# New atomic data for Ge XX

Waleed Othman Younis[a,*], Alan Hibbert[b]

[a] Physics Department, Faculty of Science, Beni-Sueif University, Beni-Sueif city, Egypt
[b] School of Mathematics and Physics, Queen's University Belfast, Belfast BT7 1NN, UK

**Highlights**

- Large-scale configuration interaction (CI) including 726 configurations.
- New atomic data for highly ionized ion Ge XX.
- Using the "fine-tuning" process within CIV3.

**Abstract**

We have performed large-scale configuration interaction (CI) calculations using CIV3 for the lowest (in energy) 155 fine-structure levels of aluminum-like germanium ion. We have calculated the energy levels, lifetimes, oscillator strengths, and transition probabilities for the electric-dipole allowed and intercombination transitions among the levels of ground state $3s^23p(^2P)$ and higher energy levels of states $3s3p^2$, $3s^23d$, $3p^3$, $3s3p3d$, $3p^23d$, $3s3d^2$, $3p3d^2$, $3d^3$, $3s^2(4s, 4p, 4d, 4f)$ of Ge XX in the LSJ coupling scheme. The present results include relativistic effects through the Breit-Pauli operator. In order to keep our calculated energy splittings as close as possible to the experimental and theoretical results complied by NIST, we attempt to correct the inaccuracies in the CI coefficients in the wavefunctions, which would lead to inaccuracies in transition probabilities, by applying a "fine-tuning" technique. Fine-tuning of the *ab initio* energies was done through adjusting, by a small amount, some diagonal elements of the Hamiltonian matrix. Comparisons are made with other available experimental and theoretical results and the accuracy of the present results is assessed.

[*] Corresponding author.
*E-mail address*: waleedegy2005@yahoo.com (W.O. Younis).

# 1. Introduction

The spectra of Al-like ions with Z > 30 have received a great deal of attention both experimentally and theoretically. In previous experimental works, which mainly make use of high-energy lasers or tokomak discharges, the spectra of some Al-like ions in plasmas have been measured, though for the heavier ions like germanium, selenium, molybdenum and silver only the spectra from the transitions of $3s3p^2$ and $3s^23d$ configurations have been identified by Hinnov et al [1] and by Sugar et al [2] for copper through molybdenum. Transitions in Ge XX between the $3s^23p$, $3s3p^2$, $3s^23d$, $3p^3$ and $3s3p3d$ configurations were identified in the extreme-ultraviolet spectra emitted from linear plasmas, produced using line-focused laser beams [3].

Large-scale theoretical studies of Al-like ions have been performed using different methods: systematic studies of oscillator strengths for fine-structure transitions in the aluminum isoelectronic sequence [4]; relativistic many-body calculations of electric-dipole lifetimes, transitions rates and oscillator strengths for $n = 3$ sates in Al-like ions [5].

Critical data compilations based on available theoretical and experimental sources are given in National Institute of Standards and Technology (NIST), Atomic Spectra Database (ASD) [6].

In this work, we have chosen to use Hibbert's program CIV3 to calculate energy levels and oscillator strengths. In a wide range of previous calculations it has demonstrated its ability to obtain results more accurate than those produced by many other programs, especially when strong mixing occurs.

The present study will include extensive calculations and give results for allowed and intercombination electric-dipole transitions within the $n = 4$ complex of states in Ge XX ion by including configuration-interaction and valence correlation effects via configuration interaction.

# 2. Theoretical method

The wavefunctions describing the atomic states included in these calculations were obtained using configuration interaction code CIV3 of Hibbert [7, 8]. The CI wavefunctions are represented as

$$\Psi(J) = \sum_{j=1}^{M} a_{ij}\phi_j(\alpha_j L_j S_j J), \qquad (1)$$

where ($\phi_j$) denotes a set of single-configuration wavefunctions, ($\alpha_j$) defines the coupling of the angular momenta of the electrons, and the orbital ($L_j$) and the spin ($S_j$) angular momenta are coupled to give the total angular momentum ($J$). The mixing coefficients ($a_{ij}$) are obtained by diagonalizing the Breit-Pauli Hamiltonian with respect to the basis ($\phi_j$).

The Hamiltonian used in this work consists of the non-relativistic electrostatic terms plus the one-body mass correction, Darwin term, and spin-orbit, spin-other-orbit, and spin-spin operators of the Breit-Pauli Hamiltonian. The inclusion of the mass correction and Darwin terms shift the energy of a configuration as a whole, while the spin-orbit, spin-other-orbit and spin-spin terms give rise to the fine-structure splitting. The general configuration-interaction code CIV3 uses orthonormal orbitals. Their radial functions $P_{nl}(r)$ are expressed as superposition of normalized Slater-type Orbitals (STOs) of the form

$$P_{nl}(r) = \sum_{j=1}^{k} C_{jnl} \chi_{jnl}(r), \qquad (2)$$

where $C_{jnl}$ are the Clementi-type coefficients and

$$\chi_{jnl}(r) = \frac{(2\xi_{jnl})^{I_{jnl}+\frac{1}{2}}}{\left[(2I_{jnl})!\right]^{\frac{1}{2}}} r^{I_{jnl}} \exp(-\xi_{jnl} r) \qquad (3)$$

with the integer

$$I_{jnl} \geq l+1$$

The parameters $C_{jnl}$ and $\xi_{jnl}$ are determined variationally while the parameters $I_{jnl}$ are normally kept fixed in any optimization process. The STOs are chosen to satisfy the orthonormality condition

$$\int_0^\infty P_{nl}(r)P_{n'l}(r)\,dr = \delta_{nn'} \;;\; l+1 < n' \leq n \quad (4)$$

In this calculations we used the 14 orthonormal one-electron orbitals: 1s, 2s, 2p, 3s, 3p, 3d, 4s, 4p, 4d, 4f, 5s, 5p, 5d, and 5f.

The optimized parameters of the radial functions are recorded and the method of optimizing the radial functions is given. All the optimizations were undertaken in LS coupling, that is, without the inclusion of the Breit-Pauli effects. In our calculation the 1s, 2s, 2p, 3s and 3p radial functions are taken as the Hartree-Fock orbitals of the ground state $1s^2 2s^2 2p^6 3s^2 3p\,(^2P^o)$ of the Ge XX ion given by Clementi and Roetti [9], whereas for the other basis radial functions 3d, 4s, 4p, 4d and 4f we initially took $k = n - l$ in Eq. (1), so that the coefficients $C_{jnl}$ are uniquely specified by the orthogonality condition on $P_{nl}$.

However, to increase the flexibility of the form of these radial functions, we then took $k \succ n - l$ in Eq. (1), thus adding extra basis functions into the $n = 4$ orbital's and also 3d.

Finally, we optimized 5s, 5p, 5d and 5f on different sets of configurations in order to improve the representation of valence correlation in various states.

The configurations included in the CI calculations for each parity are shown. The key of configurations used in our CIV3 calculation are displayed in Table *.

## 3. Results and discussion

The energy levels and lifetimes for 155 fine structure levels of Al-like germanium have been tabulated in Table ** and arranged in ascending order of energy. Our calculated fine-structure excitation energies relative to the ground level are compared with the compiled values of the National Institute of Standard and Technology (NIST) [6], as far as is possible. The mixing among some of the relativistic levels is very strong. Our *ab initio* calculation, 'Present (*a*)', is in good agreement with the experimental and theoretical results complied by NIST, except for the levels $3s3p^2$ ($^2P_{1/2}$, $^2S_{1/2}$), for which our calculated positions of these two levels to be reversed compared with those of NIST. This was previously noted by Gupta and Msezane [10] when comparing the *ab initio* MCHF levels of Froese Fisher et al. [11] with NIST. The wave functions for the levels $3s3p^2(^2P_{1/2})$ and $3s3p^2$ ($^2S_{1/2}$) exhibit very strong configuration mixing (55% and 42%), and in a simple *ab initio* calculation it is difficult to represent that mixing sufficiently accurately.

An *ab initio* CI calculation is necessarily approximate, and even with such a relative extensive set of CSFs our theoretical energies will not, in general, be obtained to the same level of accuracy as could be achieved experimentally. Thus, prior to undertaking the calculation of the Breit-Pauli oscillator strengths, we make further small modifications where by certain diagonal matrix elements are adjusted so that the theoretical energy differences coincide with the relevant experimental values. These corrections to the calculated energy levels result in improvements to the CI mixing coefficients $a_{ij}$, and therefore to the *ab initio* oscillator strengths. This so-called fine-tuning process has been described and justified by Hibbert [12]. In a way, these adjustments correct the *ab initio* approach for the neglected core-valence correlation [13].

In general, our adjusted theoretical energies, 'Present (*b*)', are also listed in Table ** and are in excellent agreement with the NIST data. We included in column 4 of Table ** the leading percentage composition of the various levels (corresponding to Present (*b*)). The first number of each entry in this column represents the leading percentage of the level corresponding to the level number under the first column, followed by a set of numbers of the form *X(Y)*. These mean that the next leading percentage is *X%* of the level number *Y* in the first column and so on. As can be seen from the column 4 of Table **, the mixing among several fine-structure levels is very strong, with most of the strongly mixed levels belonging to the $3p^2 3d$ configurations. No experimental or other theoretical results are available for these levels.

The last column in Table ** represents the radiative lifetime of an excited state which is related to the radiative transition probabilities $A_{ji}$ through

$$\tau_j = \left(\sum_i A_{ji}\right)^{-1} \quad (5)$$

where the sum over *i* is over all accessible final states and $\Delta E = (E_j - E_i)$, is the transition energy. Eq. (5) gives us a means of comparing theory (transition probabilities) with experiment (lifetimes), but in drawing conclusions from such a comparison, it is important to realize that the experimental lifetimes are not measured directly [14]. Estimation of the accuracy of our results has been done by comparing them with data available from other calculations [5], see Table A.

Table A: Lifetimes (in *ns*) of some exited levels in Ge XX in comparison with theoretical calculations of Safronova et al. [5]

| Configuration | τ (Present) | τ ([5]) | Configuration | τ (Present) | τ ([5]) |
|---|---|---|---|---|---|
| $3s^2(^1S)3d\ ^2D_{3/2}$ | 1.42E-02 | 1.48E-02 | $3s3p(^3P)3d\ ^4D_{3/2}$ | 1.69E-02 | 2.11E-02 |
| $3s^2(^1S)3d\ ^2D_{5/2}$ | 1.69E-02 | 1.78E-02 | $3s3p(^3P)3d\ ^4D_{5/2}$ | 1.62E-02 | 1.82E-02 |
| $3s3p(^3P)3d\ ^4F_{3/2}$ | 5.96E-01 | 6.34E-01 | $3s3p(^3P)3d\ ^4D_{7/2}$ | 1.64E-02 | 1.79E-02 |
| $3s3p(^3P)3d\ ^4F_{5/2}$ | 5.85E-01 | 5.87E-01 | $3s3p(^1P)3d\ ^2D_{3/2}$ | 8.87E-03 | 1.00E-02 |
| $3s3p(^3P)3d\ ^4F_{7/2}$ | 6.08E-01 | 6.07E-01 | $3s3p(^1P)3d\ ^2D_{5/2}$ | 8.38E-03 | 8.73E-03 |
| $3s3p(^3P)3d\ ^4P_{1/2}$ | 2.22E-02 | 1.62E-02 | $3s3p(^1P)3d\ ^2F_{5/2}$ | 1.15E-02 | 1.25E-02 |
| $3s3p(^3P)3d\ ^4P_{3/2}$ | 1.99E-02 | 1.88E-02 | $3s3p(^1P)3d\ ^2F_{7/2}$ | 1.24E-02 | 1.39E-02 |
| $3s3p(^3P)3d\ ^4P_{5/2}$ | 2.30E-02 | 2.41E-02 | $3s3p(^1P)3d\ ^2P_{1/2}$ | 8.99E-03 | 1.23E-02 |
| $3s3p(^3P)3d\ ^4D_{1/2}$ | 1.49E-02 | 2.37E-02 | $3s3p(^1P)3d\ ^2P_{3/2}$ | 9.49E-03 | 2.42E-02 |

In Table *** we have tabulated our calculated wavelengths (λ) in Ă, oscillator strengths in length form $(f_l)$ and velocity form $(f_v)$, and transition probabilities in length form $(A_l)$ for all allowed dipole transitions among the 155 *LS* levels of aluminum-like germanium ion. In calculating these parameters we used our adjusted theoretical energy splitting, corresponding to the Present (*b*) in Table **. The keys of the lower and upper levels involved in a transition are given in Table *. The value $f_l/f_v$ gives an indication of calculation precision, a good agreement between the length and velocity values of the oscillator strengths indicates to some extent the accuracy of the wavefunctions used in our CIV3 calculation. The magnitudes of the oscillator strengths for most intercombination transitions are smaller by several orders of magnitude than those for allowed transitions, and there is a significant difference between the $f_l$ and $f_v$ values for several of these transitions.

In Table B, to assess our results for oscillator strengthsfor selected transitions, we have compared the present calculations of ($f_l$) with theoretical results which available since 1986 until now. Huang [15] who used the multiconfiguration Dirac-Fock method (MCDF) to calculate oscillator strengths for transition [$3s^2$ ($^1S$) $3p\ ^2P_J$ - $3s^2$ ($^1S$) $3d\ ^2D_{J'}$] with $J = 1/2, 3/2$ and $J' = 3/2, 5/2$ .Lavin et al. [4] used relativistic quantum defect orbital (RQDO) method, without explicit account for polarization effects.

For oscillator strengths a few calculations within the $n = 3$ complex of states have been performed by Safronova et al. [5], relativistic many-body perturbation theory (MBPT), while we have performed these calculations up to $n = 4$ without including core-valence correlation effects.

Since we lack any experimental measurements of oscillator strengths and any compiled data by NIST the comparison will be limited only to the energy levels.

Table B: Comparison of present oscillator strengths with theoretical calculations of Safronova et al. [5] (MBPT), Lavin et al. [4](RQDO), and Huang[15](MCDF) for some transitions of Ge XX

| Lower State | Upper State | $f_l$ (Present) | $f_l$ ([5]) | $f_l$ (RQDO [4]) | $f_l$ (MCDF[11]) |
|---|---|---|---|---|---|
| $3s^2(^1S)3d\ ^2D_{5/2}$ | $3s3p(^1P)3d\ ^2F_{7/2}$ | 2.8690E-01 | 3.19E-01 | ----- | ----- |
| $3s^2(^1S)3d\ ^2D_{3/2}$ | $3s3p(^1P)3d\ ^2F_{5/2}$ | 3.5347E-01 | 3.56E-01 | ----- | ----- |
| $3s^2(^1S)3d\ ^2D_{5/2}$ | $3s3p(^1P)3d\ ^2P_{3/2}$ | 1.2920 E-01 | 5.46E-02 | ----- | ----- |
| $3s^2(^1S)3d\ ^2D_{3/2}$ | $3s3p(^1P)3d\ ^2P_{1/2}$ | 1.2822 E-01 | 1.14E-01 | ----- | ----- |
| $3s^2(^1S)3d\ ^2D_{5/2}$ | $3s3p(^1P)3d\ ^2D_{5/2}$ | 1.8325 E-01 | 1.73E-01 | ----- | ----- |
| $3s^2(^1S)3d\ ^2D_{5/2}$ | $3s3p(^1P)3d\ ^2D_{3/2}$ | 4.6306 E-02 | 1.06E-01 | ----- | ----- |
| $3s^2(^1S)3d\ ^2D_{3/2}$ | $3s3p(^1P)3d\ ^2P_{3/2}$ | 1.0250 E-01 | 8.21E-02 | ----- | ----- |
| $3s^2(^1S)3d\ ^2D_{3/2}$ | $3s3p(^1P)3d\ ^2D_{3/2}$ | 9.4828 E-02 | 1.08E-01 | ----- | ----- |
| $3s^2(^1S)3p\ ^2P_{1/2}$ | $3s^2(^1S)3d\ ^2D_{3/2}$ | 3.5033E-01 | ----- | 3.089E-01 | 3.585E-01 |
| $3s^2(^1S)3p\ ^2P_{3/2}$ | $3s^2(^1S)3d\ ^2D_{3/2}$ | 6.0455 E-02 | 5.71E-02 | 2.847 E-02 | 6.046 E-02 |
| $3s^2(^1S)3p\ ^2P_{3/2}$ | $3s^2(^1S)3d\ ^2D_{5/2}$ | 3.2928 E-01 | 3.15E-01 | 2.597E-01 | 3.307E-01 |

## 4. Summary and conclusion

In summary, we have presented our calculated energy splittings of 155 fine-structure levels as well as oscillator strengths and transition probabilities for 3540 transitions among the fine-structure levels of the terms belonging to the $(1s^22s^22p^6)$ $3s^23p$, $3s3p^2$, $3s^23d$, $3p^3$, $3s3p3d$, $3p^23d$, $3s3d^2$, $3p3d^2$, $3d^3$, $3s^24s$, $3s^24p$, $3s^24d$, and $3s^24f$ configurations of GeXX. In this calculation, we have used an extensive set of CI wavefunctions and included valence-correlation effects in the excitation up to the 5f orbital. Our calculated excitation energies are in excellent agreement with the values from the NIST Atomic Spectra Database. Our new predicted levels, for which no experimental or theoretical data is available, can guide experimentalists in identifying strongly mixed fine-structure levels [16]. Finally, we believe that the present results are the most extensive and definitive to date and should be useful in astrophysical applications and technical plasma modeling.

## References

[1] E. Hinnov, F. Boody, S. Cohen, U. Feldman J. Hosea, K. Sato, J. L. Schwob, S. Suckewer and A. Wouters, J. Opt. Soc. Am. B **3** (1986) 1288.
[2] J. Sugar, V. Kaufman and W. L. Rowan, J. Opt. Soc. Am. B **5** (1988)2183.
[3] J. O. Ekberg and A. Redfors, Phys. Scr. **44** (1991)539.
[4] C. Lavin, A.B. Alvarez, I. Martin, J. Quant. Spectrosc. Radiat. Transfer 57 (1997) 831.
[5] U.I. Safronova, M. Sataka, J.R. Albritton, W.R. Johnson, M.S. Safronova, At. Data Nucl. Data Tables 84 (2003) 1.
[6] A. Kramida, Yu. Ralchenko, J. Reader, and NIST ASD Team (2014). NIST Atomic Spectra Database (ver. 5.2),[Online].Available: http://physics.nist.gov/asd (2014) December 7.
[7] R. Glass, A. Hibbert, Comput. Phys. Commun. 16 (1978) 19.
[8] A. Hibbert, Comput. Phys. Commun. 9 (1975) 141.
[9] E. Clementi and C. Roetti, At.Data Nucl. Data Tables 14(1974)177.
[10] G. P. Gupta and A. Z. Msezane, Phys. Scr. 76 (2007) 225.
[11] C. Froese Fischer, G. Tachiev and A. Irimia, At.Data Nucl. Data Tables 92 (2006) 607.
[12] A. Hibbert, Phys. Scr. T65 (1996) 104.
[13] P. Jonsson, C. Froese Fischer and M. R. Godefroid, J. Phys. B: Atom. Molec. Opt. Phys. 32 (1999) 123.
[14] A. Hibbert, Rep. Progr. Phys. 38 (1975) 1217.
[15] K. N. Huang, At.Data Nucl. Data Tables 34 (1986) 1.
[16] Jagjit Singh, A. K. S. Jha, N. Verma and M. Mohan, At.Data Nucl. Data Tables 96 (2010) 759.

Table * States (Eigen value) used in Ge XX for fine-tuning calculations.

| Key | State | Key | State | Key | State | Key | State |
|---|---|---|---|---|---|---|---|
| 370 | $3s^23p\ (^2P_{1/2})$ | 212 | $3p^23d\ (^4F_{5/2})$ | 561 | $3p3d^2(^3F)(^4G_{5/2})$ | 551 | $3p3d^2(^3P)(^2D_{5/2})$ |
| 468 | $3s^23p(^2P_{3/2})$ | 269 | $3p^2(^1D)3d\ (^2F_{7/2})$ | 650 | $3p3d^2(^3F)(^4G_{7/2})$ | 358 | $3p3d^2(^3P)(^2P_{1/2})$ |
| 53 | $3s3p^2\ (^4P_{1/2})$ | 268 | $3p^23d\ (^4F_{7/2})$ | 703 | $3p3d^2(^3F)(^4G_{9/2})$ | 447 | $3p3d^2(^3P)\ (^2P_{3/2})$ |
| 134 | $3s3p^2\ (^4P_{3/2})$ | 129 | $3p^2(^3P)3d\ (^2P_{3/2})$ | 560 | $3p3d^2(^1D)(^2F_{5/2})$ | 118 | $3d^3\ (^4F_{3/2})$ |
| 216 | $3s3p^2\ (^4P_{5/2})$ | 50 | $3p^23d\ (^4D_{1/2})$ | 457 | $3p3d^2(^3F)\ (^2D_{3/2})$ | 201 | $3d^3\ (^4F_{5/2})$ |
| 133 | $3s3p^2\ (^2D_{3/2})$ | 302 | $3p^23d\ (^4F_{9/2})$ | 726 | $3p3d^2(^3F)(^4G_{11/2})$ | 261 | $3d^3\ (^4F_{7/2})$ |
| 215 | $3s3p^2\ (^2D_{5/2})$ | 128 | $3p^2(^3P)3d\ (^4D_{3/2})$ | 649 | $3p3d^2(^3F)\ (^2F_{7/2})$ | 298 | $3d^3\ (^4F_{9/2})$ |
| 52 | $3s3p^2\ (^2P_{1/2})$ | 211 | $3p^23d\ (^4D_{5/2})$ | 364 | $3p3d^2(^3P)(^4D_{1/2})$ | 42 | $3d^3\ (^4P_{1/2})$ |
| 51 | $3s3p^2\ (^2S_{1/2})$ | 49 | $3p^2(^3P)3d\ (^2P_{1/2})$ | 363 | $3p3d^2(^3P)\ (^2S_{1/2})$ | 117 | $3d^3\ (^4P_{3/2})$ |
| 132 | $3s3p^2\ (^2P_{3/2})$ | 267 | $3p^23d\ (^4D_{7/2})$ | 559 | $3p3d^2(^1D)(^2F_{5/2})$ | 260 | $3d^3\ (^2G_{7/2})$ |
| 131 | $3s^23d\ (^2D_{3/2})$ | 266 | $3p^23d\ (^2G_{7/2})$ | 456 | $3p3d^2(^3F)\ (^4F_{3/2})$ | 41 | $3s^24s\ (^2S_{1/2})$ |
| 214 | $3s^23d\ (^2D_{5/2})$ | 210 | $3p^23d\ (^4P_{5/2})$ | 455 | $3p3d^2(^3P)\ (^4D_{3/2})$ | 297 | $3d^3\ (^2G_{9/2})$ |
| 467 | $3p^3\ (^2D_{3/2})$ | 127 | $3p^2(^3P)3d\ (^4P_{3/2})$ | 558 | $3p3d^2(^3F)(^4F_{5/2})$ | 200 | $3d^3\ (^4P_{5/2})$ |
| 569 | $3p^3\ (^2D_{5/2})$ | 301 | $3p^23d\ (^2G_{9/2})$ | 648 | $3p3d^2(^3F)\ (^4F_{7/2})$ | 311 | $3d^3\ (^2H_{11/2})$ |
| 466 | $3p^3\ (^4S_{3/2})$ | 48 | $3p^23d\ (^4P_{1/2})$ | 702 | $3p3d^2(^1G)(^2G_{9/2})$ | 296 | $3d^3\ (^2H_{9/2})$ |
| 465 | $3s3p(^3P)3d(^4F_{3/2})$ | 126 | $3p^2(^3P)3d\ (^4P_{3/2})$ | 557 | $3p3d^2(^3P)(^4D_{5/2})$ | 116 | $3d^3\ (^2D_{3/2})$ |
| 568 | $3s3p(^3P)3d(^4F_{5/2})$ | 209 | $3p^2(^3P)3d\ (^4P_{5/2})$ | 647 | $3p3d^2(^1G)\ (^2G_{7/2})$ | 40 | $3d^3\ (^2P_{1/2})$ |
| 369 | $3p^3\ (^2P_{1/2})$ | 47 | $3p^2(^1D)3d\ (^2P_{1/2})$ | 454 | $3p3d^2(^3P)(^4P_{3/2})$ | 199 | $3d^3\ (^2D_{5/2})$ |
| 654 | $3s3p(^3P)3d(^4F_{7/2})$ | 125 | $3p^2(^1S)3d\ (^2D_{3/2})$ | 701 | $3p3d^2(^3F)(^4F_{9/2})$ | 115 | $3d^3\ (^2P_{3/2})$ |
| 464 | $3p^3\ (^2P_{3/2})$ | 208 | $3p^23d\ (^2F_{5/2})$ | 646 | $3p3d^2(^3P)\ (^4D_{7/2})$ | 259 | $3d^3\ (^2F_{7/2})$ |
| 704 | $3s3p(^3P)3d(^4F_{9/2})$ | 124 | $3p^2(^3P)3d\ (^2P_{3/2})$ | 362 | $3p3d^2(^3P)\ (^4P_{1/2})$ | 198 | $3d^3\ (^2F_{5/2})$ |
| 567 | $3s3p(^3P)3d(^4P_{5/2})$ | 46 | $3p^2(^1D)3d\ (^2S_{1/2})$ | 556 | $3p3d^2(^3P)(^4P_{5/2})$ | 197 | $3d^3\ (^2D_{5/2})$ |
| 463 | $3s3p(^3P)3d(^4D_{3/2})$ | 207 | $3p^2(^1S)3d\ (^2D_{5/2})$ | 453 | $3p3d^2(^3F)\ (^4D_{3/2})$ | 114 | $3d^3\ (^2D_{3/2})$ |
| 368 | $3s3p(^3P)3d(^4D_{1/2})$ | 265 | $3p^2(^3P)3d\ (^2F_{7/2})$ | 700 | $3p3d^2(^1G)(^2H_{9/2})$ | 357 | $3s^24p(^2P_{1/2})$ |
| 653 | $3s3p(^3P)3d(^4D_{7/2})$ | 123 | $3s3d^2\ (^4F_{3/2})$ | 361 | $3p3d^2(^3F)(^4D_{1/2})$ | 446 | $3s^24p(^2P_{3/2})$ |
| 566 | $3s3p(^3P)3d(^4D_{5/2})$ | 206 | $3s3d^2\ (^4F_{5/2})$ | 555 | $3p3d^2(^3F)(^4D_{5/2})$ | 113 | $3s^24d(^2D_{3/2})$ |
| 367 | $3s3p(^3P)3d(^4P_{1/2})$ | 264 | $3s3d^2\ (^4F_{7/2})$ | 645 | $3p3d^2(^3P)\ (^4D_{7/2})$ | 196 | $3s^24d\ (^2D_{5/2})$ |
| 462 | $3s3p(^3P)3d(^4P_{3/2})$ | 300 | $3s3d^2\ (^4F_{9/2})$ | 452 | $3p3d^2(^1D)(^2P_{3/2})$ | 550 | $3s^24f(^2F_{5/2})$ |
| 565 | $3s3p(^3P)3d(^2D_{5/2})$ | 45 | $3s3d^2\ (^4P_{1/2})$ | 725 | $3p3d^2(^1G)(^2H_{11/2})$ | 641 | $3s^24f(^2F_{7/2})$ |
| 461 | $3s3p(^3P)3d(^2D_{3/2})$ | 122 | $3s3d^2\ (^4P_{3/2})$ | 360 | $3p3d^2(^1D)(^2P_{1/2})$ | | |
| 564 | $3s3p(^3P)3d(^2F_{5/2})$ | 205 | $3s3d^2\ (^4P_{5/2})$ | 451 | $3p3d^2(^3P)\ (^4S_{3/2})$ | | |
| 652 | $3s3p(^3P)3d(^2F_{7/2})$ | 204 | $3p^23d\ (^2D_{5/2})$ | 554 | $3p3d^2(^3F)(^2F_{5/2})$ | | |
| 460 | $3s3p(^3P)3d(^2P_{3/2})$ | 121 | $3p^2(^3P)3d\ (^2D_{3/2})$ | 644 | $3p3d^2(^1G)\ (^2F_{7/2})$ | | |
| 651 | $3s3p(^1P)3d(^2F_{7/2})$ | 120 | $3s3d^2\ (^2D_{3/2})$ | 450 | $3p3d^2(^1D)\ (^2D_{3/2})$ | | |
| 366 | $3s3p(^3P)3d(^2P_{1/2})$ | 203 | $3s3d^2\ (^2D_{5/2})$ | 359 | $3p3d^2(^1S)(^2P_{1/2})$ | | |
| 563 | $3s3p(^1P)3d(^2F_{5/2})$ | 263 | $3s3d^2\ (^2G_{7/2})$ | 553 | $3p3d^2(^1D)(^2D_{5/2})$ | | |
| 365 | $3s3p(^1P)3d(^2P_{1/2})$ | 299 | $3s3d^2\ (^2G_{9/2})$ | 643 | $3p3d^2(^3F)\ (^2F_{7/2})$ | | |
| 459 | $3s3p(^1P)3d(^2D_{3/2})$ | 202 | $3s3d^2\ (^2F_{5/2})$ | 552 | $3p3d^2(^1G)(^2F_{5/2})$ | | |
| 562 | $3s3p(^1P)3d(^2D_{5/2})$ | 262 | $3s3d^2\ (^2F_{7/2})$ | 449 | $3p3d^2(^1S)\ (^2P_{3/2})$ | | |
| 458 | $3s3p(^1P)3d(^2P_{3/2})$ | 44 | $3s3d^2\ (^2P_{1/2})$ | 699 | $3p3d^2(^3F)(^2G_{9/2})$ | | |
| 130 | $3p^23d\ (^4F_{3/2})$ | 119 | $3s3d^2\ (^2P_{3/2})$ | 642 | $3p3d^2(^3F)\ (^2G_{7/2})$ | | |

| 213 | 3p$^2$($^1$D)3d($^2$F$_{5/2}$) | 43 | 3s3d$^2$ ($^2$S$_{1/2}$) | 448 | 3p3d$^2$($^3$P) ($^2$D$_{3/2}$) | | |

Table ** Fine-Structure Energy Levels (in cm$^{-1}$) of Ge XX relative to the ground level and their lifetimes (in nanoseconds).

| Key. Eigen value. | Present calculations | | NIST | Leading (%) | τ (*ns*) |
|---|---|---|---|---|---|
| | (*a*) | (*b*) | | | |
| 370 | .00 | .00 | .00 | 97 | ---------- |
| 468 | 53926.44 | 54564.43 | 54564.0 | 97 | ---------- |
| 53 | 327166.93 | 327449.10 | | 96 | 0.447E+01 |
| 134 | 351132.37 | 351281.18 | | 99 | 0.295E+02 |
| 216 | 374358.93 | 374970.81 | | 93 | 0.643E+01 |
| 133 | 450626.56 | 453869.52 | 453869.0 | 83, 13(131) | 0.182E+00 |
| 215 | 461269.59 | 466407.60 | 466407.0 | 79, 13(214) | 0.324E+00 |
| 52 | 584308.20 | 532728.61 | 530521.0 | 55, 40(51) | 0.207E-01 |
| 51 | 532290.42 | 585257.99 | 585257.0 | 55, 42(52) | 0.225E-01 |
| 132 | 596355.95 | 597020.86 | 597020.0 | 93 | 0.139E-01 |
| 131 | 692091.31 | 682246.15 | 682246.0 | 83, 14(133) | 0.142E-01 |
| 214 | 700660.60 | 689533.56 | 689533.0 | 85, 13(215) | 0.169E-01 |
| 467 | 839821.55 | 844871.42 | | 47, 19(461), 15(464) | 0.595E-01 |
| 569 | 859856.26 | 859880.48 | | 68, 30(565) | 0.140E+00 |
| 466 | 867619.91 | 875850.58 | 875850 | 74, 15(467) | 0.196E-01 |
| 465 | 928770.39 | 928745.55 | | 96 | 0.596E+00 |
| 568 | 940037.76 | 940014.48 | | 96 | 0.585E+00 |
| 369 | 948573.01 | 946764.20 | | 80, 13(366) | 0.349E-01 |
| 654 | 956973.30 | 956880.74 | | 95 | 0.608E+00 |
| 464 | 966827.57 | 966609.31 | | 58, 14(460) | 0.339E-01 |
| 704 | 982532.57 | 982651.18 | | 100 | ---------- |
| 567 | 1000164.81 | 996202.66 | 996200.0 | 64, 24(566) | 0.230E-01 |
| 463 | 1003830.97 | 1000789.76 | 1000790.0 | 70, 25(462) | 0.169E-01 |
| 368 | 1005265.80 | 1002640.78 | 1002640.0 | 90 | 0.149E-01 |
| 653 | 1034859.52 | 1033390.68 | 1033390.0 | 95 | 0.164E-01 |
| 566 | 1035220.18 | 1033974.10 | | 72, 16(567) | 0.162E-01 |
| 367 | 1036586.05 | 1035630.83 | 1035630.0 | 92 | 0.222E-01 |
| 462 | 1036710.55 | 1035971.17 | | 65, 27(463) | 0.199E-01 |
| 565 | 1053698.49 | 1050330.72 | | 35, 18(567), 16(562), 12(569), 9(564) | 0.212E-01 |
| 461 | 1051245.48 | 1050589.59 | 1050590.0 | 44, 26(459), 15(467), 9(462) | 0.178E-01 |
| 564 | 1078155.69 | 1073969.58 | | 53, 29(563) | 0.337E-01 |
| 652 | 1116449.41 | 1112259.48 | 1114490.0 | 53, 46(651) | 0.370E-01 |
| 460 | 1166611.25 | 1157626.35 | 1157630.0 | 61, 14(464) | 0.113E-01 |
| 651 | 1192900.10 | 1184190.35 | 1184190.0 | 51, 46(652) | 0.124E-01 |
| 366 | 1195281.84 | 1184750.81 | 1184750.0 | 86, 11(369) | 0.150E-01 |
| 563 | 1197726.50 | 1190041.58 | 1190040.0 | 60, 33(564) | 0.115E-01 |
| 365 | 1220330.08 | 1211960.40 | 1211960.0 | 90 | 0.899E-02 |
| 459 | 1225696.48 | 1217454.54 | 1217450.0 | 43, 20(460), 13(461), 11(467) | 0.887E-02 |
| 562 | 1229958.10 | 1222807.60 | 1222780.0 | 62, 17(565), 12(569) | 0.838E-02 |
| 458 | 1233178.31 | 1223699.54 | 1223700.0 | 80, 4(464), 5(459), 4(461) | 0.949E-02 |
| 130 | 1393488.69 | 1393587.73 | | 84, 6(128) | 0.339E-01 |
| 213 | 1399427.12 | 1399543.57 | | 59, 25(208), 11(202) | 0.850E-01 |
| 212 | 1408641.57 | 1408748.49 | | 86, 10(211) | 0.330E-01 |
| 269 | 1415434.50 | 1415553.10 | | 53, 16(265), 10(268), 8(267) | 0.617E-01 |
| 268 | 1428670.11 | 1428788.72 | | 79, 15(267) | 0.333E-01 |
| 129 | 1428973.61 | 1429089.82 | | 43, 23(128), 10(130) | 0.253E-01 |
| 50 | 1436541.68 | 1436660.25 | | 79, 13(49), 7(47) | 0.269E-01 |
| 302 | 1444939.03 | 1445057.63 | | 84, 15(301) | 0.348E-01 |
| 128 | 1455343.08 | 1455461.60 | | 65, 17(129) | 0.246E-01 |
| 211 | 1456536.37 | 1456653.36 | | 76, 10(212) | 0.260E-01 |
| 49 | 1480892.17 | 1481010.91 | | 42, 27(47), 19(50) | 0.219E-01 |
| 267 | 1482137.04 | 1482255.64 | | 77, 9(268), 7(269) | 0.280E-01 |
| 266 | 1491402.48 | 1491521.09 | | 82, 9(263) | 0.612E-01 |
| 210 | 1500540.63 | 1500632.14 | | 51, 36($3p^2\ ^1D\ 3d\ ^2D_{5/2}$) | 0.113E-01 |
| 127 | 1506793.10 | 1506888.82 | | 40, 34($3p^2\ ^1D\ 3d\ ^2D_{3/2}$) | 0.113E-01 |
| 301 | 1508334.80 | 1508453.41 | | 74, 16(302), 10(299) | 0.558E-01 |
| 48 | 1522569.67 | 1522683.41 | | 85, 7(49) | 0.913E-02 |
| 126 | 1532111.43 | 1532215.76 | | 48, 33($3p^2\ ^1D\ 3d\ ^2D_{3/2}$) | 0.107E-01 |
| 209 | 1533143.54 | 1533252.19 | | 39, 37($3p^2\ ^1D\ 3d\ ^2D_{5/2}$), 8(203), 7(204) | 0.116E-01 |
| 47 | 1565722.25 | 1565776.32 | | 37, 31(49), 19(44), 8(46) | 0.158E-01 |
| 125 | 1567987.65 | 1568075.24 | | 26, 24(121), 13($3p^2\ ^1D\ 3d\ ^2P_{3/2}$) | 0.207E-01 |
| 208 | 1596374.66 | 1596472.57 | | 60, 14(202), 13(213), 8(207) | 0.100E-01 |
| 124 | 1597246.55 | 1597331.49 | | 26, 22($3p^2\ ^1D\ 3d\ ^2P_{3/2}$), 20(125), 15(119) | 0.156E-01 |
| 46 | 1597273.12 | 1597395.23 | | 75, 9(48), 9(43), 5(49) | 0.135E-01 |
| 207 | 1617030.22 | 1617072.52 | | 51, 29(204), 6(203) | 0.201E-01 |
| 265 | 1617482.06 | 1617600.67 | | 75, 12(262), 10(269) | 0.968E-02 |
| 123 | 1629102.67 | 1629221.17 | | 100 | 0.114E-01 |
| 206 | 1632639.40 | 1632757.58 | | 99 | 0.117E-01 |
| 264 | 1637333.06 | 1637451.66 | | 100 | 0.123E-01 |
| 300 | 1643202.73 | 1643321.34 | | 100 | 0.130E-01 |
| 45 | 1680433.44 | 1680552.10 | | 99 | 0.951E-02 |
| 122 | 1682716.51 | 1682835.12 | | 99 | 0.965E-02 |
| 205 | 1685349.18 | 1685462.27 | | 89, 5(204) | 0.970E-02 |
| 204 | 1686681.62 | 1686785.29 | | 51, 23(207), 10(205) | 0.856E-02 |
| 121 | 1695501.29 | 1695581.36 | | 56, 35(125) | 0.842E-02 |
| 120 | 1746788.62 | 1746882.99 | | 78, 14($3p^2\ ^1D\ 3d\ ^2D_{3/2}$) | 0.828E-02 |
| 203 | 1746868.61 | 1746966.13 | | 80, 16($3p^2\ ^1D\ 3d\ ^2D_{5/2}$) | 0.838E-02 |
| 263 | 1751957.59 | 1752076.19 | | 89, 10(266) | 0.138E-01 |
| 299 | 1752746.27 | 1752864.88 | | 89, 11(301) | 0.134E-01 |
| 202 | 1806498.53 | 1806617.28 | | 70, 24(213) | 0.696E-02 |
| 262 | 1811665.93 | 1811784.53 | | 74, 23(269) | 0.733E-02 |
| 44 | 1848192.30 | 1848283.51 | | 56, 20(47), 18(43) | 0.571E-02 |
| 119 | 1857779.32 | 1857908.84 | | 69, 26($3p^2\,^1D\,3d\,^2P_{3/2}$) | 0.582E-02 |
| 43 | 1861303.95 | 1861469.81 | | 70, 15(44), 8(46), 5(47) | 0.652E-02 |
| 561 | 1963102.40 | 1963212.06 | | 89, 5(554), 4(559) | 0.457E-01 |
| 650 | 1978617.90 | 1978733.29 | | 93 | 0.445E-01 |
| 703 | 1997800.40 | 1997919.00 | | 93 | 0.416E-01 |
| 560 | 2011208.03 | 2011291.64 | | 30, 16(554), 12(553), 10(551) | 0.258E-01 |
| 457 | 2024056.05 | 2024167.81 | | 24, 20(448), 20(450), 15(455) | 0.190E-01 |
| 726 | 2025001.42 | 2025120.03 | | 100 | 0.409E-01 |
| 649 | 2041232.10 | 2041300.85 | | 36, 30($3p\,^2P\,3d^2\,^2F_{7/2}$), 13($3p\,^2P\,3d^2\,^4D_{7/2}$) | 0.225E-01 |
| 364 | 2041854.27 | 2041969.71 | | 55, 21(363), 19(362) | 0.176E-01 |
| 363 | 2045431.07 | 2045545.45 | | 41, 35(364), 18(362) | 0.148E-01 |
| 559 | 2045480.99 | 2045572.11 | | 24, 16(557), 13(554), 10(551) | 0.219E-01 |
| 456 | 2053949.78 | 2054060.84 | | 62, 31(455) | 0.118E-01 |
| 455 | 2060818.70 | 2060940.20 | | 38, 19(450), 17(456), 8(448), 7(457), 6(453) | 0.167E-01 |
| 558 | 2061206.03 | 2061322.79 | | 77, 6(553), 5(557), 5(551) | 0.112E-01 |
| 648 | 2064252.39 | 2064372.17 | | 49, 17(647), 10($3p\,^2P\,3d^2\,^4D_{7/2}$) | 0.137E-01 |
| 702 | 2065024.46 | 2065143.07 | | 38, 24(700), 19(701), 15(699) | 0.189E-01 |
| 557 | 2078400.72 | 2078512.91 | | 55, 13(553), 11($3p\,^2P\,3d^2\,^2D_{5/2}$) | 0.192E-01 |
| 647 | 2080319.92 | 2080415.86 | | 54, 17(642), 12(648) | 0.166E-01 |
| 454 | 2084115.79 | 2084237.17 | | 85, 5(451), 5(455) | 0.147E-01 |
| 701 | 2084541.07 | 2084659.68 | | 71, 22(700) | 0.140E-01 |
| 646 | 2088483.67 | 2088599.42 | | 46, 26($3p\,^2P\,3d^2\,^4D_{7/2}$), 24(648) | 0.165E-01 |
| 362 | 2097644.31 | 2097760.98 | | 61, 37(363) | 0.130E-01 |
| 556 | 2105211.98 | 2105318.62 | | 68, 24(555) | 0.131E-01 |
| 453 | 2113948.32 | 2114050.96 | | 66, 16(452), 7(455) | 0.837E-02 |
| 700 | 2116227.57 | 2116346.18 | | 53, 27(702), 14(699) | 0.228E-01 |
| 361 | 2116349.89 | 2116462.90 | | 88, 6(364) | 0.837E-02 |
| 555 | 2118755.53 | 2118869.93 | | 51, 22(556), 19(557) | 0.935E-02 |
| 645 | 2119028.48 | 2119146.58 | | 48, 45($3p\,^2P\,3d^2\,^4D_{7/2}$) | 0.877E-02 |
| 452 | 2130506.23 | 2130591.94 | | 62, 20(453) | 0.841E-02 |
| 725 | 2136900.48 | 2137019.08 | | 100 | 0.338E-01 |
| 360 | 2151917.46 | 2152031.32 | | 59, 22(359) | 0.865E-02 |
| 451 | 2161417.89 | 2161606.14 | | 91, 6(454) | 0.617E-02 |
| 554 | 2171618.50 | 2171720.44 | | 35, 30(560), 21(552) | 0.982E-02 |
| 644 | 2181895.58 | 2181924.88 | | 54, 34($3p\,^2P\,3d^2\,^2F_{7/2}$) | 0.133E-01 |
| 450 | 2192666.17 | 2192690.96 | | 46, 35(457) | 0.854E-02 |
| 359 | 2204026.35 | 2204006.88 | | 55, 35(360) | 0.139E-01 |
| 553 | 2208257.43 | 2208302.51 | | 52, 29($3p\,^2P\,3d^2\,^2D_{5/2}$) | 0.858E-02 |
| 643 | 2218771.82 | 2218810.79 | | 34, 24(642), 20(644) | 0.888E-02 |
| 552 | 2237420.42 | 2237411.41 | | 70, 24(554) | 0.109E-01 |
| 449 | 2239573.35 | 2239516.71 | | 54, 27(447), 12(452) | 0.138E-01 |
| 699 | 2242673.77 | 2242792.37 | | 70, 28(702) | 0.917E-02 |
| 642 | 2253482.41 | 2253560.67 | | 49, 21(647), 13(644), 12(643) | 0.942E-02 |
| 448 | 2294085.02 | 2294114.28 | | 61, 22(457) | 0.707E-02 |
| 551 | 2294166.16 | 2294163.26 | | 60, 31(3p $^2$P 3d$^2$ $^2$D$_{5/2}$) | 0.758E-02 |
| 358 | 2321634.61 | 2321656.44 | | 76, 19(359) | 0.616E-02 |
| 447 | 2329996.51 | 2330046.66 | | 55, 34(449) | 0.612E-02 |
| 118 | 2672778.66 | 2672897.21 | | 99 | 0.837E-02 |
| 201 | 2676587.37 | 2676705.95 | | 99 | 0.859E-02 |
| 261 | 2681264.70 | 2681383.30 | | 99 | 0.885E-02 |
| 298 | 2686429.98 | 2686548.59 | | 98 | 0.915E-02 |
| 42 | 2724117.67 | 2724236.26 | | 97 | 0.759E-02 |
| 117 | 2724768.05 | 2724886.60 | | 92 | 0.755E-02 |
| 260 | 2726108.44 | 2726227.04 | | 99 | 0.102E-01 |
| 41 | 2726259.56 | 2726378.47 | | 98 | 0.140E-02 |
| 297 | 2727541.66 | 2727660.26 | | 63, 35(296) | 0.110E-01 |
| 200 | 2729946.39 | 2730064.89 | | 99 | 0.780E-02 |
| 311 | 2740754.44 | 2740873.05 | | 100 | 0.136E-01 |
| 296 | 2741415.04 | 2741533.65 | | 65, 34(297) | 0.125E-01 |
| 116 | 2749679.00 | 2749795.44 | | 54, 26(115), 13(114) | 0.773E-02 |
| 40 | 2756167.86 | 2756285.89 | | 95 | 0.755E-02 |
| 199 | 2759423.14 | 2759539.20 | | 84, 14(197) | 0.823E-02 |
| 115 | 2766760.04 | 2766877.71 | | 66, 24(116) | 0.814E-02 |
| 259 | 2790060.83 | 2790179.43 | | 99 | 0.967E-02 |
| 198 | 2792062.32 | 2792180.82 | | 98 | 0.962E-02 |
| 197 | 2860704.45 | 2860806.90 | | 83, 14(199) | 0.807E-02 |
| 114 | 2862314.45 | 2862419.48 | | 78, 20(116) | 0.822E-02 |
| 357 | 2894912.00 | 2895029.28 | | 98 | 0.495E-02 |
| 446 | 2915395.79 | 2915512.75 | | 98 | 0.538E-02 |
| 113 | 3128824.69 | 3128943.13 | | 98 | 0.784E-03 |
| 196 | 3132763.28 | 3132881.88 | | 98 | 0.782E-03 |
| 550 | 3267541.23 | 3267650.16 | | 98 | 0.317E-03 |
| 641 | 3268942.46 | 3269047.81 | | 98 | 0.318E-03 |

Table *** Wavelengths, oscillator strengths, and transitions probabilities in Ge XX.

| Transitions | | λ (Å) | $f_l$ | $f_v$ | $A_l$ (s$^{-1}$) | $f_l/f_v$ |
|---|---|---|---|---|---|---|
| i | j | | | | | |
| 370 | 53 | 305.388 | .26582E-02 | .56627E-02 | .19011E+09 | 4.69E-01 |
| 53 | 369 | 161.467 | .16061E-03 | .17408E-03 | .41091E+08 | 9.23E-01 |
| 53 | 368 | 148.105 | .21190E+00 | .18302E+00 | .64437E+11 | 1.16E+00 |
| 53 | 367 | 141.205 | .42872E-03 | .16992E-03 | .14342E+09 | 2.52E+00 |
| 53 | 366 | 116.644 | .80054E-03 | .37730E-03 | .39246E+09 | 2.12E+00 |
| 53 | 365 | 113.056 | .11523E-04 | .39612E-04 | .60133E+07 | 2.91E-01 |
| 53 | 364 | 58.325 | .64202E-05 | .12594E-06 | .12589E+08 | 5.10E+01 |
| 53 | 363 | 58.203 | .20732E-04 | .11564E-06 | .40821E+08 | 1.79E+02 |
| 53 | 362 | 56.487 | .76703E-05 | .41271E-06 | .16034E+08 | 1.86E+01 |
| 53 | 361 | 55.896 | .72714E-05 | .26509E-05 | .15523E+08 | 2.74E+00 |
| 53 | 360 | 54.806 | .68658E-06 | .39725E-05 | .15246E+07 | 1.73E-01 |
| 53 | 359 | 53.288 | .70977E-05 | .89860E-06 | .16672E+08 | 7.90E+00 |
| 53 | 358 | 50.145 | .39777E-06 | .18787E-05 | .10551E+07 | 2.12E-01 |
| 53 | 357 | 38.947 | .13672E-03 | .11225E-03 | .60121E+09 | 1.22E+00 |
| 370 | 52 | 187.711 | .25310E+00 | .20503E+00 | .47913E+11 | 1.23E+00 |
| 52 | 369 | 241.523 | .36710E-02 | .34383E-02 | .41977E+09 | 1.07E+00 |
| 52 | 368 | 212.803 | .49634E-03 | .12274E-02 | .73106E+08 | 4.04E-01 |
| 52 | 367 | 198.844 | .31767E-09 | .12580E-06 | .53591E+02 | 2.53E-03 |
| 52 | 366 | 153.367 | .13026E-01 | .11256E-01 | .36939E+10 | 1.16E+00 |
| 52 | 365 | 147.224 | .12384E+00 | .98081E-01 | .38111E+11 | 1.26E+00 |
| 52 | 364 | 66.258 | .49691E-05 | .13088E-06 | .75498E+07 | 3.80E+01 |
| 52 | 363 | 66.101 | .15499E-04 | .42102E-05 | .23660E+08 | 3.68E+00 |
| 52 | 362 | 63.896 | .17827E-04 | .19565E-06 | .29124E+08 | 9.11E+01 |
| 52 | 361 | 63.141 | .44710E-05 | .84911E-06 | .74802E+07 | 5.27E+00 |
| 52 | 360 | 61.754 | .29107E-03 | .86606E-06 | .50908E+09 | 3.36E+02 |
| 52 | 359 | 59.834 | .30419E-03 | .82101E-04 | .56675E+09 | 3.71E+00 |
| 52 | 358 | 55.899 | .60736E-04 | .17478E-07 | .12965E+09 | 3.47E+03 |
| 52 | 357 | 42.331 | .18189E-02 | .79850E-03 | .67705E+10 | 2.28E+00 |
| 370 | 51 | 170.863 | .27065E-01 | .22174E-01 | .61836E+10 | 1.22E+00 |
| 51 | 369 | 276.618 | .62707E-01 | .61238E-01 | .54662E+10 | 1.02E+00 |
| 51 | 368 | 239.586 | .34872E-06 | .62792E-04 | .40522E+05 | 5.55E-03 |
| 51 | 367 | 222.036 | .34540E-04 | .72230E-04 | .46732E+07 | 4.78E-01 |
| 51 | 366 | 166.806 | .23245E+00 | .22490E+00 | .55725E+11 | 1.03E+00 |
| 51 | 365 | 159.564 | .14065E-01 | .12314E-01 | .36847E+10 | 1.14E+00 |
| 51 | 364 | 68.647 | .29302E-05 | .16944E-09 | .41475E+07 | 1.73E+04 |
| 51 | 363 | 68.479 | .77549E-05 | .14894E-05 | .11030E+08 | 5.21E+00 |
| 51 | 362 | 66.115 | .10154E-04 | .99066E-07 | .15494E+08 | 1.02E+02 |
| 51 | 361 | 65.307 | .10158E-05 | .25952E-06 | .15885E+07 | 3.91E+00 |
| 51 | 360 | 63.825 | .13774E-03 | .16959E-04 | .22554E+09 | 8.12E+00 |
| 51 | 359 | 61.775 | .71668E-05 | .25174E-04 | .12526E+08 | 2.85E-01 |
| 51 | 358 | 57.590 | .19824E-03 | .51605E-04 | .39868E+09 | 3.84E+00 |
| 51  | 357 | 43.294  | .56321E-03 | .13209E-02 | .20043E+10 | 4.26E-01 |
| 370 | 50  | 69.605  | .42895E-05 | .15392E-06 | .59054E+07 | 2.79E+01 |
| 369 | 50  | 204.123 | .36842E-02 | .36898E-02 | .58979E+09 | 9.98E-01 |
| 368 | 50  | 230.402 | .13686E-05 | .56376E-07 | .17196E+06 | 2.43E+01 |
| 367 | 50  | 249.356 | .61819E-01 | .70741E-01 | .66316E+10 | 8.74E-01 |
| 366 | 50  | 396.964 | .68663E-03 | .78613E-03 | .29064E+08 | 8.73E-01 |
| 365 | 50  | 445.033 | .21866E-02 | .37789E-02 | .73639E+08 | 5.79E-01 |
| 50  | 364 | 165.203 | .10528E+00 | .83466E-01 | .25729E+11 | 1.26E+00 |
| 50  | 363 | 164.233 | .71371E-02 | .48182E-02 | .17649E+10 | 1.48E+00 |
| 50  | 362 | 151.261 | .18383E-01 | .12230E-01 | .53592E+10 | 1.50E+00 |
| 50  | 361 | 147.100 | .60347E-01 | .51373E-01 | .18602E+11 | 1.17E+00 |
| 50  | 360 | 139.786 | .80747E-02 | .60495E-02 | .27563E+10 | 1.33E+00 |
| 50  | 359 | 130.318 | .37229E-02 | .38800E-02 | .14622E+10 | 9.60E-01 |
| 50  | 358 | 112.994 | .24916E-05 | .95132E-05 | .13017E+07 | 2.62E-01 |
| 50  | 357 | 68.569  | .89623E-06 | .18055E-09 | .12714E+07 | 4.96E+03 |
| 370 | 49  | 67.521  | .27931E-04 | .19740E-06 | .40864E+08 | 1.41E+02 |
| 369 | 49  | 187.177 | .12357E-01 | .10844E-01 | .23525E+10 | 1.14E+00 |
| 368 | 49  | 209.041 | .21308E-02 | .17628E-02 | .32525E+09 | 1.21E+00 |
| 367 | 49  | 224.525 | .11376E-01 | .10156E-01 | .15052E+10 | 1.12E+00 |
| 366 | 49  | 337.538 | .33048E-02 | .25835E-02 | .19348E+09 | 1.28E+00 |
| 365 | 49  | 371.674 | .11080E-01 | .11972E-01 | .53499E+09 | 9.25E-01 |
| 49  | 364 | 178.264 | .54809E-04 | .23198E-04 | .11504E+08 | 2.36E+00 |
| 49  | 363 | 177.135 | .25753E-01 | .23255E-01 | .54747E+10 | 1.11E+00 |
| 49  | 362 | 162.139 | .73952E-01 | .54129E-01 | .18763E+11 | 1.37E+00 |
| 49  | 361 | 157.367 | .73732E-02 | .64235E-02 | .19859E+10 | 1.15E+00 |
| 49  | 360 | 149.025 | .90596E-01 | .78571E-01 | .27209E+11 | 1.15E+00 |
| 49  | 359 | 138.312 | .13226E-01 | .15422E-01 | .46116E+10 | 8.58E-01 |
| 49  | 358 | 118.955 | .46498E-06 | .44490E-04 | .21918E+06 | 1.05E-02 |
| 49  | 357 | 70.720  | .62160E-05 | .39126E-07 | .82901E+07 | 1.59E+02 |
| 370 | 48  | 65.673  | .85784E-09 | .10741E-05 | .13267E+04 | 7.99E-04 |
| 369 | 48  | 173.634 | .21607E-02 | .16719E-02 | .47804E+09 | 1.29E+00 |
| 368 | 48  | 192.290 | .69193E-01 | .52714E-01 | .12482E+11 | 1.31E+00 |
| 367 | 48  | 205.314 | .46543E-01 | .40175E-01 | .73645E+10 | 1.16E+00 |
| 366 | 48  | 295.914 | .10200E-03 | .12562E-03 | .77695E+07 | 8.12E-01 |
| 365 | 48  | 321.827 | .16333E-02 | .31226E-02 | .10519E+09 | 5.23E-01 |
| 48  | 364 | 192.570 | .28486E-03 | .24543E-03 | .51238E+08 | 1.16E+00 |
| 48  | 363 | 191.253 | .13759E-05 | .87929E-04 | .25091E+06 | 1.56E-02 |
| 48  | 362 | 173.888 | .17139E-05 | .26071E-04 | .37808E+06 | 6.57E-02 |
| 48  | 361 | 168.411 | .18583E+00 | .16111E+00 | .43703E+11 | 1.15E+00 |
| 48  | 360 | 158.893 | .52328E-02 | .35632E-02 | .13825E+10 | 1.47E+00 |
| 48  | 359 | 146.772 | .11438E-02 | .71949E-03 | .35417E+09 | 1.59E+00 |
| 48  | 358 | 125.159 | .95514E-04 | .28208E-03 | .40670E+08 | 3.39E-01 |
| 48  | 357 | 72.867  | .51348E-05 | .11397E-06 | .64504E+07 | 4.51E+01 |
| 370 | 47  | 63.865  | .11683E-03 | .27829E-04 | .19106E+09 | 4.20E+00 |
| 369 | 47 | 161.546 | .17232E+00 | .14659E+00 | .44042E+11 | 1.18E+00 |
| 368 | 47 | 177.575 | .91930E-03 | .89855E-03 | .19446E+09 | 1.02E+00 |
| 367 | 47 | 188.625 | .95371E-03 | .66140E-03 | .17879E+09 | 1.44E+00 |
| 366 | 47 | 262.447 | .67015E-02 | .52877E-02 | .64896E+09 | 1.27E+00 |
| 365 | 47 | 282.630 | .78897E-02 | .75005E-02 | .65881E+09 | 1.05E+00 |
| 47 | 364 | 209.997 | .16842E-02 | .20926E-02 | .25474E+09 | 8.05E-01 |
| 47 | 363 | 208.431 | .46979E-02 | .37398E-02 | .72130E+09 | 1.26E+00 |
| 47 | 362 | 187.973 | .42819E-02 | .43451E-02 | .80831E+09 | 9.85E-01 |
| 47 | 361 | 181.590 | .10410E-02 | .16451E-02 | .21058E+09 | 6.33E-01 |
| 47 | 360 | 170.572 | .47338E-01 | .30994E-01 | .10852E+11 | 1.53E+00 |
| 47 | 359 | 156.682 | .11604E-03 | .50427E-03 | .31530E+08 | 2.30E-01 |
| 47 | 358 | 132.295 | .79301E-01 | .43318E-01 | .30222E+11 | 1.83E+00 |
| 47 | 357 | 75.229 | .30377E-06 | .53743E-05 | .35801E+06 | 5.65E-02 |
| 370 | 46 | 62.601 | .74421E-04 | .35200E-06 | .12667E+09 | 2.11E+02 |
| 369 | 46 | 153.695 | .20081E-01 | .15480E-01 | .56700E+10 | 1.30E+00 |
| 368 | 46 | 168.135 | .40874E-02 | .26100E-02 | .96441E+09 | 1.57E+00 |
| 367 | 46 | 178.009 | .33101E-02 | .21320E-02 | .69678E+09 | 1.55E+00 |
| 366 | 46 | 242.337 | .57630E-03 | .18935E-02 | .65455E+08 | 3.04E-01 |
| 365 | 46 | 259.445 | .34471E-01 | .37645E-01 | .34158E+10 | 9.16E-01 |
| 46 | 364 | 224.932 | .22231E-03 | .36418E-03 | .29308E+08 | 6.10E-01 |
| 46 | 363 | 223.137 | .19051E-02 | .25056E-02 | .25521E+09 | 7.60E-01 |
| 46 | 362 | 199.852 | .55165E-05 | .29663E-05 | .92125E+06 | 1.86E+00 |
| 46 | 361 | 192.651 | .26573E-02 | .36846E-02 | .47756E+09 | 7.21E-01 |
| 46 | 360 | 180.297 | .10611E+00 | .96955E-01 | .21772E+11 | 1.09E+00 |
| 46 | 359 | 164.848 | .17135E-02 | .97227E-04 | .42058E+09 | 1.76E+01 |
| 46 | 358 | 138.070 | .47791E-01 | .40437E-01 | .16722E+11 | 1.18E+00 |
| 46 | 357 | 77.063 | .10082E-03 | .20966E-04 | .11324E+09 | 4.81E+00 |
| 370 | 45 | 59.504 | .22755E-05 | .65374E-06 | .42867E+07 | 3.48E+00 |
| 369 | 45 | 136.278 | .31920E-06 | .10589E-05 | .11464E+06 | 3.01E-01 |
| 368 | 45 | 147.510 | .76162E-01 | .42640E-01 | .23347E+11 | 1.79E+00 |
| 367 | 45 | 155.056 | .16611E-01 | .14794E-01 | .46085E+10 | 1.12E+00 |
| 366 | 45 | 201.692 | .15979E-03 | .18810E-03 | .26200E+08 | 8.49E-01 |
| 365 | 45 | 213.403 | .43647E-04 | .62168E-04 | .63928E+07 | 7.02E-01 |
| 45 | 364 | 276.685 | .24319E-01 | .25788E-01 | .21189E+10 | 9.43E-01 |
| 45 | 363 | 273.975 | .53971E-01 | .48435E-01 | .47959E+10 | 1.11E+00 |
| 45 | 362 | 239.686 | .17922E-01 | .12805E-01 | .20809E+10 | 1.40E+00 |
| 45 | 361 | 229.402 | .54337E-02 | .52639E-02 | .68871E+09 | 1.03E+00 |
| 45 | 360 | 212.096 | .24480E-02 | .22295E-02 | .36297E+09 | 1.10E+00 |
| 45 | 359 | 191.036 | .28196E-03 | .24453E-03 | .51533E+08 | 1.15E+00 |
| 45 | 358 | 155.979 | .17010E-02 | .12623E-02 | .46634E+09 | 1.35E+00 |
| 45 | 357 | 82.339 | .42218E-06 | .10174E-05 | .41535E+06 | 4.15E-01 |
| 370 | 44 | 54.104 | .24027E-03 | .36139E-05 | .54749E+09 | 6.65E+01 |
| 369 | 44 | 110.923 | .25973E-03 | .29450E-03 | .14081E+09 | 8.82E-01 |
| 368 | 44 | 118.252 | .25846E-02 | .10089E-02 | .12328E+10 | 2.56E+00 |
| 367 | 44 | 123.053 | .11417E-03 | .25782E-04 | .50294E+08 | 4.43E+00 |
| 366 | 44 | 150.707 | .18694E-03 | .31455E-03 | .54901E+08 | 5.94E-01 |
| 365 | 44 | 157.151 | .32028E+00 | .23380E+00 | .86503E+11 | 1.37E+00 |
| 44 | 364 | 516.294 | .91130E-03 | .34484E-05 | .22804E+08 | 2.64E+02 |
| 44 | 363 | 506.935 | .53919E-02 | .69614E-03 | .13995E+09 | 7.75E+00 |
| 44 | 362 | 400.834 | .26623E-02 | .10828E-03 | .11053E+09 | 2.46E+01 |
| 44 | 361 | 372.881 | .43094E-03 | .85919E-03 | .20673E+08 | 5.02E-01 |
| 44 | 360 | 329.217 | .62646E-02 | .48231E-02 | .38554E+09 | 1.30E+00 |
| 44 | 359 | 281.114 | .33499E-01 | .19269E-01 | .28274E+10 | 1.74E+00 |
| 44 | 358 | 211.248 | .63701E-01 | .38684E-01 | .95213E+10 | 1.65E+00 |
| 44 | 357 | 95.533 | .24672E-04 | .21242E-03 | .18031E+08 | 1.16E-01 |
| 370 | 43 | 53.720 | .54639E-04 | .93416E-05 | .12629E+09 | 5.85E+00 |
| 369 | 43 | 109.324 | .10380E-03 | .11706E-03 | .57928E+08 | 8.87E-01 |
| 368 | 43 | 116.436 | .58020E-03 | .14863E-03 | .28545E+09 | 3.90E+00 |
| 367 | 43 | 121.088 | .23403E-06 | .14744E-06 | .10647E+06 | 1.59E+00 |
| 366 | 43 | 147.770 | .13613E+00 | .71523E-01 | .41582E+11 | 1.90E+00 |
| 365 | 43 | 153.961 | .12470E-03 | .18407E-02 | .35090E+08 | 6.77E-02 |
| 43 | 364 | 554.011 | .20902E-03 | .47932E-04 | .45423E+07 | 4.36E+00 |
| 43 | 363 | 543.249 | .74091E-03 | .22578E-03 | .16746E+08 | 3.28E+00 |
| 43 | 362 | 423.202 | .52707E-03 | .99508E-06 | .19629E+08 | 5.30E+02 |
| 43 | 361 | 392.163 | .10550E-03 | .27616E-04 | .45754E+07 | 3.82E+00 |
| 43 | 360 | 344.158 | .32595E-02 | .14207E-02 | .18356E+09 | 2.29E+00 |
| 43 | 359 | 291.936 | .42857E-01 | .36074E-01 | .33541E+10 | 1.19E+00 |
| 43 | 358 | 217.301 | .15388E+00 | .14396E+00 | .21737E+11 | 1.07E+00 |
| 43 | 357 | 96.752 | .40761E-03 | .47896E-03 | .29044E+09 | 8.51E-01 |
| 370 | 42 | 36.707 | .26058E-08 | .20394E-08 | .12900E+05 | 1.28E+00 |
| 369 | 42 | 56.259 | .31106E-07 | .16422E-06 | .65553E+05 | 1.89E-01 |
| 368 | 42 | 58.085 | .40942E-04 | .74793E-05 | .80942E+08 | 5.47E+00 |
| 367 | 42 | 59.220 | .97144E-05 | .23094E-05 | .18476E+08 | 4.21E+00 |
| 366 | 42 | 64.956 | .83345E-07 | .62603E-07 | .13176E+06 | 1.33E+00 |
| 365 | 42 | 66.125 | .74583E-05 | .10655E-05 | .11377E+08 | 7.00E+00 |
| 364 | 42 | 146.569 | .37421E-01 | .33329E-01 | .11619E+11 | 1.12E+00 |
| 363 | 42 | 147.341 | .13194E-03 | .10923E-03 | .40538E+08 | 1.21E+00 |
| 362 | 42 | 159.622 | .11475E-02 | .11045E-02 | .30039E+09 | 1.04E+00 |
| 361 | 42 | 164.533 | .10590E+00 | .11667E+00 | .26093E+11 | 9.08E-01 |
| 360 | 42 | 174.761 | .98716E-03 | .11068E-02 | .21559E+09 | 8.92E-01 |
| 359 | 42 | 192.221 | .11222E-02 | .15114E-02 | .20257E+09 | 7.42E-01 |
| 358 | 42 | 248.395 | .11778E-02 | .19271E-02 | .12732E+09 | 6.11E-01 |
| 42 | 357 | 585.498 | .75891E-10 | .58733E-08 | .14766E+01 | 1.29E-02 |
| 370 | 41 | 36.678 | .46300E-01 | .52155E-01 | .22956E+12 | 8.88E-01 |
| 369 | 41 | 56.191 | .10755E-05 | .43463E-05 | .22719E+07 | 2.47E-01 |
| 368 | 41 | 58.013 | .10715E-04 | .24735E-04 | .21236E+08 | 4.33E-01 |
| 367 | 41 | 59.145 | .54513E-07 | .19622E-06 | .10394E+06 | 2.78E-01 |
| 366 | 41 | 64.866 | .21099E-04 | .28580E-03 | .33448E+08 | 7.38E-02 |
| 365 | 41 | 66.031 | .38533E-03 | .67776E-03 | .58948E+09 | 5.69E-01 |
| 364 | 41 | 146.110 | .62410E-07 | .10433E-08 | .19500E+05 | 5.98E+01 |
| 363 | 41 | 146.877 | .50317E-06 | .66487E-07 | .15557E+06 | 7.57E+00 |
| 362 | 41 | 159.078 | .13583E-08 | .57068E-09 | .35802E+03 | 2.38E+00 |
| 361 | 41 | 163.955 | .51820E-06 | .13127E-06 | .12858E+06 | 3.95E+00 |
| 360 | 41 | 174.109 | .12040E-04 | .33360E-05 | .26492E+07 | 3.61E+00 |
| 359 | 41 | 191.433 | .30921E-04 | .14934E-04 | .56281E+07 | 2.07E+00 |
| 358 | 41 | 247.081 | .97029E-05 | .12193E-04 | .10601E+07 | 7.96E-01 |
| 41 | 357 | 592.935 | .16866E+00 | .11726E+00 | .31998E+10 | 1.44E+00 |
| 370 | 40 | 36.280 | .11526E-06 | .75855E-07 | .58410E+06 | 1.52E+00 |
| 369 | 40 | 55.263 | .64415E-05 | .73948E-05 | .14069E+08 | 8.71E-01 |
| 368 | 40 | 57.023 | .58902E-05 | .75088E-06 | .12083E+08 | 7.84E+00 |
| 367 | 40 | 58.117 | .24082E-05 | .38527E-06 | .47558E+07 | 6.25E+00 |
| 366 | 40 | 63.631 | .82184E-07 | .54403E-05 | .13539E+06 | 1.51E-02 |
| 365 | 40 | 64.753 | .51505E-03 | .62026E-04 | .81935E+09 | 8.30E+00 |
| 364 | 40 | 139.993 | .80029E-02 | .68648E-02 | .27238E+10 | 1.17E+00 |
| 363 | 40 | 140.697 | .23216E-01 | .20640E-01 | .78226E+10 | 1.12E+00 |
| 362 | 40 | 151.853 | .31253E-01 | .30664E-01 | .90402E+10 | 1.02E+00 |
| 361 | 40 | 156.292 | .66440E-03 | .56481E-03 | .18142E+09 | 1.18E+00 |
| 360 | 40 | 165.491 | .37645E-01 | .43116E-01 | .91682E+10 | 8.73E-01 |
| 359 | 40 | 181.066 | .63477E-01 | .77658E-01 | .12915E+11 | 8.17E-01 |
| 358 | 40 | 230.079 | .45298E-01 | .69397E-01 | .57077E+10 | 6.53E-01 |
| 40 | 357 | 720.748 | .39163E-08 | .27823E-06 | .50286E+02 | 1.41E-02 |
| 468 | 53 | 366.451 | .33925E-03 | .69579E-03 | .33701E+08 | 4.88E-01 |
| 53 | 467 | 193.264 | .27519E-01 | .23833E-01 | .24572E+10 | 1.15E+00 |
| 53 | 466 | 182.346 | .96421E-01 | .78924E-01 | .96712E+10 | 1.22E+00 |
| 53 | 465 | 166.306 | .62834E-03 | .85421E-03 | .75768E+08 | 7.36E-01 |
| 53 | 464 | 156.454 | .15175E-01 | .11332E-01 | .20675E+10 | 1.34E+00 |
| 53 | 463 | 148.512 | .28179E+00 | .24132E+00 | .42610E+11 | 1.17E+00 |
| 53 | 462 | 141.137 | .16306E-02 | .64907E-03 | .27301E+09 | 2.51E+00 |
| 53 | 461 | 138.284 | .13125E-01 | .92596E-02 | .22891E+10 | 1.42E+00 |
| 53 | 460 | 120.455 | .47622E-02 | .27856E-02 | .10946E+10 | 1.71E+00 |
| 53 | 459 | 112.358 | .80385E-04 | .12553E-04 | .21236E+08 | 6.40E+00 |
| 53 | 458 | 111.575 | .27225E-04 | .10079E-05 | .72935E+07 | 2.70E+01 |
| 53 | 457 | 58.937 | .38513E-06 | .10379E-05 | .36978E+06 | 3.71E-01 |
| 53 | 456 | 57.916 | .30195E-05 | .16074E-06 | .30022E+07 | 1.88E+01 |
| 53 | 455 | 57.686 | .16084E-04 | .15942E-05 | .16119E+08 | 1.01E+01 |
| 53 | 454 | 56.921 | .11745E-03 | .10557E-04 | .12089E+09 | 1.11E+01 |
| 53 | 453 | 55.972 | .49322E-05 | .18567E-05 | .52505E+07 | 2.66E+00 |
| 53 | 452 | 55.458 | .25173E-04 | .95325E-05 | .27297E+08 | 2.64E+00 |
| 53 | 451 | 54.520 | .67823E-04 | .24764E-04 | .76096E+08 | 2.74E+00 |
| 53 | 450 | 53.612 | .52891E-06 | .34016E-07 | .61371E+06 | 1.55E+01 |
| 53 | 449 | 52.299 | .16851E-05 | .32123E-06 | .20547E+07 | 5.25E+00 |
| 53 | 448 | 50.847 | .13640E-05 | .15488E-05 | .17594E+07 | 8.81E-01 |
| 53 | 447 | 49.935 | .10424E-05 | .19627E-05 | .13943E+07 | 5.31E-01 |
| 53 | 446 | 38.639 | .16795E-03 | .21840E-03 | .37517E+09 | 7.69E-01 |
| 468 | 52 | 209.131 | .14497E-02 | .30774E-02 | .44218E+09 | 4.71E-01 |
| 52 | 467 | 320.363 | .51745E-01 | .45833E-01 | .16815E+10 | 1.13E+00 |
| 52 | 466 | 291.438 | .43610E-02 | .28159E-02 | .17124E+09 | 1.55E+00 |
| 52 | 465 | 252.512 | .63223E-03 | .21196E-03 | .33069E+08 | 2.98E+00 |
| 52 | 464 | 230.476 | .36619E-01 | .32824E-01 | .22991E+10 | 1.12E+00 |
| 52 | 463 | 213.645 | .34421E-03 | .88322E-03 | .25150E+08 | 3.90E-01 |
| 52 | 462 | 198.709 | .52006E-03 | .48511E-03 | .43926E+08 | 1.07E+00 |
| 52 | 461 | 193.100 | .18150E-01 | .19096E-01 | .16234E+10 | 9.50E-01 |
| 52 | 460 | 160.024 | .53742E+00 | .50031E+00 | .69991E+11 | 1.07E+00 |
| 52 | 459 | 146.042 | .46806E-01 | .45506E-01 | .73190E+10 | 1.03E+00 |
| 52 | 458 | 144.722 | .36986E-01 | .31500E-01 | .58894E+10 | 1.17E+00 |
| 52 | 457 | 67.049 | .15457E-03 | .66065E-04 | .11467E+09 | 2.34E+00 |
| 52 | 456 | 65.731 | .39179E-06 | .33546E-07 | .30242E+06 | 1.17E+01 |
| 52 | 455 | 65.435 | .13248E-03 | .59347E-04 | .10319E+09 | 2.23E+00 |
| 52 | 454 | 64.453 | .12395E-05 | .27607E-12 | .99508E+06 | 4.49E+06 |
| 52 | 453 | 63.238 | .40692E-04 | .73870E-05 | .33936E+08 | 5.51E+00 |
| 52 | 452 | 62.583 | .14651E-03 | .32871E-04 | .12475E+09 | 4.46E+00 |
| 52 | 451 | 61.391 | .25477E-07 | .17692E-07 | .22545E+05 | 1.44E+00 |
| 52 | 450 | 60.242 | .21440E-03 | .47675E-04 | .19703E+09 | 4.50E+00 |
| 52 | 449 | 58.589 | .23362E-04 | .14937E-04 | .22697E+08 | 1.56E+00 |
| 52 | 448 | 56.773 | .22507E-03 | .15575E-03 | .23288E+09 | 1.45E+00 |
| 52 | 447 | 55.638 | .92258E-04 | .15633E-04 | .99395E+08 | 5.90E+00 |
| 52 | 446 | 41.967 | .12415E-02 | .18826E-02 | .23509E+10 | 6.59E-01 |
| 468 | 51 | 188.431 | .10208E+00 | .80960E-01 | .38355E+11 | 1.26E+00 |
| 51 | 467 | 385.184 | .97879E-02 | .69694E-02 | .22002E+09 | 1.40E+00 |
| 51 | 466 | 344.121 | .30023E-02 | .18814E-02 | .84553E+08 | 1.60E+00 |
| 51 | 465 | 291.128 | .10958E-04 | .62186E-05 | .43119E+06 | 1.76E+00 |
| 51 | 464 | 262.223 | .91238E-02 | .64495E-02 | .44253E+09 | 1.41E+00 |
| 51 | 463 | 240.653 | .12932E-03 | .42487E-06 | .74468E+07 | 3.04E+02 |
| 51 | 462 | 221.868 | .10630E-02 | .12473E-02 | .72020E+08 | 8.52E-01 |
| 51 | 461 | 214.898 | .83815E-02 | .82478E-02 | .60529E+09 | 1.02E+00 |
| 51 | 460 | 174.711 | .10836E-02 | .54155E-03 | .11839E+09 | 2.00E+00 |
| 51 | 459 | 158.177 | .45674E+00 | .43934E+00 | .60881E+11 | 1.04E+00 |
| 51 | 458 | 156.630 | .17592E-01 | .14751E-01 | .23915E+10 | 1.19E+00 |
| 51 | 457 | 69.496 | .92885E-04 | .48143E-04 | .64139E+08 | 1.93E+00 |
| 51 | 456 | 68.082 | .30938E-07 | .45359E-06 | .22261E+05 | 6.82E-02 |
| 51 | 455 | 67.765 | .71310E-04 | .40259E-04 | .51790E+08 | 1.77E+00 |
| 51 | 454 | 66.711 | .15476E-05 | .37018E-06 | .11597E+07 | 4.18E+00 |
| 51 | 453 | 65.410 | .13479E-04 | .33998E-06 | .10507E+08 | 3.96E+01 |
| 51 | 452 | 64.710 | .26071E-04 | .23542E-07 | .20764E+08 | 1.11E+03 |
| 51 | 451 | 63.437 | .22479E-05 | .24116E-08 | .18629E+07 | 9.32E+02 |
| 51 | 450 | 62.210 | .87566E-04 | .64876E-04 | .75460E+08 | 1.35E+00 |
| 51 | 449 | 60.449 | .77170E-03 | .13925E-04 | .70432E+09 | 5.54E+01 |
| 51 | 448 | 58.518 | .11607E-03 | .55688E-04 | .11304E+09 | 2.08E+00 |
| 51 | 447 | 57.313 | .15388E-07 | .83043E-04 | .15624E+05 | 1.85E-04 |
| 51 | 446 | 42.913 | .17825E-02 | .21552E-02 | .32282E+10 | 8.27E-01 |
| 468 | 50 | 72.353 | .68118E-06 | .68281E-07 | .17358E+07 | 9.98E+00 |
| 467 | 50 | 168.977 | .85864E-02 | .68423E-02 | .40116E+10 | 1.25E+00 |
| 466 | 50 | 178.312 | .78403E-02 | .69556E-02 | .32896E+10 | 1.13E+00 |
| 465 | 50 | 196.881 | .57087E-01 | .45099E-01 | .19647E+11 | 1.27E+00 |
| 464 | 50 | 212.741 | .13410E-02 | .13827E-02 | .39528E+09 | 9.70E-01 |
| 463 | 50 | 229.424 | .58134E-05 | .29970E-05 | .14734E+07 | 1.94E+00 |
| 462 | 50 | 249.567 | .11251E-01 | .12359E-01 | .24099E+10 | 9.10E-01 |
| 461 | 50 | 259.017 | .11675E-03 | .66987E-05 | .23214E+08 | 1.74E+01 |
| 460 | 50 | 358.376 | .20934E-03 | .29579E-03 | .21744E+08 | 7.08E-01 |
| 459 | 50 | 456.188 | .13230E-03 | .17332E-03 | .84806E+07 | 7.63E-01 |
| 458 | 50 | 469.565 | .40955E-03 | .57144E-03 | .24779E+08 | 7.17E-01 |
| 50 | 457 | 170.209 | .38246E-01 | .32669E-01 | .44027E+10 | 1.17E+00 |
| 50 | 456 | 161.968 | .36891E-02 | .43715E-02 | .46899E+09 | 8.44E-01 |
| 50 | 455 | 160.183 | .50169E-01 | .43297E-01 | .65209E+10 | 1.16E+00 |
| 50 | 454 | 154.420 | .10080E-01 | .73972E-02 | .14098E+10 | 1.36E+00 |
| 50 | 453 | 147.624 | .62878E-01 | .53495E-01 | .96225E+10 | 1.18E+00 |
| 50 | 452 | 144.105 | .33383E-04 | .14472E-06 | .53613E+07 | 2.31E+02 |
| 50 | 451 | 137.940 | .13702E-06 | .43341E-08 | .24017E+05 | 3.16E+01 |
| 50 | 450 | 132.268 | .83395E-02 | .73857E-02 | .15898E+10 | 1.13E+00 |
| 50 | 449 | 124.554 | .49937E-03 | .62579E-03 | .10735E+09 | 7.98E-01 |
| 50 | 448 | 116.623 | .49004E-02 | .28033E-02 | .12016E+10 | 1.75E+00 |
| 50 | 447 | 111.932 | .52034E-05 | .16812E-04 | .13851E+07 | 3.10E-01 |
| 50 | 446 | 67.619 | .66748E-08 | .41070E-07 | .48686E+04 | 1.63E-01 |
| 468 | 49 | 70.104 | .17992E-05 | .21450E-06 | .48839E+07 | 8.39E+00 |
| 467 | 49 | 157.197 | .41650E-01 | .32085E-01 | .22485E+11 | 1.30E+00 |
| 466 | 49 | 165.244 | .31452E-02 | .19032E-02 | .15366E+10 | 1.65E+00 |
| 465 | 49 | 181.070 | .10269E-01 | .76700E-02 | .41781E+10 | 1.34E+00 |
| 464 | 49 | 194.399 | .13304E-01 | .11857E-01 | .46963E+10 | 1.12E+00 |
| 463 | 49 | 208.235 | .99881E-03 | .69127E-03 | .30728E+09 | 1.44E+00 |
| 462 | 49 | 224.697 | .20951E-02 | .25916E-02 | .55357E+09 | 8.08E-01 |
| 461 | 49 | 232.328 | .26831E-01 | .22818E-01 | .66313E+10 | 1.18E+00 |
| 460 | 49 | 309.226 | .14522E-02 | .18856E-02 | .20260E+09 | 7.70E-01 |
| 459 | 49 | 379.421 | .19748E-03 | .38953E-04 | .18300E+08 | 5.07E+00 |
| 458 | 49 | 388.630 | .85940E-03 | .23223E-03 | .75907E+08 | 3.70E+00 |
| 49 | 457 | 184.107 | .33665E-02 | .41299E-02 | .33124E+09 | 8.15E-01 |
| 49 | 456 | 174.503 | .20412E-02 | .23185E-02 | .22355E+09 | 8.80E-01 |
| 49 | 455 | 172.433 | .15172E-01 | .11644E-01 | .17018E+10 | 1.30E+00 |
| 49 | 454 | 165.774 | .56595E-02 | .51340E-02 | .68684E+09 | 1.10E+00 |
| 49 | 453 | 157.966 | .22164E-02 | .12693E-02 | .29623E+09 | 1.75E+00 |
| 49 | 452 | 153.944 | .26610E-01 | .23807E-01 | .37448E+10 | 1.12E+00 |
| 49 | 451 | 146.929 | .59582E-02 | .40043E-02 | .92046E+09 | 1.49E+00 |
| 49 | 450 | 140.511 | .61267E-01 | .54290E-01 | .10349E+11 | 1.13E+00 |
| 49 | 449 | 131.837 | .19977E-02 | .18409E-02 | .38332E+09 | 1.09E+00 |
| 49 | 448 | 122.984 | .24874E-01 | .15865E-01 | .54847E+10 | 1.57E+00 |
| 49 | 447 | 117.779 | .41226E-03 | .13394E-03 | .99115E+08 | 3.08E+00 |
| 49 | 446 | 69.710 | .63177E-06 | .53162E-06 | .43358E+06 | 1.19E+00 |
| 468 | 48 | 68.114 | .63414E-05 | .11596E-07 | .18234E+08 | 5.47E+02 |
| 467 | 48 | 147.532 | .18790E-01 | .15490E-01 | .11516E+11 | 1.21E+00 |
| 466 | 48 | 154.598 | .10408E+00 | .91478E-01 | .58090E+11 | 1.14E+00 |
| 465 | 48 | 168.366 | .34673E-03 | .31645E-03 | .16317E+09 | 1.10E+00 |
| 464 | 48 | 179.830 | .83882E-04 | .58413E-06 | .34603E+08 | 1.44E+02 |
| 463 | 48 | 191.608 | .86962E-02 | .66711E-02 | .31599E+10 | 1.30E+00 |
| 462 | 48 | 205.458 | .50116E-01 | .44363E-01 | .15838E+11 | 1.13E+00 |
| 461 | 48 | 211.820 | .10617E-06 | .17600E-04 | .31568E+05 | 6.03E-03 |
| 460 | 48 | 273.927 | .24325E-03 | .17921E-03 | .43247E+08 | 1.36E+00 |
| 459 | 48 | 327.620 | .96529E-03 | .11119E-02 | .11997E+09 | 8.68E-01 |
| 458 | 48 | 334.463 | .12665E-02 | .20482E-02 | .15103E+09 | 6.18E-01 |
| 48 | 457 | 199.406 | .80706E-02 | .71605E-02 | .67691E+09 | 1.13E+00 |
| 48 | 456 | 188.188 | .19273E-02 | .22504E-02 | .18149E+09 | 8.56E-01 |
| 48 | 455 | 185.783 | .18383E-02 | .14498E-02 | .17763E+09 | 1.27E+00 |
| 48 | 454 | 178.075 | .15511E-02 | .30430E-02 | .16313E+09 | 5.10E-01 |
| 48 | 453 | 169.098 | .17771E+00 | .16044E+00 | .20727E+11 | 1.11E+00 |
| 48 | 452 | 164.497 | .49354E-01 | .44785E-01 | .60829E+10 | 1.10E+00 |
| 48 | 451 | 156.512 | .13940E+00 | .95414E-01 | .18979E+11 | 1.46E+00 |
| 48 | 450 | 149.250 | .26218E-02 | .28730E-02 | .39253E+09 | 9.13E-01 |
| 48 | 449 | 139.501 | .35778E-03 | .56851E-04 | .61314E+08 | 6.29E+00 |
| 48 | 448 | 129.628 | .15024E-02 | .47605E-03 | .29819E+09 | 3.16E+00 |
| 48 | 447 | 123.859 | .67901E-02 | .40208E-02 | .14761E+10 | 1.69E+00 |
| 48 | 446 | 71.796 | .62411E-05 | .13095E-05 | .40380E+07 | 4.77E+00 |
| 468 | 47 | 66.171 | .10269E-06 | .37112E-05 | .31285E+06 | 2.77E-02 |
| 467 | 47 | 138.713 | .66714E-03 | .35934E-03 | .46254E+09 | 1.86E+00 |
| 466 | 47 | 144.942 | .93426E-03 | .80583E-03 | .59326E+09 | 1.16E+00 |
| 465 | 47 | 156.977 | .43083E-03 | .33562E-03 | .23324E+09 | 1.28E+00 |
| 464 | 47 | 166.897 | .12040E-01 | .11184E-01 | .57660E+10 | 1.08E+00 |
| 463 | 47 | 176.993 | .14038E-02 | .72675E-03 | .59781E+09 | 1.93E+00 |
| 462 | 47 | 188.747 | .12608E-03 | .10480E-03 | .47212E+08 | 1.20E+00 |
| 461 | 47 | 194.102 | .13562E-01 | .10989E-01 | .48020E+10 | 1.23E+00 |
| 460 | 47 | 245.005 | .23380E-02 | .18009E-02 | .51957E+09 | 1.30E+00 |
| 459 | 47 | 287.088 | .22163E-01 | .28144E-01 | .35873E+10 | 7.87E-01 |
| 458 | 47 | 292.329 | .49068E-02 | .59051E-02 | .76598E+09 | 8.31E-01 |
| 47 | 457 | 218.152 | .74320E-01 | .71674E-01 | .52082E+10 | 1.04E+00 |
| 47 | 456 | 204.796 | .21115E-03 | .23940E-04 | .16790E+08 | 8.82E+00 |
| 47 | 455 | 201.951 | .43750E-01 | .35670E-01 | .35776E+10 | 1.23E+00 |
| 47 | 454 | 192.877 | .15634E-02 | .16929E-02 | .14016E+09 | 9.24E-01 |
| 47 | 453 | 182.388 | .21565E-01 | .17913E-01 | .21620E+10 | 1.20E+00 |
| 47 | 452 | 177.047 | .15554E-01 | .10028E-01 | .16549E+10 | 1.55E+00 |
| 47 | 451 | 167.831 | .45392E-02 | .37309E-02 | .53745E+09 | 1.22E+00 |
| 47 | 450 | 159.510 | .42115E-01 | .40990E-01 | .55203E+10 | 1.03E+00 |
| 47 | 449 | 148.424 | .12583E-02 | .76045E-03 | .19049E+09 | 1.65E+00 |
| 47 | 448 | 137.297 | .54351E-01 | .29079E-01 | .96158E+10 | 1.87E+00 |
| 47 | 447 | 130.842 | .21601E-03 | .68941E-03 | .42081E+08 | 3.13E-01 |
| 47 | 446 | 74.088 | .39443E-04 | .31304E-06 | .23965E+08 | 1.26E+02 |
| 468 | 46 | 64.815 | .12588E-03 | .88060E-06 | .39973E+09 | 1.43E+02 |
| 467 | 46 | 132.885 | .46093E-02 | .25805E-02 | .34821E+10 | 1.79E+00 |
| 466 | 46 | 138.590 | .11977E-02 | .11088E-02 | .83188E+09 | 1.08E+00 |
| 465 | 46 | 149.554 | .61238E-03 | .46759E-03 | .36525E+09 | 1.31E+00 |
| 464 | 46 | 158.531 | .96271E-01 | .76486E-01 | .51101E+11 | 1.26E+00 |
| 463 | 46 | 167.613 | .60460E-04 | .70255E-05 | .28709E+08 | 8.61E+00 |
| 462 | 46 | 178.117 | .62754E-02 | .43377E-02 | .26387E+10 | 1.45E+00 |
| 461 | 46 | 182.878 | .57682E-03 | .31571E-03 | .23008E+09 | 1.83E+00 |
| 460 | 46 | 227.390 | .12032E-02 | .51545E-04 | .31043E+09 | 2.33E+01 |
| 459 | 46 | 263.196 | .67784E-05 | .48371E-03 | .13054E+07 | 1.40E-02 |
| 458 | 46 | 267.595 | .19945E-01 | .22668E-01 | .37158E+10 | 8.80E-01 |
| 46 | 457 | 234.314 | .87549E-03 | .57220E-03 | .53181E+08 | 1.53E+00 |
| 46 | 456 | 218.976 | .17831E-03 | .30955E-03 | .12401E+08 | 5.76E-01 |
| 46 | 455 | 215.727 | .23215E-02 | .25894E-02 | .16637E+09 | 8.97E-01 |
| 46 | 454 | 205.403 | .71316E-03 | .10469E-02 | .56374E+08 | 6.81E-01 |
| 46 | 453 | 193.550 | .22310E-05 | .95494E-04 | .19862E+06 | 2.34E-02 |
| 46 | 452 | 187.546 | .68401E-01 | .61313E-01 | .64856E+10 | 1.12E+00 |
| 46 | 451 | 177.237 | .52241E-02 | .30206E-02 | .55464E+09 | 1.73E+00 |
| 46 | 450 | 167.982 | .35838E-02 | .25264E-02 | .42357E+09 | 1.42E+00 |
| 46 | 449 | 155.732 | .16740E-01 | .31835E-01 | .23020E+10 | 5.26E-01 |
| 46 | 448 | 143.528 | .20875E-01 | .14145E-01 | .33795E+10 | 1.48E+00 |
| 46 | 447 | 136.489 | .12483E+00 | .89717E-01 | .22346E+11 | 1.39E+00 |
| 46 | 446 | 75.865 | .18947E-03 | .45250E-04 | .10979E+09 | 4.19E+00 |
| 468 | 45 | 61.500 | .15656E-05 | .61640E-06 | .55220E+07 | 2.54E+00 |
| 467 | 45 | 119.662 | .26816E-04 | .18392E-04 | .24983E+08 | 1.46E+00 |
| 466 | 45 | 124.268 | .10569E-04 | .83141E-05 | .91301E+07 | 1.27E+00 |
| 465 | 45 | 133.012 | .27391E-03 | .11730E-03 | .20654E+09 | 2.34E+00 |
| 464 | 45 | 140.066 | .50915E-03 | .26947E-03 | .34622E+09 | 1.89E+00 |
| 463 | 45 | 147.109 | .89508E-01 | .55049E-01 | .55176E+11 | 1.63E+00 |
| 462 | 45 | 155.138 | .30200E-01 | .26763E-01 | .16739E+11 | 1.13E+00 |
| 461 | 45 | 158.738 | .85851E-02 | .70962E-02 | .45451E+10 | 1.21E+00 |
| 460 | 45 | 191.230 | .31816E-03 | .35118E-03 | .11606E+09 | 9.06E-01 |
| 459 | 45 | 215.935 | .11505E-04 | .23751E-04 | .32916E+07 | 4.84E-01 |
| 458 | 45 | 218.887 | .10375E-03 | .21232E-03 | .28889E+08 | 4.89E-01 |
| 45 | 457 | 291.020 | .58814E-02 | .65551E-02 | .23160E+09 | 8.97E-01 |
| 45 | 456 | 267.729 | .10893E-01 | .11785E-01 | .50683E+09 | 9.24E-01 |
| 45 | 455 | 262.887 | .22227E-01 | .21786E-01 | .10726E+10 | 1.02E+00 |
| 45 | 454 | 247.715 | .10127E+00 | .73830E-01 | .55041E+10 | 1.37E+00 |
| 45 | 453 | 230.679 | .69022E-02 | .62531E-02 | .43259E+09 | 1.10E+00 |
| 45 | 452 | 222.200 | .28066E-02 | .26592E-02 | .18958E+09 | 1.06E+00 |
| 45 | 451 | 207.875 | .95456E-01 | .65459E-01 | .73672E+10 | 1.46E+00 |
| 45 | 450 | 195.257 | .13595E-03 | .96520E-04 | .11893E+08 | 1.41E+00 |
| 45 | 449 | 178.900 | .28620E-03 | .29658E-03 | .29823E+08 | 9.65E-01 |
| 45 | 448 | 162.981 | .67979E-03 | .55601E-03 | .85350E+08 | 1.22E+00 |
| 45 | 447 | 153.964 | .71734E-03 | .55291E-03 | .10092E+09 | 1.30E+00 |
| 45 | 446 | 80.973 | .80994E-06 | .18129E-05 | .41198E+06 | 4.47E-01 |
| 468 | 44 | 55.750 | .71611E-05 | .21160E-05 | .30737E+08 | 3.38E+00 |
| 467 | 44 | 99.659 | .24783E-03 | .13593E-05 | .33288E+09 | 1.82E+02 |
| 466 | 44 | 102.834 | .78107E-04 | .50240E-05 | .98533E+08 | 1.55E+01 |
| 465 | 44 | 108.749 | .91917E-04 | .16577E-04 | .10368E+09 | 5.54E+00 |
| 464 | 44 | 113.419 | .77073E-04 | .22278E-05 | .79926E+08 | 3.46E+01 |
| 463 | 44 | 117.994 | .79938E-04 | .21590E-04 | .76594E+08 | 3.70E+00 |
| 462 | 44 | 123.104 | .19685E-04 | .75119E-06 | .17328E+08 | 2.62E+01 |
| 461 | 44 | 125.360 | .56312E-03 | .32973E-03 | .47802E+09 | 1.71E+00 |
| 460 | 44 | 144.788 | .11841E+00 | .69601E-01 | .75350E+11 | 1.70E+00 |
| 459 | 44 | 158.520 | .19184E-01 | .15772E-01 | .10184E+11 | 1.22E+00 |
| 458 | 44 | 160.105 | .29767E-05 | .61862E-03 | .15491E+07 | 4.81E-03 |
| 44 | 457 | 568.550 | .48297E-02 | .22169E-02 | .49829E+08 | 2.18E+00 |
| 44 | 456 | 485.957 | .82290E-03 | .19071E-02 | .11621E+08 | 4.31E-01 |
| 44 | 455 | 470.237 | .16141E-02 | .16566E-03 | .24344E+08 | 9.74E+00 |
| 44 | 454 | 423.808 | .38852E-03 | .66523E-03 | .72140E+07 | 5.84E-01 |
| 44 | 453 | 376.265 | .78197E-04 | .64076E-04 | .18421E+07 | 1.22E+00 |
| 44 | 452 | 354.219 | .60394E-03 | .36629E-03 | .16053E+08 | 1.65E+00 |
| 44 | 451 | 319.156 | .27432E-03 | .41996E-03 | .89814E+07 | 6.53E-01 |
| 44 | 450 | 290.351 | .28176E-02 | .58517E-02 | .11146E+09 | 4.82E-01 |
| 44 | 449 | 255.599 | .28000E-02 | .14309E-02 | .14294E+09 | 1.96E+00 |
| 44 | 448 | 224.298 | .36717E+00 | .36301E+00 | .24340E+11 | 1.01E+00 |
| 44 | 447 | 207.569 | .76977E-01 | .60622E-01 | .59586E+10 | 1.27E+00 |
| 44 | 446 | 93.700 | .24645E-03 | .20994E-03 | .93617E+08 | 1.17E+00 |
| 468 | 43 | 55.343 | .24277E-03 | .11598E-04 | .10574E+10 | 2.09E+01 |
| 467 | 43 | 98.366 | .64107E-04 | .53471E-04 | .88384E+08 | 1.20E+00 |
| 466 | 43 | 101.458 | .13404E-04 | .18352E-05 | .17371E+08 | 7.30E+00 |
| 465 | 43 | 107.212 | .35326E-04 | .17183E-04 | .40999E+08 | 2.06E+00 |
| 464 | 43 | 111.748 | .20976E-05 | .24870E-03 | .22408E+07 | 8.43E-03 |
| 463 | 43 | 116.186 | .12399E-02 | .43154E-03 | .12253E+10 | 2.87E+00 |
| 462 | 43 | 121.138 | .17183E-04 | .64821E-05 | .15621E+08 | 2.65E+00 |
| 461 | 43 | 123.321 | .60424E-04 | .44352E-08 | .53002E+08 | 1.36E+04 |
| 460 | 43 | 142.076 | .63454E-02 | .15617E-02 | .41936E+10 | 4.06E+00 |
| 459 | 43 | 155.274 | .10388E+00 | .59851E-01 | .57475E+11 | 1.74E+00 |
| 458 | 43 | 156.795 | .86877E-01 | .50759E-01 | .47142E+11 | 1.71E+00 |
| 43 | 457 | 614.629 | .84126E-03 | .11055E-03 | .74269E+07 | 7.61E+00 |
| 43 | 456 | 519.230 | .15743E-03 | .18964E-03 | .19475E+07 | 8.30E-01 |
| 43 | 455 | 501.322 | .22479E-03 | .38785E-04 | .29829E+07 | 5.80E+00 |
| 43 | 454 | 448.894 | .22547E-03 | .14872E-04 | .37316E+07 | 1.52E+01 |
| 43 | 453 | 395.908 | .11900E-03 | .59685E-05 | .25320E+07 | 1.99E+01 |
| 43 | 452 | 371.575 | .20372E-03 | .90156E-03 | .49209E+07 | 2.26E-01 |
| 43 | 451 | 333.178 | .38968E-03 | .98198E-04 | .11707E+08 | 3.97E+00 |
| 43 | 450 | 301.910 | .70219E-02 | .61385E-02 | .25692E+09 | 1.14E+00 |
| 43 | 449 | 264.515 | .13369E+00 | .80203E-01 | .63722E+10 | 1.67E+00 |
| 43 | 448 | 231.134 | .27402E-01 | .28016E-01 | .17106E+10 | 9.78E-01 |
| 43 | 447 | 213.410 | .10251E+00 | .10831E+00 | .75067E+10 | 9.46E-01 |
| 43 | 446 | 94.872 | .49948E-03 | .11358E-02 | .18507E+09 | 4.40E-01 |
| 468 | 42 | 37.457 | .10029E-08 | .98980E-09 | .95354E+04 | 1.01E+00 |
| 467 | 42 | 53.209 | .84877E-07 | .51450E-06 | .39993E+06 | 1.65E-01 |
| 466 | 42 | 54.101 | .26021E-06 | .13015E-05 | .11860E+07 | 2.00E-01 |
| 465 | 42 | 55.694 | .14694E-06 | .74635E-07 | .63195E+06 | 1.97E+00 |
| 464 | 42 | 56.894 | .50060E-07 | .12986E-06 | .20631E+06 | 3.85E-01 |
| 463 | 42 | 58.023 | .61145E-05 | .10578E-05 | .24229E+08 | 5.78E+00 |
| 462 | 42 | 59.232 | .12856E-04 | .34739E-05 | .48884E+08 | 3.70E+00 |
| 461 | 42 | 59.749 | .50907E-06 | .21994E-06 | .19023E+07 | 2.31E+00 |
| 460 | 42 | 63.831 | .97351E-06 | .80576E-08 | .31874E+07 | 1.21E+02 |
| 459 | 42 | 66.366 | .16102E-05 | .47244E-07 | .48769E+07 | 3.41E+01 |
| 458 | 42 | 66.642 | .33749E-06 | .70767E-08 | .10137E+07 | 4.77E+01 |
| 457 | 42 | 142.842 | .20618E-01 | .17976E-01 | .13481E+11 | 1.15E+00 |
| 456 | 42 | 149.213 | .58829E-02 | .53753E-02 | .35248E+10 | 1.09E+00 |
| 455 | 42 | 150.761 | .82059E-02 | .75809E-02 | .48163E+10 | 1.08E+00 |
| 454 | 42 | 156.249 | .44122E-01 | .43316E-01 | .24109E+11 | 1.02E+00 |
| 453 | 42 | 163.883 | .41495E-01 | .44993E-01 | .20611E+11 | 9.22E-01 |
| 452 | 42 | 168.449 | .19217E-01 | .21532E-01 | .90348E+10 | 8.92E-01 |
| 451 | 42 | 177.735 | .40620E-01 | .49766E-01 | .17154E+11 | 8.16E-01 |
| 450 | 42 | 188.129 | .15362E-03 | .21634E-03 | .57902E+08 | 7.10E-01 |
| 449 | 42 | 206.303 | .24720E-03 | .42269E-03 | .77481E+08 | 5.85E-01 |
| 448 | 42 | 232.490 | .26979E-03 | .30827E-03 | .66585E+08 | 8.75E-01 |
| 447 | 42 | 253.682 | .22128E-03 | .44938E-03 | .45870E+08 | 4.92E-01 |
| 42 | 446 | 522.798 | .15706E-08 | .62961E-08 | .19164E+02 | 2.49E-01 |
| 468 | 41 | 37.427 | .50514E-01 | .55808E-01 | .48105E+12 | 9.05E-01 |
| 467 | 41 | 53.148 | .17750E-06 | .48592E-06 | .83829E+06 | 3.65E-01 |
| 466 | 41 | 54.038 | .30193E-08 | .11825E-06 | .13793E+05 | 2.55E-02 |
| 465 | 41 | 55.628 | .63149E-08 | .24470E-07 | .27223E+05 | 2.58E-01 |
| 464 | 41 | 56.825 | .44017E-06 | .87802E-06 | .18185E+07 | 5.01E-01 |
| 463 | 41 | 57.951 | .10198E-04 | .21049E-04 | .40510E+08 | 4.84E-01 |
| 462 | 41 | 59.157 | .66599E-06 | .11201E-05 | .25388E+07 | 5.95E-01 |
| 461 | 41 | 59.673 | .10679E-05 | .12167E-06 | .40006E+07 | 8.78E+00 |
| 460 | 41 | 63.744 | .71130E-06 | .92281E-04 | .23352E+07 | 7.71E-03 |
| 459 | 41 | 66.272 | .66733E-04 | .25536E-03 | .20270E+09 | 2.61E-01 |
| 458 | 41 | 66.547 | .42047E-03 | .75638E-03 | .12666E+10 | 5.56E-01 |
| 457 | 41 | 142.406 | .14022E-07 | .27767E-08 | .92243E+04 | 5.05E+00 |
| 456 | 41 | 148.738 | .22753E-07 | .15530E-08 | .13720E+05 | 1.47E+01 |
| 455 | 41 | 150.275 | .13748E-07 | .62525E-08 | .81213E+04 | 2.20E+00 |
| 454 | 41 | 155.727 | .22466E-06 | .24570E-07 | .12358E+06 | 9.14E+00 |
| 453 | 41 | 163.310 | .42649E-06 | .65271E-07 | .21333E+06 | 6.53E+00 |
| 452 | 41 | 167.844 | .11529E-05 | .16007E-06 | .54594E+06 | 7.20E+00 |
| 451 | 41 | 177.061 | .12964E-07 | .27096E-08 | .55166E+04 | 4.78E+00 |
| 450 | 41 | 187.374 | .17475E-05 | .51387E-06 | .66401E+06 | 3.40E+00 |
| 449 | 41 | 205.395 | .25938E-04 | .17687E-04 | .82021E+07 | 1.47E+00 |
| 448 | 41 | 231.338 | .15435E-05 | .18230E-05 | .38476E+06 | 8.47E-01 |
| 447 | 41 | 252.311 | .15136E-04 | .21253E-04 | .31717E+07 | 7.12E-01 |
| 41 | 446 | 528.719 | .38107E+00 | .26599E+00 | .45463E+10 | 1.43E+00 |
| 468 | 40 | 37.013 | .16751E-07 | .40978E-07 | .16312E+06 | 4.09E-01 |
| 467 | 40 | 52.317 | .92241E-05 | .16266E-05 | .44958E+08 | 5.67E+00 |
| 466 | 40 | 53.179 | .18715E-05 | .23072E-06 | .88285E+07 | 8.11E+00 |
| 465 | 40 | 54.718 | .90747E-06 | .10438E-06 | .40433E+07 | 8.69E+00 |
| 464 | 40 | 55.875 | .32874E-06 | .10372E-05 | .14047E+07 | 3.17E-01 |
| 463 | 40 | 56.963 | .17720E-05 | .35896E-06 | .72853E+07 | 4.94E+00 |
| 462 | 40 | 58.128 | .18643E-05 | .26562E-06 | .73604E+07 | 7.02E+00 |
| 461 | 40 | 58.626 | .13868E-06 | .58231E-07 | .53825E+06 | 2.38E+00 |
| 460 | 40 | 62.552 | .86975E-04 | .26689E-05 | .29654E+09 | 3.26E+01 |
| 459 | 40 | 64.984 | .13375E-03 | .11619E-04 | .42253E+09 | 1.15E+01 |
| 458 | 40 | 65.248 | .34862E-04 | .61391E-05 | .10924E+09 | 5.68E+00 |
| 457 | 40 | 136.589 | .49669E-01 | .41671E-01 | .35516E+11 | 1.19E+00 |
| 456 | 40 | 142.403 | .14316E-03 | .84322E-04 | .94177E+08 | 1.70E+00 |
| 455 | 40 | 143.812 | .40459E-01 | .36253E-01 | .26097E+11 | 1.12E+00 |
| 454 | 40 | 148.797 | .11686E-02 | .11398E-02 | .70410E+09 | 1.03E+00 |
| 453 | 40 | 155.705 | .51250E-02 | .52380E-02 | .28201E+10 | 9.78E-01 |
| 452 | 40 | 159.821 | .35750E-02 | .34756E-02 | .18671E+10 | 1.03E+00 |
| 451 | 40 | 168.156 | .35043E-03 | .45126E-03 | .16533E+09 | 7.77E-01 |
| 450 | 40 | 177.431 | .24544E-02 | .35711E-02 | .10400E+10 | 6.87E-01 |
| 449 | 40 | 193.508 | .13594E-01 | .17285E-01 | .48431E+10 | 7.86E-01 |
| 448 | 40 | 216.368 | .32779E-01 | .41750E-01 | .93405E+10 | 7.85E-01 |
| 447 | 40 | 234.608 | .27304E-02 | .46734E-02 | .66176E+09 | 5.84E-01 |
| 40 | 446 | 628.028 | .39483E-07 | .64934E-08 | .33385E+03 | 6.08E+00 |
| 370 | 134 | 284.669 | .67478E-04 | .25158E-04 | .27770E+07 | 2.68E+00 |
| 134 | 369 | 167.929 | .24502E-03 | .16240E-03 | .11591E+09 | 1.51E+00 |
| 134 | 368 | 153.523 | .39820E-02 | .41490E-02 | .22538E+10 | 9.60E-01 |
| 134 | 367 | 146.123 | .71623E-01 | .56733E-01 | .44749E+11 | 1.26E+00 |
| 134 | 366 | 119.979 | .69049E-04 | .15333E-04 | .63990E+08 | 4.50E+00 |
| 134 | 365 | 116.186 | .27415E-05 | .20493E-05 | .27092E+07 | 1.34E+00 |
| 134 | 364 | 59.147 | .74548E-05 | .52457E-08 | .28427E+08 | 1.42E+03 |
| 134 | 363 | 59.022 | .70322E-06 | .27952E-08 | .26929E+07 | 2.52E+02 |
| 134 | 362 | 57.257 | .13806E-04 | .73327E-06 | .56178E+08 | 1.88E+01 |
| 134 | 361 | 56.651 | .70314E-07 | .20733E-06 | .29228E+06 | 3.39E-01 |
| 134 | 360 | 55.532 | .67190E-08 | .66916E-10 | .29066E+05 | 1.00E+02 |
| 134 | 359 | 53.974 | .41836E-06 | .12282E-07 | .19158E+07 | 3.41E+01 |
| 134 | 358 | 50.751 | .28124E-06 | .10459E-09 | .14566E+07 | 2.69E+03 |
| 134 | 357 | 39.312 | .92495E-05 | .11729E-04 | .79843E+08 | 7.89E-01 |
| 370 | 133 | 220.325 | .79834E-01 | .48698E-01 | .54848E+10 | 1.64E+00 |
| 133 | 369 | 202.881 | .67253E-01 | .54361E-01 | .21797E+11 | 1.24E+00 |
| 133 | 368 | 182.223 | .11309E-04 | .50043E-04 | .45432E+07 | 2.26E-01 |
| 133 | 367 | 171.890 | .39849E-03 | .43453E-03 | .17992E+09 | 9.17E-01 |
| 133 | 366 | 136.820 | .89299E-03 | .29930E-02 | .63638E+09 | 2.98E-01 |
| 133 | 365 | 131.909 | .16213E-03 | .41857E-04 | .12431E+09 | 3.87E+00 |
| 133 | 364 | 62.968 | .55397E-06 | .34169E-07 | .18639E+07 | 1.62E+01 |
| 133 | 363 | 62.826 | .52970E-06 | .36427E-06 | .17902E+07 | 1.45E+00 |
| 133 | 362 | 60.831 | .40560E-06 | .10777E-08 | .14622E+07 | 3.76E+02 |
| 133 | 361 | 60.146 | .71259E-06 | .66318E-08 | .26277E+07 | 1.07E+02 |
| 133 | 360 | 58.887 | .19307E-04 | .21353E-06 | .74277E+08 | 9.04E+01 |
| 133 | 359 | 57.138 | .19400E-03 | .15414E-05 | .79273E+09 | 1.26E+02 |
| 133 | 358 | 53.539 | .11987E-03 | .14562E-05 | .55788E+09 | 8.23E+01 |
| 133 | 357 | 40.964 | .27368E-02 | .39913E-02 | .21757E+11 | 6.86E-01 |
| 370 | 132 | 167.497 | .13833E+00 | .10879E+00 | .16444E+11 | 1.27E+00 |
| 132 | 369 | 285.921 | .50098E-02 | .50721E-02 | .81751E+09 | 9.88E-01 |
| 132 | 368 | 246.534 | .16883E-04 | .28416E-04 | .37057E+07 | 5.94E-01 |
| 132 | 367 | 227.991 | .10666E-03 | .25817E-03 | .27373E+08 | 4.13E-01 |
| 132 | 366 | 170.144 | .11711E-01 | .11600E-01 | .53964E+10 | 1.01E+00 |
| 132 | 365 | 162.616 | .42127E-01 | .35583E-01 | .21252E+11 | 1.18E+00 |
| 132 | 364 | 69.206 | .68548E-05 | .64383E-06 | .19093E+08 | 1.06E+01 |
| 132 | 363 | 69.035 | .10151E-04 | .68467E-07 | .28413E+08 | 1.48E+02 |
| 132 | 362 | 66.633 | .14837E-04 | .22897E-06 | .44578E+08 | 6.48E+01 |
| 132 | 361 | 65.813 | .33675E-05 | .23578E-07 | .10372E+08 | 1.43E+02 |
| 132 | 360 | 64.308 | .11642E-03 | .13768E-05 | .37553E+09 | 8.46E+01 |
| 132 | 359 | 62.228 | .12807E-04 | .31215E-04 | .44122E+08 | 4.10E-01 |
| 132 | 358 | 57.983 | .45825E-04 | .86340E-05 | .18183E+09 | 5.31E+00 |
| 132 | 357 | 43.515 | .95346E-03 | .48422E-03 | .67171E+10 | 1.97E+00 |
| 370 | 131 | 146.573 | .35033E+00 | .29867E+00 | .54384E+11 | 1.17E+00 |
| 131 | 369 | 378.042 | .13235E-04 | .19285E-03 | .12354E+07 | 6.86E-02 |
| 131 | 368 | 312.112 | .10910E-02 | .13854E-02 | .14941E+09 | 7.87E-01 |
| 131 | 367 | 282.975 | .79826E-06 | .13774E-04 | .13299E+06 | 5.80E-02 |
| 131 | 366 | 199.001 | .25737E-02 | .14265E-03 | .86698E+09 | 1.80E+01 |
| 131 | 365 | 188.779 | .12822E+00 | .66512E-01 | .47997E+11 | 1.93E+00 |
| 131 | 364 | 73.544 | .53673E-07 | .42848E-08 | .13238E+06 | 1.25E+01 |
| 131 | 363 | 73.351 | .10023E-05 | .30364E-07 | .24852E+07 | 3.30E+01 |
| 131 | 362 | 70.645 | .77801E-08 | .27229E-08 | .20796E+05 | 2.86E+00 |
| 131 | 361 | 69.724 | .26035E-06 | .94264E-10 | .71443E+06 | 2.76E+03 |
| 131 | 360 | 68.036 | .68563E-06 | .28701E-08 | .19759E+07 | 2.39E+02 |
| 131 | 359 | 65.713 | .18325E-06 | .21426E-04 | .56611E+06 | 8.55E-03 |
| 131 | 358 | 60.997 | .91863E-04 | .22260E-04 | .32937E+09 | 4.13E+00 |
| 131 | 357 | 45.191 | .24511E-01 | .22251E-01 | .16011E+12 | 1.10E+00 |
| 370 | 130 | 71.756 | .46318E-06 | .49073E-05 | .30001E+06 | 9.44E-02 |
| 369 | 130 | 223.800 | .72289E-03 | .15033E-02 | .48134E+08 | 4.81E-01 |
| 368 | 130 | 255.786 | .95589E-01 | .10663E+00 | .48725E+10 | 8.96E-01 |
| 367 | 130 | 279.360 | .73043E-03 | .92054E-03 | .31214E+08 | 7.93E-01 |
| 366 | 130 | 478.837 | .10661E-04 | .14631E-03 | .15507E+06 | 7.29E-02 |
| 365 | 130 | 550.572 | .38471E-03 | .61910E-03 | .42326E+07 | 6.21E-01 |
| 130 | 364 | 154.228 | .13939E-01 | .80324E-02 | .78177E+10 | 1.74E+00 |
| 130 | 363 | 153.383 | .41817E-01 | .25887E-01 | .23712E+11 | 1.62E+00 |
| 130 | 362 | 142.009 | .76790E-03 | .39204E-03 | .50797E+09 | 1.96E+00 |
| 130 | 361 | 138.335 | .34335E-04 | .13841E-03 | .23935E+08 | 2.48E-01 |
| 130 | 360 | 131.848 | .50040E-03 | .24025E-03 | .38400E+09 | 2.08E+00 |
| 130 | 359 | 123.392 | .15774E-02 | .46651E-03 | .13820E+10 | 3.38E+00 |
| 130 | 358 | 107.750 | .40791E-04 | .37284E-04 | .46870E+08 | 1.09E+00 |
| 130 | 357 | 66.602 | .60322E-06 | .53324E-09 | .18141E+07 | 1.13E+03 |
| 370 | 129 | 69.974 | .59445E-05 | .17257E-09 | .40490E+07 | 3.44E+04 |
| 369 | 129 | 207.327 | .37745E-02 | .34545E-02 | .29286E+09 | 1.09E+00 |
| 368 | 129 | 234.492 | .37939E-03 | .34480E-03 | .23011E+08 | 1.10E+00 |
| 367 | 129 | 254.153 | .17103E-01 | .20180E-01 | .88304E+09 | 8.48E-01 |
| 366 | 129 | 409.263 | .65028E-03 | .88643E-03 | .12948E+08 | 7.34E-01 |
| 365 | 129 | 460.550 | .44012E-02 | .84204E-02 | .69202E+08 | 5.23E-01 |
| 129 | 364 | 163.162 | .23997E-01 | .17111E-01 | .12025E+11 | 1.40E+00 |
| 129 | 363 | 162.216 | .53983E-01 | .42524E-01 | .27367E+11 | 1.27E+00 |
| 129 | 362 | 149.549 | .10110E-01 | .75089E-02 | .60304E+10 | 1.35E+00 |
| 129 | 361 | 145.480 | .15659E-01 | .13334E-01 | .98700E+10 | 1.17E+00 |
| 129 | 360 | 138.322 | .88098E-02 | .60426E-02 | .61425E+10 | 1.46E+00 |
| 129 | 359 | 129.045 | .14244E-02 | .19451E-02 | .11410E+10 | 7.32E-01 |
| 129 | 358 | 112.035 | .14049E-04 | .13195E-06 | .14932E+08 | 1.06E+02 |
| 129 | 357 | 68.215 | .10718E-05 | .14925E-10 | .30725E+07 | 7.18E+04 |
| 370 | 128 | 68.706 | .20989E-06 | .24254E-08 | .14829E+06 | 8.65E+01 |
| 369 | 128 | 196.578 | .54945E-05 | .60334E-04 | .47420E+06 | 9.11E-02 |
| 368 | 128 | 220.836 | .70191E-02 | .60741E-02 | .48000E+09 | 1.16E+00 |
| 367 | 128 | 238.189 | .31360E-01 | .31735E-01 | .18435E+10 | 9.88E-01 |
| 366 | 128 | 369.394 | .35811E-05 | .22660E-03 | .87528E+05 | 1.58E-02 |
| 365 | 128 | 410.671 | .14225E-05 | .16991E-03 | .28130E+05 | 8.37E-03 |
| 128 | 364 | 170.499 | .59311E-02 | .57470E-02 | .27218E+10 | 1.03E+00 |
| 128 | 363 | 169.466 | .30477E-03 | .24821E-03 | .14157E+09 | 1.23E+00 |
| 128 | 362 | 155.689 | .58533E-01 | .39718E-01 | .32214E+11 | 1.47E+00 |
| 128 | 361 | 151.284 | .27158E-01 | .23118E-01 | .15830E+11 | 1.17E+00 |
| 128 | 360 | 143.559 | .85586E-02 | .65089E-02 | .55399E+10 | 1.31E+00 |
| 128 | 359 | 133.591 | .61759E-03 | .46388E-03 | .46164E+09 | 1.33E+00 |
| 128 | 358 | 115.446 | .29589E-03 | .12291E-03 | .29616E+09 | 2.41E+00 |
| 128 | 357 | 69.465 | .63529E-06 | .17787E-06 | .17563E+07 | 3.57E+00 |
| 370 | 127 | 66.361 | .21269E-04 | .19742E-05 | .16107E+08 | 1.08E+01 |
| 369 | 127 | 178.530 | .19982E-01 | .20890E-01 | .20909E+10 | 9.57E-01 |
| 368 | 127 | 198.313 | .33009E-01 | .24794E-01 | .27992E+10 | 1.33E+00 |
| 367 | 127 | 212.196 | .13470E-01 | .12947E-01 | .99766E+09 | 1.04E+00 |
| 366 | 127 | 310.423 | .44813E-02 | .65053E-02 | .15509E+09 | 6.89E-01 |
| 365 | 127 | 339.062 | .87877E-02 | .10263E-01 | .25493E+09 | 8.56E-01 |
| 127 | 364 | 186.886 | .19322E-03 | .38379E-03 | .73801E+08 | 5.03E-01 |
| 127 | 363 | 185.645 | .40656E-02 | .44328E-02 | .15737E+10 | 9.17E-01 |
| 127 | 362 | 169.240 | .81859E-03 | .48603E-03 | .38126E+09 | 1.68E+00 |
| 127 | 361 | 164.047 | .66082E-02 | .51339E-02 | .32757E+10 | 1.29E+00 |
| 127 | 360 | 155.003 | .39033E-02 | .51041E-02 | .21673E+10 | 7.65E-01 |
| 127 | 359 | 143.446 | .52838E-02 | .47657E-02 | .34255E+10 | 1.11E+00 |
| 127 | 358 | 122.733 | .22253E-03 | .23445E-03 | .19708E+09 | 9.49E-01 |
| 127 | 357 | 72.038 | .11868E-04 | .16654E-05 | .30509E+08 | 7.13E+00 |
| 370 | 126 | 65.264 | .55665E-04 | .53053E-05 | .43585E+08 | 1.05E+01 |
| 369 | 126 | 170.807 | .52892E-01 | .48155E-01 | .60462E+10 | 1.10E+00 |
| 368 | 126 | 188.829 | .23523E-01 | .18444E-01 | .22002E+10 | 1.28E+00 |
| 367 | 126 | 201.373 | .36741E-01 | .29629E-01 | .30217E+10 | 1.24E+00 |
| 366 | 126 | 287.796 | .48368E-02 | .60675E-02 | .19476E+09 | 7.97E-01 |
| 365 | 126 | 312.248 | .93565E-02 | .96460E-02 | .32005E+09 | 9.70E-01 |
| 126 | 364 | 196.171 | .44519E-04 | .10041E-04 | .15433E+08 | 4.43E+00 |
| 126 | 363 | 194.805 | .40881E-04 | .46164E-05 | .14371E+08 | 8.86E+00 |
| 126 | 362 | 176.819 | .53494E-02 | .58742E-02 | .22825E+10 | 9.11E-01 |
| 126 | 361 | 171.159 | .75993E-02 | .61977E-02 | .34605E+10 | 1.23E+00 |
| 126 | 360 | 161.337 | .92607E-02 | .95776E-02 | .47461E+10 | 9.67E-01 |
| 126 | 359 | 148.854 | .29682E-02 | .35030E-02 | .17870E+10 | 8.47E-01 |
| 126 | 358 | 126.671 | .21627E-02 | .11417E-02 | .17980E+10 | 1.89E+00 |
| 126 | 357 | 73.377 | .11841E-04 | .33391E-05 | .29337E+08 | 3.55E+00 |
| 370 | 125 | 63.772 | .33668E-03 | .15544E-04 | .27610E+09 | 2.17E+01 |
| 369 | 125 | 160.948 | .67335E-01 | .55350E-01 | .86690E+10 | 1.22E+00 |
| 368 | 125 | 176.853 | .65953E-04 | .81646E-05 | .70325E+07 | 8.08E+00 |
| 367 | 125 | 187.811 | .79611E-03 | .61936E-03 | .75272E+08 | 1.29E+00 |
| 366 | 125 | 260.873 | .24295E-01 | .29438E-01 | .11906E+10 | 8.25E-01 |
| 365 | 125 | 280.805 | .19324E-01 | .16998E-01 | .81733E+09 | 1.14E+00 |
| 125 | 364 | 211.015 | .51750E-03 | .46254E-03 | .15504E+09 | 1.12E+00 |
| 125 | 363 | 209.435 | .17034E-02 | .24506E-02 | .51806E+09 | 6.95E-01 |
| 125 | 362 | 188.789 | .57259E-03 | .79484E-03 | .21432E+09 | 7.20E-01 |
| 125 | 361 | 182.351 | .47661E-03 | .71545E-04 | .19121E+09 | 6.66E+00 |
| 125 | 360 | 171.244 | .48826E-01 | .26584E-01 | .22212E+11 | 1.84E+00 |
| 125 | 359 | 157.248 | .46838E-01 | .24294E-01 | .25269E+11 | 1.93E+00 |
| 125 | 358 | 132.698 | .10433E-01 | .34859E-02 | .79040E+10 | 2.99E+00 |
| 125 | 357 | 75.360 | .31202E-04 | .26641E-05 | .73293E+08 | 1.17E+01 |
| 370 | 124 | 62.604 | .85013E-05 | .68544E-05 | .72341E+07 | 1.24E+00 |
| 369 | 124 | 153.710 | .55863E-01 | .46721E-01 | .78854E+10 | 1.20E+00 |
| 368 | 124 | 168.153 | .26400E-02 | .21265E-02 | .31139E+09 | 1.24E+00 |
| 367 | 124 | 178.029 | .15828E-02 | .11064E-02 | .16655E+09 | 1.43E+00 |
| 366 | 124 | 242.374 | .29632E-01 | .24827E-01 | .16822E+10 | 1.19E+00 |
| 365 | 124 | 259.487 | .40389E-01 | .35445E-01 | .20005E+10 | 1.14E+00 |
| 124 | 364 | 224.900 | .15645E-02 | .15246E-02 | .41264E+09 | 1.03E+00 |
| 124 | 363 | 223.105 | .79995E-03 | .69968E-03 | .21439E+09 | 1.14E+00 |
| 124 | 362 | 199.826 | .21864E-02 | .22510E-02 | .73044E+09 | 9.71E-01 |
| 124 | 361 | 192.627 | .90984E-05 | .10782E-05 | .32711E+07 | 8.44E+00 |
| 124 | 360 | 180.276 | .54495E-02 | .69185E-02 | .22369E+10 | 7.88E-01 |
| 124 | 359 | 164.831 | .24180E-01 | .18525E-01 | .11873E+11 | 1.31E+00 |
| 124 | 358 | 138.058 | .57176E-02 | .41861E-02 | .40018E+10 | 1.37E+00 |
| 124 | 357 | 77.059 | .14090E-04 | .16864E-07 | .31655E+08 | 8.36E+02 |
| 370 | 123 | 61.378 | .41925E-05 | .77198E-06 | .37115E+07 | 5.43E+00 |
| 369 | 123 | 146.528 | .51004E-04 | .39212E-04 | .79226E+07 | 1.30E+00 |
| 368 | 123 | 159.595 | .25091E+00 | .23845E+00 | .32854E+11 | 1.05E+00 |
| 367 | 123 | 168.465 | .21354E-01 | .22644E-01 | .25094E+10 | 9.43E-01 |
| 366 | 123 | 224.985 | .17490E-03 | .58549E-03 | .11523E+08 | 2.99E-01 |
| 365 | 123 | 239.656 | .46059E-03 | .16193E-02 | .26745E+08 | 2.84E-01 |
| 123 | 364 | 242.276 | .31660E-02 | .12960E-02 | .71954E+09 | 2.44E+00 |
| 123 | 363 | 240.195 | .21402E-02 | .89993E-03 | .49486E+09 | 2.38E+00 |
| 123 | 362 | 213.427 | .89140E-03 | .43799E-03 | .26106E+09 | 2.04E+00 |
| 123 | 361 | 205.235 | .65797E-01 | .34384E-01 | .20839E+11 | 1.91E+00 |
| 123 | 360 | 191.272 | .27987E-02 | .15487E-02 | .10205E+10 | 1.81E+00 |
| 123 | 359 | 173.976 | .37058E-04 | .27727E-04 | .16333E+08 | 1.34E+00 |
| 123 | 358 | 144.416 | .29695E-04 | .94051E-05 | .18994E+08 | 3.16E+00 |
| 123 | 357 | 79.000 | .29738E-08 | .13399E-06 | .63566E+04 | 2.22E-02 |
| 370 | 122 | 59.423 | .33210E-07 | .47731E-07 | .31366E+05 | 6.96E-01 |
| 369 | 122 | 135.855 | .12923E-03 | .10366E-03 | .23351E+08 | 1.25E+00 |
| 368 | 122 | 147.015 | .11013E-02 | .15084E-02 | .16994E+09 | 7.30E-01 |
| 367 | 122 | 154.509 | .20874E+00 | .15946E+00 | .29161E+11 | 1.31E+00 |
| 366 | 122 | 200.767 | .32487E-03 | .30322E-03 | .26880E+08 | 1.07E+00 |
| 365 | 122 | 212.368 | .16657E-05 | .13990E-04 | .12317E+06 | 1.19E-01 |
| 122 | 364 | 278.444 | .26236E-01 | .21995E-01 | .45143E+10 | 1.19E+00 |
| 122 | 363 | 275.699 | .13650E-02 | .63047E-03 | .23956E+09 | 2.17E+00 |
| 122 | 362 | 241.004 | .29699E-01 | .20483E-01 | .68212E+10 | 1.45E+00 |
| 122 | 361 | 230.610 | .82953E-05 | .40718E-04 | .20808E+07 | 2.04E-01 |
| 122 | 360 | 213.128 | .47401E-04 | .54911E-04 | .13921E+08 | 8.63E-01 |
| 122 | 359 | 191.873 | .72349E-05 | .33765E-05 | .26216E+07 | 2.14E+00 |
| 122 | 358 | 156.537 | .98885E-04 | .57621E-04 | .53834E+08 | 1.72E+00 |
| 122 | 357 | 82.494 | .14344E-07 | .94318E-07 | .28118E+05 | 1.52E-01 |
| 370 | 121 | 58.976 | .50194E-04 | .24141E-04 | .48128E+08 | 2.08E+00 |
| 369 | 121 | 133.543 | .19269E+00 | .15304E+00 | .36035E+11 | 1.26E+00 |
| 368 | 121 | 144.311 | .15162E-02 | .86680E-03 | .24280E+09 | 1.75E+00 |
| 367 | 121 | 151.525 | .17834E-04 | .10336E-04 | .25905E+07 | 1.73E+00 |
| 366 | 121 | 195.758 | .32547E+00 | .33264E+00 | .28326E+11 | 9.78E-01 |
| 365 | 121 | 206.771 | .65704E-01 | .60871E-01 | .51252E+10 | 1.08E+00 |
| 121 | 364 | 288.690 | .30386E-06 | .63323E-07 | .48638E+05 | 4.80E+00 |
| 121 | 363 | 285.741 | .13457E-06 | .26761E-04 | .21987E+05 | 5.03E-03 |
| 121 | 362 | 248.643 | .77703E-04 | .22391E-04 | .16767E+08 | 3.47E+00 |
| 121 | 361 | 237.594 | .20258E-03 | .29445E-04 | .47872E+08 | 6.88E+00 |
| 121 | 360 | 219.080 | .12753E-03 | .10120E-02 | .35446E+08 | 1.26E-01 |
| 121 | 359 | 196.684 | .75664E-04 | .11477E-03 | .26093E+08 | 6.59E-01 |
| 121 | 358 | 159.724 | .10059E+00 | .70125E-01 | .52601E+11 | 1.43E+00 |
| 121 | 357 | 83.371 | .23042E-05 | .38382E-05 | .44224E+07 | 6.00E-01 |
| 370 | 120 | 57.244 | .16402E-03 | .92699E-07 | .16693E+09 | 1.77E+03 |
| 369 | 120 | 124.980 | .33825E-02 | .23003E-02 | .72219E+09 | 1.47E+00 |
| 368 | 120 | 134.363 | .16162E-02 | .63380E-03 | .29857E+09 | 2.55E+00 |
| 367 | 120 | 140.596 | .14013E-03 | .49575E-04 | .23642E+08 | 2.83E+00 |
| 366 | 120 | 177.892 | .67617E-01 | .67465E-01 | .71260E+10 | 1.00E+00 |
| 365 | 120 | 186.941 | .16095E+00 | .14426E+00 | .15359E+11 | 1.12E+00 |
| 120 | 364 | 338.880 | .50031E-04 | .26920E-04 | .58118E+07 | 1.86E+00 |
| 120 | 363 | 334.823 | .10225E-03 | .20134E-04 | .12167E+08 | 5.08E+00 |
| 120 | 362 | 284.996 | .75046E-04 | .25922E-06 | .12326E+08 | 2.90E+02 |
| 120 | 361 | 270.575 | .58888E-03 | .24226E-03 | .10730E+09 | 2.43E+00 |
| 120 | 360 | 246.821 | .22354E-01 | .76815E-02 | .48950E+10 | 2.91E+00 |
| 120 | 359 | 218.757 | .43228E-01 | .18601E-01 | .12050E+11 | 2.32E+00 |
| 120 | 358 | 173.980 | .82845E-02 | .55747E-02 | .36512E+10 | 1.49E+00 |
| 120 | 357 | 87.096 | .40942E-05 | .62600E-04 | .72000E+07 | 6.54E-02 |
| 370 | 119 | 53.823 | .53179E-04 | .17071E-08 | .61221E+08 | 3.12E+04 |
| 369 | 119 | 109.751 | .12043E-03 | .36637E-04 | .33343E+08 | 3.29E+00 |
| 368 | 119 | 116.921 | .38268E-03 | .17453E-03 | .93359E+08 | 2.19E+00 |
| 367 | 119 | 121.612 | .33388E-04 | .88117E-05 | .75291E+07 | 3.79E+00 |
| 366 | 119 | 148.552 | .10791E-01 | .94088E-02 | .16308E+10 | 1.15E+00 |
| 365 | 119 | 154.809 | .11612E+00 | .88535E-01 | .16158E+11 | 1.31E+00 |
| 119 | 364 | 543.293 | .16294E-02 | .24892E-04 | .73641E+08 | 6.55E+01 |
| 119 | 363 | 532.939 | .17503E-02 | .62104E-04 | .82208E+08 | 2.82E+01 |
| 119 | 362 | 416.919 | .33687E-02 | .16617E-03 | .25853E+09 | 2.03E+01 |
| 119 | 361 | 386.762 | .34138E-03 | .17752E-03 | .30445E+08 | 1.92E+00 |
| 119 | 360 | 339.991 | .15697E-02 | .47734E-03 | .18116E+09 | 3.29E+00 |
| 119 | 359 | 288.932 | .90834E-03 | .44999E-03 | .14515E+09 | 2.02E+00 |
| 119 | 358 | 215.632 | .39944E-01 | .29407E-01 | .11460E+11 | 1.36E+00 |
| 119 | 357 | 96.420 | .64884E-05 | .12726E-05 | .93104E+07 | 5.10E+00 |
| 370 | 118 | 37.412 | .73164E-08 | .11377E-09 | .17433E+05 | 6.43E+01 |
| 369 | 118 | 57.932 | .33788E-07 | .18620E-07 | .33576E+05 | 1.81E+00 |
| 368 | 118 | 59.870 | .17592E-04 | .38489E-05 | .16368E+08 | 4.57E+00 |
| 367 | 118 | 61.077 | .19188E-05 | .43241E-06 | .17154E+07 | 4.44E+00 |
| 366 | 118 | 67.197 | .21083E-07 | .88646E-08 | .15572E+05 | 2.38E+00 |
| 365 | 118 | 68.449 | .19992E-06 | .13273E-06 | .14231E+06 | 1.51E+00 |
| 364 | 118 | 158.495 | .33590E-01 | .33259E-01 | .44594E+10 | 1.01E+00 |
| 363 | 118 | 159.399 | .24213E-01 | .24149E-01 | .31782E+10 | 1.00E+00 |
| 362 | 118 | 173.870 | .55044E-03 | .67414E-03 | .60725E+08 | 8.17E-01 |
| 361 | 118 | 179.714 | .19075E+00 | .23641E+00 | .19697E+11 | 8.07E-01 |
| 360 | 118 | 191.986 | .60597E-02 | .82528E-02 | .54829E+09 | 7.34E-01 |
| 359 | 118 | 213.267 | .26419E-04 | .59404E-04 | .19372E+07 | 4.45E-01 |
| 358 | 118 | 284.702 | .12156E-03 | .30918E-03 | .50018E+07 | 3.93E-01 |
| 118 | 357 | 450.178 | .16621E-08 | .22094E-10 | .10941E+03 | 7.52E+01 |
| 370 | 117 | 36.698 | .15207E-07 | .23489E-08 | .37658E+05 | 6.47E+00 |
| 369 | 117 | 56.239 | .18497E-07 | .81822E-07 | .19505E+05 | 2.26E-01 |
| 368 | 117 | 58.063 | .22742E-04 | .42372E-05 | .22497E+08 | 5.37E+00 |
| 367 | 117 | 59.197 | .50018E-05 | .16861E-05 | .47602E+07 | 2.97E+00 |
| 366 | 117 | 64.929 | .12082E-06 | .37642E-09 | .95578E+05 | 3.21E+02 |
| 365 | 117 | 66.096 | .15024E-04 | .22847E-05 | .11469E+08 | 6.58E+00 |
| 364 | 117 | 146.429 | .14242E-01 | .12746E-01 | .22152E+10 | 1.12E+00 |
| 363 | 117 | 147.200 | .51334E-01 | .45953E-01 | .79012E+10 | 1.12E+00 |
| 362 | 117 | 159.456 | .37510E-01 | .38140E-01 | .49201E+10 | 9.83E-01 |
| 361 | 117 | 164.357 | .16394E-01 | .18394E-01 | .20240E+10 | 8.91E-01 |
| 360 | 117 | 174.562 | .73653E-03 | .83169E-03 | .80610E+08 | 8.86E-01 |
| 359 | 117 | 191.981 | .90603E-03 | .11614E-02 | .81984E+08 | 7.80E-01 |
| 358 | 117 | 247.995 | .88473E-03 | .14762E-02 | .47976E+08 | 5.99E-01 |
| 117 | 357 | 587.736 | .45382E-09 | .11974E-08 | .17526E+02 | 3.79E-01 |
| 370 | 116 | 36.366 | .25107E-06 | .10571E-07 | .63315E+06 | 2.38E+01 |
| 369 | 116 | 55.462 | .20195E-05 | .14930E-05 | .21896E+07 | 1.35E+00 |
| 368 | 116 | 57.235 | .32675E-05 | .28532E-06 | .33265E+07 | 1.15E+01 |
| 367 | 116 | 58.337 | .29597E-05 | .44614E-06 | .29005E+07 | 6.63E+00 |
| 366 | 116 | 63.895 | .42202E-05 | .18420E-05 | .34474E+07 | 2.29E+00 |
| 365 | 116 | 65.026 | .24546E-03 | .53924E-04 | .19360E+09 | 4.55E+00 |
| 364 | 116 | 141.276 | .42689E-02 | .37604E-02 | .71331E+09 | 1.14E+00 |
| 363 | 116 | 141.994 | .13971E-01 | .12009E-01 | .23110E+10 | 1.16E+00 |
| 362 | 116 | 153.365 | .29530E-01 | .28976E-01 | .41872E+10 | 1.02E+00 |
| 361 | 116 | 157.893 | .16143E-02 | .15321E-02 | .21595E+09 | 1.05E+00 |
| 360 | 116 | 167.288 | .69815E-01 | .76551E-01 | .83199E+10 | 9.12E-01 |
| 359 | 116 | 183.219 | .14506E-01 | .16930E-01 | .14412E+10 | 8.57E-01 |
| 358 | 116 | 233.567 | .22884E-01 | .40707E-01 | .13990E+10 | 5.62E-01 |
| 116 | 357 | 688.538 | .20777E-07 | .26673E-07 | .58466E+03 | 7.79E-01 |
| 370 | 115 | 36.141 | .46204E-08 | .45595E-07 | .11797E+05 | 1.01E-01 |
| 369 | 115 | 54.941 | .23701E-06 | .15574E-05 | .26187E+06 | 1.52E-01 |
| 368 | 115 | 56.681 | .27038E-08 | .59156E-10 | .28068E+04 | 4.57E+01 |
| 367 | 115 | 57.761 | .66649E-06 | .12787E-06 | .66623E+06 | 5.21E+00 |
| 366 | 115 | 63.205 | .11124E-04 | .10647E-04 | .92870E+07 | 1.04E+00 |
| 365 | 115 | 64.311 | .68895E-04 | .19490E-05 | .55554E+08 | 3.53E+01 |
| 364 | 115 | 137.947 | .12342E-01 | .10420E-01 | .21630E+10 | 1.18E+00 |
| 363 | 115 | 138.631 | .16654E-01 | .14022E-01 | .28901E+10 | 1.19E+00 |
| 362 | 115 | 149.449 | .31838E-01 | .30329E-01 | .47540E+10 | 1.05E+00 |
| 361 | 115 | 153.746 | .39514E-02 | .40253E-02 | .55750E+09 | 9.82E-01 |
| 360 | 115 | 162.641 | .15372E+00 | .16729E+00 | .19381E+11 | 9.19E-01 |
| 359 | 115 | 177.659 | .16802E-01 | .20622E-01 | .17754E+10 | 8.15E-01 |
| 358 | 115 | 224.605 | .48618E-02 | .55855E-02 | .32141E+09 | 8.70E-01 |
| 115 | 357 | 780.318 | .32099E-07 | .23785E-07 | .70326E+03 | 1.35E+00 |
| 370 | 114 | 34.935 | .44942E-06 | .83880E-06 | .12281E+07 | 5.36E-01 |
| 369 | 114 | 52.201 | .58610E-06 | .27952E-04 | .71733E+06 | 2.10E-02 |
| 368 | 114 | 53.769 | .14298E-09 | .79231E-10 | .16493E+03 | 1.80E+00 |
| 367 | 114 | 54.740 | .28612E-07 | .51502E-11 | .31845E+05 | 5.56E+03 |
| 366 | 114 | 59.606 | .13312E-04 | .30806E-07 | .12496E+08 | 4.32E+02 |
| 365 | 114 | 60.589 | .15279E-03 | .21762E-05 | .13881E+09 | 7.02E+01 |
| 364 | 114 | 121.883 | .54320E-04 | .36597E-04 | .12195E+08 | 1.48E+00 |
| 363 | 114 | 122.417 | .71862E-03 | .51783E-03 | .15993E+09 | 1.39E+00 |
| 362 | 114 | 130.776 | .14943E-04 | .92485E-05 | .29140E+07 | 1.62E+00 |
| 361 | 114 | 134.055 | .23058E-04 | .32320E-04 | .42792E+07 | 7.13E-01 |
| 360 | 114 | 140.767 | .21891E-02 | .16801E-02 | .36844E+09 | 1.30E+00 |
| 359 | 114 | 151.879 | .22901E-01 | .22298E-01 | .33110E+10 | 1.03E+00 |
| 358 | 114 | 184.922 | .35410E+00 | .46108E+00 | .34535E+11 | 7.68E-01 |
| 114 | 357 | 3066.530 | .12500E-06 | .12968E-05 | .17733E+03 | 9.64E-02 |
| 370 | 113 | 31.959 | .31992E+00 | .30774E+00 | .10446E+13 | 1.04E+00 |
| 369 | 113 | 45.825 | .40311E-05 | .70999E-05 | .64020E+07 | 5.68E-01 |
| 368 | 113 | 47.030 | .13822E-03 | .16127E-03 | .20842E+09 | 8.57E-01 |
| 367 | 113 | 47.771 | .94946E-06 | .12285E-05 | .13876E+07 | 7.73E-01 |
| 366 | 113 | 51.435 | .14848E-02 | .20843E-02 | .18718E+10 | 7.12E-01 |
| 365 | 113 | 52.165 | .37190E-02 | .41037E-02 | .45580E+10 | 9.06E-01 |
| 364 | 113 | 91.998 | .71844E-10 | .20289E-08 | .28310E+02 | 3.54E-02 |
| 363 | 113 | 92.301 | .47759E-09 | .27334E-07 | .18696E+03 | 1.75E-02 |
| 362 | 113 | 96.975 | .53054E-09 | .46524E-08 | .18815E+03 | 1.14E-01 |
| 361 | 113 | 98.766 | .84991E-08 | .34165E-07 | .29057E+04 | 2.49E-01 |
| 360 | 113 | 102.362 | .13781E-07 | .94495E-06 | .43863E+04 | 1.46E-02 |
| 359 | 113 | 108.114 | .18864E-09 | .75962E-06 | .53822E+02 | 2.48E-04 |
| 358 | 113 | 123.870 | .22102E-05 | .69366E-06 | .48040E+06 | 3.19E+00 |
| 357 | 113 | 427.503 | .44835E+00 | .43386E+00 | .81817E+10 | 1.03E+00 |
| 468 | 134 | 337.018 | .52960E-03 | .16587E-02 | .31101E+08 | 3.19E-01 |
| 134 | 467 | 202.595 | .13887E-01 | .12349E-01 | .22567E+10 | 1.12E+00 |
| 134 | 466 | 190.631 | .99433E-01 | .82996E-01 | .18251E+11 | 1.20E+00 |
| 134 | 465 | 173.169 | .28547E-02 | .29838E-02 | .63498E+09 | 9.57E-01 |
| 134 | 464 | 162.513 | .84339E-02 | .62176E-02 | .21300E+10 | 1.36E+00 |
| 134 | 463 | 153.961 | .49498E-01 | .47934E-01 | .13928E+11 | 1.03E+00 |
| 134 | 462 | 146.050 | .90349E-01 | .74360E-01 | .28252E+11 | 1.22E+00 |
| 134 | 461 | 142.997 | .16493E-03 | .13188E-03 | .53798E+08 | 1.25E+00 |
| 134 | 460 | 124.015 | .11990E-04 | .95489E-05 | .51999E+07 | 1.26E+00 |
| 134 | 459 | 115.449 | .30171E-03 | .23782E-03 | .15099E+09 | 1.27E+00 |
| 134 | 458 | 114.623 | .24932E-06 | .18267E-05 | .12657E+06 | 1.36E-01 |
| 134 | 457 | 59.776 | .17849E-05 | .20477E-06 | .33318E+07 | 8.72E+00 |
| 134 | 456 | 58.727 | .10651E-04 | .44954E-06 | .20599E+08 | 2.37E+01 |
| 134 | 455 | 58.491 | .11312E-04 | .67892E-06 | .22055E+08 | 1.67E+01 |
| 134 | 454 | 57.704 | .13456E-04 | .27797E-05 | .26955E+08 | 4.84E+00 |
| 134 | 453 | 56.728 | .92688E-06 | .68752E-06 | .19211E+07 | 1.35E+00 |
| 134 | 452 | 56.201 | .49716E-06 | .84128E-06 | .10499E+07 | 5.91E-01 |
| 134 | 451 | 55.238 | .76442E-04 | .23638E-04 | .16710E+09 | 3.23E+00 |
| 134 | 450 | 54.306 | .48203E-05 | .12543E-05 | .10902E+08 | 3.84E+00 |
| 134 | 449 | 52.959 | .22240E-05 | .19160E-06 | .52891E+07 | 1.16E+01 |
| 134 | 448 | 51.471 | .15266E-05 | .12119E-05 | .38437E+07 | 1.26E+00 |
| 134 | 447 | 50.536 | .25508E-06 | .22665E-06 | .66621E+06 | 1.13E+00 |
| 134 | 446 | 38.998 | .11781E-05 | .21384E-05 | .51671E+07 | 5.51E-01 |
| 468 | 133 | 250.432 | .16756E-05 | .10305E-02 | .17820E+06 | 1.63E-03 |
| 133 | 467 | 255.751 | .23968E-01 | .20309E-01 | .24442E+10 | 1.18E+00 |
| 133 | 466 | 236.975 | .65735E-02 | .47192E-02 | .78077E+09 | 1.39E+00 |
| 133 | 465 | 210.579 | .39485E-02 | .54334E-02 | .59393E+09 | 7.27E-01 |
| 133 | 464 | 195.029 | .10777E-01 | .67421E-02 | .18900E+10 | 1.60E+00 |
| 133 | 463 | 182.840 | .89961E-02 | .10013E-01 | .17949E+10 | 8.98E-01 |
| 133 | 462 | 171.789 | .11343E-01 | .99947E-02 | .25637E+10 | 1.13E+00 |
| 133 | 461 | 167.581 | .19225E+00 | .17167E+00 | .45662E+11 | 1.12E+00 |
| 133 | 460 | 142.093 | .71669E-03 | .94846E-03 | .23677E+09 | 7.56E-01 |
| 133 | 459 | 130.960 | .74005E-03 | .54352E-04 | .28782E+09 | 1.36E+01 |
| 133 | 458 | 129.897 | .17798E-02 | .88124E-03 | .70357E+09 | 2.02E+00 |
| 133 | 457 | 63.681 | .66470E-05 | .41990E-06 | .10933E+08 | 1.58E+01 |
| 133 | 456 | 62.492 | .10165E-06 | .92950E-07 | .17361E+06 | 1.09E+00 |
| 133 | 455 | 62.224 | .32001E-05 | .11646E-05 | .55129E+07 | 2.75E+00 |
| 133 | 454 | 61.335 | .29565E-07 | .22002E-07 | .52419E+05 | 1.34E+00 |
| 133 | 453 | 60.234 | .37756E-05 | .96015E-06 | .69412E+07 | 3.93E+00 |
| 133 | 452 | 59.640 | .27162E-04 | .44358E-05 | .50936E+08 | 6.12E+00 |
| 133 | 451 | 58.556 | .91159E-07 | .11087E-06 | .17733E+06 | 8.22E-01 |
| 133 | 450 | 57.510 | .18056E-04 | .21884E-04 | .36414E+08 | 8.25E-01 |
| 133 | 449 | 56.002 | .47008E-05 | .34118E-06 | .99978E+07 | 1.38E+01 |
| 133 | 448 | 54.340 | .55784E-04 | .10385E-04 | .12601E+09 | 5.37E+00 |
| 133 | 447 | 53.299 | .26940E-04 | .25488E-05 | .63254E+08 | 1.06E+01 |
| 133 | 446 | 40.623 | .69964E-03 | .73665E-03 | .28279E+10 | 9.50E-01 |
| 468 | 132 | 184.345 | .28172E+00 | .22653E+00 | .55296E+11 | 1.24E+00 |
| 132 | 467 | 403.465 | .15613E-03 | .50536E-03 | .63974E+07 | 3.09E-01 |
| 132 | 466 | 358.638 | .51681E-02 | .54498E-02 | .26801E+09 | 9.48E-01 |
| 132 | 465 | 301.452 | .22741E-04 | .29559E-04 | .16692E+07 | 7.69E-01 |
| 132 | 464 | 270.568 | .49692E-01 | .42549E-01 | .45276E+10 | 1.17E+00 |
| 132 | 463 | 247.664 | .11811E-02 | .11735E-02 | .12844E+09 | 1.01E+00 |
| 132 | 462 | 227.814 | .44741E-03 | .20673E-03 | .57501E+08 | 2.16E+00 |
| 132 | 461 | 220.471 | .66422E-02 | .56691E-02 | .91147E+09 | 1.17E+00 |
| 132 | 460 | 178.377 | .54338E-01 | .50800E-01 | .11391E+11 | 1.07E+00 |
| 132 | 459 | 161.176 | .48645E-01 | .50484E-01 | .12490E+11 | 9.64E-01 |
| 132 | 458 | 159.570 | .14555E+00 | .14151E+00 | .38128E+11 | 1.03E+00 |
| 132 | 457 | 70.069 | .75948E-05 | .79646E-05 | .10318E+08 | 9.54E-01 |
| 132 | 456 | 68.632 | .35201E-06 | .40271E-08 | .49847E+06 | 8.74E+01 |
| 132 | 455 | 68.309 | .38114E-05 | .16688E-05 | .54483E+07 | 2.28E+00 |
| 132 | 454 | 67.239 | .74368E-06 | .28021E-05 | .10972E+07 | 2.65E-01 |
| 132 | 453 | 65.918 | .23420E-04 | .94793E-06 | .35952E+08 | 2.47E+01 |
| 132 | 452 | 65.207 | .15550E-03 | .15418E-04 | .24393E+09 | 1.01E+01 |
| 132 | 451 | 63.914 | .86450E-05 | .65725E-06 | .14116E+08 | 1.32E+01 |
| 132 | 450 | 62.669 | .28020E-03 | .16515E-04 | .47587E+09 | 1.70E+01 |
| 132 | 449 | 60.882 | .21229E-03 | .13350E-03 | .38202E+09 | 1.59E+00 |
| 132 | 448 | 58.924 | .25861E-06 | .19948E-04 | .49682E+06 | 1.30E-02 |
| 132 | 447 | 57.702 | .46467E-03 | .20535E-04 | .93088E+09 | 2.26E+01 |
| 132 | 446 | 43.131 | .41869E-04 | .14322E-03 | .15012E+09 | 2.92E-01 |
| 468 | 131 | 159.315 | .60455E-01 | .52681E-01 | .15888E+11 | 1.15E+00 |
| 131 | 467 | 614.904 | .54706E-03 | .21388E-03 | .96505E+07 | 2.56E+00 |
| 131 | 466 | 516.512 | .49960E-03 | .39452E-03 | .12491E+08 | 1.27E+00 |
| 131 | 465 | 405.676 | .41254E-03 | .90651E-03 | .16720E+08 | 4.55E-01 |
| 131 | 464 | 351.659 | .14699E-02 | .13063E-02 | .79282E+08 | 1.13E+00 |
| 131 | 463 | 313.925 | .10766E-02 | .11869E-02 | .72871E+08 | 9.07E-01 |
| 131 | 462 | 282.702 | .96350E-03 | .52487E-03 | .80413E+08 | 1.84E+00 |
| 131 | 461 | 271.483 | .90276E-02 | .25421E-02 | .81700E+09 | 3.55E+00 |
| 131 | 460 | 210.356 | .21406E-01 | .15594E-01 | .32267E+10 | 1.37E+00 |
| 131 | 459 | 186.841 | .94828E-01 | .84333E-01 | .18119E+11 | 1.12E+00 |
| 131 | 458 | 184.686 | .10250E+00 | .71647E-01 | .20045E+11 | 1.43E+00 |
| 131 | 457 | 74.519 | .38054E-06 | .64682E-05 | .45708E+06 | 5.88E-02 |
| 131 | 456 | 72.895 | .11828E-06 | .39758E-07 | .14847E+06 | 2.97E+00 |
| 131 | 455 | 72.532 | .70149E-07 | .37687E-05 | .88940E+05 | 1.86E-02 |
| 131 | 454 | 71.326 | .29244E-06 | .15269E-06 | .38342E+06 | 1.92E+00 |
| 131 | 453 | 69.841 | .10844E-06 | .22273E-06 | .14829E+06 | 4.87E-01 |
| 131 | 452 | 69.044 | .14418E-05 | .16043E-05 | .20173E+07 | 8.99E-01 |
| 131 | 451 | 67.596 | .25391E-07 | .65457E-09 | .37065E+05 | 3.88E+01 |
| 131 | 450 | 66.205 | .29791E-04 | .93372E-05 | .45335E+08 | 3.19E+00 |
| 131 | 449 | 64.214 | .69785E-05 | .69177E-05 | .11288E+08 | 1.01E+00 |
| 131 | 448 | 62.039 | .11044E-03 | .59476E-04 | .19139E+09 | 1.86E+00 |
| 131 | 447 | 60.686 | .84608E-04 | .18053E-04 | .15324E+09 | 4.69E+00 |
| 131 | 446 | 44.777 | .43222E-02 | .43596E-02 | .14379E+11 | 9.91E-01 |
| 468 | 130 | 74.681 | .16585E-05 | .98981E-06 | .19836E+07 | 1.68E+00 |
| 467 | 130 | 182.242 | .47246E-02 | .50244E-02 | .94886E+09 | 9.40E-01 |
| 466 | 130 | 193.146 | .10548E-03 | .88167E-04 | .18860E+08 | 1.20E+00 |
| 465 | 130 | 215.125 | .80644E-01 | .74147E-01 | .11623E+11 | 1.09E+00 |
| 464 | 130 | 234.201 | .17832E-03 | .23889E-03 | .21685E+08 | 7.46E-01 |
| 463 | 130 | 254.581 | .22966E-01 | .26238E-01 | .23635E+10 | 8.75E-01 |
| 462 | 130 | 279.626 | .49504E-03 | .65705E-03 | .42230E+08 | 7.53E-01 |
| 461 | 130 | 291.544 | .80849E-04 | .10505E-03 | .63445E+07 | 7.70E-01 |
| 460 | 130 | 423.794 | .85606E-03 | .96285E-03 | .31793E+08 | 8.89E-01 |
| 459 | 130 | 567.746 | .27531E-03 | .43397E-03 | .56969E+07 | 6.34E-01 |
| 458 | 130 | 588.616 | .39026E-06 | .20056E-05 | .75131E+04 | 1.95E-01 |
| 130 | 457 | 158.582 | .39927E-01 | .33230E-01 | .10590E+11 | 1.20E+00 |
| 130 | 456 | 151.405 | .12384E+00 | .10835E+00 | .36035E+11 | 1.14E+00 |
| 130 | 455 | 149.844 | .55587E-01 | .42848E-01 | .16513E+11 | 1.30E+00 |
| 130 | 454 | 144.790 | .29472E-02 | .14207E-02 | .93769E+09 | 2.07E+00 |
| 130 | 453 | 138.798 | .24267E-02 | .22414E-02 | .84019E+09 | 1.08E+00 |
| 130 | 452 | 135.683 | .42016E-02 | .26733E-02 | .15223E+10 | 1.57E+00 |
| 130 | 451 | 130.204 | .10690E-04 | .82767E-05 | .42061E+07 | 1.29E+00 |
| 130 | 450 | 125.139 | .16159E-02 | .87581E-03 | .68829E+09 | 1.85E+00 |
| 130 | 449 | 118.212 | .18913E-03 | .16073E-03 | .90277E+08 | 1.18E+00 |
| 130 | 448 | 111.045 | .50168E-04 | .35238E-04 | .27137E+08 | 1.42E+00 |
| 130 | 447 | 106.784 | .21577E-05 | .14275E-07 | .12621E+07 | 1.51E+02 |
| 130 | 446 | 65.706 | .31077E-07 | .52077E-08 | .48014E+05 | 5.97E+00 |
| 468 | 129 | 72.752 | .13044E-04 | .58271E-06 | .16438E+08 | 2.24E+01 |
| 467 | 129 | 171.167 | .24427E-01 | .19834E-01 | .55611E+10 | 1.23E+00 |
| 466 | 129 | 180.752 | .25873E-03 | .12530E-03 | .52822E+08 | 2.06E+00 |
| 465 | 129 | 199.860 | .32468E-01 | .25800E-01 | .54218E+10 | 1.26E+00 |
| 464 | 129 | 216.223 | .67687E-02 | .67440E-02 | .96568E+09 | 1.00E+00 |
| 463 | 129 | 233.479 | .87327E-03 | .73120E-03 | .10685E+09 | 1.19E+00 |
| 462 | 129 | 254.373 | .15510E-01 | .19289E-01 | .15988E+10 | 8.04E-01 |
| 461 | 129 | 264.198 | .51252E-02 | .43958E-02 | .48976E+09 | 1.17E+00 |
| 460 | 129 | 368.370 | .62551E-02 | .65425E-02 | .30747E+09 | 9.56E-01 |
| 459 | 129 | 472.506 | .49627E-04 | .65489E-05 | .14827E+07 | 7.58E+00 |
| 458 | 129 | 486.873 | .24466E-02 | .36477E-02 | .68842E+08 | 6.71E-01 |
| 129 | 457 | 168.043 | .96255E-02 | .94937E-02 | .22736E+10 | 1.01E+00 |
| 129 | 456 | 160.006 | .74160E-02 | .52977E-02 | .19321E+10 | 1.40E+00 |
| 129 | 455 | 158.264 | .67121E-02 | .37321E-02 | .17874E+10 | 1.80E+00 |
| 129 | 454 | 152.636 | .73870E-02 | .50790E-02 | .21149E+10 | 1.45E+00 |
| 129 | 453 | 145.992 | .71986E-01 | .59612E-01 | .22528E+11 | 1.21E+00 |
| 129 | 452 | 142.550 | .70472E-01 | .57611E-01 | .23132E+11 | 1.22E+00 |
| 129 | 451 | 136.514 | .18405E-04 | .79333E-05 | .65872E+07 | 2.32E+00 |
| 129 | 450 | 130.957 | .22162E-05 | .22970E-04 | .86195E+06 | 9.65E-02 |
| 129 | 449 | 123.390 | .57448E-03 | .12979E-02 | .25168E+09 | 4.43E-01 |
| 129 | 448 | 115.602 | .39875E-03 | .19450E-03 | .19902E+09 | 2.05E+00 |
| 129 | 447 | 110.992 | .53849E-03 | .16541E-03 | .29156E+09 | 3.26E+00 |
| 129 | 446 | 67.275 | .11720E-05 | .87745E-08 | .17273E+07 | 1.34E+02 |
| 468 | 128 | 71.382 | .89427E-05 | .14183E-06 | .11706E+08 | 6.31E+01 |
| 467 | 128 | 163.774 | .13409E-01 | .10727E-01 | .33345E+10 | 1.25E+00 |
| 466 | 128 | 172.528 | .13245E-01 | .13627E-01 | .29680E+10 | 9.72E-01 |
| 465 | 128 | 189.854 | .31435E-01 | .24544E-01 | .58171E+10 | 1.28E+00 |
| 464 | 128 | 204.559 | .16511E-02 | .18748E-02 | .26318E+09 | 8.81E-01 |
| 463 | 128 | 219.937 | .29906E-02 | .27266E-02 | .41237E+09 | 1.10E+00 |
| 462 | 128 | 238.382 | .39038E-01 | .39819E-01 | .45822E+10 | 9.80E-01 |
| 461 | 128 | 246.989 | .79672E-03 | .18322E-02 | .87113E+08 | 4.35E-01 |
| 460 | 128 | 335.753 | .50726E-02 | .46927E-02 | .30014E+09 | 1.08E+00 |
| 459 | 128 | 420.151 | .16669E-03 | .17043E-04 | .62985E+07 | 9.78E+00 |
| 458 | 128 | 431.472 | .16878E-02 | .21709E-02 | .60472E+08 | 7.77E-01 |
| 128 | 457 | 175.836 | .49395E-02 | .27548E-02 | .10656E+10 | 1.79E+00 |
| 128 | 456 | 167.055 | .35114E-01 | .30014E-01 | .83925E+10 | 1.17E+00 |
| 128 | 455 | 165.157 | .78701E-02 | .76639E-02 | .19245E+10 | 1.03E+00 |
| 128 | 454 | 159.038 | .30626E-01 | .22235E-01 | .80764E+10 | 1.38E+00 |
| 128 | 453 | 151.838 | .33425E-02 | .27203E-02 | .96705E+09 | 1.23E+00 |
| 128 | 452 | 148.118 | .98375E-01 | .84626E-01 | .29909E+11 | 1.16E+00 |
| 128 | 451 | 141.613 | .16189E-02 | .94020E-03 | .53845E+09 | 1.72E+00 |
| 128 | 450 | 135.642 | .67525E-03 | .47098E-03 | .24480E+09 | 1.43E+00 |
| 128 | 449 | 127.541 | .89545E-04 | .44809E-03 | .36718E+08 | 2.00E-01 |
| 128 | 448 | 119.238 | .28578E-02 | .18543E-02 | .13407E+10 | 1.54E+00 |
| 128 | 447 | 114.339 | .52217E-03 | .26885E-03 | .26642E+09 | 1.94E+00 |
| 128 | 446 | 68.490 | .20357E-05 | .10063E-07 | .28947E+07 | 2.02E+02 |
| 468 | 127 | 68.854 | .43414E-05 | .19443E-05 | .61080E+07 | 2.23E+00 |
| 467 | 127 | 151.052 | .16371E+00 | .13605E+00 | .47858E+11 | 1.20E+00 |
| 466 | 127 | 158.467 | .37646E-01 | .34682E-01 | .99995E+10 | 1.09E+00 |
| 465 | 127 | 172.966 | .24446E-03 | .18198E-03 | .54503E+08 | 1.34E+00 |
| 464 | 127 | 185.087 | .34481E-03 | .72801E-05 | .67137E+08 | 4.74E+01 |
| 463 | 127 | 197.588 | .34789E-01 | .27819E-01 | .59437E+10 | 1.25E+00 |
| 462 | 127 | 212.349 | .97508E-04 | .67442E-04 | .14424E+08 | 1.45E+00 |
| 461 | 127 | 219.152 | .19947E-01 | .19222E-01 | .27703E+10 | 1.04E+00 |
| 460 | 127 | 286.315 | .99748E-02 | .97313E-02 | .81162E+09 | 1.03E+00 |
| 459 | 127 | 345.498 | .67659E-02 | .55331E-02 | .37806E+09 | 1.22E+00 |
| 458 | 127 | 353.117 | .46315E-02 | .50143E-02 | .24775E+09 | 9.24E-01 |
| 127 | 457 | 193.317 | .20865E-01 | .15660E-01 | .37240E+10 | 1.33E+00 |
| 127 | 456 | 182.756 | .63361E-04 | .49192E-04 | .12654E+08 | 1.29E+00 |
| 127 | 455 | 180.487 | .74528E-02 | .50160E-02 | .15260E+10 | 1.49E+00 |
| 127 | 454 | 173.204 | .61499E-02 | .60687E-02 | .13674E+10 | 1.01E+00 |
| 127 | 453 | 164.699 | .76275E-01 | .61564E-01 | .18756E+11 | 1.24E+00 |
| 127 | 452 | 160.331 | .56509E-02 | .38130E-02 | .14663E+10 | 1.48E+00 |
| 127 | 451 | 152.736 | .78501E-01 | .52365E-01 | .22445E+11 | 1.50E+00 |
| 127 | 450 | 145.813 | .19389E-01 | .16526E-01 | .60825E+10 | 1.17E+00 |
| 127 | 449 | 136.494 | .19480E-02 | .85582E-03 | .69744E+09 | 2.28E+00 |
| 127 | 448 | 127.027 | .17443E-01 | .12652E-01 | .72104E+10 | 1.38E+00 |
| 127 | 447 | 121.482 | .26907E-02 | .57395E-03 | .12161E+10 | 4.69E+00 |
| 127 | 446 | 70.991  | .14520E-08 | .19166E-05 | .19218E+04 | 7.58E-04 |
| 468 | 126 | 67.674  | .14535E-05 | .17591E-06 | .21170E+07 | 8.26E+00 |
| 467 | 126 | 145.486 | .66094E-02 | .49799E-02 | .20828E+10 | 1.33E+00 |
| 466 | 126 | 152.353 | .17097E+00 | .14889E+00 | .49131E+11 | 1.15E+00 |
| 465 | 126 | 165.707 | .18747E-04 | .53153E-04 | .45539E+07 | 3.53E-01 |
| 464 | 126 | 176.800 | .54977E-02 | .60502E-02 | .11731E+10 | 9.09E-01 |
| 463 | 126 | 188.171 | .26461E-01 | .20013E-01 | .49846E+10 | 1.32E+00 |
| 462 | 126 | 201.511 | .21750E-01 | .16964E-01 | .35726E+10 | 1.28E+00 |
| 461 | 126 | 207.628 | .26910E-01 | .23531E-01 | .41636E+10 | 1.14E+00 |
| 460 | 126 | 266.956 | .64160E-02 | .48495E-02 | .60051E+09 | 1.32E+00 |
| 459 | 126 | 317.698 | .64877E-02 | .41407E-02 | .42874E+09 | 1.57E+00 |
| 458 | 126 | 324.129 | .23363E-03 | .52363E-04 | .14833E+08 | 4.46E+00 |
| 126 | 457 | 203.270 | .53469E-02 | .38484E-02 | .86316E+09 | 1.39E+00 |
| 126 | 456 | 191.626 | .23829E-03 | .39463E-03 | .43284E+08 | 6.04E-01 |
| 126 | 455 | 189.132 | .15031E-01 | .11420E-01 | .28028E+10 | 1.32E+00 |
| 126 | 454 | 181.150 | .16131E-02 | .18970E-02 | .32788E+09 | 8.50E-01 |
| 126 | 453 | 171.868 | .26226E-01 | .23492E-01 | .59221E+10 | 1.12E+00 |
| 126 | 452 | 167.117 | .19318E-01 | .16745E-01 | .46138E+10 | 1.15E+00 |
| 126 | 451 | 158.882 | .86094E-01 | .60917E-01 | .22749E+11 | 1.41E+00 |
| 126 | 450 | 151.405 | .28469E-01 | .22472E-01 | .82838E+10 | 1.27E+00 |
| 126 | 449 | 141.381 | .80687E-04 | .26837E-04 | .26925E+08 | 3.01E+00 |
| 126 | 448 | 131.250 | .21349E-01 | .16112E-01 | .82662E+10 | 1.33E+00 |
| 126 | 447 | 125.339 | .51245E-03 | .38695E-03 | .21758E+09 | 1.32E+00 |
| 126 | 446 | 72.290  | .24077E-05 | .46561E-06 | .30730E+07 | 5.17E+00 |
| 468 | 125 | 66.071  | .59372E-06 | .55860E-05 | .90719E+06 | 1.06E-01 |
| 467 | 125 | 138.272 | .18276E-02 | .15469E-02 | .63758E+09 | 1.18E+00 |
| 466 | 125 | 144.460 | .30172E-03 | .30540E-03 | .96436E+08 | 9.88E-01 |
| 465 | 125 | 156.412 | .10706E-03 | .13008E-03 | .29189E+08 | 8.23E-01 |
| 464 | 125 | 166.259 | .35971E-01 | .33549E-01 | .86800E+10 | 1.07E+00 |
| 463 | 125 | 176.276 | .24835E-02 | .22865E-02 | .53311E+09 | 1.09E+00 |
| 462 | 125 | 187.931 | .78961E-03 | .52483E-03 | .14912E+09 | 1.50E+00 |
| 461 | 125 | 193.240 | .11215E-01 | .85984E-02 | .20032E+10 | 1.30E+00 |
| 460 | 125 | 243.633 | .41421E-02 | .36874E-02 | .46546E+09 | 1.12E+00 |
| 459 | 125 | 285.205 | .45396E-02 | .67099E-02 | .37225E+09 | 6.77E-01 |
| 458 | 125 | 290.377 | .23831E-01 | .23305E-01 | .18852E+10 | 1.02E+00 |
| 125 | 457 | 219.251 | .28730E-02 | .14034E-02 | .39864E+09 | 2.05E+00 |
| 125 | 456 | 205.765 | .34491E-06 | .57865E-04 | .54337E+05 | 5.96E-03 |
| 125 | 455 | 202.893 | .23194E-02 | .11076E-02 | .37581E+09 | 2.09E+00 |
| 125 | 454 | 193.736 | .94393E-03 | .81444E-03 | .16775E+09 | 1.16E+00 |
| 125 | 453 | 183.156 | .47708E-03 | .43198E-03 | .94858E+08 | 1.10E+00 |
| 125 | 452 | 177.771 | .31433E-01 | .30216E-01 | .66343E+10 | 1.04E+00 |
| 125 | 451 | 168.481 | .38026E-02 | .29235E-02 | .89352E+09 | 1.30E+00 |
| 125 | 450 | 160.097 | .93815E-01 | .82465E-01 | .24414E+11 | 1.14E+00 |
| 125 | 449 | 148.932 | .21822E-04 | .12978E-04 | .65624E+07 | 1.68E+00 |
| 125 | 448 | 137.732 | .32140E-01 | .28697E-01 | .11301E+11 | 1.12E+00 |
| 125 | 447 | 131.237 | .14476E-01 | .92850E-02 | .56063E+10 | 1.56E+00 |
| 125 | 446 | 74.214 | .23946E-04 | .10439E-06 | .28999E+08 | 2.29E+02 |
| 468 | 124 | 64.818 | .76302E-04 | .21761E-04 | .12114E+09 | 3.51E+00 |
| 467 | 124 | 132.896 | .62691E-02 | .35788E-02 | .23676E+10 | 1.75E+00 |
| 466 | 124 | 138.602 | .20299E-03 | .25898E-03 | .70480E+08 | 7.84E-01 |
| 465 | 124 | 149.568 | .39385E-03 | .30878E-03 | .11743E+09 | 1.28E+00 |
| 464 | 124 | 158.547 | .91854E-01 | .77458E-01 | .24374E+11 | 1.19E+00 |
| 463 | 124 | 167.631 | .60134E-02 | .44943E-02 | .14274E+10 | 1.34E+00 |
| 462 | 124 | 178.137 | .22991E-02 | .19647E-02 | .48326E+09 | 1.17E+00 |
| 461 | 124 | 182.900 | .45763E-01 | .34765E-01 | .91248E+10 | 1.32E+00 |
| 460 | 124 | 227.423 | .71683E-04 | .78287E-04 | .92444E+07 | 9.16E-01 |
| 459 | 124 | 263.240 | .18675E-02 | .46908E-03 | .17976E+09 | 3.98E+00 |
| 458 | 124 | 267.640 | .85471E-02 | .92733E-02 | .79589E+09 | 9.22E-01 |
| 124 | 457 | 234.279 | .27843E-02 | .19887E-02 | .33836E+09 | 1.40E+00 |
| 124 | 456 | 218.946 | .14304E-03 | .27363E-03 | .19903E+08 | 5.23E-01 |
| 124 | 455 | 215.697 | .20923E-02 | .10941E-02 | .29997E+09 | 1.91E+00 |
| 124 | 454 | 205.376 | .51669E-03 | .44232E-03 | .81707E+08 | 1.17E+00 |
| 124 | 453 | 193.527 | .12798E-02 | .43413E-03 | .22793E+09 | 2.95E+00 |
| 124 | 452 | 187.524 | .38806E-01 | .30223E-01 | .73608E+10 | 1.28E+00 |
| 124 | 451 | 177.217 | .69672E-02 | .54888E-02 | .14797E+10 | 1.27E+00 |
| 124 | 450 | 167.964 | .94739E-01 | .76446E-01 | .22399E+11 | 1.24E+00 |
| 124 | 449 | 155.717 | .50876E-02 | .65623E-03 | .13995E+10 | 7.75E+00 |
| 124 | 448 | 143.515 | .48096E-02 | .41290E-02 | .15576E+10 | 1.16E+00 |
| 124 | 447 | 136.477 | .39185E-01 | .17350E-01 | .14032E+11 | 2.26E+00 |
| 124 | 446 | 75.861 | .93789E-06 | .16226E-05 | .10870E+07 | 5.78E-01 |
| 468 | 123 | 63.505 | .37645E-06 | .73821E-07 | .62262E+06 | 5.10E+00 |
| 467 | 123 | 127.493 | .74908E-04 | .45239E-04 | .30739E+08 | 1.66E+00 |
| 466 | 123 | 132.735 | .15930E-03 | .10456E-03 | .60307E+08 | 1.52E+00 |
| 465 | 123 | 142.759 | .10116E+00 | .76847E-01 | .33107E+11 | 1.32E+00 |
| 464 | 123 | 150.916 | .88646E-03 | .74858E-03 | .25961E+09 | 1.18E+00 |
| 463 | 123 | 159.125 | .31782E-01 | .29914E-01 | .83722E+10 | 1.06E+00 |
| 462 | 123 | 168.561 | .10526E-01 | .11112E-01 | .24711E+10 | 9.47E-01 |
| 461 | 123 | 172.820 | .49874E-03 | .56312E-03 | .11138E+09 | 8.86E-01 |
| 460 | 123 | 212.044 | .19955E-04 | .26464E-04 | .29603E+07 | 7.54E-01 |
| 459 | 123 | 242.853 | .54986E-05 | .13418E-05 | .62186E+06 | 4.10E+00 |
| 458 | 123 | 246.593 | .51694E-04 | .14768E-03 | .56703E+07 | 3.50E-01 |
| 123 | 457 | 253.196 | .49976E-02 | .48304E-02 | .51998E+09 | 1.03E+00 |
| 123 | 456 | 235.380 | .62816E-01 | .45180E-01 | .75625E+10 | 1.39E+00 |
| 123 | 455 | 231.630 | .74982E-02 | .67573E-02 | .93218E+09 | 1.11E+00 |
| 123 | 454 | 219.770 | .27318E-02 | .16323E-02 | .37726E+09 | 1.67E+00 |
| 123 | 453 | 206.256 | .46014E-01 | .26748E-01 | .72145E+10 | 1.72E+00 |
| 123 | 452 | 199.451 | .49994E-02 | .26041E-02 | .83826E+09 | 1.92E+00 |
| 123 | 451 | 187.832 | .35592E-04 | .23571E-04 | .67289E+07 | 1.51E+00 |
| 123 | 450 | 177.470 | .78953E-03 | .56801E-03 | .16721E+09 | 1.39E+00 |
| 123 | 449 | 163.853 | .10591E-06 | .20408E-07 | .26313E+05 | 5.19E+00 |
| 123 | 448 | 150.399 | .64068E-03 | .52701E-03 | .18892E+09 | 1.22E+00 |
| 123 | 447 | 142.687 | .36224E-05 | .76301E-05 | .11867E+07 | 4.75E-01 |
| 123 | 446 | 77.742 | .57655E-10 | .27296E-07 | .63629E+02 | 2.11E-03 |
| 468 | 122 | 61.414 | .24191E-06 | .31812E-07 | .42780E+06 | 7.60E+00 |
| 467 | 122 | 119.336 | .21064E-04 | .11284E-04 | .98658E+07 | 1.87E+00 |
| 466 | 122 | 123.917 | .16803E-04 | .13497E-04 | .72987E+07 | 1.24E+00 |
| 465 | 122 | 132.609 | .40503E-03 | .17777E-03 | .15363E+09 | 2.28E+00 |
| 464 | 122 | 139.619 | .83959E-04 | .92074E-04 | .28728E+08 | 9.12E-01 |
| 463 | 122 | 146.616 | .37326E-02 | .10750E-02 | .11582E+10 | 3.47E+00 |
| 462 | 122 | 154.590 | .58313E-01 | .40536E-01 | .16275E+11 | 1.44E+00 |
| 461 | 122 | 158.165 | .17470E-02 | .13342E-02 | .46581E+09 | 1.31E+00 |
| 460 | 122 | 190.398 | .43188E-05 | .15138E-04 | .79464E+06 | 2.85E-01 |
| 459 | 122 | 214.876 | .16038E-03 | .18370E-03 | .23168E+08 | 8.73E-01 |
| 458 | 122 | 217.798 | .15980E-05 | .22333E-04 | .22470E+06 | 7.16E-02 |
| 122 | 457 | 292.966 | .10832E-01 | .10451E-01 | .84182E+09 | 1.04E+00 |
| 122 | 456 | 269.375 | .21862E-01 | .20203E-01 | .20096E+10 | 1.08E+00 |
| 122 | 455 | 264.474 | .25550E-01 | .23301E-01 | .24364E+10 | 1.10E+00 |
| 122 | 454 | 249.124 | .30907E-03 | .59293E-04 | .33217E+08 | 5.21E+00 |
| 122 | 453 | 231.900 | .29210E-02 | .29620E-02 | .36229E+09 | 9.86E-01 |
| 122 | 452 | 223.333 | .14772E-02 | .13349E-02 | .19755E+09 | 1.11E+00 |
| 122 | 451 | 208.866 | .52541E-01 | .35656E-01 | .80334E+10 | 1.47E+00 |
| 122 | 450 | 196.132 | .76571E-03 | .56854E-03 | .13277E+09 | 1.35E+00 |
| 122 | 449 | 179.634 | .59805E-03 | .47308E-03 | .12362E+09 | 1.26E+00 |
| 122 | 448 | 163.590 | .11913E-02 | .86227E-03 | .29692E+09 | 1.38E+00 |
| 122 | 447 | 154.507 | .42820E-03 | .26094E-03 | .11964E+09 | 1.64E+00 |
| 122 | 446 | 81.123 | .25412E-10 | .35996E-07 | .25756E+02 | 7.06E-04 |
| 468 | 121 | 60.937 | .46063E-05 | .30909E-05 | .82742E+07 | 1.49E+00 |
| 467 | 121 | 117.548 | .22572E-03 | .19015E-03 | .10896E+09 | 1.19E+00 |
| 466 | 121 | 121.990 | .28472E-02 | .19442E-02 | .12761E+10 | 1.46E+00 |
| 465 | 121 | 130.405 | .20553E-03 | .88486E-04 | .80618E+08 | 2.32E+00 |
| 464 | 121 | 137.178 | .28109E-01 | .23093E-01 | .99634E+10 | 1.22E+00 |
| 463 | 121 | 143.927 | .42427E-03 | .34166E-03 | .13661E+09 | 1.24E+00 |
| 462 | 121 | 151.603 | .22368E-04 | .50331E-04 | .64914E+07 | 4.44E-01 |
| 461 | 121 | 155.039 | .48075E-02 | .24171E-02 | .13340E+10 | 1.99E+00 |
| 460 | 121 | 185.887 | .16507E-02 | .29937E-02 | .31863E+09 | 5.51E-01 |
| 459 | 121 | 209.147 | .83764E-01 | .82094E-01 | .12773E+11 | 1.02E+00 |
| 458 | 121 | 211.915 | .23156E-01 | .19722E-01 | .34393E+10 | 1.17E+00 |
| 121 | 457 | 304.331 | .36901E-04 | .23069E-04 | .26575E+07 | 1.60E+00 |
| 121 | 456 | 278.953 | .11435E-03 | .18968E-03 | .98015E+07 | 6.03E-01 |
| 121 | 455 | 273.701 | .13748E-04 | .19378E-05 | .12241E+07 | 7.09E+00 |
| 121 | 454 | 257.294 | .72147E-05 | .65897E-05 | .72692E+06 | 1.09E+00 |
| 121 | 453 | 238.963 | .14546E-04 | .63256E-04 | .16990E+07 | 2.30E-01 |
| 121 | 452 | 229.877 | .70404E-03 | .24206E-03 | .88867E+08 | 2.91E+00 |
| 121 | 451 | 214.579 | .42425E-05 | .77121E-05 | .61458E+06 | 5.50E-01 |
| 121 | 450 | 201.161 | .18562E-02 | .51741E-02 | .30597E+09 | 3.59E-01 |
| 121 | 449 | 183.843 | .35985E-04 | .11535E-04 | .71017E+07 | 3.12E+00 |
| 121 | 448 | 167.073 | .37817E-01 | .35310E-01 | .90366E+10 | 1.07E+00 |
| 121 | 447 | 157.611 | .41841E-01 | .29814E-01 | .11235E+11 | 1.40E+00 |
| 121 | 446 | 81.971 | .34991E-06 | .74239E-06 | .34735E+06 | 4.71E-01 |
| 468 | 120 | 59.090 | .23532E-04 | .40579E-06 | .44953E+08 | 5.80E+01 |
| 467 | 120 | 110.862 | .41164E-03 | .15116E-03 | .22340E+09 | 2.72E+00 |
| 466 | 120 | 114.805 | .19627E-03 | .83344E-04 | .99328E+08 | 2.35E+00 |
| 465 | 120 | 122.228 | .31119E-02 | .17962E-02 | .13894E+10 | 1.73E+00 |
| 464 | 120 | 128.159 | .64901E-03 | .40369E-03 | .26356E+09 | 1.61E+00 |
| 463 | 120 | 134.030 | .46003E-02 | .32566E-02 | .17081E+10 | 1.41E+00 |
| 462 | 120 | 140.663 | .12755E-01 | .94321E-02 | .43000E+10 | 1.35E+00 |
| 461 | 120 | 143.616 | .18724E+00 | .14773E+00 | .60552E+11 | 1.27E+00 |
| 460 | 120 | 169.704 | .35352E-04 | .26688E-04 | .81878E+07 | 1.32E+00 |
| 459 | 120 | 188.881 | .62283E-02 | .59621E-02 | .11645E+10 | 1.04E+00 |
| 458 | 120 | 191.136 | .14876E-01 | .14794E-01 | .27160E+10 | 1.01E+00 |
| 120 | 457 | 360.636 | .57042E-02 | .32460E-02 | .29254E+09 | 1.76E+00 |
| 120 | 456 | 325.541 | .97590E-04 | .42621E-04 | .61422E+07 | 2.29E+00 |
| 120 | 455 | 318.410 | .81614E-02 | .60273E-02 | .53694E+09 | 1.35E+00 |
| 120 | 454 | 296.421 | .17020E-06 | .59750E-05 | .12920E+05 | 2.85E-02 |
| 120 | 453 | 272.352 | .25029E-04 | .28127E-05 | .22507E+07 | 8.90E+00 |
| 120 | 452 | 260.611 | .10844E-02 | .32402E-04 | .10650E+09 | 3.35E+01 |
| 120 | 451 | 241.122 | .14210E-03 | .92464E-04 | .16302E+08 | 1.54E+00 |
| 120 | 450 | 224.309 | .78744E-01 | .50880E-01 | .10439E+11 | 1.55E+00 |
| 120 | 449 | 202.988 | .12723E-01 | .61825E-02 | .20595E+10 | 2.06E+00 |
| 120 | 448 | 182.736 | .61624E-01 | .66510E-01 | .12309E+11 | 9.27E-01 |
| 120 | 447 | 171.477 | .76073E-02 | .71971E-02 | .17256E+10 | 1.06E+00 |
| 120 | 446 | 85.569 | .49279E-06 | .11713E-04 | .44891E+06 | 4.21E-02 |
| 468 | 119 | 55.452 | .15765E-03 | .69892E-08 | .34197E+09 | 2.26E+04 |
| 467 | 119 | 98.712 | .17849E-08 | .29323E-05 | .12218E+04 | 6.09E-04 |
| 466 | 119 | 101.826 | .89133E-10 | .17990E-05 | .57340E+02 | 4.95E-05 |
| 465 | 119 | 107.623 | .86035E-05 | .19903E-05 | .49545E+07 | 4.32E+00 |
| 464 | 119 | 112.195 | .53072E-03 | .12805E-03 | .28123E+09 | 4.14E+00 |
| 463 | 119 | 116.669 | .12326E-02 | .49066E-03 | .60401E+09 | 2.51E+00 |
| 462 | 119 | 121.662 | .42588E-03 | .15094E-03 | .19191E+09 | 2.82E+00 |
| 461 | 119 | 123.865 | .21502E-02 | .99582E-03 | .93477E+09 | 2.16E+00 |
| 460 | 119 | 142.798 | .49093E-01 | .36154E-01 | .16059E+11 | 1.36E+00 |
| 459 | 119 | 156.138 | .28806E-02 | .22000E-02 | .78815E+09 | 1.31E+00 |
| 458 | 119 | 157.675 | .33046E+00 | .25493E+00 | .88659E+11 | 1.30E+00 |
| 119 | 457 | 601.465 | .17327E-03 | .11096E-04 | .31947E+07 | 1.56E+01 |
| 119 | 456 | 509.803 | .24920E-05 | .22617E-04 | .63956E+05 | 1.10E-01 |
| 119 | 455 | 492.530 | .23223E-03 | .22367E-04 | .63855E+07 | 1.04E+01 |
| 119 | 454 | 441.831 | .87757E-03 | .10570E-02 | .29985E+08 | 8.30E-01 |
| 119 | 453 | 390.404 | .72261E-03 | .11809E-02 | .31623E+08 | 6.12E-01 |
| 119 | 452 | 366.722 | .11243E-02 | .12965E-02 | .55764E+08 | 8.67E-01 |
| 119 | 451 | 329.272 | .85627E-03 | .55190E-03 | .52679E+08 | 1.55E+00 |
| 119 | 450 | 298.699 | .64596E-03 | .15071E-03 | .48291E+08 | 4.29E+00 |
| 119 | 449 | 262.046 | .28029E-01 | .17691E-01 | .27226E+10 | 1.58E+00 |
| 119 | 448 | 229.247 | .72142E-02 | .98246E-02 | .91562E+09 | 7.34E-01 |
| 119 | 447 | 211.800 | .22488E+00 | .17590E+00 | .33437E+11 | 1.28E+00 |
| 119 | 446 | 94.552 | .38599E-04 | .16267E-04 | .28798E+08 | 2.37E+00 |
| 468 | 118 | 38.192 | .57141E-09 | .66344E-10 | .26130E+04 | 8.61E+00 |
| 467 | 118 | 54.703 | .33508E-06 | .15936E-06 | .74690E+06 | 2.10E+00 |
| 466 | 118 | 55.646 | .15183E-06 | .48566E-07 | .32706E+06 | 3.13E+00 |
| 465 | 118 | 57.334 | .10581E-04 | .27182E-05 | .21470E+08 | 3.89E+00 |
| 464 | 118 | 58.606 | .23007E-08 | .22802E-08 | .44680E+04 | 1.01E+00 |
| 463 | 118 | 59.804 | .22874E-05 | .47197E-06 | .42658E+07 | 4.85E+00 |
| 462 | 118 | 61.089 | .88540E-06 | .30477E-06 | .15825E+07 | 2.91E+00 |
| 461 | 118 | 61.640 | .14329E-07 | .61551E-07 | .25155E+05 | 2.33E-01 |
| 460 | 118 | 65.994 | .24693E-06 | .33116E-07 | .37818E+06 | 7.46E+00 |
| 459 | 118 | 68.707 | .10350E-05 | .30796E-06 | .14624E+07 | 3.36E+00 |
| 458 | 118 | 69.003 | .10707E-06 | .59129E-07 | .15000E+06 | 1.81E+00 |
| 457 | 118 | 154.146 | .20901E-01 | .20608E-01 | .58673E+10 | 1.01E+00 |
| 456 | 118 | 161.592 | .19307E+00 | .20274E+00 | .49320E+11 | 9.52E-01 |
| 455 | 118 | 163.408 | .17840E-01 | .19503E-01 | .44564E+10 | 9.15E-01 |
| 454 | 118 | 169.876 | .18715E-02 | .20843E-02 | .43257E+09 | 8.98E-01 |
| 453 | 118 | 178.938 | .92794E-02 | .11308E-01 | .19331E+10 | 8.21E-01 |
| 452 | 118 | 184.396 | .59664E-02 | .76598E-02 | .11704E+10 | 7.79E-01 |
| 451 | 118 | 195.581 | .10813E-03 | .15011E-03 | .18855E+08 | 7.20E-01 |
| 450 | 118 | 208.242 | .35225E-03 | .60359E-03 | .54182E+08 | 5.84E-01 |
| 449 | 118 | 230.742 | .53775E-04 | .10420E-03 | .67369E+07 | 5.16E-01 |
| 448 | 118 | 264.001 | .81109E-03 | .17890E-02 | .77624E+08 | 4.53E-01 |
| 447 | 118 | 291.669 | .21467E-06 | .11640E-05 | .16831E+05 | 1.84E-01 |
| 118 | 446 | 412.170 | .31017E-09 | .69866E-10 | .12178E+02 | 4.44E+00 |
| 468 | 117 | 37.448 | .55534E-08 | .79723E-08 | .26414E+05 | 6.97E-01 |
| 467 | 117 | 53.191 | .28514E-07 | .10570E-05 | .67224E+05 | 2.70E-02 |
| 466 | 117 | 54.082 | .92229E-07 | .30906E-05 | .21033E+06 | 2.98E-02 |
| 465 | 117 | 55.674 | .19532E-06 | .32673E-07 | .42032E+06 | 5.98E+00 |
| 464 | 117 | 56.873 | .12846E-09 | .10824E-06 | .26490E+03 | 1.19E-03 |
| 463 | 117 | 58.001 | .32239E-04 | .60344E-05 | .63922E+08 | 5.34E+00 |
| 462 | 117 | 59.209 | .13411E-05 | .22374E-06 | .25517E+07 | 5.99E+00 |
| 461 | 117 | 59.726 | .18727E-05 | .62919E-07 | .35016E+07 | 2.98E+01 |
| 460 | 117 | 63.805 | .89470E-08 | .31152E-06 | .14659E+05 | 2.87E-02 |
| 459 | 117 | 66.337 | .19733E-06 | .27509E-08 | .29909E+06 | 7.17E+01 |
| 458 | 117 | 66.613 | .16910E-04 | .13586E-05 | .25419E+08 | 1.24E+01 |
| 457 | 117 | 142.709 | .65486E-02 | .57580E-02 | .21448E+10 | 1.14E+00 |
| 456 | 117 | 149.068 | .16797E-03 | .17603E-03 | .50418E+08 | 9.54E-01 |
| 455 | 117 | 150.613 | .55376E-02 | .50817E-02 | .16283E+10 | 1.09E+00 |
| 454 | 117 | 156.090 | .63284E-01 | .62581E-01 | .17325E+11 | 1.01E+00 |
| 453 | 117 | 163.708 | .57386E-01 | .63238E-01 | .14282E+11 | 9.07E-01 |
| 452 | 117 | 168.265 | .14872E-03 | .17695E-03 | .35035E+08 | 8.40E-01 |
| 451 | 117 | 177.530 | .10715E+00 | .13112E+00 | .22677E+11 | 8.17E-01 |
| 450 | 117 | 187.899 | .90407E-05 | .16195E-05 | .17080E+07 | 5.58E+00 |
| 449 | 117 | 206.026 | .23780E-04 | .98465E-05 | .37368E+07 | 2.42E+00 |
| 448 | 117 | 232.139 | .10887E-03 | .15653E-03 | .13476E+08 | 6.96E-01 |
| 447 | 117 | 253.265 | .41512E-03 | .46391E-03 | .43167E+08 | 8.95E-01 |
| 117 | 446 | 524.581 | .10874E-08 | .99064E-11 | .26356E+02 | 1.10E+02 |
| 468 | 116 | 37.102 | .12536E-08 | .36606E-07 | .60741E+04 | 3.42E-02 |
| 467 | 116 | 52.495 | .22027E-07 | .14761E-05 | .53314E+05 | 1.49E-02 |
| 466 | 116 | 53.363 | .14158E-06 | .35980E-06 | .33164E+06 | 3.93E-01 |
| 465 | 116 | 54.913 | .32599E-05 | .11301E-06 | .72109E+07 | 2.88E+01 |
| 464 | 116 | 56.079 | .40411E-06 | .84257E-06 | .85710E+06 | 4.80E-01 |
| 463 | 116 | 57.175 | .37102E-05 | .27891E-06 | .75705E+07 | 1.33E+01 |
| 462 | 116 | 58.348 | .21412E-04 | .18853E-05 | .41951E+08 | 1.14E+01 |
| 461 | 116 | 58.850 | .11432E-03 | .79838E-05 | .22016E+09 | 1.43E+01 |
| 460 | 116 | 62.807 | .32565E-04 | .18573E-04 | .55065E+08 | 1.75E+00 |
| 459 | 116 | 65.259 | .73024E-04 | .12400E-04 | .11437E+09 | 5.89E+00 |
| 458 | 116 | 65.526 | .16162E-03 | .17745E-04 | .25107E+09 | 9.11E+00 |
| 457 | 116 | 137.810 | .34889E-02 | .31883E-02 | .12253E+10 | 1.09E+00 |
| 456 | 116 | 143.731 | .91934E-03 | .79184E-03 | .29683E+09 | 1.16E+00 |
| 455 | 116 | 145.167 | .93054E-04 | .14395E-03 | .29453E+08 | 6.46E-01 |
| 454 | 116 | 150.248 | .20033E-02 | .19302E-02 | .59190E+09 | 1.04E+00 |
| 453 | 116 | 157.294 | .16776E-01 | .17246E-01 | .45228E+10 | 9.73E-01 |
| 452 | 116 | 161.496 | .14281E+00 | .15230E+00 | .36524E+11 | 9.38E-01 |
| 451 | 116 | 170.012 | .48591E-02 | .57832E-02 | .11213E+10 | 8.40E-01 |
| 450 | 116 | 179.498 | .10622E+00 | .12614E+00 | .21989E+11 | 8.42E-01 |
| 449 | 116 | 195.969 | .35686E-03 | .35187E-03 | .61980E+08 | 1.01E+00 |
| 448 | 116 | 219.449 | .11770E-04 | .13548E-04 | .16302E+07 | 8.69E-01 |
| 447 | 116 | 238.235 | .76746E-02 | .11143E-01 | .90194E+09 | 6.89E-01 |
| 116 | 446 | 603.431 | .21279E-07 | .81348E-08 | .38978E+03 | 2.62E+00 |
| 468 | 115 | 36.869 | .15909E-06 | .79482E-07 | .78065E+06 | 2.00E+00 |
| 467 | 115 | 52.028 | .18304E-07 | .80259E-08 | .45103E+05 | 2.28E+00 |
| 466 | 115 | 52.881 | .37987E-06 | .40631E-06 | .90608E+06 | 9.35E-01 |
| 465 | 115 | 54.402 | .10184E-05 | .15696E-06 | .22952E+07 | 6.49E+00 |
| 464 | 115 | 55.547 | .11096E-04 | .81172E-05 | .23988E+08 | 1.37E+00 |
| 463 | 115 | 56.622 | .84600E-06 | .21277E-06 | .17601E+07 | 3.98E+00 |
| 462 | 115 | 57.773 | .70720E-06 | .12796E-07 | .14133E+07 | 5.53E+01 |
| 461 | 115 | 58.265 | .19176E-04 | .12970E-05 | .37677E+08 | 1.48E+01 |
| 460 | 115 | 62.140 | .63792E-04 | .37609E-05 | .11019E+09 | 1.70E+01 |
| 459 | 115 | 64.539 | .70966E-04 | .28490E-04 | .11364E+09 | 2.49E+00 |
| 458 | 115 | 64.801 | .42758E-03 | .49154E-04 | .67919E+09 | 8.70E+00 |
| 457 | 115 | 134.641 | .62442E-03 | .46877E-03 | .22975E+09 | 1.33E+00 |
| 456 | 115 | 140.287 | .56851E-05 | .47512E-05 | .19268E+07 | 1.20E+00 |
| 455 | 115 | 141.654 | .16117E-02 | .13490E-02 | .53575E+09 | 1.19E+00 |
| 454 | 115 | 146.488 | .12520E-03 | .15276E-03 | .38917E+08 | 8.20E-01 |
| 453 | 115 | 153.178 | .19457E-03 | .26625E-03 | .55312E+08 | 7.31E-01 |
| 452 | 115 | 157.160 | .13297E-01 | .14174E-01 | .35908E+10 | 9.38E-01 |
| 451 | 115 | 165.213 | .85940E-03 | .10781E-02 | .21001E+09 | 7.97E-01 |
| 450 | 115 | 174.158 | .55688E-01 | .64936E-01 | .12246E+11 | 8.58E-01 |
| 449 | 115 | 189.621 | .76190E-01 | .96055E-01 | .14134E+11 | 7.93E-01 |
| 448 | 115 | 211.520 | .13372E-02 | .15947E-02 | .19935E+09 | 8.39E-01 |
| 447 | 115 | 228.919 | .43659E-01 | .66276E-01 | .55570E+10 | 6.59E-01 |
| 115 | 446 | 672.782 | .76896E-08 | .20856E-06 | .11332E+03 | 3.69E-02 |
| 468 | 114 | 35.614 | .51998E-07 | .84229E-07 | .27345E+06 | 6.17E-01 |
| 467 | 114 | 49.565 | .40841E-06 | .89393E-07 | .11089E+07 | 4.57E+00 |
| 466 | 114 | 50.338 | .19663E-06 | .71337E-06 | .51760E+06 | 2.76E-01 |
| 465 | 114 | 51.714 | .32726E-06 | .80272E-08 | .81620E+06 | 4.08E+01 |
| 464 | 114 | 52.747 | .30577E-06 | .33809E-05 | .73304E+06 | 9.04E-02 |
| 463 | 114 | 53.716 | .24135E-06 | .75805E-07 | .55791E+06 | 3.18E+00 |
| 462 | 114 | 54.750 | .80007E-06 | .18637E-06 | .17803E+07 | 4.29E+00 |
| 461 | 114 | 55.192 | .14794E-04 | .20779E-05 | .32395E+08 | 7.12E+00 |
| 460 | 114 | 58.658 | .92209E-05 | .15416E-05 | .17876E+08 | 5.98E+00 |
| 459 | 114 | 60.791 | .59963E-04 | .32557E-05 | .10823E+09 | 1.84E+01 |
| 458 | 114 | 61.023 | .42685E-04 | .81536E-06 | .76459E+08 | 5.24E+01 |
| 457 | 114 | 119.295 | .87441E-04 | .63616E-04 | .40983E+08 | 1.37E+00 |
| 456 | 114 | 123.706 | .73944E-03 | .55402E-03 | .32229E+09 | 1.33E+00 |
| 455 | 114 | 124.768 | .20697E-04 | .14860E-04 | .88683E+07 | 1.39E+00 |
| 454 | 114 | 128.503 | .79554E-04 | .57269E-04 | .32134E+08 | 1.39E+00 |
| 453 | 114 | 133.623 | .20409E-03 | .15718E-03 | .76243E+08 | 1.30E+00 |
| 452 | 114 | 136.643 | .10624E-03 | .66108E-04 | .37953E+08 | 1.61E+00 |
| 451 | 114 | 142.690 | .11105E-03 | .10767E-03 | .36379E+08 | 1.03E+00 |
| 450 | 114 | 149.313 | .12676E-01 | .11412E-01 | .37924E+10 | 1.11E+00 |
| 449 | 114 | 160.537 | .15578E-02 | .15448E-02 | .40317E+09 | 1.01E+00 |
| 448 | 114 | 175.960 | .44872E-01 | .52956E-01 | .96667E+10 | 8.47E-01 |
| 447 | 114 | 187.836 | .63190E-01 | .83800E-01 | .11946E+11 | 7.54E-01 |
| 114 | 446 | 1883.458 | .36507E-07 | .42356E-07 | .68644E+02 | 8.62E-01 |
| 468 | 113 | 32.527 | .33694E-01 | .32055E-01 | .21243E+12 | 1.05E+00 |
| 467 | 113 | 43.781 | .33766E-05 | .70826E-06 | .11750E+08 | 4.77E+00 |
| 466 | 113 | 44.383 | .57647E-06 | .69168E-08 | .19520E+07 | 8.33E+01 |
| 465 | 113 | 45.450 | .40065E-07 | .11746E-07 | .12937E+06 | 3.41E+00 |
| 464 | 113 | 46.246 | .49893E-06 | .38750E-07 | .15561E+07 | 1.29E+01 |
| 463 | 113 | 46.989 | .13161E-04 | .12588E-04 | .39759E+08 | 1.05E+00 |
| 462 | 113 | 47.778 | .11448E-05 | .57564E-06 | .33451E+07 | 1.99E+00 |
| 461 | 113 | 48.115 | .18864E-05 | .90504E-08 | .54351E+07 | 2.08E+02 |
| 460 | 113 | 50.727 | .72598E-04 | .67310E-04 | .18818E+09 | 1.08E+00 |
| 459 | 113 | 52.315 | .74578E-04 | .17578E-03 | .18176E+09 | 4.24E-01 |
| 458 | 113 | 52.486 | .45151E-03 | .44267E-03 | .10932E+10 | 1.02E+00 |
| 457 | 113 | 90.515 | .18825E-06 | .54989E-07 | .15326E+06 | 3.42E+00 |
| 456 | 113 | 93.032 | .63405E-07 | .17290E-08 | .48864E+05 | 3.67E+01 |
| 455 | 113 | 93.632 | .20768E-06 | .31304E-07 | .15801E+06 | 6.63E+00 |
| 454 | 113 | 95.720 | .16319E-09 | .24066E-09 | .11880E+03 | 6.78E-01 |
| 453 | 113 | 98.532 | .15716E-07 | .29856E-07 | .10797E+05 | 5.26E-01 |
| 452 | 113 | 100.164 | .65397E-08 | .10578E-06 | .43478E+04 | 6.18E-02 |
| 451 | 113 | 103.376 | .36901E-09 | .54590E-09 | .23032E+03 | 6.76E-01 |
| 450 | 113 | 106.808 | .12969E-05 | .13746E-08 | .75826E+06 | 9.43E+02 |
| 449 | 113 | 112.431 | .13291E-07 | .79262E-07 | .70131E+04 | 1.68E-01 |
| 448 | 113 | 119.784 | .56523E-06 | .25244E-06 | .26276E+06 | 2.24E+00 |
| 447 | 113 | 125.171 | .24611E-06 | .77656E-07 | .10478E+06 | 3.17E+00 |
| 446 | 113 | 468.532 | .41034E-01 | .41528E-01 | .12468E+10 | 9.88E-01 |
| 134 | 569 | 196.616 | .14644E-03 | .18589E-03 | .16844E+08 | 7.88E-01 |
| 134 | 568 | 169.854 | .44696E-02 | .52512E-02 | .68891E+09 | 8.51E-01 |
| 134 | 567 | 155.056 | .18458E+00 | .17054E+00 | .34139E+11 | 1.08E+00 |
| 134 | 566 | 146.477 | .99971E-01 | .88691E-01 | .20719E+11 | 1.13E+00 |
| 134 | 565 | 143.050 | .79509E-03 | .48464E-03 | .17278E+09 | 1.64E+00 |
| 134 | 564 | 138.371 | .25383E-02 | .18946E-02 | .58951E+09 | 1.34E+00 |
| 134 | 563 | 119.222 | .17278E-03 | .15748E-03 | .54052E+08 | 1.10E+00 |
| 134 | 562 | 114.740 | .42750E-03 | .29422E-03 | .14439E+09 | 1.45E+00 |
| 134 | 561 | 62.037 | .68870E-07 | .17080E-07 | .79574E+05 | 4.03E+00 |
| 134 | 560 | 60.240 | .15430E-06 | .41346E-07 | .18908E+06 | 3.73E+00 |
| 134 | 559 | 59.021 | .44290E-07 | .75733E-06 | .56538E+05 | 5.85E-02 |
| 134 | 558 | 58.477 | .48893E-06 | .19770E-06 | .63579E+06 | 2.47E+00 |
| 134 | 557 | 57.896 | .31228E-04 | .24524E-05 | .41428E+08 | 1.27E+01 |
| 134 | 556 | 57.011 | .20486E-04 | .15282E-06 | .28028E+08 | 1.34E+02 |
| 134 | 555 | 56.574 | .49625E-04 | .10436E-04 | .68947E+08 | 4.76E+00 |
| 134 | 554 | 54.931 | .32866E-06 | .21004E-06 | .48434E+06 | 1.56E+00 |
| 134 | 553 | 53.849 | .18934E-05 | .64891E-06 | .29035E+07 | 2.92E+00 |
| 134 | 552 | 53.018 | .19571E-07 | .12706E-06 | .30961E+05 | 1.54E-01 |
| 134 | 551 | 51.469 | .86865E-07 | .18549E-06 | .14581E+06 | 4.68E-01 |
| 134 | 550 | 34.289 | .31198E-03 | .28626E-03 | .11799E+10 | 1.09E+00 |
| 133 | 569 | 246.296 | .92970E-02 | .63790E-02 | .68150E+09 | 1.46E+00 |
| 133 | 568 | 205.698 | .36072E-03 | .24758E-03 | .37910E+08 | 1.46E+00 |
| 133 | 567 | 184.387 | .16198E-02 | .15684E-02 | .21186E+09 | 1.03E+00 |
| 133 | 566 | 172.381 | .70566E-02 | .68761E-02 | .10560E+10 | 1.03E+00 |
| 133 | 565 | 167.654 | .74328E-01 | .73826E-01 | .11759E+11 | 1.01E+00 |
| 133 | 564 | 161.263 | .67378E-01 | .70612E-01 | .11521E+11 | 9.54E-01 |
| 133 | 563 | 135.836 | .16595E+00 | .17337E+00 | .39992E+11 | 9.57E-01 |
| 133 | 562 | 130.048 | .55816E-03 | .70939E-03 | .14675E+09 | 7.87E-01 |
| 133 | 561 | 66.253 | .79904E-05 | .57560E-05 | .80946E+07 | 1.39E+00 |

| 133 | 560 | 64.208 | .10188E-03 | .59594E-04 | .10989E+09 | 1.71E+00 |
| --- | --- | --- | --- | --- | --- | --- |
| 133 | 559 | 62.825 | .90469E-04 | .32754E-04 | .10192E+09 | 2.76E+00 |
| 133 | 558 | 62.210 | .46047E-06 | .11628E-08 | .52909E+06 | 3.96E+02 |
| 133 | 557 | 61.551 | .52214E-06 | .24937E-08 | .61285E+06 | 2.09E+02 |
| 133 | 556 | 60.552 | .13313E-05 | .10642E-06 | .16146E+07 | 1.25E+01 |
| 133 | 555 | 60.059 | .24211E-06 | .15664E-06 | .29846E+06 | 1.55E+00 |
| 133 | 554 | 58.212 | .44915E-04 | .70883E-05 | .58940E+08 | 6.34E+00 |
| 133 | 553 | 56.998 | .16046E-04 | .11500E-04 | .21962E+08 | 1.40E+00 |
| 133 | 552 | 56.068 | .40664E-05 | .23474E-04 | .57521E+07 | 1.73E-01 |
| 133 | 551 | 54.339 | .31826E-04 | .15996E-04 | .47930E+08 | 1.99E+00 |
| 133 | 550 | 35.539 | .11315E+00 | .10344E+00 | .39837E+12 | 1.09E+00 |
| 132 | 569 | 380.427 | .36809E-01 | .34693E-01 | .11310E+10 | 1.06E+00 |
| 132 | 568 | 291.548 | .44117E-04 | .46073E-05 | .23079E+07 | 9.58E+00 |
| 132 | 567 | 250.510 | .11019E-02 | .99007E-03 | .78082E+08 | 1.11E+00 |
| 132 | 566 | 228.855 | .57801E-03 | .54476E-03 | .49074E+08 | 1.06E+00 |
| 132 | 565 | 220.597 | .80765E-02 | .93568E-02 | .73801E+09 | 8.63E-01 |
| 132 | 564 | 209.664 | .97627E-03 | .54437E-03 | .98756E+08 | 1.79E+00 |
| 132 | 563 | 168.626 | .31249E-01 | .23921E-01 | .48867E+10 | 1.31E+00 |
| 132 | 562 | 159.797 | .48115E+00 | .49181E+00 | .83789E+11 | 9.78E-01 |
| 132 | 561 | 73.195 | .11098E-06 | .81512E-07 | .92115E+05 | 1.36E+00 |
| 132 | 560 | 70.707 | .51306E-04 | .27626E-04 | .45633E+08 | 1.86E+00 |
| 132 | 559 | 69.034 | .52893E-04 | .25929E-04 | .49354E+08 | 2.04E+00 |
| 132 | 558 | 68.291 | .20157E-04 | .68883E-05 | .19220E+08 | 2.93E+00 |
| 132 | 557 | 67.499 | .12097E-03 | .62386E-04 | .11807E+09 | 1.94E+00 |
| 132 | 556 | 66.299 | .56037E-05 | .49480E-05 | .56689E+07 | 1.13E+00 |
| 132 | 555 | 65.709 | .90436E-05 | .10651E-05 | .93140E+07 | 8.49E+00 |
| 132 | 554 | 63.504 | .41301E-04 | .78637E-05 | .45541E+08 | 5.25E+00 |
| 132 | 553 | 62.062 | .41489E-03 | .96480E-04 | .47899E+09 | 4.30E+00 |
| 132 | 552 | 60.960 | .78889E-07 | .37542E-06 | .94398E+05 | 2.10E-01 |
| 132 | 551 | 58.922 | .13736E-03 | .17050E-03 | .17593E+09 | 8.06E-01 |
| 132 | 550 | 37.444 | .19381E-01 | .17622E-01 | .61469E+11 | 1.10E+00 |
| 131 | 569 | 562.948 | .30865E-04 | .28317E-03 | .43309E+06 | 1.09E-01 |
| 131 | 568 | 387.941 | .14717E-02 | .42766E-02 | .43485E+08 | 3.44E-01 |
| 131 | 567 | 318.512 | .53355E-03 | .59410E-03 | .23387E+08 | 8.98E-01 |
| 131 | 566 | 284.308 | .16825E-02 | .16522E-02 | .92561E+08 | 1.02E+00 |
| 131 | 565 | 271.674 | .14473E-01 | .10714E-01 | .87196E+09 | 1.35E+00 |
| 131 | 564 | 255.279 | .28407E-01 | .21823E-01 | .19384E+10 | 1.30E+00 |
| 131 | 563 | 196.928 | .35347E+00 | .27125E+00 | .40530E+11 | 1.30E+00 |
| 131 | 562 | 184.991 | .20832E-02 | .33678E-02 | .27069E+09 | 6.19E-01 |
| 131 | 561 | 78.065 | .79467E-08 | .46219E-06 | .57984E+04 | 1.72E-02 |
| 131 | 560 | 75.241 | .18991E-05 | .29752E-05 | .14917E+07 | 6.38E-01 |
| 131 | 559 | 73.349 | .55112E-08 | .14317E-05 | .45551E+04 | 3.85E-03 |
| 131 | 558 | 72.512 | .34807E-10 | .33526E-07 | .29437E+02 | 1.04E-03 |
| 131 | 557 | 71.619 | .80958E-06 | .76217E-09 | .70185E+06 | 1.06E+03 |
| 131 | 556 | 70.270 | .43622E-08 | .34536E-09 | .39283E+04 | 1.26E+01 |
| 131 | 555 | 69.607 | .35990E-06 | .14584E-08 | .33031E+06 | 2.47E+02 |
| 131 | 554 | 67.137 | .77691E-06 | .12165E-05 | .76645E+06 | 6.39E-01 |
| 131 | 553 | 65.528 | .12294E-04 | .16696E-05 | .12731E+08 | 7.36E+00 |
| 131 | 552 | 64.301 | .12798E-03 | .85842E-04 | .13764E+09 | 1.49E+00 |
| 131 | 551 | 62.037 | .18724E-04 | .42102E-05 | .21634E+08 | 4.45E+00 |
| 131 | 550 | 38.678 | .83336E+00 | .76123E+00 | .24771E+13 | 1.09E+00 |
| 569 | 130 | 187.367 | .28506E-02 | .25512E-02 | .81241E+09 | 1.12E+00 |
| 568 | 130 | 220.469 | .39546E-01 | .38273E-01 | .81402E+10 | 1.03E+00 |
| 567 | 130 | 251.642 | .17042E-02 | .18503E-02 | .26926E+09 | 9.21E-01 |
| 566 | 130 | 278.073 | .92553E-06 | .12569E-04 | .11976E+06 | 7.36E-02 |
| 565 | 130 | 291.324 | .16659E-02 | .19878E-02 | .19639E+09 | 8.38E-01 |
| 564 | 130 | 312.870 | .85732E-03 | .13301E-02 | .87627E+08 | 6.45E-01 |
| 563 | 130 | 491.284 | .13665E-03 | .13294E-03 | .56647E+07 | 1.03E+00 |
| 562 | 130 | 585.542 | .30250E-06 | .42850E-05 | .88274E+04 | 7.06E-02 |
| 130 | 561 | 175.552 | .81200E-01 | .69672E-01 | .11716E+11 | 1.17E+00 |
| 130 | 560 | 161.888 | .14393E-02 | .21748E-02 | .24420E+09 | 6.62E-01 |
| 130 | 559 | 153.376 | .22397E-03 | .13824E-03 | .42336E+08 | 1.62E+00 |
| 130 | 558 | 149.758 | .10487E-01 | .93042E-02 | .20792E+10 | 1.13E+00 |
| 130 | 557 | 146.000 | .18999E-03 | .22586E-03 | .39635E+08 | 8.41E-01 |
| 130 | 556 | 140.501 | .45665E-04 | .28455E-04 | .10286E+08 | 1.60E+00 |
| 130 | 555 | 137.876 | .42757E-08 | .43209E-05 | .10002E+04 | 9.90E-04 |
| 130 | 554 | 128.511 | .50338E-03 | .64189E-03 | .13554E+09 | 7.84E-01 |
| 130 | 553 | 122.741 | .51258E-04 | .88500E-04 | .15129E+08 | 5.79E-01 |
| 130 | 552 | 118.507 | .68039E-03 | .37346E-03 | .21543E+09 | 1.82E+00 |
| 130 | 551 | 111.039 | .33021E-03 | .19637E-03 | .11909E+09 | 1.68E+00 |
| 130 | 550 | 53.359 | .75777E-05 | .50414E-06 | .11835E+08 | 1.50E+01 |
| 569 | 129 | 175.680 | .54252E-01 | .46980E-01 | .17587E+11 | 1.15E+00 |
| 568 | 129 | 204.465 | .91422E-02 | .73028E-02 | .21879E+10 | 1.25E+00 |
| 567 | 129 | 231.005 | .27265E-02 | .21286E-02 | .51119E+09 | 1.28E+00 |
| 566 | 129 | 253.088 | .96920E-02 | .10209E-01 | .15139E+10 | 9.49E-01 |
| 565 | 129 | 264.017 | .11572E-01 | .11148E-01 | .16610E+10 | 1.04E+00 |
| 564 | 129 | 281.592 | .15141E-02 | .21610E-02 | .19105E+09 | 7.01E-01 |
| 563 | 129 | 418.321 | .17501E-04 | .38599E-04 | .10006E+07 | 4.53E-01 |
| 562 | 129 | 484.768 | .20815E-03 | .60795E-04 | .88620E+07 | 3.42E+00 |
| 129 | 561 | 187.221 | .86123E-02 | .95674E-02 | .10926E+10 | 9.00E-01 |
| 129 | 560 | 171.760 | .14255E-01 | .98802E-02 | .21486E+10 | 1.44E+00 |
| 129 | 559 | 162.209 | .99256E-02 | .82025E-02 | .16774E+10 | 1.21E+00 |
| 129 | 558 | 158.168 | .15971E-01 | .16194E-01 | .28388E+10 | 9.86E-01 |
| 129 | 557 | 153.981 | .25169E-02 | .25559E-02 | .47204E+09 | 9.85E-01 |
| 129 | 556 | 147.877 | .64749E-02 | .56849E-02 | .13167E+10 | 1.14E+00 |
| 129 | 555 | 144.972 | .78451E-02 | .61769E-02 | .16598E+10 | 1.27E+00 |
| 129 | 554 | 134.655 | .33825E-02 | .32646E-02 | .82954E+09 | 1.04E+00 |
| 129 | 553 | 128.333 | .16245E-01 | .14591E-01 | .43861E+10 | 1.11E+00 |
| 129 | 552 | 123.712 | .87756E-05 | .10182E-05 | .25497E+07 | 8.62E+00 |
| 129 | 551 | 115.596 | .10317E-01 | .60271E-02 | .34332E+10 | 1.71E+00 |
| 129 | 550 | 54.390 | .35231E-06 | .70083E-06 | .52958E+06 | 5.03E-01 |
| 569 | 128 | 167.901 | .14002E-01 | .11299E-01 | .49694E+10 | 1.24E+00 |
| 568 | 128 | 194.004 | .41137E-01 | .33218E-01 | .10935E+11 | 1.24E+00 |
| 567 | 128 | 217.740 | .68587E-03 | .41997E-03 | .14474E+09 | 1.63E+00 |
| 566 | 128 | 237.252 | .24140E-02 | .29387E-02 | .42908E+09 | 8.21E-01 |
| 565 | 128 | 246.831 | .21155E-01 | .21975E-01 | .34740E+10 | 9.63E-01 |
| 564 | 128 | 262.126 | .35000E-02 | .41640E-02 | .50965E+09 | 8.41E-01 |
| 563 | 128 | 376.757 | .19259E-04 | .51869E-04 | .13575E+07 | 3.71E-01 |
| 562 | 128 | 429.818 | .17650E-05 | .10425E-03 | .95585E+05 | 1.69E-02 |
| 128 | 561 | 196.945 | .35274E-02 | .41975E-02 | .40439E+09 | 8.40E-01 |
| 128 | 560 | 179.909 | .26643E-02 | .31523E-02 | .36603E+09 | 8.45E-01 |
| 128 | 559 | 169.458 | .21132E-01 | .18740E-01 | .32723E+10 | 1.13E+00 |
| 128 | 558 | 165.053 | .21026E-01 | .18872E-01 | .34321E+10 | 1.11E+00 |
| 128 | 557 | 160.499 | .77038E-02 | .45308E-02 | .13298E+10 | 1.70E+00 |
| 128 | 556 | 153.878 | .12653E-01 | .11329E-01 | .23762E+10 | 1.12E+00 |
| 128 | 555 | 150.735 | .12666E-01 | .12662E-01 | .24789E+10 | 1.00E+00 |
| 128 | 554 | 139.613 | .11791E-01 | .83979E-02 | .26898E+10 | 1.40E+00 |
| 128 | 553 | 132.829 | .58504E-02 | .60074E-02 | .14745E+10 | 9.74E-01 |
| 128 | 552 | 127.884 | .48417E-06 | .14365E-04 | .13165E+06 | 3.37E-02 |
| 128 | 551 | 119.231 | .33998E-02 | .20294E-02 | .10635E+10 | 1.68E+00 |
| 128 | 550 | 55.181 | .16254E-05 | .12164E-04 | .23736E+07 | 1.34E-01 |
| 569 | 127 | 154.556 | .97986E-02 | .74868E-02 | .41041E+10 | 1.31E+00 |
| 568 | 127 | 176.404 | .66641E-04 | .33825E-04 | .21426E+08 | 1.97E+00 |
| 567 | 127 | 195.813 | .17947E-04 | .30833E-04 | .46831E+07 | 5.82E-01 |
| 566 | 127 | 211.452 | .44788E-01 | .42236E-01 | .10022E+11 | 1.06E+00 |
| 565 | 127 | 219.028 | .70588E-04 | .10777E-03 | .14722E+08 | 6.55E-01 |
| 564 | 127 | 230.988 | .12668E-03 | .15297E-03 | .23754E+08 | 8.28E-01 |
| 563 | 127 | 315.606 | .32165E-04 | .10612E-02 | .32308E+07 | 3.03E-02 |
| 562 | 127 | 352.008 | .72187E-04 | .25407E-03 | .58288E+07 | 2.84E-01 |
| 127 | 561 | 219.141 | .21671E-02 | .27240E-02 | .20067E+09 | 7.96E-01 |
| 127 | 560 | 198.252 | .41425E-01 | .42283E-01 | .46867E+10 | 9.80E-01 |
| 127 | 559 | 185.636 | .45396E-02 | .44547E-02 | .58578E+09 | 1.02E+00 |
| 127 | 558 | 180.362 | .10265E-01 | .11442E-01 | .14032E+10 | 8.97E-01 |
| 127 | 557 | 174.938 | .66198E-03 | .58750E-03 | .96187E+08 | 1.13E+00 |
| 127 | 556 | 167.102 | .30554E-01 | .24648E-01 | .48657E+10 | 1.24E+00 |
| 127 | 555 | 163.402 | .91659E-01 | .81817E-01 | .15265E+11 | 1.12E+00 |
| 127 | 554 | 150.412 | .26423E-01 | .22517E-01 | .51934E+10 | 1.17E+00 |
| 127 | 553 | 142.568 | .21392E-01 | .14196E-01 | .46800E+10 | 1.51E+00 |
| 127 | 552 | 136.887 | .76613E-02 | .69615E-02 | .18181E+10 | 1.10E+00 |
| 127 | 551 | 127.019 | .11389E-01 | .63928E-02 | .31391E+10 | 1.78E+00 |
| 127 | 550 | 56.793 | .87264E-04 | .13811E-03 | .12031E+09 | 6.32E-01 |
| 569 | 126 | 148.734 | .12041E-01 | .91118E-02 | .54459E+10 | 1.32E+00 |
| 568 | 126 | 168.860 | .41457E-03 | .24892E-03 | .14547E+09 | 1.67E+00 |
| 567 | 126 | 186.561 | .95027E-03 | .72570E-03 | .27317E+09 | 1.31E+00 |
| 566 | 126 | 200.704 | .35522E-01 | .30159E-01 | .88229E+10 | 1.18E+00 |
| 565 | 126 | 207.516 | .42030E-02 | .45097E-02 | .97652E+09 | 9.32E-01 |
| 564 | 126 | 218.221 | .80824E-04 | .26776E-04 | .16981E+08 | 3.02E+00 |
| 563 | 126 | 292.246 | .69820E-04 | .17099E-03 | .81792E+07 | 4.08E-01 |
| 562 | 126 | 323.194 | .16488E-02 | .20532E-02 | .15793E+09 | 8.03E-01 |
| 126 | 561 | 232.018 | .12627E-02 | .17167E-02 | .10431E+09 | 7.36E-01 |
| 126 | 560 | 208.733 | .48658E-02 | .59949E-02 | .49661E+09 | 8.12E-01 |
| 126 | 559 | 194.794 | .29233E-01 | .28585E-01 | .34258E+10 | 1.02E+00 |
| 126 | 558 | 188.996 | .22745E-04 | .24848E-04 | .28316E+07 | 9.15E-01 |
| 126 | 557 | 183.049 | .47512E-02 | .39185E-02 | .63053E+09 | 1.21E+00 |
| 126 | 556 | 174.487 | .47373E-01 | .40722E-01 | .69191E+10 | 1.16E+00 |
| 126 | 555 | 170.456 | .99049E-01 | .94393E-01 | .15159E+11 | 1.05E+00 |
| 126 | 554 | 156.369 | .58614E-01 | .49801E-01 | .10660E+11 | 1.18E+00 |
| 126 | 553 | 147.908 | .16274E-01 | .14357E-01 | .33078E+10 | 1.13E+00 |
| 126 | 552 | 141.803 | .91629E-02 | .89830E-02 | .20263E+10 | 1.02E+00 |
| 126 | 551 | 131.241 | .31587E-02 | .26580E-02 | .81546E+09 | 1.19E+00 |
| 126 | 550 | 57.622 | .93591E-04 | .19179E-03 | .12534E+09 | 4.88E-01 |
| 569 | 125 | 141.203 | .39632E-04 | .40612E-04 | .19888E+08 | 9.76E-01 |
| 568 | 125 | 159.219 | .82652E-03 | .39709E-03 | .32620E+09 | 2.08E+00 |
| 567 | 125 | 174.862 | .12912E-02 | .95927E-03 | .42251E+09 | 1.35E+00 |
| 566 | 125 | 187.228 | .76706E-04 | .41211E-04 | .21893E+08 | 1.86E+00 |
| 565 | 125 | 193.143 | .39887E-01 | .30765E-01 | .10698E+11 | 1.30E+00 |
| 564 | 125 | 202.384 | .43091E-01 | .30685E-01 | .10526E+11 | 1.40E+00 |
| 563 | 125 | 264.524 | .87311E-03 | .10742E-02 | .12484E+09 | 8.13E-01 |
| 562 | 125 | 289.627 | .30208E-02 | .40445E-02 | .36030E+09 | 7.47E-01 |
| 125 | 561 | 253.074 | .21757E-03 | .13008E-02 | .15106E+08 | 1.67E-01 |
| 125 | 560 | 225.621 | .21398E-01 | .14643E-01 | .18692E+10 | 1.46E+00 |
| 125 | 559 | 209.423 | .27730E-03 | .69513E-03 | .28115E+08 | 3.99E-01 |
| 125 | 558 | 202.736 | .70069E-02 | .62764E-02 | .75807E+09 | 1.12E+00 |
| 125 | 557 | 195.908 | .15221E-01 | .13261E-01 | .17635E+10 | 1.15E+00 |
| 125 | 556 | 186.133 | .27840E-03 | .26851E-03 | .35732E+08 | 1.04E+00 |
| 125 | 555 | 181.554 | .87761E-02 | .82759E-02 | .11839E+10 | 1.06E+00 |
| 125 | 554 | 165.658 | .16262E-01 | .12231E-01 | .26351E+10 | 1.33E+00 |
| 125 | 553 | 156.193 | .32256E-03 | .16821E-03 | .58793E+08 | 1.92E+00 |
| 125 | 552 | 149.400 | .15377E-01 | .98547E-02 | .30635E+10 | 1.56E+00 |
| 125 | 551 | 137.723 | .74502E-02 | .31745E-02 | .17466E+10 | 2.35E+00 |
| 125 | 550 | 58.838 | .20088E-04 | .23467E-03 | .25803E+08 | 8.56E-02 |
| 569 | 124 | 135.601 | .89994E-04 | .11879E-03 | .48968E+08 | 7.58E-01 |
| 568 | 124 | 152.132 | .39578E-03 | .22508E-03 | .17109E+09 | 1.76E+00 |
| 567 | 124 | 166.352 | .66448E-03 | .45002E-03 | .24024E+09 | 1.48E+00 |
| 566 | 124 | 177.505 | .22919E-02 | .13894E-02 | .72777E+09 | 1.65E+00 |
| 565 | 124 | 182.813 | .26050E-03 | .30544E-03 | .77985E+08 | 8.53E-01 |
| 564 | 124 | 191.070 | .30671E-01 | .21902E-01 | .84054E+10 | 1.40E+00 |
| 563 | 124 | 245.523 | .67827E-02 | .57790E-02 | .11258E+10 | 1.17E+00 |
| 562 | 124 | 267.003 | .15542E-01 | .14916E-01 | .21813E+10 | 1.04E+00 |
| 124 | 561 | 273.310 | .83607E-04 | .29532E-03 | .49771E+07 | 2.83E-01 |
| 124 | 560 | 241.567 | .31775E-02 | .36599E-02 | .24213E+09 | 8.68E-01 |
| 124 | 559 | 223.092 | .34417E-01 | .32502E-01 | .30750E+10 | 1.06E+00 |
| 124 | 558 | 215.519 | .68444E-02 | .57248E-02 | .65525E+09 | 1.20E+00 |
| 124 | 557 | 207.820 | .28815E-01 | .25808E-01 | .29668E+10 | 1.12E+00 |
| 124 | 556 | 196.853 | .31523E-04 | .76051E-05 | .36173E+07 | 4.14E+00 |
| 124 | 555 | 191.738 | .74320E-02 | .82057E-02 | .89893E+09 | 9.06E-01 |
| 124 | 554 | 174.096 | .72067E-02 | .64500E-02 | .10573E+10 | 1.12E+00 |
| 124 | 553 | 163.672 | .23005E-01 | .29556E-01 | .38187E+10 | 7.78E-01 |
| 124 | 552 | 156.229 | .74889E-05 | .10463E-03 | .13644E+07 | 7.16E-02 |
| 124 | 551 | 143.505 | .43877E-01 | .30594E-01 | .94742E+10 | 1.43E+00 |
| 124 | 550 | 59.868 | .39203E-05 | .37330E-04 | .48638E+07 | 1.05E-01 |
| 569 | 123 | 129.980 | .15771E-04 | .10506E-04 | .93396E+07 | 1.50E+00 |
| 568 | 123 | 145.093 | .16022E-01 | .12557E-01 | .76149E+10 | 1.28E+00 |
| 567 | 123 | 157.972 | .34714E-03 | .32926E-03 | .13918E+09 | 1.05E+00 |
| 566 | 123 | 167.996 | .56160E-03 | .61433E-03 | .19909E+09 | 9.14E-01 |
| 565 | 123 | 172.742 | .23417E-03 | .25711E-03 | .78516E+08 | 9.11E-01 |
| 564 | 123 | 180.097 | .62700E-04 | .70747E-04 | .19341E+08 | 8.86E-01 |
| 563 | 123 | 227.695 | .18120E-03 | .22305E-03 | .34968E+08 | 8.12E-01 |
| 562 | 123 | 246.052 | .10347E-04 | .86086E-05 | .17100E+07 | 1.20E+00 |
| 123 | 561 | 299.406 | .95432E-01 | .11957E+00 | .47339E+10 | 7.98E-01 |
| 123 | 560 | 261.729 | .36262E-02 | .45974E-02 | .23539E+09 | 7.89E-01 |
| 123 | 559 | 240.179 | .90622E-02 | .91843E-02 | .69856E+09 | 9.87E-01 |
| 123 | 558 | 231.425 | .25371E-01 | .19687E-01 | .21065E+10 | 1.29E+00 |
| 123 | 557 | 222.570 | .13372E-02 | .95225E-03 | .12003E+09 | 1.40E+00 |
| 123 | 556 | 210.039 | .43262E-02 | .28139E-02 | .43606E+09 | 1.54E+00 |
| 123 | 555 | 204.226 | .62644E-02 | .40498E-02 | .66789E+09 | 1.55E+00 |
| 123 | 554 | 184.330 | .23346E-02 | .22016E-02 | .30554E+09 | 1.06E+00 |
| 123 | 553 | 172.685 | .37250E-04 | .36568E-04 | .55547E+07 | 1.02E+00 |
| 123 | 552 | 164.421 | .11242E-02 | .10207E-02 | .18491E+09 | 1.10E+00 |
| 123 | 551 | 150.387 | .54581E-04 | .46433E-04 | .10732E+08 | 1.18E+00 |
| 123 | 550 | 61.033 | .40468E-05 | .35965E-05 | .48308E+07 | 1.13E+00 |
| 569 | 122 | 121.512 | .76955E-04 | .38163E-04 | .52146E+08 | 2.02E+00 |
| 568 | 122 | 134.621 | .10684E-02 | .46449E-03 | .58984E+09 | 2.30E+00 |
| 567 | 122 | 145.637 | .10711E+00 | .67010E-01 | .50528E+11 | 1.60E+00 |
| 566 | 122 | 154.115 | .43984E-02 | .13556E-02 | .18528E+10 | 3.24E+00 |
| 565 | 122 | 158.100 | .69202E-02 | .61592E-02 | .27700E+10 | 1.12E+00 |
| 564 | 122 | 164.238 | .81042E-03 | .67994E-03 | .30060E+09 | 1.19E+00 |
| 563 | 122 | 202.923 | .28480E-04 | .34413E-04 | .69201E+07 | 8.28E-01 |
| 562 | 122 | 217.376 | .86011E-04 | .14031E-03 | .18212E+08 | 6.13E-01 |
| 122 | 561 | 356.659 | .12926E-04 | .89826E-05 | .45185E+06 | 1.44E+00 |
| 122 | 560 | 304.451 | .27977E-02 | .30754E-02 | .13422E+09 | 9.10E-01 |
| 122 | 559 | 275.679 | .68292E-02 | .74096E-02 | .39958E+09 | 9.22E-01 |
| 122 | 558 | 264.207 | .89531E-03 | .12397E-02 | .57033E+08 | 7.22E-01 |
| 122 | 557 | 252.728 | .53444E-01 | .49027E-01 | .37207E+10 | 1.09E+00 |
| 122 | 556 | 236.693 | .44516E-01 | .30260E-01 | .35334E+10 | 1.47E+00 |
| 122 | 555 | 229.337 | .67296E-01 | .54244E-01 | .56896E+10 | 1.24E+00 |
| 122 | 554 | 204.545 | .18326E-04 | .27773E-04 | .19477E+07 | 6.60E-01 |
| 122 | 553 | 190.305 | .84392E-06 | .17581E-05 | .10362E+06 | 4.80E-01 |
| 122 | 552 | 180.316 | .30100E-03 | .29060E-03 | .41167E+08 | 1.04E+00 |
| 122 | 551 | 163.577 | .42891E-03 | .36698E-03 | .71279E+08 | 1.17E+00 |
| 122 | 550 | 63.098 | .11909E-05 | .31260E-05 | .13301E+07 | 3.81E-01 |
| 569 | 121 | 119.659 | .32373E-03 | .23664E-03 | .22621E+09 | 1.37E+00 |
| 568 | 121 | 132.350 | .10015E-04 | .22146E-04 | .57203E+07 | 4.52E-01 |
| 567 | 121 | 142.983 | .12231E-04 | .17227E-04 | .59859E+07 | 7.10E-01 |
| 566 | 121 | 151.145 | .25619E-04 | .28730E-04 | .11220E+08 | 8.92E-01 |
| 565 | 121 | 154.977 | .31945E-04 | .82609E-05 | .13307E+08 | 3.87E+00 |
| 564 | 121 | 160.870 | .60395E-02 | .58548E-02 | .23349E+10 | 1.03E+00 |
| 563 | 121 | 197.806 | .66118E-01 | .51612E-01 | .16907E+11 | 1.28E+00 |
| 562 | 121 | 211.515 | .16789E-04 | .10110E-04 | .37547E+07 | 1.66E+00 |
| 121 | 561 | 373.645 | .10632E-03 | .67392E-06 | .33863E+07 | 1.58E+02 |
| 121 | 560 | 316.743 | .38792E-02 | .29652E-02 | .17194E+09 | 1.31E+00 |
| 121 | 559 | 285.719 | .67735E-02 | .61513E-02 | .36896E+09 | 1.10E+00 |
| 121 | 558 | 273.414 | .53211E-04 | .62004E-04 | .31652E+07 | 8.58E-01 |
| 121 | 557 | 261.141 | .27264E-03 | .27343E-03 | .17778E+08 | 9.97E-01 |
| 121 | 556 | 244.056 | .31977E-03 | .32162E-03 | .23872E+08 | 9.94E-01 |
| 121 | 555 | 236.243 | .22307E-05 | .10402E-04 | .17773E+06 | 2.14E-01 |
| 121 | 554 | 210.020 | .57250E-02 | .73385E-02 | .57716E+09 | 7.80E-01 |
| 121 | 553 | 195.036 | .10643E-04 | .55804E-04 | .12442E+07 | 1.91E-01 |
| 121 | 552 | 184.558 | .38873E+00 | .36206E+00 | .50749E+11 | 1.07E+00 |
| 121 | 551 | 167.060 | .75925E-02 | .65360E-02 | .12097E+10 | 1.16E+00 |
| 121 | 550 | 63.610 | .97922E-04 | .25832E-03 | .10762E+09 | 3.79E-01 |
| 569 | 120 | 112.738 | .32846E-04 | .86563E-05 | .25856E+08 | 3.79E+00 |
| 568 | 120 | 123.935 | .71081E-05 | .27319E-04 | .46301E+07 | 2.60E-01 |
| 567 | 120 | 133.211 | .12207E-02 | .81296E-03 | .68829E+09 | 1.50E+00 |
| 566 | 120 | 140.269 | .81834E-03 | .69730E-03 | .41614E+09 | 1.17E+00 |
| 565 | 120 | 143.563 | .10690E-01 | .72144E-02 | .51896E+10 | 1.48E+00 |
| 564 | 120 | 148.606 | .40838E-03 | .17759E-02 | .18502E+09 | 2.30E-01 |
| 563 | 120 | 179.582 | .56431E-01 | .23826E-01 | .17507E+11 | 2.37E+00 |
| 562 | 120 | 190.810 | .23427E-02 | .69556E-03 | .64379E+09 | 3.37E+00 |
| 120 | 561 | 462.254 | .77050E-03 | .26609E-03 | .16034E+08 | 2.90E+00 |
| 120 | 560 | 378.198 | .16078E-01 | .18302E-01 | .49986E+09 | 8.78E-01 |
| 120 | 559 | 334.793 | .21486E-01 | .19215E-01 | .85239E+09 | 1.12E+00 |
| 120 | 558 | 318.023 | .27617E-04 | .72058E-04 | .12142E+07 | 3.83E-01 |
| 120 | 557 | 301.538 | .15513E-02 | .84788E-03 | .75867E+08 | 1.83E+00 |
| 120 | 556 | 278.987 | .10030E-02 | .99475E-03 | .57301E+08 | 1.01E+00 |
| 120 | 555 | 268.824 | .30743E-03 | .58894E-03 | .18917E+08 | 5.22E-01 |
| 120 | 554 | 235.382 | .10012E+00 | .12368E+00 | .80358E+10 | 8.10E-01 |
| 120 | 553 | 216.720 | .67548E-01 | .61088E-01 | .63952E+10 | 1.11E+00 |
| 120 | 552 | 203.860 | .96450E-06 | .16964E-03 | .10320E+06 | 5.69E-03 |
| 120 | 551 | 182.720 | .11051E-01 | .11268E-01 | .14719E+10 | 9.81E-01 |
| 120 | 550 | 65.756 | .19928E-02 | .16859E-02 | .20495E+10 | 1.18E+00 |
| 569 | 119 | 100.196 | .30884E-03 | .44766E-05 | .30779E+09 | 6.90E+01 |
| 568 | 119 | 108.944 | .72656E-04 | .19187E-04 | .61248E+08 | 3.79E+00 |
| 567 | 119 | 116.048 | .39961E-03 | .13229E-03 | .29688E+09 | 3.02E+00 |
| 566 | 119 | 121.368 | .20504E-03 | .55393E-04 | .13927E+09 | 3.70E+00 |
| 565 | 119 | 123.826 | .12008E-05 | .21368E-05 | .78357E+06 | 5.62E-01 |
| 564 | 119 | 127.560 | .10028E-02 | .51082E-03 | .61660E+09 | 1.96E+00 |
| 563 | 119 | 149.729 | .38169E-02 | .27474E-02 | .17034E+10 | 1.39E+00 |
| 562 | 119 | 157.454 | .10605E+00 | .70958E-01 | .42798E+11 | 1.49E+00 |
| 119 | 561 | 949.629 | .65323E-05 | .45176E-06 | .32211E+05 | 1.45E+01 |
| 119 | 560 | 651.957 | .18693E-02 | .54423E-03 | .19556E+08 | 3.43E+00 |
| 119 | 559 | 532.864 | .41587E-02 | .38906E-02 | .65128E+08 | 1.07E+00 |
| 119 | 558 | 491.603 | .24033E-02 | .24242E-02 | .44221E+08 | 9.91E-01 |
| 119 | 557 | 453.296 | .16861E-02 | .10090E-03 | .36488E+08 | 1.67E+01 |
| 119 | 556 | 404.183 | .14113E-05 | .85239E-04 | .38416E+05 | 1.66E-02 |
| 119 | 555 | 383.195 | .25036E-03 | .40910E-04 | .75817E+07 | 6.12E+00 |
| 119 | 554 | 318.659 | .10424E-02 | .18221E-02 | .45648E+08 | 5.72E-01 |
| 119 | 553 | 285.390 | .42089E-02 | .69101E-02 | .22979E+09 | 6.09E-01 |
| 119 | 552 | 263.500 | .46205E-03 | .69989E-03 | .29592E+08 | 6.60E-01 |
| 119 | 551 | 229.222 | .27431E+00 | .28462E+00 | .23215E+11 | 9.64E-01 |
| 119 | 550 | 70.934 | .55527E-05 | .41643E-05 | .49072E+07 | 1.33E+00 |
| 569 | 118 | 55.156 | .29144E-07 | .16450E-07 | .95850E+05 | 1.77E+00 |
| 568 | 118 | 57.707 | .99600E-06 | .38058E-06 | .29925E+07 | 2.62E+00 |
| 567 | 118 | 59.641 | .14415E-07 | .19974E-08 | .40547E+05 | 7.22E+00 |
| 566 | 118 | 61.015 | .76582E-07 | .30226E-07 | .20582E+06 | 2.53E+00 |
| 565 | 118 | 61.630 | .10949E-07 | .14575E-07 | .28841E+05 | 7.51E-01 |
| 564 | 118 | 62.541 | .72300E-07 | .21872E-07 | .18494E+06 | 3.31E+00 |
| 563 | 118 | 67.437 | .36295E-06 | .22824E-07 | .79852E+06 | 1.59E+01 |
| 562 | 118 | 68.961 | .10329E-06 | .30190E-07 | .21731E+06 | 3.42E+00 |
| 561 | 118 | 140.906 | .32671E-01 | .27803E-01 | .16464E+11 | 1.18E+00 |
| 560 | 118 | 151.146 | .36412E-02 | .34270E-02 | .15947E+10 | 1.06E+00 |
| 559 | 118 | 159.405 | .97479E-03 | .10040E-02 | .38382E+09 | 9.71E-01 |
| 558 | 118 | 163.511 | .24920E-01 | .26566E-01 | .93255E+10 | 9.38E-01 |
| 557 | 118 | 168.240 | .55691E-03 | .60391E-03 | .19686E+09 | 9.22E-01 |
| 556 | 118 | 176.185 | .34173E-03 | .40475E-03 | .11015E+09 | 8.44E-01 |
| 555 | 118 | 180.495 | .33140E-04 | .41125E-04 | .10178E+08 | 8.06E-01 |
| 554 | 118 | 199.528 | .14698E-03 | .17431E-03 | .36939E+08 | 8.43E-01 |
| 553 | 118 | 215.239 | .56149E-06 | .22339E-09 | .12126E+06 | 2.51E+03 |
| 552 | 118 | 229.626 | .27869E-05 | .15975E-05 | .52882E+06 | 1.74E+00 |
| 551 | 118 | 264.035 | .78138E-04 | .16322E-03 | .11214E+08 | 4.79E-01 |
| 118 | 550 | 168.135 | .22042E-07 | .71056E-07 | .34672E+04 | 3.10E-01 |
| 569 | 117 | 53.619 | .51247E-06 | .12535E-06 | .17835E+07 | 4.09E+00 |
| 568 | 117 | 56.026 | .53595E-06 | .14059E-06 | .17083E+07 | 3.81E+00 |
| 567 | 117 | 57.847 | .88625E-06 | .50883E-07 | .26499E+07 | 1.74E+01 |
| 566 | 117 | 59.139 | .19865E-04 | .50423E-05 | .56830E+08 | 3.94E+00 |
| 565 | 117 | 59.717 | .13229E-05 | .63318E-06 | .37116E+07 | 2.09E+00 |
| 564 | 117 | 60.572 | .22037E-06 | .64284E-09 | .60095E+06 | 3.43E+02 |
| 563 | 117 | 65.152 | .11885E-05 | .92337E-07 | .28014E+07 | 1.29E+01 |
| 562 | 117 | 66.574 | .30951E-05 | .12301E-07 | .69870E+07 | 2.52E+02 |
| 561 | 117 | 131.288 | .27847E-03 | .21810E-03 | .16164E+09 | 1.28E+00 |
| 560 | 117 | 140.134 | .23221E-01 | .19805E-01 | .11831E+11 | 1.17E+00 |
| 559 | 117 | 147.206 | .22252E-01 | .20273E-01 | .10274E+11 | 1.10E+00 |
| 558 | 117 | 150.700 | .23074E-04 | .22474E-04 | .10165E+08 | 1.03E+00 |
| 557 | 117 | 154.708 | .12040E-01 | .11532E-01 | .50331E+10 | 1.04E+00 |
| 556 | 117 | 161.401 | .54395E-03 | .65308E-03 | .20892E+09 | 8.33E-01 |
| 555 | 117 | 165.010 | .77680E-01 | .84546E-01 | .28544E+11 | 9.19E-01 |
| 554 | 117 | 180.776 | .65262E-03 | .83313E-03 | .19980E+09 | 7.83E-01 |
| 553 | 117 | 193.577 | .81795E-03 | .10302E-02 | .21840E+09 | 7.94E-01 |
| 552 | 117 | 205.136 | .44765E-04 | .77259E-04 | .10643E+08 | 5.79E-01 |
| 551 | 117 | 232.165 | .16313E-02 | .21822E-02 | .30281E+09 | 7.48E-01 |
| 117 | 550 | 184.240 | .23539E-07 | .54969E-07 | .30836E+04 | 4.28E-01 |
| 569 | 116 | 52.912 | .20857E-05 | .81614E-06 | .74538E+07 | 2.56E+00 |
| 568 | 116 | 55.255 | .25247E-06 | .33261E-07 | .82736E+06 | 7.59E+00 |
| 567 | 116 | 57.025 | .67784E-06 | .68479E-07 | .20855E+07 | 9.90E+00 |
| 566 | 116 | 58.281 | .14767E-05 | .36538E-06 | .43499E+07 | 4.04E+00 |
| 565 | 116 | 58.841 | .86180E-05 | .46924E-06 | .24904E+08 | 1.84E+01 |
| 564 | 116 | 59.671 | .35608E-05 | .27555E-07 | .10006E+08 | 1.29E+02 |
| 563 | 116 | 64.112 | .96950E-05 | .75368E-07 | .23599E+08 | 1.29E+02 |
| 562 | 116 | 65.488 | .69872E-05 | .19466E-07 | .16301E+08 | 3.59E+02 |
| 561 | 116 | 127.131 | .33701E-02 | .25403E-02 | .20863E+10 | 1.33E+00 |
| 560 | 116 | 135.407 | .38842E-01 | .32136E-01 | .21195E+11 | 1.21E+00 |
| 559 | 116 | 141.999 | .86406E-02 | .75048E-02 | .42875E+10 | 1.15E+00 |
| 558 | 116 | 145.248 | .21323E-03 | .22958E-03 | .10112E+09 | 9.29E-01 |
| 557 | 116 | 148.967 | .82320E-02 | .76183E-02 | .37115E+10 | 1.08E+00 |
| 556 | 116 | 155.163 | .97769E-04 | .75249E-04 | .40630E+08 | 1.30E+00 |
| 555 | 116 | 158.496 | .44705E-02 | .47181E-02 | .17805E+10 | 9.48E-01 |
| 554 | 116 | 172.986 | .98849E-02 | .10806E-01 | .33050E+10 | 9.15E-01 |
| 553 | 116 | 184.673 | .14581E-01 | .16834E-01 | .42776E+10 | 8.66E-01 |
| 552 | 116 | 195.164 | .60997E-04 | .19162E-04 | .16023E+08 | 3.18E+00 |
| 551 | 116 | 219.473 | .85335E-02 | .10891E-01 | .17725E+10 | 7.84E-01 |
| 116 | 550 | 193.102 | .68953E-06 | .31029E-05 | .82228E+05 | 2.22E-01 |
| 569 | 115 | 52.438 | .69353E-05 | .51416E-06 | .25235E+08 | 1.35E+01 |
| 568 | 115 | 54.738 | .59777E-06 | .47945E-07 | .19961E+07 | 1.25E+01 |
| 567 | 115 | 56.475 | .39221E-05 | .29970E-06 | .12303E+08 | 1.31E+01 |
| 566 | 115 | 57.706 | .63406E-05 | .76292E-06 | .19051E+08 | 8.31E+00 |
| 565 | 115 | 58.256 | .39456E-05 | .21649E-06 | .11632E+08 | 1.82E+01 |
| 564 | 115 | 59.069 | .39878E-06 | .23093E-06 | .11435E+07 | 1.73E+00 |
| 563 | 115 | 63.417 | .39539E-05 | .80814E-06 | .98363E+07 | 4.89E+00 |
| 562 | 115 | 64.763 | .18654E-03 | .13793E-04 | .44497E+09 | 1.35E+01 |
| 561 | 115 | 124.429 | .38814E-03 | .28297E-03 | .25083E+09 | 1.37E+00 |
| 560 | 115 | 132.346 | .20612E-02 | .16847E-02 | .11774E+10 | 1.22E+00 |
| 559 | 115 | 138.636 | .33403E-01 | .29079E-01 | .17389E+11 | 1.15E+00 |
| 558 | 115 | 141.731 | .65420E-02 | .60254E-02 | .32584E+10 | 1.09E+00 |
| 557 | 115 | 145.270 | .37459E-01 | .34185E-01 | .17759E+11 | 1.10E+00 |
| 556 | 115 | 151.156 | .23737E-02 | .22523E-02 | .10395E+10 | 1.05E+00 |
| 555 | 115 | 154.318 | .35429E-02 | .35490E-02 | .14885E+10 | 9.98E-01 |
| 554 | 115 | 168.021 | .79343E-03 | .63271E-03 | .28119E+09 | 1.25E+00 |
| 553 | 115 | 179.025 | .10865E-01 | .13793E-01 | .33918E+10 | 7.88E-01 |
| 552 | 115 | 188.867 | .10208E-05 | .62053E-05 | .28633E+06 | 1.65E-01 |
| 551 | 115 | 211.542 | .29384E-01 | .37078E-01 | .65697E+10 | 7.92E-01 |
| 115 | 550 | 199.689 | .28817E-06 | .15570E-05 | .32135E+05 | 1.85E-01 |
| 569 | 114 | 49.936 | .28484E-07 | .12337E-06 | .11429E+06 | 2.31E-01 |
| 568 | 114 | 52.018 | .90342E-06 | .84203E-07 | .33405E+07 | 1.07E+01 |
| 567 | 114 | 53.584 | .12897E-06 | .15224E-07 | .44940E+06 | 8.47E+00 |
| 566 | 114 | 54.691 | .39653E-06 | .35691E-07 | .13264E+07 | 1.11E+01 |
| 565 | 114 | 55.184 | .34213E-05 | .18346E-06 | .11241E+08 | 1.86E+01 |
| 564 | 114 | 55.914 | .14933E-05 | .67294E-08 | .47789E+07 | 2.22E+02 |
| 563 | 114 | 59.794 | .19953E-03 | .18566E-04 | .55835E+09 | 1.07E+01 |
| 562 | 114 | 60.989 | .29317E-07 | .66075E-07 | .78855E+05 | 4.44E-01 |
| 561 | 114 | 111.208 | .14697E-03 | .88566E-04 | .11890E+09 | 1.66E+00 |
| 560 | 114 | 117.490 | .15317E-03 | .79941E-04 | .11102E+09 | 1.92E+00 |
| 559 | 114 | 122.421 | .28494E-04 | .10862E-04 | .19023E+08 | 2.62E+00 |
| 558 | 114 | 124.828 | .14872E-05 | .56988E-06 | .95496E+06 | 2.61E+00 |
| 557 | 114 | 127.565 | .60315E-05 | .37195E-05 | .37084E+07 | 1.62E+00 |
| 556 | 114 | 132.081 | .64761E-05 | .41729E-05 | .37141E+07 | 1.55E+00 |
| 555 | 114 | 134.489 | .17216E-04 | .13777E-04 | .95234E+07 | 1.25E+00 |
| 554 | 114 | 144.779 | .70491E-02 | .60456E-02 | .33647E+10 | 1.17E+00 |
| 553 | 114 | 152.876 | .82107E-03 | .78383E-03 | .35150E+09 | 1.05E+00 |
| 552 | 114 | 159.996 | .12752E+00 | .13075E+00 | .49840E+11 | 9.75E-01 |
| 551 | 114 | 175.975 | .63898E-02 | .76860E-02 | .20645E+10 | 8.31E-01 |
| 114 | 550 | 246.770 | .56163E-06 | .53200E-05 | .41011E+05 | 1.06E-01 |
| 569 | 113 | 44.071 | .18817E-06 | .23667E-07 | .96934E+06 | 7.95E+00 |
| 568 | 113 | 45.684 | .60998E-06 | .10773E-05 | .29242E+07 | 5.66E-01 |
| 567 | 113 | 46.888 | .90843E-07 | .32720E-09 | .41343E+06 | 2.78E+02 |
| 566 | 113 | 47.733 | .18570E-06 | .57926E-06 | .81543E+06 | 3.21E-01 |
| 565 | 113 | 48.109 | .73509E-07 | .19344E-05 | .31778E+06 | 3.80E-02 |
| 564 | 113 | 48.662 | .95474E-07 | .79618E-05 | .40340E+06 | 1.20E-02 |
| 563 | 113 | 51.575 | .24640E-03 | .25445E-03 | .92681E+09 | 9.68E-01 |
| 562 | 113 | 52.462 | .22784E-04 | .10915E-04 | .82826E+08 | 2.09E+00 |
| 561 | 113 | 85.782 | .33621E-09 | .18228E-07 | .45713E+03 | 1.84E-02 |
| 560 | 113 | 89.472 | .10102E-07 | .52228E-07 | .12625E+05 | 1.93E-01 |
| 559 | 113 | 92.304 | .85010E-07 | .73924E-07 | .99829E+05 | 1.15E+00 |
| 558 | 113 | 93.665 | .13461E-07 | .16677E-08 | .15351E+05 | 8.07E+00 |
| 557 | 113 | 95.198 | .19891E-07 | .13486E-07 | .21959E+05 | 1.47E+00 |
| 556 | 113 | 97.691 | .22848E-08 | .52763E-09 | .23953E+04 | 4.33E+00 |
| 555 | 113 | 99.002 | .90672E-09 | .99417E-10 | .92558E+03 | 9.12E+00 |
| 554 | 113 | 104.468 | .38779E-07 | .54472E-09 | .35552E+05 | 7.12E+01 |
| 553 | 113 | 108.619 | .20570E-06 | .13415E-09 | .17444E+06 | 1.53E+03 |
| 552 | 113 | 112.165 | .29533E-06 | .16626E-07 | .23486E+06 | 1.78E+01 |
| 551 | 113 | 119.791 | .71082E-07 | .14246E-07 | .49561E+05 | 4.99E+00 |
| 113 | 550 | 720.936 | .16057E+00 | .15282E+00 | .13738E+10 | 1.05E+00 |
| 468 | 216 | 312.100 | .34093E-02 | .57054E-02 | .15564E+09 | 5.98E-01 |
| 216 | 467 | 212.809 | .26173E-01 | .23068E-01 | .57823E+10 | 1.13E+00 |
| 216 | 466 | 199.647 | .84305E-01 | .73541E-01 | .21162E+11 | 1.15E+00 |
| 216 | 465 | 180.577 | .25983E-08 | .49711E-05 | .79723E+03 | 5.23E-04 |
| 216 | 464 | 169.020 | .23524E-02 | .21120E-02 | .82388E+09 | 1.11E+00 |
| 216 | 463 | 159.789 | .10861E-02 | .74309E-03 | .42559E+09 | 1.46E+00 |
| 216 | 462 | 151.284 | .42694E-01 | .36516E-01 | .18664E+11 | 1.17E+00 |
| 216 | 461 | 148.011 | .11719E-02 | .91914E-03 | .53521E+09 | 1.27E+00 |
| 216 | 460 | 127.769 | .72061E-04 | .25071E-06 | .44165E+08 | 2.87E+02 |
| 216 | 459 | 118.695 | .11168E-04 | .28396E-05 | .79308E+07 | 3.93E+00 |
| 216 | 458 | 117.822 | .22023E-03 | .20709E-03 | .15873E+09 | 1.06E+00 |
| 216 | 457 | 60.635 | .42279E-06 | .76689E-07 | .11505E+07 | 5.51E+00 |
| 216 | 456 | 59.555 | .10001E-05 | .25843E-09 | .28212E+07 | 3.87E+03 |
| 216 | 455 | 59.312 | .41655E-06 | .12838E-07 | .11847E+07 | 3.24E+01 |
| 216 | 454 | 58.504 | .22203E-05 | .32841E-06 | .64902E+07 | 6.76E+00 |
| 216 | 453 | 57.501 | .35538E-06 | .31177E-06 | .10754E+07 | 1.14E+00 |
| 216 | 452 | 56.959 | .28948E-06 | .35553E-07 | .89273E+06 | 8.14E+00 |
| 216 | 451 | 55.971 | .74262E-04 | .16703E-04 | .23718E+09 | 4.45E+00 |
| 216 | 450 | 55.013 | .60664E-07 | .36382E-06 | .20055E+06 | 1.67E-01 |
| 216 | 449 | 53.632 | .14056E-04 | .24393E-07 | .48892E+08 | 5.76E+02 |
| 216 | 448 | 52.106 | .13007E-05 | .10532E-08 | .47932E+07 | 1.23E+03 |
| 216 | 447 | 51.148 | .34085E-05 | .96444E-06 | .13035E+08 | 3.53E+00 |
| 216 | 446 | 39.361 | .13856E-03 | .16199E-03 | .89478E+09 | 8.55E-01 |
| 468 | 215 | 242.808 | .40901E-01 | .26719E-01 | .30850E+10 | 1.53E+00 |
| 215 | 467 | 264.223 | .13515E-01 | .12735E-01 | .19368E+10 | 1.06E+00 |
| 215 | 466 | 244.232 | .31384E-02 | .31807E-02 | .52642E+09 | 9.87E-01 |
| 215 | 465 | 216.290 | .14937E-02 | .15739E-02 | .31947E+09 | 9.49E-01 |
| 215 | 464 | 199.917 | .60866E-01 | .48363E-01 | .15237E+11 | 1.26E+00 |
| 215 | 463 | 187.130 | .19310E-03 | .21045E-03 | .55171E+08 | 9.18E-01 |
| 215 | 462 | 175.571 | .10506E-02 | .12404E-02 | .34101E+09 | 8.47E-01 |
| 215 | 461 | 171.178 | .10684E-01 | .11164E-01 | .36481E+10 | 9.57E-01 |
| 215 | 460 | 144.670 | .36480E-05 | .52675E-03 | .17439E+07 | 6.93E-03 |
| 215 | 459 | 133.146 | .86516E-03 | .12116E-02 | .48828E+09 | 7.14E-01 |
| 215 | 458 | 132.048 | .19559E-02 | .12186E-02 | .11223E+10 | 1.61E+00 |
| 215 | 457 | 64.194 | .22550E-06 | .15320E-06 | .54749E+06 | 1.47E+00 |
| 215 | 456 | 62.985 | .48867E-08 | .64658E-11 | .12324E+05 | 7.56E+02 |
| 215 | 455 | 62.714 | .52649E-06 | .38442E-06 | .13393E+07 | 1.37E+00 |
| 215 | 454 | 61.811 | .98759E-07 | .98235E-08 | .25863E+06 | 1.01E+01 |
| 215 | 453 | 60.692 | .16377E-06 | .14127E-07 | .44483E+06 | 1.16E+01 |
| 215 | 452 | 60.089 | .14411E-05 | .55719E-08 | .39932E+07 | 2.59E+02 |
| 215 | 451 | 58.990 | .40585E-05 | .10277E-05 | .11669E+08 | 3.95E+00 |
| 215 | 450 | 57.927 | .35344E-09 | .17990E-05 | .10538E+04 | 1.96E-04 |
| 215 | 449 | 56.398 | .15838E-03 | .42170E-06 | .49819E+09 | 3.76E+02 |
| 215 | 448 | 54.713 | .22243E-04 | .47884E-08 | .74343E+08 | 4.65E+03 |
| 215 | 447 | 53.658 | .11767E-03 | .14907E-06 | .40890E+09 | 7.89E+02 |
| 215 | 446 | 40.831 | .39559E-02 | .44347E-02 | .23741E+11 | 8.92E-01 |
| 468 | 214 | 157.486 | .32928E+00 | .31601E+00 | .59037E+11 | 1.04E+00 |
| 214 | 467 | 643.751 | .85122E-04 | .95626E-05 | .20551E+07 | 8.90E+00 |
| 214 | 466 | 536.714 | .12992E-04 | .17313E-05 | .45126E+06 | 7.50E+00 |
| 214 | 465 | 418.035 | .17296E-04 | .42466E-04 | .99027E+06 | 4.07E-01 |
| 214 | 464 | 360.908 | .18291E-04 | .12268E-03 | .14049E+07 | 1.49E-01 |
| 214 | 463 | 321.275 | .48273E-03 | .58920E-03 | .46792E+08 | 8.19E-01 |
| 214 | 462 | 288.649 | .44362E-05 | .66771E-05 | .53271E+06 | 6.64E-01 |
| 214 | 461 | 276.962 | .18893E-05 | .27427E-03 | .24642E+06 | 6.89E-03 |
| 214 | 460 | 213.631 | .10080E-01 | .14812E-01 | .22098E+10 | 6.81E-01 |
| 214 | 459 | 189.420 | .46306E-01 | .23927E-01 | .12913E+11 | 1.94E+00 |
| 214 | 458 | 187.206 | .12920E+00 | .63601E-01 | .36884E+11 | 2.03E+00 |
| 214 | 457 | 74.926 | .46929E-07 | .10887E-05 | .83638E+05 | 4.31E-02 |
| 214 | 456 | 73.285 | .83038E-07 | .58102E-07 | .15469E+06 | 1.43E+00 |
| 214 | 455 | 72.917 | .45023E-06 | .10841E-05 | .84722E+06 | 4.15E-01 |
| 214 | 454 | 71.699 | .35484E-06 | .33683E-07 | .69061E+06 | 1.05E+01 |
| 214 | 453 | 70.198 | .20767E-06 | .45340E-06 | .42164E+06 | 4.58E-01 |
| 214 | 452 | 69.393 | .20006E-06 | .24421E-05 | .41568E+06 | 8.19E-02 |
| 214 | 451 | 67.931 | .44643E-08 | .59984E-07 | .96794E+04 | 7.44E-02 |
| 214 | 450 | 66.526 | .22737E-05 | .23058E-07 | .51401E+07 | 9.86E+01 |
| 214 | 449 | 64.516 | .10840E-04 | .46363E-04 | .26057E+08 | 2.34E-01 |
| 214 | 448 | 62.321 | .15832E-04 | .42451E-05 | .40786E+08 | 3.73E+00 |
| 214 | 447 | 60.956 | .63982E-04 | .86276E-05 | .17229E+09 | 7.42E+00 |
| 214 | 446 | 44.924 | .26780E-01 | .27488E-01 | .13277E+12 | 9.74E-01 |
| 468 | 213 | 74.350 | .32861E-06 | .99715E-06 | .26434E+06 | 3.30E-01 |
| 467 | 213 | 180.285 | .16365E-01 | .16573E-01 | .22389E+10 | 9.87E-01 |
| 466 | 213 | 190.950 | .10755E-01 | .12059E-01 | .13117E+10 | 8.92E-01 |
| 465 | 213 | 212.403 | .46668E-02 | .30159E-02 | .45998E+09 | 1.55E+00 |
| 464 | 213 | 230.980 | .47890E-02 | .66878E-02 | .39915E+09 | 7.16E-01 |
| 463 | 213 | 250.779 | .28608E-02 | .22751E-02 | .20228E+09 | 1.26E+00 |
| 462 | 213 | 275.045 | .11766E-01 | .13934E-01 | .69160E+09 | 8.44E-01 |
| 461 | 213 | 286.568 | .41321E-01 | .48641E-01 | .22375E+10 | 8.50E-01 |
| 460 | 213 | 413.360 | .34544E-02 | .27612E-02 | .89898E+08 | 1.25E+00 |
| 459 | 213 | 549.176 | .32956E-02 | .15808E-02 | .48591E+08 | 2.08E+00 |
| 458 | 213 | 568.680 | .65599E-03 | .37883E-03 | .90200E+07 | 1.73E+00 |
| 213 | 457 | 160.095 | .42059E-01 | .28840E-01 | .16418E+11 | 1.46E+00 |
| 213 | 456 | 152.783 | .48892E-03 | .21863E-03 | .20956E+09 | 2.24E+00 |
| 213 | 455 | 151.194 | .21732E-01 | .13822E-01 | .95116E+10 | 1.57E+00 |
| 213 | 454 | 146.049 | .18749E-02 | .10813E-02 | .87946E+09 | 1.73E+00 |
| 213 | 453 | 139.955 | .30261E-02 | .23594E-02 | .15457E+10 | 1.28E+00 |
| 213 | 452 | 136.788 | .42333E-03 | .22859E-03 | .22636E+09 | 1.85E+00 |
| 213 | 451 | 131.221 | .19464E-04 | .75307E-05 | .11309E+08 | 2.58E+00 |
| 213 | 450 | 126.079 | .63645E-04 | .35942E-03 | .40060E+08 | 1.77E-01 |
| 213 | 449 | 119.050 | .27164E-03 | .14459E-03 | .19176E+09 | 1.88E+00 |
| 213 | 448 | 111.784 | .29453E-04 | .43478E-03 | .23582E+08 | 6.77E-02 |
| 213 | 447 | 107.468 | .33918E-06 | .12927E-04 | .29383E+06 | 2.62E-02 |
| 213 | 446 | 65.964 | .32100E-06 | .62220E-08 | .73810E+06 | 5.16E+01 |
| 468 | 212 | 73.844 | .78095E-06 | .25842E-05 | .63684E+06 | 3.02E-01 |
| 467 | 212 | 177.342 | .25282E-02 | .27767E-02 | .35747E+09 | 9.11E-01 |
| 466 | 212 | 187.651 | .11089E-02 | .14554E-02 | .14003E+09 | 7.62E-01 |
| 465 | 212 | 208.330 | .31920E-01 | .26481E-01 | .32704E+10 | 1.21E+00 |
| 464 | 212 | 226.171 | .22169E-02 | .26239E-02 | .19271E+09 | 8.45E-01 |
| 463 | 212 | 245.120 | .75042E-01 | .76958E-01 | .55538E+10 | 9.75E-01 |
| 462 | 212 | 268.254 | .34162E-02 | .39039E-02 | .21110E+09 | 8.75E-01 |
| 461 | 212 | 279.203 | .46414E-04 | .14441E-03 | .26476E+07 | 3.21E-01 |
| 460 | 212 | 398.208 | .54530E-03 | .19686E-02 | .15292E+08 | 2.77E-01 |
| 459 | 212 | 522.750 | .36443E-03 | .13414E-02 | .59302E+07 | 2.72E-01 |
| 458 | 212 | 540.392 | .14036E-04 | .11335E-05 | .21374E+06 | 1.24E+01 |
| 212 | 457 | 162.489 | .33289E-03 | .43421E-04 | .12615E+09 | 7.67E+00 |
| 212 | 456 | 154.962 | .58070E-01 | .43569E-01 | .24195E+11 | 1.33E+00 |
| 212 | 455 | 153.328 | .65896E-02 | .26178E-02 | .28044E+10 | 2.52E+00 |
| 212 | 454 | 148.039 | .29744E-01 | .18164E-01 | .13579E+11 | 1.64E+00 |
| 212 | 453 | 141.782 | .77566E-03 | .98691E-03 | .38606E+09 | 7.86E-01 |
| 212 | 452 | 138.533 | .49482E-03 | .19044E-03 | .25797E+09 | 2.60E+00 |
| 212 | 451 | 132.826 | .46241E-04 | .26747E-04 | .26223E+08 | 1.73E+00 |
| 212 | 450 | 127.559 | .38392E-04 | .51584E-05 | .23607E+08 | 7.44E+00 |
| 212 | 449 | 120.369 | .84399E-03 | .31448E-03 | .58282E+09 | 2.68E+00 |
| 212 | 448 | 112.946 | .72256E-09 | .27853E-07 | .56670E+03 | 2.59E-02 |
| 212 | 447 | 108.541 | .56022E-05 | .12522E-04 | .47576E+07 | 4.47E-01 |
| 212 | 446 | 66.367 | .49069E-06 | .27408E-09 | .11146E+07 | 1.79E+03 |
| 468 | 211 | 71.321 | .78220E-06 | .31077E-07 | .68379E+06 | 2.52E+01 |
| 467 | 211 | 163.455 | .74841E-02 | .75579E-02 | .12456E+10 | 9.90E-01 |
| 466 | 211 | 172.174 | .10211E-01 | .10754E-01 | .15317E+10 | 9.50E-01 |
| 465 | 211 | 189.425 | .72864E-02 | .53388E-02 | .90298E+09 | 1.36E+00 |
| 464 | 211 | 204.061 | .53279E-02 | .67901E-02 | .56896E+09 | 7.85E-01 |
| 463 | 211 | 219.362 | .98568E-02 | .91715E-02 | .91087E+09 | 1.07E+00 |
| 462 | 211 | 237.707 | .38073E-01 | .39199E-01 | .29962E+10 | 9.71E-01 |
| 461 | 211 | 246.264 | .14883E-01 | .14669E-01 | .10913E+10 | 1.01E+00 |
| 460 | 211 | 334.414 | .25430E-02 | .39427E-02 | .10111E+09 | 6.45E-01 |
| 459 | 211 | 418.058 | .11910E-03 | .25875E-04 | .30303E+07 | 4.60E+00 |
| 458 | 211 | 429.265 | .10768E-02 | .33814E-02 | .25986E+08 | 3.18E-01 |
| 211 | 457 | 176.205 | .15015E-02 | .13142E-02 | .48385E+09 | 1.14E+00 |
| 211 | 456 | 167.388 | .19469E-03 | .44407E-04 | .69521E+08 | 4.38E+00 |
| 211 | 455 | 165.483 | .10377E-02 | .47791E-03 | .37913E+09 | 2.17E+00 |
| 211 | 454 | 159.340 | .63047E-01 | .43369E-01 | .24845E+11 | 1.45E+00 |
| 211 | 453 | 152.113 | .33685E-01 | .30620E-01 | .14565E+11 | 1.10E+00 |
| 211 | 452 | 148.380 | .10223E-01 | .73250E-02 | .46455E+10 | 1.40E+00 |
| 211 | 451 | 141.852 | .11467E-02 | .60056E-03 | .57015E+09 | 1.91E+00 |
| 211 | 450 | 135.861 | .17874E-04 | .39960E-04 | .96887E+07 | 4.47E-01 |
| 211 | 449 | 127.735 | .84060E-05 | .76148E-04 | .51546E+07 | 1.10E-01 |
| 211 | 448 | 119.407 | .14489E-04 | .87057E-06 | .10167E+08 | 1.66E+01 |
| 211 | 447 | 114.495 | .11057E-03 | .45316E-04 | .84391E+08 | 2.44E+00 |
| 211 | 446 | 68.546 | .22107E-05 | .12078E-06 | .47074E+07 | 1.83E+01 |
| 468 | 210 | 69.152 | .18970E-04 | .22309E-05 | .17640E+08 | 8.50E+00 |
| 467 | 210 | 152.493 | .46386E-01 | .39489E-01 | .88702E+10 | 1.17E+00 |
| 466 | 210 | 160.054 | .19098E+00 | .18069E+00 | .33151E+11 | 1.06E+00 |
| 465 | 210 | 174.858 | .43431E-05 | .57207E-07 | .63164E+06 | 7.59E+01 |
| 464 | 210 | 187.256 | .30107E-02 | .29625E-02 | .38180E+09 | 1.02E+00 |
| 463 | 210 | 200.061 | .31862E-01 | .24803E-01 | .35399E+10 | 1.28E+00 |
| 462 | 210 | 215.208 | .11287E-01 | .10859E-01 | .10837E+10 | 1.04E+00 |
| 461 | 210 | 222.199 | .61191E-03 | .45470E-03 | .55112E+08 | 1.35E+00 |
| 460 | 210 | 291.537 | .16243E-02 | .24351E-02 | .84981E+08 | 6.67E-01 |
| 459 | 210 | 353.132 | .26206E-02 | .36100E-02 | .93449E+08 | 7.26E-01 |
| 458 | 210 | 361.095 | .36102E-02 | .41182E-02 | .12312E+09 | 8.77E-01 |
| 210 | 457 | 191.007 | .27096E-02 | .21130E-02 | .74306E+09 | 1.28E+00 |
| 210 | 456 | 180.690 | .10251E-02 | .89308E-03 | .31415E+09 | 1.15E+00 |
| 210 | 455 | 178.471 | .15200E-03 | .12509E-03 | .47747E+08 | 1.22E+00 |
| 210 | 454 | 171.347 | .56291E-03 | .21433E-03 | .19183E+09 | 2.63E+00 |
| 210 | 453 | 163.019 | .24424E-03 | .82125E-04 | .91952E+08 | 2.97E+00 |
| 210 | 452 | 158.739 | .12254E-01 | .11258E-01 | .48658E+10 | 1.09E+00 |
| 210 | 451 | 151.290 | .94926E-01 | .62539E-01 | .41494E+11 | 1.52E+00 |
| 210 | 450 | 144.495 | .54439E-02 | .44484E-02 | .26087E+10 | 1.22E+00 |
| 210 | 449 | 135.338 | .29859E-03 | .58233E-03 | .16311E+09 | 5.13E-01 |
| 210 | 448 | 126.025 | .73188E-03 | .70409E-03 | .46105E+09 | 1.04E+00 |
| 210 | 447 | 120.566 | .25036E-02 | .13704E-02 | .17232E+10 | 1.83E+00 |
| 210 | 446 | 70.677 | .12627E-04 | .26130E-05 | .25292E+08 | 4.83E+00 |
| 468 | 209 | 67.627 | .35553E-04 | .27367E-05 | .34569E+08 | 1.30E+01 |
| 467 | 209 | 145.267 | .22953E-01 | .20044E-01 | .48367E+10 | 1.15E+00 |
| 466 | 209 | 152.112 | .76861E-01 | .68221E-01 | .14771E+11 | 1.13E+00 |
| 465 | 209 | 165.422 | .15597E-02 | .13412E-02 | .25344E+09 | 1.16E+00 |
| 464 | 209 | 176.476 | .91175E-01 | .91937E-01 | .13018E+11 | 9.92E-01 |
| 463 | 209 | 187.805 | .12301E-01 | .87162E-02 | .15509E+10 | 1.41E+00 |
| 462 | 209 | 201.091 | .96980E-02 | .72354E-02 | .10664E+10 | 1.34E+00 |
| 461 | 209 | 207.182 | .28125E-01 | .24172E-01 | .29136E+10 | 1.16E+00 |
| 460 | 209 | 266.220 | .72808E-02 | .10105E-01 | .45681E+09 | 7.21E-01 |
| 459 | 209 | 316.655 | .20376E-02 | .20449E-02 | .90361E+08 | 9.96E-01 |
| 458 | 209 | 323.043 | .89394E-02 | .89784E-02 | .38092E+09 | 9.96E-01 |
| 209 | 457 | 203.699 | .83516E-03 | .88349E-03 | .20138E+09 | 9.45E-01 |
| 209 | 456 | 192.007 | .17640E-03 | .22602E-03 | .47873E+08 | 7.80E-01 |
| 209 | 455 | 189.504 | .12960E-02 | .12492E-02 | .36108E+09 | 1.04E+00 |
| 209 | 454 | 181.491 | .51463E-04 | .18262E-03 | .15632E+08 | 2.82E-01 |
| 209 | 453 | 172.175 | .32219E-02 | .31242E-02 | .10874E+10 | 1.03E+00 |
| 209 | 452 | 167.407 | .26463E-02 | .46126E-02 | .94474E+09 | 5.74E-01 |
| 209 | 451 | 159.144 | .73597E-01 | .53954E-01 | .29074E+11 | 1.36E+00 |
| 209 | 450 | 151.643 | .40169E-02 | .38648E-02 | .17477E+10 | 1.04E+00 |
| 209 | 449 | 141.589 | .29265E-02 | .25956E-02 | .14605E+10 | 1.13E+00 |
| 209 | 448 | 131.428 | .23391E-02 | .17760E-02 | .13549E+10 | 1.32E+00 |
| 209 | 447 | 125.502 | .16268E-02 | .10779E-02 | .10334E+10 | 1.51E+00 |
| 209 | 446 | 72.344 | .16065E-04 | .47760E-05 | .30711E+08 | 3.36E+00 |
| 468 | 208 | 64.854 | .14812E-05 | .27266E-06 | .15659E+07 | 5.43E+00 |
| 467 | 208 | 133.048 | .15310E+00 | .12277E+00 | .38458E+11 | 1.25E+00 |
| 466 | 208 | 138.768 | .42760E-01 | .36967E-01 | .98743E+10 | 1.16E+00 |
| 465 | 208 | 149.760 | .74630E-02 | .59831E-02 | .14797E+10 | 1.25E+00 |
| 464 | 208 | 158.763 | .58593E-03 | .43442E-03 | .10337E+09 | 1.35E+00 |
| 463 | 208 | 167.873 | .10772E-02 | .14270E-02 | .16997E+09 | 7.55E-01 |
| 462 | 208 | 178.410 | .84916E-02 | .83460E-02 | .11863E+10 | 1.02E+00 |
| 461 | 208 | 183.188 | .10802E+00 | .11503E+00 | .14314E+11 | 9.39E-01 |
| 460 | 208 | 227.868 | .61955E-01 | .67684E-01 | .53059E+10 | 9.15E-01 |
| 459 | 208 | 263.837 | .48176E-01 | .54641E-01 | .30775E+10 | 8.82E-01 |
| 458 | 208 | 268.257 | .21840E-03 | .77784E-03 | .13496E+08 | 2.81E-01 |
| 208 | 457 | 233.809 | .47978E-03 | .11486E-02 | .87810E+08 | 4.18E-01 |
| 208 | 456 | 218.535 | .82278E-03 | .45880E-03 | .17237E+09 | 1.79E+00 |
| 208 | 455 | 215.298 | .35913E-03 | .16967E-05 | .77516E+08 | 2.12E+02 |
| 208 | 454 | 205.015 | .81991E-04 | .84873E-04 | .19517E+08 | 9.66E-01 |
| 208 | 453 | 193.205 | .15052E-02 | .40857E-03 | .40344E+09 | 3.68E+00 |
| 208 | 452 | 187.222 | .33783E-02 | .87550E-03 | .96430E+09 | 3.86E+00 |
| 208 | 451 | 176.947 | .15745E-03 | .16774E-03 | .50313E+08 | 9.39E-01 |
| 208 | 450 | 167.722 | .40045E-01 | .16109E-01 | .14243E+11 | 2.49E+00 |
| 208 | 449 | 155.509 | .15793E-02 | .14706E-02 | .65338E+09 | 1.07E+00 |
| 208 | 448 | 143.339 | .40887E-01 | .22770E-01 | .19910E+11 | 1.80E+00 |
| 208 | 447 | 136.317 | .23436E-02 | .15278E-02 | .12618E+10 | 1.53E+00 |
| 208 | 446 | 75.812 | .22428E-05 | .17938E-06 | .39044E+07 | 1.25E+01 |
| 468 | 207 | 63.999 | .25031E-03 | .18666E-05 | .27175E+09 | 1.34E+02 |
| 467 | 207 | 129.499 | .64680E-02 | .41307E-02 | .17151E+10 | 1.57E+00 |
| 466 | 207 | 134.911 | .14712E-03 | .43771E-04 | .35944E+08 | 3.36E+00 |
| 465 | 207 | 145.278 | .50994E-03 | .35902E-03 | .10744E+09 | 1.42E+00 |
| 464 | 207 | 153.735 | .22254E-02 | .97950E-03 | .41870E+09 | 2.27E+00 |
| 463 | 207 | 162.261 | .42913E-02 | .38867E-02 | .72477E+09 | 1.10E+00 |
| 462 | 207 | 172.085 | .10363E-02 | .72002E-03 | .15561E+09 | 1.44E+00 |
| 461 | 207 | 176.526 | .16239E-01 | .13755E-01 | .23173E+10 | 1.18E+00 |
| 460 | 207 | 217.651 | .96214E-02 | .78772E-02 | .90315E+09 | 1.22E+00 |
| 459 | 207 | 250.236 | .11936E-01 | .65510E-02 | .84762E+09 | 1.82E+00 |
| 458 | 207 | 254.209 | .58905E-01 | .50976E-01 | .40533E+10 | 1.16E+00 |
| 207 | 457 | 245.640 | .28372E-03 | .86390E-04 | .47045E+08 | 3.28E+00 |
| 207 | 456 | 228.837 | .17895E-03 | .46324E-04 | .34191E+08 | 3.86E+00 |
| 207 | 455 | 225.290 | .19364E-02 | .11257E-02 | .38171E+09 | 1.72E+00 |
| 207 | 454 | 214.055 | .32408E-03 | .48005E-03 | .70766E+08 | 6.75E-01 |
| 207 | 453 | 201.214 | .33717E-02 | .14169E-02 | .83322E+09 | 2.38E+00 |
| 207 | 452 | 194.733 | .73517E-02 | .36841E-02 | .19397E+10 | 2.00E+00 |
| 207 | 451 | 183.641 | .46895E-04 | .36912E-04 | .13913E+08 | 1.27E+00 |
| 207 | 450 | 173.724 | .34715E-02 | .46769E-02 | .11508E+10 | 7.42E-01 |
| 207 | 449 | 160.655 | .10703E+00 | .64164E-01 | .41491E+11 | 1.67E+00 |
| 207 | 448 | 147.700 | .34119E-03 | .15734E-04 | .15648E+09 | 2.17E+01 |
| 207 | 447 | 140.256 | .61527E-02 | .15564E-02 | .31293E+10 | 3.95E+00 |
| 207 | 446 | 77.015 | .66267E-04 | .25893E-05 | .11178E+09 | 2.56E+01 |
| 468 | 206 | 63.363 | .57590E-05 | .47600E-06 | .63785E+07 | 1.21E+01 |
| 467 | 206 | 126.921 | .46100E-03 | .30427E-03 | .12725E+09 | 1.52E+00 |
| 466 | 206 | 132.115 | .95915E-04 | .72511E-04 | .24436E+08 | 1.32E+00 |
| 465 | 206 | 142.042 | .90667E-02 | .68614E-02 | .19983E+10 | 1.32E+00 |
| 464 | 206 | 150.115 | .35493E-05 | .60063E-05 | .70038E+06 | 5.91E-01 |
| 463 | 206 | 158.234 | .16309E+00 | .15257E+00 | .28964E+11 | 1.07E+00 |
| 462 | 206 | 167.562 | .64787E-01 | .67942E-01 | .10261E+11 | 9.54E-01 |
| 461 | 206 | 171.770 | .37365E-03 | .36263E-03 | .56314E+08 | 1.03E+00 |
| 460 | 206 | 210.466 | .28524E-04 | .64570E-07 | .28634E+07 | 4.42E+02 |
| 459 | 206 | 240.785 | .17063E-04 | .11852E-03 | .13087E+07 | 1.44E-01 |
| 458 | 206 | 244.462 | .24851E-04 | .60284E-04 | .18491E+07 | 4.12E-01 |
| 206 | 457 | 255.484 | .10752E-01 | .61710E-02 | .16481E+10 | 1.74E+00 |
| 206 | 456 | 237.356 | .83199E-02 | .70057E-02 | .14775E+10 | 1.19E+00 |
| 206 | 455 | 233.543 | .15069E-01 | .85656E-02 | .27642E+10 | 1.76E+00 |
| 206 | 454 | 221.492 | .94660E-03 | .37030E-03 | .19305E+09 | 2.56E+00 |
| 206 | 453 | 207.771 | .38461E-01 | .19457E-01 | .89141E+10 | 1.98E+00 |
| 206 | 452 | 200.868 | .17072E-01 | .90087E-02 | .42334E+10 | 1.90E+00 |
| 206 | 451 | 189.088 | .64484E-04 | .45260E-04 | .18045E+08 | 1.42E+00 |
| 206 | 450 | 178.591 | .17533E-02 | .10592E-02 | .54999E+09 | 1.66E+00 |
| 206 | 449 | 164.808 | .10472E-02 | .63157E-03 | .38573E+09 | 1.66E+00 |
| 206 | 448 | 151.203 | .12440E-03 | .10742E-03 | .54440E+08 | 1.16E+00 |
| 206 | 447 | 143.411 | .24064E-03 | .10075E-03 | .11707E+09 | 2.39E+00 |
| 206 | 446 | 77.956 | .22061E-06 | .41871E-07 | .36320E+06 | 5.27E+00 |
| 468 | 205 | 61.315 | .41748E-05 | .99707E-06 | .49379E+07 | 4.19E+00 |
| 467 | 205 | 118.963 | .42352E-03 | .27246E-03 | .13307E+09 | 1.55E+00 |
| 466 | 205 | 123.515 | .14747E-02 | .98173E-03 | .42983E+09 | 1.50E+00 |
| 465 | 205 | 132.148 | .86079E-04 | .57862E-04 | .21919E+08 | 1.49E+00 |
| 464 | 205 | 139.109 | .19437E-01 | .16303E-01 | .44665E+10 | 1.19E+00 |
| 463 | 205 | 146.054 | .93827E-02 | .71142E-02 | .19559E+10 | 1.32E+00 |
| 462 | 205 | 153.965 | .70920E-01 | .52802E-01 | .13304E+11 | 1.34E+00 |
| 461 | 205 | 157.510 | .11429E-01 | .95282E-02 | .20485E+10 | 1.20E+00 |
| 460 | 205 | 189.451 | .26613E-01 | .25589E-01 | .32972E+10 | 1.04E+00 |
| 459 | 205 | 213.669 | .11271E-02 | .12857E-02 | .10977E+09 | 8.77E-01 |
| 458 | 205 | 216.559 | .64623E-02 | .65258E-02 | .61274E+09 | 9.90E-01 |
| 205 | 457 | 295.239 | .30591E-02 | .22622E-02 | .35113E+09 | 1.35E+00 |
| 205 | 456 | 271.295 | .49920E-02 | .40985E-02 | .67860E+09 | 1.22E+00 |
| 205 | 455 | 266.324 | .24711E-02 | .23259E-02 | .34857E+09 | 1.06E+00 |
| 205 | 454 | 250.765 | .37348E-01 | .26775E-01 | .59424E+10 | 1.39E+00 |
| 205 | 453 | 233.321 | .13037E-03 | .32691E-04 | .23961E+08 | 3.99E+00 |
| 205 | 452 | 224.651 | .13144E-02 | .16832E-03 | .26057E+09 | 7.81E+00 |
| 205 | 451 | 210.018 | .21129E-01 | .14891E-01 | .47929E+10 | 1.42E+00 |
| 205 | 450 | 197.148 | .11737E-02 | .35510E-03 | .30213E+09 | 3.31E+00 |
| 205 | 449 | 180.486 | .13299E-03 | .11462E-03 | .40848E+08 | 1.16E+00 |
| 205 | 448 | 164.296 | .21544E-02 | .16233E-02 | .79852E+09 | 1.33E+00 |
| 205 | 447 | 155.137 | .60659E-02 | .39831E-02 | .25217E+10 | 1.52E+00 |
| 205 | 446 | 81.297 | .81593E-09 | .36680E-06 | .12352E+04 | 2.22E-03 |
| 468 | 204 | 61.266 | .82903E-04 | .19768E-04 | .98216E+08 | 4.19E+00 |
| 467 | 204 | 118.776 | .38773E-02 | .23594E-02 | .12221E+10 | 1.64E+00 |
| 466 | 204 | 123.313 | .55185E-02 | .37520E-02 | .16138E+10 | 1.47E+00 |
| 465 | 204 | 131.918 | .96647E-04 | .12189E-03 | .24696E+08 | 7.93E-01 |
| 464 | 204 | 138.853 | .14026E+00 | .12025E+00 | .32349E+11 | 1.17E+00 |
| 463 | 204 | 145.772 | .99998E-02 | .77681E-02 | .20926E+10 | 1.29E+00 |
| 462 | 204 | 153.652 | .18824E-01 | .14925E-01 | .35456E+10 | 1.26E+00 |
| 461 | 204 | 157.183 | .12528E-02 | .15094E-02 | .22548E+09 | 8.30E-01 |
| 460 | 204 | 188.977 | .24384E+00 | .23201E+00 | .30362E+11 | 1.05E+00 |
| 459 | 204 | 213.067 | .19328E-01 | .21300E-01 | .18932E+10 | 9.07E-01 |
| 458 | 204 | 215.940 | .26294E-01 | .28770E-01 | .25075E+10 | 9.14E-01 |
| 204 | 457 | 296.396 | .37327E-04 | .22396E-04 | .42511E+07 | 1.67E+00 |
| 204 | 456 | 272.272 | .75298E-03 | .67606E-03 | .10163E+09 | 1.11E+00 |
| 204 | 455 | 267.266 | .16392E-02 | .10514E-02 | .22960E+09 | 1.56E+00 |
| 204 | 454 | 251.600 | .48012E-02 | .30751E-02 | .75885E+09 | 1.56E+00 |
| 204 | 453 | 234.044 | .54679E-03 | .23570E-04 | .99874E+08 | 2.32E+01 |
| 204 | 452 | 225.321 | .31374E-03 | .41365E-03 | .61828E+08 | 7.58E-01 |
| 204 | 451 | 210.604 | .43872E-02 | .34432E-02 | .98964E+09 | 1.27E+00 |
| 204 | 450 | 197.663 | .39637E-02 | .70209E-03 | .10150E+10 | 5.65E+00 |
| 204 | 449 | 180.918 | .37708E-02 | .34144E-02 | .11527E+10 | 1.10E+00 |
| 204 | 448 | 164.654 | .14738E-01 | .99773E-02 | .54390E+10 | 1.48E+00 |
| 204 | 447 | 155.456 | .10496E+00 | .66571E-01 | .43452E+11 | 1.58E+00 |
| 204 | 446 | 81.384 | .72055E-07 | .38162E-06 | .10885E+06 | 1.89E-01 |
| 468 | 203 | 59.087 | .64006E-04 | .70987E-06 | .81523E+08 | 9.02E+01 |
| 467 | 203 | 110.852 | .33768E-03 | .19413E-03 | .12220E+09 | 1.74E+00 |
| 466 | 203 | 114.794 | .18689E-03 | .96529E-04 | .63066E+08 | 1.94E+00 |
| 465 | 203 | 122.215 | .34593E-03 | .19114E-03 | .10299E+09 | 1.81E+00 |
| 464 | 203 | 128.145 | .35182E-03 | .14862E-03 | .95269E+08 | 2.37E+00 |
| 463 | 203 | 134.015 | .25527E-02 | .13272E-02 | .63203E+09 | 1.92E+00 |
| 462 | 203 | 140.646 | .41318E-03 | .44530E-03 | .92880E+08 | 9.28E-01 |
| 461 | 203 | 143.599 | .18041E-01 | .13562E-01 | .38905E+10 | 1.33E+00 |
| 460 | 203 | 169.680 | .76291E-01 | .69681E-01 | .11783E+11 | 1.09E+00 |
| 459 | 203 | 188.851 | .37959E-01 | .37413E-01 | .47328E+10 | 1.01E+00 |
| 458 | 203 | 191.105 | .12317E+00 | .11432E+00 | .14997E+11 | 1.08E+00 |
| 203 | 457 | 360.744 | .10417E-02 | .63391E-03 | .80086E+08 | 1.64E+00 |
| 203 | 456 | 325.629 | .17492E-03 | .42259E-04 | .16505E+08 | 4.14E+00 |
| 203 | 455 | 318.494 | .26426E-02 | .16973E-02 | .26065E+09 | 1.56E+00 |
| 203 | 454 | 296.494 | .29865E-03 | .38185E-03 | .33990E+08 | 7.82E-01 |
| 203 | 453 | 272.414 | .88520E-02 | .40977E-02 | .11935E+10 | 2.16E+00 |
| 203 | 452 | 260.668 | .40047E-01 | .17682E-01 | .58968E+10 | 2.26E+00 |
| 203 | 451 | 241.171 | .22738E-07 | .62932E-05 | .39113E+04 | 3.61E-03 |
| 203 | 450 | 224.351 | .12723E-02 | .11054E-02 | .25291E+09 | 1.15E+00 |
| 203 | 449 | 203.023 | .29741E-01 | .13619E-01 | .72191E+10 | 2.18E+00 |
| 203 | 448 | 182.764 | .69160E-03 | .16540E-02 | .20716E+09 | 4.18E-01 |
| 203 | 447 | 171.501 | .26622E-04 | .49271E-04 | .90559E+07 | 5.40E-01 |
| 203 | 446 | 85.575 | .13470E-05 | .84559E-04 | .18403E+07 | 1.59E-02 |
| 468 | 202 | 57.075 | .58001E-06 | .44234E-06 | .79174E+06 | 1.31E+00 |
| 467 | 202 | 103.976 | .93507E-03 | .56701E-03 | .38461E+09 | 1.65E+00 |
| 466 | 202 | 107.437 | .77184E-03 | .46240E-03 | .29735E+09 | 1.67E+00 |
| 465 | 202 | 113.911 | .47395E-03 | .19268E-03 | .16242E+09 | 2.46E+00 |
| 464 | 202 | 119.045 | .17933E-02 | .12278E-02 | .56269E+09 | 1.46E+00 |
| 463 | 202 | 124.095 | .29188E-03 | .11438E-03 | .84284E+08 | 2.55E+00 |
| 462 | 202 | 129.760 | .25427E-05 | .18614E-05 | .67152E+06 | 1.37E+00 |
| 461 | 202 | 132.269 | .57545E-04 | .12039E-03 | .14626E+08 | 4.78E-01 |
| 460 | 202 | 154.084 | .73081E-01 | .61577E-01 | .13688E+11 | 1.19E+00 |
| 459 | 202 | 169.731 | .41373E+00 | .39778E+00 | .63862E+11 | 1.04E+00 |
| 458 | 202 | 171.549 | .52157E-01 | .52651E-01 | .78810E+10 | 9.91E-01 |
| 202 | 457 | 459.658 | .57214E-02 | .39068E-04 | .27093E+09 | 1.46E+02 |
| 202 | 456 | 404.128 | .42998E-03 | .81514E-04 | .26341E+08 | 5.27E+00 |
| 202 | 455 | 393.197 | .21620E-02 | .90464E-04 | .13992E+09 | 2.39E+01 |
| 202 | 454 | 360.201 | .17991E-04 | .63563E-06 | .13874E+07 | 2.83E+01 |
| 202 | 453 | 325.270 | .65873E-03 | .14703E-03 | .62294E+08 | 4.48E+00 |
| 202 | 452 | 308.663 | .85042E-03 | .76808E-06 | .89308E+08 | 1.11E+03 |
| 202 | 451 | 281.696 | .28892E-06 | .32620E-07 | .36429E+05 | 8.86E+00 |
| 202 | 450 | 259.015 | .16389E-01 | .37547E-02 | .24441E+10 | 4.36E+00 |
| 202 | 449 | 230.998 | .19262E-02 | .63147E-03 | .36117E+09 | 3.05E+00 |
| 202 | 448 | 205.127 | .56326E-01 | .25080E-01 | .13393E+11 | 2.25E+00 |
| 202 | 447 | 191.046 | .91680E-02 | .53560E-02 | .25132E+10 | 1.71E+00 |
| 202 | 446 | 90.179 | .30659E-09 | .52027E-06 | .37720E+03 | 5.89E-04 |
| 468 | 201 | 38.136 | .25606E-08 | .68231E-10 | .78290E+04 | 3.75E+01 |
| 467 | 201 | 54.590 | .18719E-07 | .80140E-08 | .27932E+05 | 2.34E+00 |
| 466 | 201 | 55.529 | .48291E-07 | .20085E-07 | .69643E+05 | 2.40E+00 |
| 465 | 201 | 57.209 | .65195E-05 | .14331E-05 | .88578E+07 | 4.55E+00 |
| 464 | 201 | 58.476 | .22122E-07 | .30265E-08 | .28769E+05 | 7.31E+00 |
| 463 | 201 | 59.668 | .11074E-04 | .22109E-05 | .13831E+08 | 5.01E+00 |
| 462 | 201 | 60.948 | .54219E-05 | .10755E-05 | .64905E+07 | 5.04E+00 |
| 461 | 201 | 61.496 | .60139E-08 | .77369E-08 | .70715E+04 | 7.77E-01 |
| 460 | 201 | 65.829 | .54490E-10 | .12269E-08 | .55915E+02 | 4.44E-02 |
| 459 | 201 | 68.528 | .71544E-07 | .75026E-08 | .67746E+05 | 9.54E+00 |
| 458 | 201 | 68.822 | .59279E-08 | .39994E-07 | .55653E+04 | 1.48E-01 |
| 457 | 201 | 153.246 | .17901E-01 | .16921E-01 | .33895E+10 | 1.06E+00 |
| 456 | 201 | 160.603 | .12286E-04 | .34310E-04 | .21180E+07 | 3.58E-01 |
| 455 | 201 | 162.398 | .38976E-01 | .40841E-01 | .65718E+10 | 9.54E-01 |
| 454 | 201 | 168.783 | .63115E-05 | .20462E-05 | .98518E+06 | 3.08E+00 |
| 453 | 201 | 177.727 | .16554E+00 | .20242E+00 | .23304E+11 | 8.18E-01 |
| 452 | 201 | 183.110 | .29063E-01 | .36819E-01 | .38544E+10 | 7.89E-01 |
| 451 | 201 | 194.135 | .43266E-03 | .60230E-03 | .51049E+08 | 7.18E-01 |
| 450 | 201 | 206.603 | .38907E-05 | .32212E-05 | .40531E+06 | 1.21E+00 |
| 449 | 201 | 228.731 | .92270E-04 | .17774E-03 | .78424E+07 | 5.19E-01 |
| 448 | 201 | 261.373 | .17649E-03 | .41569E-03 | .11488E+08 | 4.25E-01 |
| 447 | 201 | 288.465 | .29538E-04 | .69252E-04 | .15785E+07 | 4.27E-01 |
| 201 | 446 | 418.744 | .81857E-09 | .49271E-11 | .46707E+02 | 1.66E+02 |
| 468 | 200 | 37.376 | .38345E-08 | .26914E-09 | .12206E+05 | 1.42E+01 |
| 467 | 200 | 53.044 | .98848E-07 | .91705E-06 | .15622E+06 | 1.08E-01 |
| 466 | 200 | 53.931 | .66458E-07 | .49507E-05 | .10161E+06 | 1.34E-02 |
| 465 | 200 | 55.514 | .27186E-07 | .38483E-07 | .39226E+05 | 7.06E-01 |
| 464 | 200 | 56.706 | .14313E-06 | .32336E-06 | .19793E+06 | 4.43E-01 |
| 463 | 200 | 57.827 | .12952E-04 | .27590E-05 | .17223E+08 | 4.69E+00 |
| 462 | 200 | 59.028 | .37118E-05 | .11052E-05 | .47371E+07 | 3.36E+00 |
| 461 | 200 | 59.542 | .17779E-06 | .73230E-07 | .22300E+06 | 2.43E+00 |
| 460 | 200 | 63.595 | .66807E-06 | .16235E-06 | .73454E+06 | 4.11E+00 |
| 459 | 200 | 66.110 | .15522E-07 | .32313E-08 | .15793E+05 | 4.80E+00 |
| 458 | 200 | 66.384 | .11004E-05 | .46388E-06 | .11104E+07 | 2.37E+00 |
| 457 | 200 | 141.662 | .53986E-04 | .49173E-04 | .11962E+08 | 1.10E+00 |
| 456 | 200 | 147.927 | .44774E-03 | .41228E-03 | .90986E+08 | 1.09E+00 |
| 455 | 200 | 149.447 | .29352E-04 | .26744E-04 | .58439E+07 | 1.10E+00 |
| 454 | 200 | 154.838 | .12073E-01 | .11505E-01 | .22392E+10 | 1.05E+00 |
| 453 | 200 | 162.332 | .34633E-02 | .38011E-02 | .58442E+09 | 9.11E-01 |
| 452 | 200 | 166.811 | .59270E-02 | .68292E-02 | .94716E+09 | 8.68E-01 |
| 451 | 200 | 175.912 | .23872E+00 | .29024E+00 | .34304E+11 | 8.22E-01 |
| 450 | 200 | 186.088 | .13817E-02 | .18268E-02 | .17742E+09 | 7.56E-01 |
| 449 | 200 | 203.851 | .12624E-02 | .19694E-02 | .13508E+09 | 6.41E-01 |
| 448 | 200 | 229.381 | .63742E-04 | .12530E-03 | .53871E+07 | 5.09E-01 |
| 447 | 200 | 249.986 | .23328E-03 | .50568E-03 | .16599E+08 | 4.61E-01 |
| 200 | 446 | 539.229 | .14989E-08 | .33329E-08 | .51576E+02 | 4.50E-01 |
| 468 | 199 | 36.969 | .15998E-06 | .15250E-09 | .52053E+06 | 1.05E+03 |
| 467 | 199 | 52.228 | .33826E-06 | .53500E-06 | .55143E+06 | 6.32E-01 |
| 466 | 199 | 53.087 | .37790E-06 | .37387E-07 | .59627E+06 | 1.01E+01 |
| 465 | 199 | 54.621 | .50477E-06 | .33814E-07 | .75236E+06 | 1.49E+01 |
| 464 | 199 | 55.774 | .65107E-07 | .17621E-06 | .93069E+05 | 3.69E-01 |
| 463 | 199 | 56.858 | .37624E-05 | .43041E-06 | .51751E+07 | 8.74E+00 |
| 462 | 199 | 58.019 | .84070E-06 | .13627E-06 | .11106E+07 | 6.17E+00 |
| 461 | 199 | 58.515 | .43845E-04 | .59053E-05 | .56942E+08 | 7.42E+00 |
| 460 | 199 | 62.425 | .26400E-05 | .38421E-05 | .30125E+07 | 6.87E-01 |
| 459 | 199 | 64.847 | .10998E-04 | .38435E-05 | .11630E+08 | 2.86E+00 |
| 458 | 199 | 65.110 | .95319E-04 | .32806E-04 | .99982E+08 | 2.91E+00 |
| 457 | 199 | 135.984 | .28963E-02 | .24000E-02 | .69647E+09 | 1.21E+00 |
| 456 | 199 | 141.746 | .23987E-03 | .21334E-03 | .53087E+08 | 1.12E+00 |
| 455 | 199 | 143.142 | .13541E-02 | .12584E-02 | .29387E+09 | 1.08E+00 |
| 454 | 199 | 148.080 | .12457E-02 | .11349E-02 | .25262E+09 | 1.10E+00 |
| 453 | 199 | 154.920 | .17116E-01 | .17328E-01 | .31712E+10 | 9.88E-01 |
| 452 | 199 | 158.994 | .82342E-01 | .87106E-01 | .14484E+11 | 9.45E-01 |
| 451 | 199 | 167.241 | .68663E-02 | .75499E-02 | .10916E+10 | 9.09E-01 |
| 450 | 199 | 176.412 | .83832E-01 | .10055E+00 | .11978E+11 | 8.34E-01 |
| 449 | 199 | 192.297 | .14144E-01 | .18757E-01 | .17009E+10 | 7.54E-01 |
| 448 | 199 | 214.855 | .34798E-03 | .63829E-03 | .33520E+08 | 5.45E-01 |
| 447 | 199 | 232.830 | .46388E-02 | .94741E-02 | .38051E+09 | 4.90E-01 |
| 199 | 446 | 641.128 | .40158E-07 | .15597E-07 | .97747E+03 | 2.57E+00 |
| 468 | 198 | 36.528 | .17666E-09 | .12003E-08 | .58875E+03 | 1.47E-01 |
| 467 | 198 | 51.352 | .16404E-05 | .11189E-04 | .27662E+07 | 1.47E-01 |
| 466 | 198 | 52.183 | .47039E-06 | .28011E-05 | .76815E+06 | 1.68E-01 |
| 465 | 198 | 53.664 | .79728E-06 | .48618E-06 | .12311E+07 | 1.64E+00 |
| 464 | 198 | 54.777 | .18336E-07 | .22518E-06 | .27173E+05 | 8.14E-02 |
| 463 | 198 | 55.822 | .19183E-05 | .48693E-06 | .27375E+07 | 3.94E+00 |
| 462 | 198 | 56.940 | .46879E-07 | .38227E-06 | .64296E+05 | 1.23E-01 |
| 461 | 198 | 57.418 | .20360E-05 | .52188E-05 | .27462E+07 | 3.90E-01 |
| 460 | 198 | 61.178 | .72840E-04 | .65716E-05 | .86540E+08 | 1.11E+01 |
| 459 | 198 | 63.502 | .18159E-03 | .14604E-04 | .20024E+09 | 1.24E+01 |
| 458 | 198 | 63.755 | .26569E-04 | .30231E-05 | .29066E+08 | 8.79E+00 |

| 457 | 198 | 130.205 | .16246E-02 | .12842E-02 | .42612E+09 | 1.27E+00 |
| 456 | 198 | 135.478 | .50216E-03 | .36986E-03 | .12166E+09 | 1.36E+00 |
| 455 | 198 | 136.752 | .28196E-05 | .96512E-06 | .67044E+06 | 2.92E+00 |
| 454 | 198 | 141.253 | .24087E-04 | .20861E-04 | .53682E+07 | 1.15E+00 |
| 453 | 198 | 147.463 | .94765E-04 | .99714E-04 | .19379E+08 | 9.50E-01 |
| 452 | 198 | 151.150 | .40332E-02 | .39549E-02 | .78502E+09 | 1.02E+00 |
| 451 | 198 | 158.584 | .99346E-04 | .10751E-03 | .17566E+08 | 9.24E-01 |
| 450 | 198 | 166.807 | .25669E-01 | .30095E-01 | .41022E+10 | 8.53E-01 |
| 449 | 198 | 180.940 | .19065E-04 | .36823E-04 | .25895E+07 | 5.18E-01 |
| 448 | 198 | 200.774 | .27752E+00 | .41068E+00 | .30614E+11 | 6.76E-01 |
| 447 | 198 | 216.385 | .86371E-02 | .13986E-01 | .82027E+09 | 6.18E-01 |
| 198 | 446 | 810.812 | .62808E-09 | .10011E-07 | .95587E+01 | 6.27E-02 |
| 468 | 197 | 35.634 | .51133E-06 | .71017E-06 | .17906E+07 | 7.20E-01 |
| 467 | 197 | 49.604 | .55734E-06 | .48712E-05 | .10072E+07 | 1.14E-01 |
| 466 | 197 | 50.378 | .21098E-07 | .54643E-06 | .36966E+05 | 3.86E-02 |
| 465 | 197 | 51.758 | .29001E-07 | .13660E-06 | .48140E+05 | 2.12E-01 |
| 464 | 197 | 52.792 | .13152E-05 | .19810E-04 | .20984E+07 | 6.64E-02 |
| 463 | 197 | 53.762 | .10213E-06 | .48949E-07 | .15712E+06 | 2.09E+00 |
| 462 | 197 | 54.799 | .50975E-07 | .15571E-07 | .75484E+05 | 3.27E+00 |
| 461 | 197 | 55.241 | .39900E-08 | .19467E-07 | .58141E+04 | 2.05E-01 |
| 460 | 197 | 58.713 | .42235E-04 | .12921E-07 | .54481E+08 | 3.27E+03 |
| 459 | 197 | 60.851 | .18430E-05 | .27738E-06 | .22133E+07 | 6.64E+00 |
| 458 | 197 | 61.083 | .57959E-04 | .10669E-08 | .69076E+08 | 5.43E+04 |
| 457 | 197 | 119.525 | .45490E-05 | .23269E-05 | .14159E+07 | 1.95E+00 |
| 456 | 197 | 123.953 | .25462E-03 | .19189E-03 | .73690E+08 | 1.33E+00 |
| 455 | 197 | 125.020 | .31639E-04 | .22904E-04 | .90014E+07 | 1.38E+00 |
| 454 | 197 | 128.770 | .10433E-02 | .80516E-03 | .27977E+09 | 1.30E+00 |
| 453 | 197 | 133.911 | .43763E-03 | .39732E-03 | .10852E+09 | 1.10E+00 |
| 452 | 197 | 136.945 | .34066E-04 | .33487E-04 | .80774E+07 | 1.02E+00 |
| 451 | 197 | 143.019 | .68613E-03 | .66683E-03 | .14916E+09 | 1.03E+00 |
| 450 | 197 | 149.673 | .18066E-02 | .15661E-02 | .35860E+09 | 1.15E+00 |
| 449 | 197 | 160.954 | .85436E-03 | .10167E-02 | .14665E+09 | 8.40E-01 |
| 448 | 197 | 176.461 | .36498E-01 | .44661E-01 | .52121E+10 | 8.17E-01 |
| 447 | 197 | 188.407 | .28947E+00 | .38978E+00 | .36262E+11 | 7.43E-01 |
| 197 | 446 | 1827.939 | .21934E-06 | .15415E-06 | .65677E+03 | 1.42E+00 |
| 468 | 196 | 32.485 | .29939E+00 | .28449E+00 | .12616E+13 | 1.05E+00 |
| 467 | 196 | 43.706 | .53865E-09 | .57784E-06 | .12539E+04 | 9.32E-04 |
| 466 | 196 | 44.306 | .59732E-06 | .59756E-06 | .13531E+07 | 1.00E+00 |
| 465 | 196 | 45.369 | .27645E-07 | .60149E-07 | .59723E+05 | 4.60E-01 |
| 464 | 196 | 46.162 | .89033E-07 | .28515E-07 | .18579E+06 | 3.12E+00 |
| 463 | 196 | 46.902 | .95041E-04 | .11441E-03 | .19212E+09 | 8.31E-01 |
| 462 | 196 | 47.689 | .43612E-05 | .51638E-05 | .85274E+07 | 8.45E-01 |
| 461 | 196 | 48.023 | .15483E-06 | .69219E-08 | .29853E+06 | 2.24E+01 |
| 460 | 196 | 50.626 | .40965E-03 | .62538E-03 | .71073E+09 | 6.55E-01 |
| 459 | 196 | 52.207 | .11730E-02 | .14687E-02 | .19137E+10 | 7.99E-01 |
| 458 | 196 | 52.378 | .34104E-02 | .40670E-02 | .55278E+10 | 8.39E-01 |
| 457 | 196 | 90.194 | .51170E-07 | .35851E-07 | .27971E+05 | 1.43E+00 |
| 456 | 196 | 92.693 | .16591E-07 | .42748E-08 | .85866E+04 | 3.88E+00 |
| 455 | 196 | 93.288 | .79114E-07 | .15388E-07 | .40425E+05 | 5.14E+00 |
| 454 | 196 | 95.360 | .16729E-09 | .71644E-08 | .81804E+02 | 2.34E-02 |
| 453 | 196 | 98.151 | .73476E-07 | .15895E-06 | .33916E+05 | 4.62E-01 |
| 452 | 196 | 99.770 | .12439E-06 | .69844E-06 | .55566E+05 | 1.78E-01 |
| 451 | 196 | 102.956 | .79568E-09 | .33588E-08 | .33379E+03 | 2.37E-01 |
| 450 | 196 | 106.360 | .73029E-07 | .11683E-06 | .28706E+05 | 6.25E-01 |
| 449 | 196 | 111.935 | .31269E-06 | .85102E-06 | .11098E+06 | 3.67E-01 |
| 448 | 196 | 119.221 | .16091E-06 | .11720E-07 | .50339E+05 | 1.37E+01 |
| 447 | 196 | 124.557 | .88642E-06 | .15943E-06 | .25406E+06 | 5.56E+00 |
| 446 | 196 | 460.042 | .37685E+00 | .37992E+00 | .79180E+10 | 9.92E-01 |
| 216 | 569 | 206.222 | .32237E-02 | .23384E-02 | .50561E+09 | 1.38E+00 |
| 216 | 568 | 176.976 | .24764E-02 | .29383E-02 | .52738E+09 | 8.43E-01 |
| 216 | 567 | 160.969 | .23300E-01 | .19933E-01 | .59980E+10 | 1.17E+00 |
| 216 | 566 | 151.743 | .13478E+00 | .11928E+00 | .39043E+11 | 1.13E+00 |
| 216 | 565 | 148.068 | .41383E-02 | .37159E-02 | .12590E+10 | 1.11E+00 |
| 216 | 564 | 143.060 | .48637E-03 | .38603E-03 | .15851E+09 | 1.26E+00 |
| 216 | 563 | 122.687 | .29715E-04 | .33705E-04 | .13168E+08 | 8.82E-01 |
| 216 | 562 | 117.946 | .46197E-03 | .42043E-03 | .22150E+09 | 1.10E+00 |
| 216 | 561 | 62.962 | .60384E-07 | .73410E-09 | .10160E+06 | 8.23E+01 |
| 216 | 560 | 61.112 | .26873E-05 | .67022E-07 | .47996E+07 | 4.01E+01 |
| 216 | 559 | 59.858 | .44314E-05 | .87704E-07 | .82494E+07 | 5.05E+01 |
| 216 | 558 | 59.299 | .44034E-05 | .62273E-07 | .83528E+07 | 7.07E+01 |
| 216 | 557 | 58.701 | .80144E-05 | .77504E-06 | .15514E+08 | 1.03E+01 |
| 216 | 556 | 57.791 | .31135E-04 | .27918E-05 | .62182E+08 | 1.12E+01 |
| 216 | 555 | 57.342 | .30405E-05 | .16768E-07 | .61677E+07 | 1.81E+02 |
| 216 | 554 | 55.655 | .12295E-05 | .82742E-06 | .26476E+07 | 1.49E+00 |
| 216 | 553 | 54.545 | .10498E-04 | .53978E-05 | .23536E+08 | 1.94E+00 |
| 216 | 552 | 53.692 | .54670E-07 | .13516E-06 | .12649E+06 | 4.04E-01 |
| 216 | 551 | 52.105 | .69687E-05 | .24402E-05 | .17121E+08 | 2.86E+00 |
| 216 | 550 | 34.570 | .18185E-03 | .16629E-03 | .10150E+10 | 1.09E+00 |
| 215 | 569 | 254.144 | .46256E-01 | .40694E-01 | .47769E+10 | 1.14E+00 |
| 215 | 568 | 211.143 | .27247E-02 | .33921E-02 | .40766E+09 | 8.03E-01 |
| 215 | 567 | 188.750 | .15059E-01 | .16259E-01 | .28195E+10 | 9.26E-01 |
| 215 | 566 | 176.189 | .31287E-02 | .20946E-02 | .67225E+09 | 1.49E+00 |
| 215 | 565 | 171.254 | .14178E+00 | .13346E+00 | .32246E+11 | 1.06E+00 |
| 215 | 564 | 164.591 | .57068E-01 | .54122E-01 | .14051E+11 | 1.05E+00 |
| 215 | 563 | 138.190 | .45873E-02 | .51425E-02 | .16023E+10 | 8.92E-01 |
| 215 | 562 | 132.204 | .31340E-04 | .45455E-03 | .11960E+08 | 6.89E-02 |
| 215 | 561 | 66.808 | .49159E-06 | .69464E-07 | .73464E+06 | 7.08E+00 |
| 215 | 560 | 64.729 | .11621E-04 | .16277E-06 | .18499E+08 | 7.14E+01 |
| 215 | 559 | 63.324 | .21536E-07 | .11882E-05 | .35823E+05 | 1.81E-02 |
| 215 | 558 | 62.699 | .10633E-05 | .38871E-06 | .18041E+07 | 2.74E+00 |
| 215 | 557 | 62.030 | .41060E-05 | .10947E-05 | .71178E+07 | 3.75E+00 |
| 215 | 556 | 61.015 | .17218E-05 | .38178E-08 | .30848E+07 | 4.51E+02 |
| 215 | 555 | 60.515 | .74593E-06 | .60174E-07 | .13586E+07 | 1.24E+01 |
| 215 | 554 | 58.640 | .35509E-05 | .36679E-05 | .68879E+07 | 9.68E-01 |
| 215 | 553 | 57.408 | .48895E-04 | .31162E-04 | .98958E+08 | 1.57E+00 |
| 215 | 552 | 56.465 | .31309E-06 | .16038E-05 | .65501E+06 | 1.95E-01 |
| 215 | 551 | 54.711 | .57483E-04 | .11115E-04 | .12809E+09 | 5.17E+00 |
| 215 | 550 | 35.698 | .53329E-02 | .48624E-02 | .27913E+11 | 1.10E+00 |
| 214 | 569 | 587.031 | .12022E-02 | .10720E-02 | .23270E+08 | 1.12E+00 |
| 214 | 568 | 399.228 | .30530E-04 | .24379E-04 | .12777E+07 | 1.25E+00 |
| 214 | 567 | 326.081 | .19960E-02 | .21977E-02 | .12521E+09 | 9.08E-01 |
| 214 | 566 | 290.323 | .63782E-03 | .24040E-03 | .50474E+08 | 2.65E+00 |
| 214 | 565 | 277.161 | .18532E-02 | .71655E-05 | .16091E+09 | 2.59E+02 |
| 214 | 564 | 260.119 | .13686E-01 | .99367E-02 | .13491E+10 | 1.38E+00 |
| 214 | 563 | 199.795 | .12442E-02 | .55090E-03 | .20790E+09 | 2.26E+00 |
| 214 | 562 | 187.519 | .18325E+00 | .14060E+00 | .34760E+11 | 1.30E+00 |
| 214 | 561 | 78.512 | .55314E-08 | .27679E-06 | .59854E+04 | 2.00E-02 |
| 214 | 560 | 75.656 | .12035E-07 | .59369E-05 | .14024E+05 | 2.03E-03 |
| 214 | 559 | 73.743 | .51916E-06 | .38365E-06 | .63677E+06 | 1.35E+00 |
| 214 | 558 | 72.897 | .26289E-07 | .83800E-06 | .32998E+05 | 3.14E-02 |
| 214 | 557 | 71.995 | .15796E-06 | .27755E-05 | .20327E+06 | 5.69E-02 |
| 214 | 556 | 70.631 | .35053E-06 | .18067E-06 | .46866E+06 | 1.94E+00 |
| 214 | 555 | 69.962 | .13761E-06 | .52440E-08 | .18752E+06 | 2.62E+01 |
| 214 | 554 | 67.467 | .17095E-05 | .15477E-06 | .25051E+07 | 1.10E+01 |
| 214 | 553 | 65.842 | .52765E-04 | .10043E-04 | .81184E+08 | 5.25E+00 |
| 214 | 552 | 64.604 | .39597E-05 | .27949E-05 | .63281E+07 | 1.42E+00 |
| 214 | 551 | 62.319 | .15292E-03 | .66237E-04 | .26264E+09 | 2.31E+00 |
| 214 | 550 | 38.788 | .40730E-01 | .37198E-01 | .18058E+12 | 1.09E+00 |
| 569 | 213 | 185.299 | .57235E-03 | .87389E-03 | .11119E+09 | 6.55E-01 |
| 568 | 213 | 217.612 | .33819E-02 | .30781E-02 | .47636E+09 | 1.10E+00 |
| 567 | 213 | 247.927 | .10869E-02 | .11422E-02 | .11794E+09 | 9.52E-01 |
| 566 | 213 | 273.543 | .26223E-02 | .39651E-02 | .23376E+09 | 6.61E-01 |
| 565 | 213 | 286.355 | .14047E-02 | .85310E-03 | .11426E+09 | 1.65E+00 |
| 564 | 213 | 307.147 | .38317E-01 | .42071E-01 | .27092E+10 | 9.11E-01 |
| 563 | 213 | 477.317 | .16283E-03 | .57882E-04 | .47671E+07 | 2.81E+00 |
| 562 | 213 | 565.810 | .17678E-03 | .49886E-07 | .36832E+07 | 3.54E+03 |
| 213 | 561 | 177.407 | .34555E-02 | .33963E-02 | .73233E+09 | 1.02E+00 |
| 213 | 560 | 163.464 | .35821E-01 | .33289E-01 | .89418E+10 | 1.08E+00 |
| 213 | 559 | 154.790 | .40918E-02 | .51978E-02 | .11391E+10 | 7.87E-01 |
| 213 | 558 | 151.106 | .11896E-02 | .75720E-03 | .34752E+09 | 1.57E+00 |
| 213 | 557 | 147.281 | .61070E-03 | .51695E-03 | .18779E+09 | 1.18E+00 |
| 213 | 556 | 141.687 | .42274E-02 | .33263E-02 | .14046E+10 | 1.27E+00 |
| 213 | 555 | 139.017 | .62020E-02 | .50044E-02 | .21406E+10 | 1.24E+00 |
| 213 | 554 | 129.503 | .50227E-01 | .41944E-01 | .19976E+11 | 1.20E+00 |
| 213 | 553 | 123.645 | .37460E-02 | .36301E-02 | .16344E+10 | 1.03E+00 |
| 213 | 552 | 119.349 | .43610E-02 | .32922E-02 | .20421E+10 | 1.32E+00 |
| 213 | 551 | 111.778 | .25400E-04 | .34657E-06 | .13560E+08 | 7.33E+01 |
| 213 | 550 | 53.530 | .14942E-04 | .23859E-07 | .34781E+08 | 6.26E+02 |
| 569 | 212 | 182.191 | .11132E-03 | .19927E-03 | .22369E+08 | 5.59E-01 |
| 568 | 212 | 213.338 | .60377E-01 | .54670E-01 | .88484E+10 | 1.10E+00 |
| 567 | 212 | 242.395 | .17007E-01 | .18201E-01 | .19307E+10 | 9.34E-01 |
| 566 | 212 | 266.824 | .67610E-02 | .86468E-02 | .63342E+09 | 7.82E-01 |
| 565 | 212 | 279.001 | .12614E-03 | .20829E-03 | .10809E+08 | 6.06E-01 |
| 564 | 212 | 298.701 | .98422E-03 | .14657E-02 | .73579E+08 | 6.72E-01 |
| 563 | 212 | 457.228 | .77933E-07 | .69432E-05 | .24865E+04 | 1.12E-02 |
| 562 | 212 | 537.800 | .17655E-03 | .19509E-03 | .40716E+07 | 9.05E-01 |
| 212 | 561 | 180.353 | .30101E-02 | .22499E-02 | .61725E+09 | 1.34E+00 |
| 212 | 560 | 165.961 | .77486E-02 | .71249E-02 | .18765E+10 | 1.09E+00 |
| 212 | 559 | 157.028 | .24589E-01 | .21179E-01 | .66515E+10 | 1.16E+00 |
| 212 | 558 | 153.238 | .16708E+00 | .14392E+00 | .47459E+11 | 1.16E+00 |
| 212 | 557 | 149.305 | .10648E-06 | .77902E-04 | .31859E+05 | 1.37E-03 |
| 212 | 556 | 143.559 | .99782E-03 | .59310E-03 | .32294E+09 | 1.68E+00 |
| 212 | 555 | 140.820 | .51519E-04 | .12570E-04 | .17329E+08 | 4.10E+00 |
| 212 | 554 | 131.065 | .23801E-03 | .15243E-03 | .92417E+08 | 1.56E+00 |
| 212 | 553 | 125.068 | .85069E-03 | .40792E-03 | .36275E+09 | 2.09E+00 |
| 212 | 552 | 120.675 | .31919E-04 | .11090E-04 | .14620E+08 | 2.88E+00 |
| 212 | 551 | 112.940 | .33394E-05 | .74536E-06 | .17462E+07 | 4.48E+00 |
| 212 | 550 | 53.795 | .40684E-07 | .12895E-07 | .93773E+05 | 3.16E+00 |
| 569 | 211 | 167.566 | .23496E-02 | .23242E-02 | .55816E+09 | 1.01E+00 |
| 568 | 211 | 193.557 | .53360E-01 | .41812E-01 | .95002E+10 | 1.28E+00 |
| 567 | 211 | 217.176 | .86123E-02 | .79313E-02 | .12179E+10 | 1.09E+00 |
| 566 | 211 | 236.584 | .33519E-01 | .33758E-01 | .39944E+10 | 9.93E-01 |
| 565 | 211 | 246.107 | .81244E-02 | .10625E-01 | .89470E+09 | 7.65E-01 |
| 564 | 211 | 261.310 | .50528E-02 | .48215E-02 | .49357E+09 | 1.05E+00 |
| 563 | 211 | 375.073 | .21028E-08 | .28741E-05 | .99701E+02 | 7.32E-04 |
| 562 | 211 | 427.628 | .11665E-04 | .48303E-04 | .42547E+06 | 2.41E-01 |
| 211 | 561 | 197.408 | .46149E-03 | .48873E-03 | .78988E+08 | 9.44E-01 |
| 211 | 560 | 180.296 | .42025E-03 | .15258E-03 | .86232E+08 | 2.75E+00 |
| 211 | 559 | 169.801 | .20418E-01 | .16989E-01 | .47235E+10 | 1.20E+00 |
| 211 | 558 | 165.378 | .13999E-01 | .11758E-01 | .34141E+10 | 1.19E+00 |
| 211 | 557 | 160.806 | .25663E-01 | .19289E-01 | .66197E+10 | 1.33E+00 |
| 211 | 556 | 154.161 | .49957E-01 | .40565E-01 | .14021E+11 | 1.23E+00 |
| 211 | 555 | 151.006 | .38195E-01 | .32975E-01 | .11172E+11 | 1.16E+00 |
| 211 | 554 | 139.846 | .98188E-02 | .80506E-02 | .33489E+10 | 1.22E+00 |
| 211 | 553 | 133.039 | .24522E-04 | .87291E-04 | .92411E+07 | 2.81E-01 |
| 211 | 552 | 128.079 | .22704E-03 | .30079E-03 | .92316E+08 | 7.55E-01 |
| 211 | 551 | 119.400 | .12035E-02 | .91000E-03 | .56310E+09 | 1.32E+00 |
| 211 | 550 | 55.218 | .31302E-06 | .85064E-06 | .68478E+06 | 3.68E-01 |
| 569 | 210 | 156.065 | .62737E-01 | .56280E-01 | .17181E+11 | 1.11E+00 |
| 568 | 210 | 178.373 | .22663E-04 | .42483E-04 | .47511E+07 | 5.33E-01 |
| 567 | 210 | 198.242 | .60906E-01 | .50228E-01 | .10337E+11 | 1.21E+00 |
| 566 | 210 | 214.287 | .52714E-04 | .25898E-04 | .76571E+07 | 2.04E+00 |
| 565 | 210 | 222.071 | .10189E-01 | .10907E-01 | .13781E+10 | 9.34E-01 |
| 564 | 210 | 234.375 | .43887E-02 | .40636E-02 | .53290E+09 | 1.08E+00 |
| 563 | 210 | 321.964 | .12936E-02 | .16546E-02 | .83236E+08 | 7.82E-01 |
| 562 | 210 | 359.936 | .11421E-01 | .11619E-01 | .58802E+09 | 9.83E-01 |
| 210 | 561 | 216.177 | .58083E-03 | .61706E-03 | .82902E+08 | 9.41E-01 |
| 210 | 560 | 195.823 | .19533E-01 | .15949E-01 | .33976E+10 | 1.22E+00 |
| 210 | 559 | 183.505 | .72751E-02 | .53430E-02 | .14410E+10 | 1.36E+00 |
| 210 | 558 | 178.350 | .21622E-02 | .15932E-02 | .45341E+09 | 1.36E+00 |
| 210 | 557 | 173.044 | .17745E-02 | .82316E-03 | .39527E+09 | 2.16E+00 |
| 210 | 556 | 165.373 | .87771E-02 | .84195E-02 | .21407E+10 | 1.04E+00 |
| 210 | 555 | 161.748 | .24217E-01 | .17605E-01 | .61740E+10 | 1.38E+00 |
| 210 | 554 | 149.010 | .12245E-01 | .95289E-02 | .36784E+10 | 1.29E+00 |
| 210 | 553 | 141.307 | .25992E-01 | .18906E-01 | .86826E+10 | 1.37E+00 |
| 210 | 552 | 135.724 | .44335E-04 | .66162E-04 | .16053E+08 | 6.70E-01 |
| 210 | 551 | 126.018 | .18956E-01 | .14243E-01 | .79618E+10 | 1.33E+00 |
| 210 | 550 | 56.592 | .17541E-05 | .68727E-05 | .36532E+07 | 2.55E-01 |
| 569 | 209 | 148.505 | .85844E-01 | .69781E-01 | .25964E+11 | 1.23E+00 |
| 568 | 209 | 168.565 | .12699E-03 | .49909E-04 | .29809E+08 | 2.54E+00 |
| 567 | 209 | 186.201 | .65411E-02 | .49691E-02 | .12584E+10 | 1.32E+00 |
| 566 | 209 | 200.287 | .27466E-01 | .22017E-01 | .45669E+10 | 1.25E+00 |
| 565 | 209 | 207.071 | .26931E-01 | .25413E-01 | .41894E+10 | 1.06E+00 |
| 564 | 209 | 217.729 | .60273E-02 | .53444E-02 | .84805E+09 | 1.13E+00 |
| 563 | 209 | 291.363 | .62387E-03 | .45697E-03 | .49018E+08 | 1.37E+00 |
| 562 | 209 | 322.115 | .94608E-02 | .62174E-02 | .60819E+09 | 1.52E+00 |
| 209 | 561 | 232.577 | .31546E-03 | .32457E-03 | .38899E+08 | 9.72E-01 |
| 209 | 560 | 209.186 | .93369E-02 | .84797E-02 | .14232E+10 | 1.10E+00 |
| 209 | 559 | 195.188 | .22835E-03 | .18783E-03 | .39978E+08 | 1.22E+00 |
| 209 | 558 | 189.367 | .34049E-02 | .30998E-02 | .63334E+09 | 1.10E+00 |
| 209 | 557 | 183.397 | .19230E-01 | .14633E-01 | .38136E+10 | 1.31E+00 |
| 209 | 556 | 174.803 | .11115E-03 | .45183E-03 | .24263E+08 | 2.46E-01 |
| 209 | 555 | 170.758 | .11795E-01 | .85755E-02 | .26981E+10 | 1.38E+00 |
| 209 | 554 | 156.623 | .17637E-01 | .14186E-01 | .47955E+10 | 1.24E+00 |
| 209 | 553 | 148.136 | .32038E-01 | .25083E-01 | .97382E+10 | 1.28E+00 |
| 209 | 552 | 142.012 | .54137E-04 | .88811E-05 | .17905E+08 | 6.10E+00 |
| 209 | 551 | 131.420 | .30258E-01 | .22757E-01 | .11685E+11 | 1.33E+00 |
| 209 | 550 | 57.656 | .38746E-05 | .11893E-04 | .77744E+07 | 3.26E-01 |
| 569 | 208 | 135.759 | .49236E-02 | .37345E-02 | .17819E+10 | 1.32E+00 |
| 568 | 208 | 152.331 | .36751E-04 | .38672E-04 | .10564E+08 | 9.50E-01 |
| 567 | 208 | 166.590 | .80758E-04 | .69789E-05 | .19410E+08 | 1.16E+01 |
| 566 | 208 | 177.776 | .59257E-03 | .37229E-03 | .12506E+09 | 1.59E+00 |
| 565 | 208 | 183.101 | .83740E-03 | .62540E-03 | .16661E+09 | 1.34E+00 |
| 564 | 208 | 191.384 | .89197E-01 | .82112E-01 | .16243E+11 | 1.09E+00 |
| 563 | 208 | 246.042 | .33558E-01 | .31585E-01 | .36975E+10 | 1.06E+00 |
| 562 | 208 | 267.617 | .19391E-02 | .28652E-02 | .18060E+09 | 6.77E-01 |
| 208 | 561 | 272.670 | .21655E-02 | .17730E-02 | .19428E+09 | 1.22E+00 |
| 208 | 560 | 241.066 | .48471E-02 | .17476E-02 | .55634E+09 | 2.77E+00 |
| 208 | 559 | 222.665 | .67980E-02 | .38126E-02 | .91455E+09 | 1.78E+00 |
| 208 | 558 | 215.121 | .14294E-04 | .13754E-03 | .20602E+07 | 1.04E-01 |
| 208 | 557 | 207.449 | .79810E-03 | .51820E-03 | .12370E+09 | 1.54E+00 |
| 208 | 556 | 196.521 | .14530E-04 | .18004E-04 | .25094E+07 | 8.07E-01 |
| 208 | 555 | 191.423 | .59670E-03 | .55202E-03 | .10862E+09 | 1.08E+00 |
| 208 | 554 | 173.836 | .98191E-01 | .89727E-01 | .21673E+11 | 1.09E+00 |
| 208 | 553 | 163.442 | .10713E-05 | .55069E-04 | .26750E+06 | 1.95E-02 |
| 208 | 552 | 156.019 | .67250E-02 | .66640E-02 | .18428E+10 | 1.01E+00 |
| 208 | 551 | 143.328 | .65728E-02 | .55944E-02 | .21341E+10 | 1.17E+00 |
| 208 | 550 | 59.837 | .13876E-03 | .29558E-06 | .25850E+09 | 4.69E+02 |
| 569 | 207 | 132.066 | .25695E-02 | .19541E-02 | .98266E+09 | 1.31E+00 |
| 568 | 207 | 147.696 | .11423E-02 | .81779E-03 | .34927E+09 | 1.40E+00 |
| 567 | 207 | 161.063 | .22416E-02 | .13832E-02 | .57637E+09 | 1.62E+00 |
| 566 | 207 | 171.496 | .47583E-02 | .39492E-02 | .10791E+10 | 1.20E+00 |
| 565 | 207 | 176.445 | .28300E-01 | .21559E-01 | .60631E+10 | 1.31E+00 |
| 564 | 207 | 184.125 | .13079E-01 | .88594E-02 | .25733E+10 | 1.48E+00 |
| 563 | 207 | 234.173 | .24026E-02 | .30224E-02 | .29224E+09 | 7.95E-01 |
| 562 | 207 | 253.634 | .47827E-02 | .28759E-02 | .49590E+09 | 1.66E+00 |
| 207 | 561 | 288.898 | .46221E-03 | .55288E-03 | .36939E+08 | 8.36E-01 |
| 207 | 560 | 253.663 | .17385E-02 | .80576E-03 | .18021E+09 | 2.16E+00 |
| 207 | 559 | 233.370 | .83279E-03 | .33232E-03 | .10200E+09 | 2.51E+00 |
| 207 | 558 | 225.096 | .12190E-02 | .39457E-03 | .16047E+09 | 3.09E+00 |
| 207 | 557 | 216.710 | .28667E-02 | .17611E-02 | .40715E+09 | 1.63E+00 |
| 207 | 556 | 204.813 | .30046E-03 | .10431E-03 | .47776E+08 | 2.88E+00 |
| 207 | 555 | 199.282 | .21540E-04 | .21797E-04 | .36178E+07 | 9.88E-01 |
| 207 | 554 | 180.293 | .55876E-02 | .61507E-02 | .11466E+10 | 9.08E-01 |
| 207 | 553 | 169.137 | .16820E+00 | .15809E+00 | .39217E+11 | 1.06E+00 |
| 207 | 552 | 161.201 | .14758E-02 | .14312E-02 | .37882E+09 | 1.03E+00 |
| 207 | 551 | 147.689 | .40285E-01 | .41127E-01 | .12319E+11 | 9.80E-01 |
| 207 | 550 | 60.584 | .22995E-05 | .19090E-04 | .41787E+07 | 1.20E-01 |
| 569 | 206 | 129.385 | .29437E-04 | .16538E-04 | .11729E+08 | 1.78E+00 |
| 568 | 206 | 144.352 | .90920E-01 | .70756E-01 | .29104E+11 | 1.28E+00 |
| 567 | 206 | 157.094 | .95567E-02 | .86473E-02 | .25830E+10 | 1.11E+00 |
| 566 | 206 | 167.004 | .20448E-01 | .21302E-01 | .48901E+10 | 9.60E-01 |
| 565 | 206 | 171.694 | .24996E-02 | .25451E-02 | .56558E+09 | 9.82E-01 |
| 564 | 206 | 178.957 | .12687E-05 | .34899E-06 | .26425E+06 | 3.64E+00 |
| 563 | 206 | 225.876 | .45632E-04 | .21466E-03 | .59657E+07 | 2.13E-01 |
| 562 | 206 | 243.930 | .16979E-05 | .12207E-04 | .19033E+06 | 1.39E-01 |
| 206 | 561 | 302.610 | .15967E-01 | .18708E-01 | .11630E+10 | 8.53E-01 |
| 206 | 560 | 264.174 | .34634E-02 | .39011E-02 | .33102E+09 | 8.88E-01 |
| 206 | 559 | 242.237 | .39148E-03 | .41122E-03 | .44500E+08 | 9.52E-01 |
| 206 | 558 | 233.334 | .46283E-01 | .34483E-01 | .56702E+10 | 1.34E+00 |
| 206 | 557 | 224.336 | .45975E-02 | .23190E-02 | .60934E+09 | 1.98E+00 |
| 206 | 556 | 211.611 | .20399E-01 | .12242E-01 | .30386E+10 | 1.67E+00 |
| 206 | 555 | 205.712 | .30933E-01 | .17977E-01 | .48758E+10 | 1.72E+00 |
| 206 | 554 | 185.540 | .68235E-03 | .41597E-03 | .13221E+09 | 1.64E+00 |
| 206 | 553 | 173.747 | .12283E-02 | .11513E-02 | .27139E+09 | 1.07E+00 |
| 206 | 552 | 165.382 | .25639E-03 | .17218E-03 | .62526E+08 | 1.49E+00 |
| 206 | 551 | 151.192 | .57853E-03 | .58618E-03 | .16881E+09 | 9.87E-01 |
| 206 | 550 | 61.165 | .53843E-06 | .16694E-10 | .95995E+06 | 3.23E+04 |
| 569 | 205 | 121.125 | .24787E-03 | .22870E-03 | .11269E+09 | 1.08E+00 |
| 568 | 205 | 134.146 | .50786E-03 | .24954E-03 | .18825E+09 | 2.04E+00 |
| 567 | 205 | 145.082 | .79976E-01 | .58468E-01 | .25344E+11 | 1.37E+00 |
| 566 | 205 | 153.493 | .76241E-01 | .52739E-01 | .21585E+11 | 1.45E+00 |
| 565 | 205 | 157.446 | .25402E-01 | .17904E-01 | .68349E+10 | 1.42E+00 |
| 564 | 205 | 163.533 | .15838E-04 | .31963E-04 | .39502E+07 | 4.96E-01 |
| 563 | 205 | 201.847 | .23788E-03 | .49926E-04 | .38945E+08 | 4.76E+00 |
| 562 | 205 | 216.142 | .10449E-01 | .95663E-02 | .14919E+10 | 1.09E+00 |
| 205 | 561 | 360.032 | .62056E-07 | .11751E-05 | .31933E+04 | 5.28E-02 |
| 205 | 560 | 306.906 | .36245E-02 | .31823E-02 | .25667E+09 | 1.14E+00 |
| 205 | 559 | 277.690 | .16267E-01 | .12272E-01 | .14071E+10 | 1.33E+00 |
| 205 | 558 | 266.053 | .84933E-02 | .69373E-02 | .80034E+09 | 1.22E+00 |
| 205 | 557 | 254.417 | .19917E-01 | .18431E-01 | .20524E+10 | 1.08E+00 |
| 205 | 556 | 238.174 | .51412E-01 | .37184E-01 | .60452E+10 | 1.38E+00 |
| 205 | 555 | 230.727 | .27756E-02 | .95892E-03 | .34777E+09 | 2.89E+00 |
| 205 | 554 | 205.650 | .21334E-02 | .17875E-02 | .33647E+09 | 1.19E+00 |
| 205 | 553 | 191.261 | .66988E-03 | .99949E-04 | .12214E+09 | 6.70E+00 |
| 205 | 552 | 181.174 | .12195E-02 | .10608E-02 | .24781E+09 | 1.15E+00 |
| 205 | 551 | 164.283 | .19081E-01 | .15558E-01 | .47157E+10 | 1.23E+00 |
| 205 | 550 | 63.203 | .81202E-05 | .29467E-07 | .13559E+08 | 2.76E+02 |
| 569 | 204 | 120.932 | .71406E-02 | .48851E-02 | .32568E+10 | 1.46E+00 |
| 568 | 204 | 133.908 | .31989E-05 | .10539E-05 | .11899E+07 | 3.04E+00 |
| 567 | 204 | 144.804 | .30571E-02 | .22320E-02 | .97248E+09 | 1.37E+00 |
| 566 | 204 | 153.182 | .71315E-02 | .43069E-02 | .20272E+10 | 1.66E+00 |
| 565 | 204 | 157.119 | .95053E-03 | .14222E-02 | .25683E+09 | 6.68E-01 |
| 564 | 204 | 163.179 | .17376E-01 | .15615E-01 | .43527E+10 | 1.11E+00 |
| 563 | 204 | 201.309 | .25293E-02 | .10764E-02 | .41629E+09 | 2.35E+00 |
| 562 | 204 | 215.525 | .79850E-01 | .68130E-01 | .11466E+11 | 1.17E+00 |
| 204 | 561 | 361.756 | .19261E-03 | .98579E-04 | .98170E+07 | 1.95E+00 |
| 204 | 560 | 308.157 | .16516E-02 | .14839E-02 | .11601E+09 | 1.11E+00 |
| 204 | 559 | 278.714 | .54325E-05 | .28887E-03 | .46646E+06 | 1.88E-02 |
| 204 | 558 | 266.993 | .54503E-06 | .37042E-05 | .50998E+05 | 1.47E-01 |
| 204 | 557 | 255.277 | .84266E-02 | .56598E-02 | .86250E+09 | 1.49E+00 |
| 204 | 556 | 238.927 | .60866E-02 | .32630E-02 | .71118E+09 | 1.87E+00 |
| 204 | 555 | 231.434 | .22536E-02 | .18933E-02 | .28064E+09 | 1.19E+00 |
| 204 | 554 | 206.211 | .10470E-01 | .90027E-02 | .16423E+10 | 1.16E+00 |
| 204 | 553 | 191.746 | .45562E-03 | .12147E-03 | .82657E+08 | 3.75E+00 |
| 204 | 552 | 181.610 | .92401E-02 | .84368E-02 | .18687E+10 | 1.10E+00 |
| 204 | 551 | 164.640 | .41971E-01 | .35389E-01 | .10328E+11 | 1.19E+00 |
| 204 | 550 | 63.256 | .32079E-04 | .25886E-05 | .53475E+08 | 1.24E+01 |
| 569 | 203 | 112.727 | .13509E-02 | .66923E-03 | .70910E+09 | 2.02E+00 |
| 568 | 203 | 123.922 | .20100E-02 | .11313E-02 | .87302E+09 | 1.78E+00 |
| 567 | 203 | 133.196 | .18950E-01 | .12915E-01 | .71247E+10 | 1.47E+00 |
| 566 | 203 | 140.253 | .14393E-01 | .10809E-01 | .48804E+10 | 1.33E+00 |
| 565 | 203 | 143.546 | .14768E+00 | .11679E+00 | .47804E+11 | 1.26E+00 |
| 564 | 203 | 148.588 | .27233E-01 | .20887E-01 | .82275E+10 | 1.30E+00 |
| 563 | 203 | 179.556 | .49582E-02 | .20851E-02 | .10258E+10 | 2.38E+00 |
| 562 | 203 | 190.780 | .91206E-03 | .92628E-03 | .16714E+09 | 9.85E-01 |
| 203 | 561 | 462.432 | .31017E-03 | .17224E-03 | .96748E+07 | 1.80E+00 |
| 203 | 560 | 378.317 | .94115E-02 | .64774E-02 | .43861E+09 | 1.45E+00 |
| 203 | 559 | 334.886 | .25664E-03 | .92462E-03 | .15264E+08 | 2.78E-01 |
| 203 | 558 | 318.107 | .12635E-02 | .57978E-03 | .83283E+08 | 2.18E+00 |
| 203 | 557 | 301.613 | .62755E-02 | .40889E-02 | .46013E+09 | 1.53E+00 |
| 203 | 556 | 279.052 | .46233E-03 | .40378E-03 | .39602E+08 | 1.15E+00 |
| 203 | 555 | 268.884 | .11059E-02 | .66962E-03 | .10202E+09 | 1.65E+00 |
| 203 | 554 | 235.428 | .22822E-01 | .21342E-01 | .27464E+10 | 1.07E+00 |
| 203 | 553 | 216.759 | .56994E-01 | .36608E-01 | .80911E+10 | 1.56E+00 |
| 203 | 552 | 203.894 | .30258E-03 | .45578E-04 | .48548E+08 | 6.64E+00 |
| 203 | 551 | 182.748 | .50434E-01 | .57100E-01 | .10073E+11 | 8.83E-01 |
| 203 | 550 | 65.759 | .10762E-03 | .79502E-04 | .16601E+09 | 1.35E+00 |
| 569 | 202 | 105.625 | .86125E-04 | .66536E-04 | .51491E+08 | 1.29E+00 |
| 568 | 202 | 115.392 | .21531E-03 | .11942E-03 | .10786E+09 | 1.80E+00 |
| 567 | 202 | 123.392 | .55287E-04 | .20112E-04 | .24220E+08 | 2.75E+00 |
| 566 | 202 | 129.424 | .84830E-03 | .44844E-03 | .33779E+09 | 1.89E+00 |
| 565 | 202 | 132.224 | .97651E-02 | .67622E-02 | .37256E+10 | 1.44E+00 |
| 564 | 202 | 136.490 | .75453E-01 | .54500E-01 | .27015E+11 | 1.38E+00 |
| 563 | 202 | 162.184 | .46528E-01 | .37714E-01 | .11799E+11 | 1.23E+00 |
| 562 | 202 | 171.287 | .49125E-01 | .43836E-01 | .11168E+11 | 1.12E+00 |
| 202 | 561 | 638.584 | .10900E-02 | .29851E-03 | .17828E+08 | 3.65E+00 |
| 202 | 560 | 488.576 | .46944E-02 | .34215E-03 | .13118E+09 | 1.37E+01 |
| 202 | 559 | 418.485 | .17854E-02 | .29355E-03 | .67998E+08 | 6.08E+00 |
| 202 | 558 | 392.606 | .43411E-03 | .56418E-03 | .18785E+08 | 7.69E-01 |
| 202 | 557 | 367.784 | .72222E-04 | .19873E-03 | .35613E+07 | 3.63E-01 |
| 202 | 556 | 334.779 | .73712E-04 | .26112E-04 | .43868E+07 | 2.82E+00 |
| 202 | 555 | 320.250 | .51010E-03 | .40173E-03 | .33175E+08 | 1.27E+00 |
| 202 | 554 | 273.892 | .37419E-01 | .29457E-01 | .33271E+10 | 1.27E+00 |
| 202 | 553 | 248.949 | .16041E-03 | .84637E-03 | .17265E+08 | 1.90E-01 |
| 202 | 552 | 232.127 | .47890E-01 | .34265E-01 | .59282E+10 | 1.40E+00 |
| 560 | 200 | 139.124 | .69749E-03 | .59948E-03 | .24036E+09 | 1.16E+00 |
| 559 | 200 | 146.092 | .28529E-03 | .26132E-03 | .89158E+08 | 1.09E+00 |
| 558 | 200 | 149.533 | .63165E-02 | .59655E-02 | .18843E+10 | 1.06E+00 |
| 557 | 200 | 153.478 | .33220E-02 | .31646E-02 | .94069E+09 | 1.05E+00 |
| 556 | 200 | 160.063 | .10880E+00 | .11254E+00 | .28325E+11 | 9.67E-01 |
| 555 | 200 | 163.612 | .14016E-03 | .23233E-03 | .34925E+08 | 6.03E-01 |
| 554 | 200 | 179.099 | .15346E-04 | .24730E-04 | .31912E+07 | 6.21E-01 |
| 553 | 200 | 191.656 | .75310E-03 | .97586E-03 | .13675E+09 | 7.72E-01 |
| 552 | 200 | 202.980 | .80297E-04 | .11605E-03 | .12999E+08 | 6.92E-01 |
| 551 | 200 | 229.407 | .11622E-03 | .23272E-03 | .14730E+08 | 4.99E-01 |
| 200 | 550 | 186.015 | .16255E-08 | .25593E-08 | .31334E+03 | 6.35E-01 |
| 569 | 199 | 52.640 | .14761E-06 | .31735E-05 | .35532E+06 | 4.65E-02 |
| 568 | 199 | 54.959 | .31293E-05 | .17163E-06 | .69104E+07 | 1.82E+01 |
| 567 | 199 | 56.710 | .26128E-04 | .19862E-05 | .54190E+08 | 1.32E+01 |
| 566 | 199 | 57.951 | .26585E-04 | .22186E-05 | .52800E+08 | 1.20E+01 |
| 565 | 199 | 58.506 | .93131E-04 | .72267E-05 | .18148E+09 | 1.29E+01 |
| 564 | 199 | 59.327 | .50271E-04 | .55201E-05 | .95269E+08 | 9.11E+00 |
| 563 | 199 | 63.714 | .18451E-04 | .35047E-05 | .30317E+08 | 5.26E+00 |
| 562 | 199 | 65.072 | .15569E-03 | .44762E-04 | .24525E+09 | 3.48E+00 |
| 561 | 199 | 125.575 | .27322E-03 | .21073E-03 | .11557E+09 | 1.30E+00 |
| 560 | 199 | 133.644 | .47947E-02 | .41020E-02 | .17906E+10 | 1.17E+00 |
| 559 | 199 | 140.061 | .96261E-03 | .82956E-03 | .32730E+09 | 1.16E+00 |
| 558 | 199 | 143.221 | .15312E-04 | .22876E-04 | .49792E+07 | 6.69E-01 |
| 557 | 199 | 146.836 | .28080E-03 | .33546E-03 | .86869E+08 | 8.37E-01 |
| 556 | 199 | 152.852 | .44177E-05 | .15939E-05 | .12612E+07 | 2.77E+00 |
| 555 | 199 | 156.085 | .27889E-03 | .23616E-03 | .76355E+08 | 1.18E+00 |
| 554 | 199 | 170.119 | .50734E-01 | .56156E-01 | .11693E+11 | 9.03E-01 |
| 553 | 199 | 181.408 | .18698E+00 | .22865E+00 | .37897E+11 | 8.18E-01 |
| 552 | 199 | 191.522 | .15587E-02 | .22066E-02 | .28343E+09 | 7.06E-01 |
| 551 | 199 | 214.878 | .17576E-03 | .24732E-03 | .25390E+08 | 7.11E-01 |
| 199 | 550 | 196.805 | .35824E-07 | .22579E-06 | .61692E+04 | 1.59E-01 |
| 569 | 198 | 51.751 | .21067E-07 | .31167E-06 | .52467E+05 | 6.76E-02 |
| 568 | 198 | 53.990 | .61360E-06 | .62373E-07 | .14041E+07 | 9.84E+00 |
| 567 | 198 | 55.679 | .68861E-06 | .15151E-06 | .14816E+07 | 4.54E+00 |
| 566 | 198 | 56.876 | .74286E-06 | .21343E-07 | .15318E+07 | 3.48E+01 |
| 565 | 198 | 57.410 | .97272E-07 | .13221E-06 | .19686E+06 | 7.36E-01 |
| 564 | 198 | 58.199 | .45169E-05 | .10470E-06 | .88949E+07 | 4.31E+01 |
| 563 | 198 | 62.416 | .81790E-04 | .18679E-04 | .14004E+09 | 4.38E+00 |
| 562 | 198 | 63.719 | .35217E-04 | .60689E-05 | .57856E+08 | 5.80E+00 |
| 561 | 198 | 120.631 | .66143E-03 | .48173E-03 | .30318E+09 | 1.37E+00 |
| 560 | 198 | 128.058 | .10175E-02 | .84855E-03 | .41387E+09 | 1.20E+00 |
| 559 | 198 | 133.938 | .24424E-03 | .22420E-03 | .90813E+08 | 1.09E+00 |
| 558 | 198 | 136.824 | .23738E-03 | .22511E-03 | .84576E+08 | 1.05E+00 |
| 557 | 198 | 140.120 | .97199E-04 | .76292E-04 | .33021E+08 | 1.27E+00 |
| 556 | 198 | 145.588 | .35076E-05 | .36663E-05 | .11038E+07 | 9.57E-01 |
| 555 | 198 | 148.518 | .71778E-03 | .71867E-03 | .21705E+09 | 9.99E-01 |
| 554 | 198 | 161.169 | .10703E+00 | .11750E+00 | .27484E+11 | 9.11E-01 |
| 553 | 198 | 171.267 | .78340E-02 | .93795E-02 | .17814E+10 | 8.35E-01 |
| 552 | 198 | 180.253 | .24253E-01 | .27697E-01 | .49789E+10 | 8.76E-01 |
| 551 | 198 | 200.794 | .89983E-02 | .13274E-01 | .14886E+10 | 6.78E-01 |
| 198 | 550 | 210.316 | .28112E-06 | .66200E-07 | .42392E+05 | 4.25E+00 |
| 569 | 197 | 49.976 | .58751E-06 | .13559E-05 | .15690E+07 | 4.33E-01 |
| 568 | 197 | 52.061 | .64815E-07 | .19835E-07 | .15951E+06 | 3.27E+00 |
| 567 | 197 | 53.630 | .10731E-05 | .19630E-06 | .24886E+07 | 5.47E+00 |
| 566 | 197 | 54.739 | .10622E-05 | .22962E-06 | .23645E+07 | 4.63E+00 |
| 565 | 197 | 55.234 | .95262E-05 | .16793E-05 | .20828E+08 | 5.67E+00 |
| 564 | 197 | 55.964 | .96011E-05 | .65951E-06 | .20447E+08 | 1.46E+01 |
| 563 | 197 | 59.852 | .57753E-05 | .76534E-06 | .10754E+08 | 7.55E+00 |
| 562 | 197 | 61.049 | .14982E-03 | .10951E-04 | .26813E+09 | 1.37E+01 |
| 561 | 197 | 111.408 | .23246E-05 | .12073E-05 | .12492E+07 | 1.93E+00 |
| 560 | 197 | 117.713 | .15374E-05 | .57980E-06 | .74005E+06 | 2.65E+00 |
| 559 | 197 | 122.663 | .18026E-04 | .83221E-05 | .79913E+07 | 2.17E+00 |
| 558 | 197 | 125.079 | .68505E-03 | .51922E-03 | .29207E+09 | 1.32E+00 |
| 557 | 197 | 127.828 | .11700E-03 | .95546E-04 | .47759E+08 | 1.22E+00 |
| 556 | 197 | 132.363 | .91372E-04 | .69360E-04 | .34786E+08 | 1.32E+00 |
| 555 | 197 | 134.781 | .38568E-03 | .34580E-03 | .14161E+09 | 1.12E+00 |
| 554 | 197 | 145.118 | .33388E-05 | .59303E-06 | .10575E+07 | 5.63E+00 |
| 553 | 197 | 153.254 | .17862E-01 | .16490E-01 | .50728E+10 | 1.08E+00 |
| 552 | 197 | 160.410 | .11821E-01 | .12289E-01 | .30642E+10 | 9.62E-01 |
| 551 | 197 | 176.476 | .86734E-01 | .10416E+00 | .18576E+11 | 8.33E-01 |
| 197 | 550 | 245.792 | .14942E-07 | .35230E-06 | .16497E+04 | 4.24E-02 |
| 569 | 196 | 43.994 | .48943E-05 | .46085E-06 | .16867E+08 | 1.06E+01 |
| 568 | 196 | 45.602 | .56876E-07 | .75502E-07 | .18243E+06 | 7.53E-01 |
| 567 | 196 | 46.801 | .61923E-08 | .60775E-08 | .18857E+05 | 1.02E+00 |
| 566 | 196 | 47.643 | .49837E-07 | .80296E-07 | .14645E+06 | 6.21E-01 |
| 565 | 196 | 48.018 | .40127E-07 | .38337E-06 | .11608E+06 | 1.05E-01 |
| 564 | 196 | 48.569 | .59022E-07 | .27441E-06 | .16689E+06 | 2.15E-01 |
| 563 | 196 | 51.470 | .26108E-04 | .16908E-04 | .65733E+08 | 1.54E+00 |
| 562 | 196 | 52.353 | .33377E-06 | .10404E-05 | .81226E+06 | 3.21E-01 |
| 561 | 196 | 85.493 | .57909E-08 | .36474E-08 | .52847E+04 | 1.59E+00 |
| 560 | 196 | 89.158 | .23461E-06 | .74670E-07 | .19686E+06 | 3.14E+00 |
| 559 | 196 | 91.969 | .40590E-07 | .92157E-08 | .32009E+05 | 4.40E+00 |
| 558 | 196 | 93.321 | .17036E-06 | .63216E-08 | .13048E+06 | 2.69E+01 |
| 557 | 196 | 94.842 | .13666E-06 | .43331E-07 | .10134E+06 | 3.15E+00 |
| 556 | 196 | 97.317 | .75238E-09 | .47281E-08 | .52991E+03 | 1.59E-01 |
| 555 | 196 | 98.617 | .33610E-07 | .14777E-10 | .23051E+05 | 2.27E+03 |
| 554 | 196 | 104.040 | .26515E-06 | .21418E-09 | .16339E+06 | 1.24E+03 |
| 553 | 196 | 108.156 | .11698E-05 | .26311E-09 | .66705E+06 | 4.45E+03 |
| 552 | 196 | 111.672 | .10339E-07 | .22736E-08 | .55299E+04 | 4.55E+00 |
| 551 | 196 | 119.228 | .79938E-06 | .31262E-06 | .37509E+06 | 2.56E+00 |
| 196 | 550 | 742.007 | .74154E-02 | .70522E-02 | .89836E+08 | 1.05E+00 |
| 216 | 654 | 171.846 | .92964E-02 | .11358E-01 | .15748E+10 | 8.18E-01 |
| 216 | 653 | 151.877 | .27380E+00 | .25572E+00 | .59380E+11 | 1.07E+00 |
| 216 | 652 | 135.631 | .75945E-02 | .55306E-02 | .20653E+10 | 1.37E+00 |
| 216 | 651 | 123.575 | .51282E-02 | .45024E-02 | .16800E+10 | 1.14E+00 |
| 216 | 650 | 62.353 | .58359E-06 | .21093E-06 | .75092E+06 | 2.77E+00 |
| 216 | 649 | 60.011 | .19235E-04 | .32975E-05 | .26719E+08 | 5.83E+00 |
| 216 | 648 | 59.192 | .16955E-05 | .25655E-05 | .24208E+07 | 6.61E-01 |
| 216 | 647 | 58.635 | .48048E-06 | .77659E-06 | .69912E+06 | 6.19E-01 |
| 216 | 646 | 58.355 | .20472E-04 | .47717E-07 | .30075E+08 | 4.29E+02 |
| 216 | 645 | 57.333 | .46753E-04 | .12849E-04 | .71153E+08 | 3.64E+00 |
| 216 | 644 | 55.341 | .12024E-05 | .14187E-08 | .19641E+07 | 8.48E+02 |
| 216 | 643 | 54.234 | .42450E-06 | .96084E-06 | .72198E+06 | 4.42E-01 |
| 216 | 642 | 53.231 | .38728E-06 | .52047E-06 | .68374E+06 | 7.44E-01 |
| 216 | 641 | 34.553 | .36123E-02 | .33043E-02 | .15136E+11 | 1.09E+00 |
| 215 | 654 | 203.883 | .55975E-04 | .17492E-04 | .67364E+07 | 3.20E+00 |
| 215 | 653 | 176.370 | .94255E-02 | .10586E-01 | .15158E+10 | 8.90E-01 |
| 215 | 652 | 154.833 | .87418E-01 | .75995E-01 | .18242E+11 | 1.15E+00 |
| 215 | 651 | 139.316 | .17045E+00 | .18459E+00 | .43933E+11 | 9.23E-01 |
| 215 | 650 | 66.123 | .39821E-05 | .23080E-05 | .45563E+07 | 1.73E+00 |
| 215 | 649 | 63.496 | .15372E-03 | .75075E-04 | .19074E+09 | 2.05E+00 |
| 215 | 648 | 62.579 | .34751E-04 | .15817E-04 | .44392E+08 | 2.20E+00 |
| 215 | 647 | 61.957 | .18603E-04 | .51536E-05 | .24244E+08 | 3.61E+00 |
| 215 | 646 | 61.644 | .20823E-04 | .52271E-05 | .27412E+08 | 3.98E+00 |
| 215 | 645 | 60.505 | .38360E-06 | .57452E-06 | .52419E+06 | 6.68E-01 |
| 215 | 644 | 58.291 | .38134E-04 | .10848E-05 | .56144E+08 | 3.52E+01 |
| 215 | 643 | 57.064 | .16283E-04 | .23159E-04 | .25015E+08 | 7.03E-01 |
| 215 | 642 | 55.954 | .85934E-05 | .98484E-05 | .13731E+08 | 8.73E-01 |
| 215 | 641 | 35.680 | .10622E+00 | .96838E-01 | .41741E+12 | 1.10E+00 |
| 214 | 654 | 374.042 | .17745E-02 | .53183E-02 | .63449E+08 | 3.34E-01 |
| 214 | 653 | 290.815 | .10669E-02 | .15771E-02 | .63110E+08 | 6.76E-01 |
| 214 | 652 | 236.557 | .75104E-01 | .54086E-01 | .67140E+10 | 1.39E+00 |
| 214 | 651 | 202.158 | .28690E+00 | .24142E+00 | .35119E+11 | 1.19E+00 |
| 214 | 650 | 77.567 | .64281E-07 | .18719E-07 | .53447E+05 | 3.43E+00 |
| 214 | 649 | 73.976 | .28733E-07 | .67305E-06 | .26266E+05 | 4.27E-02 |
| 214 | 648 | 72.735 | .27932E-07 | .10577E-06 | .26413E+05 | 2.64E-01 |
| 214 | 647 | 71.896 | .12898E-06 | .13541E-05 | .12482E+06 | 9.53E-02 |
| 214 | 646 | 71.476 | .16493E-06 | .67706E-08 | .16151E+06 | 2.44E+01 |
| 214 | 645 | 69.948  | .34106E-07 | .10989E-06 | .34871E+05 | 3.10E-01 |
| 214 | 644 | 67.006  | .25825E-05 | .10401E-04 | .28775E+07 | 2.48E-01 |
| 214 | 643 | 65.390  | .46007E-04 | .41289E-04 | .53827E+08 | 1.11E+00 |
| 214 | 642 | 63.937  | .35638E-04 | .23168E-04 | .43612E+08 | 1.54E+00 |
| 214 | 641 | 38.767  | .81295E+00 | .74229E+00 | .27061E+13 | 1.10E+00 |
| 654 | 213 | 225.903 | .14338E-02 | .15729E-02 | .24987E+09 | 9.12E-01 |
| 653 | 213 | 273.107 | .24995E-03 | .34453E-03 | .29803E+08 | 7.25E-01 |
| 652 | 213 | 348.084 | .26096E-03 | .25287E-03 | .19155E+08 | 1.03E+00 |
| 651 | 213 | 464.349 | .21137E-04 | .89011E-04 | .87182E+06 | 2.37E-01 |
| 213 | 650 | 172.653 | .35784E-02 | .37897E-02 | .60053E+09 | 9.44E-01 |
| 213 | 649 | 155.821 | .33857E-02 | .24142E-02 | .69757E+09 | 1.40E+00 |
| 213 | 648 | 150.413 | .18433E-01 | .13840E-01 | .40760E+10 | 1.33E+00 |
| 213 | 647 | 146.869 | .28725E-01 | .19237E-01 | .66619E+10 | 1.49E+00 |
| 213 | 646 | 145.125 | .97099E-04 | .15834E-03 | .23064E+08 | 6.13E-01 |
| 213 | 645 | 138.964 | .18313E-02 | .14676E-02 | .47440E+09 | 1.25E+00 |
| 213 | 644 | 127.814 | .52619E-04 | .10316E-03 | .16113E+08 | 5.10E-01 |
| 213 | 643 | 122.059 | .13314E-01 | .87911E-02 | .44705E+10 | 1.51E+00 |
| 213 | 642 | 117.092 | .10280E-01 | .58772E-02 | .37510E+10 | 1.75E+00 |
| 213 | 641 | 53.490  | .91851E-07 | .13339E-05 | .16060E+06 | 6.89E-02 |
| 654 | 212 | 221.301 | .49262E-01 | .48786E-01 | .89458E+10 | 1.01E+00 |
| 653 | 212 | 266.410 | .25914E-03 | .39375E-03 | .32472E+08 | 6.58E-01 |
| 652 | 212 | 337.277 | .54512E-03 | .80509E-03 | .42617E+08 | 6.77E-01 |
| 651 | 212 | 445.314 | .62463E-04 | .23964E-04 | .28013E+07 | 2.61E+00 |
| 212 | 650 | 175.441 | .70876E-01 | .62424E-01 | .11519E+11 | 1.14E+00 |
| 212 | 649 | 158.088 | .60199E-03 | .87802E-03 | .12050E+09 | 6.86E-01 |
| 212 | 648 | 152.525 | .19464E-04 | .20091E-03 | .41856E+07 | 9.69E-02 |
| 212 | 647 | 148.882 | .40185E-02 | .27099E-02 | .90693E+09 | 1.48E+00 |
| 212 | 646 | 147.090 | .24820E-02 | .20339E-02 | .57388E+09 | 1.22E+00 |
| 212 | 645 | 140.765 | .22665E-04 | .15241E-05 | .57221E+07 | 1.49E+01 |
| 212 | 644 | 129.335 | .87215E-04 | .46547E-05 | .26083E+08 | 1.87E+01 |
| 212 | 643 | 123.446 | .73122E-03 | .39052E-03 | .24004E+09 | 1.87E+00 |
| 212 | 642 | 118.368 | .29935E-04 | .34962E-04 | .10688E+08 | 8.56E-01 |
| 212 | 641 | 53.754  | .55971E-05 | .58071E-06 | .96902E+07 | 9.64E+00 |
| 654 | 211 | 200.089 | .49419E-01 | .41135E-01 | .10978E+11 | 1.20E+00 |
| 653 | 211 | 236.257 | .87299E-02 | .92427E-02 | .13910E+10 | 9.45E-01 |
| 652 | 211 | 290.362 | .26103E-03 | .18023E-03 | .27535E+08 | 1.45E+00 |
| 651 | 211 | 367.018 | .10624E-04 | .88367E-05 | .70146E+06 | 1.20E+00 |
| 211 | 650 | 191.540 | .15521E-01 | .17406E-01 | .21164E+10 | 8.92E-01 |
| 211 | 649 | 171.041 | .91300E-02 | .79230E-02 | .15612E+10 | 1.15E+00 |
| 211 | 648 | 164.548 | .47864E-01 | .43190E-01 | .88434E+10 | 1.11E+00 |
| 211 | 647 | 160.316 | .28155E-01 | .23958E-01 | .54801E+10 | 1.18E+00 |
| 211 | 646 | 158.240 | .10843E-02 | .20514E-02 | .21663E+09 | 5.29E-01 |
| 211 | 645 | 150.943 | .97527E-02 | .91903E-02 | .21414E+10 | 1.06E+00 |
| 211 | 644 | 137.878 | .39058E-02 | .17574E-02 | .10278E+10 | 2.22E+00 |
| 211 | 643 | 131.205 | .75522E-03 | .49158E-03 | .21947E+09 | 1.54E+00 |
| 211 | 642 | 125.484 | .40889E-03 | .40076E-03 | .12991E+09 | 1.02E+00 |
| 211 | 641 | 55.175 | .70159E-07 | .11890E-04 | .11529E+06 | 5.90E-03 |
| 654 | 210 | 183.906 | .50651E-05 | .20547E-04 | .13319E+07 | 2.47E-01 |
| 653 | 210 | 214.020 | .56263E-01 | .53205E-01 | .10924E+11 | 1.06E+00 |
| 652 | 210 | 257.482 | .39390E-04 | .28830E-04 | .52841E+07 | 1.37E+00 |
| 651 | 210 | 316.011 | .92858E-04 | .14192E-02 | .82697E+07 | 6.54E-02 |
| 210 | 650 | 209.159 | .81185E-03 | .11424E-02 | .92837E+08 | 7.11E-01 |
| 210 | 649 | 184.954 | .93227E-01 | .95134E-01 | .13633E+11 | 9.80E-01 |
| 210 | 648 | 177.385 | .12936E-02 | .11959E-02 | .20566E+09 | 1.08E+00 |
| 210 | 647 | 172.476 | .98592E-05 | .38230E-04 | .16580E+07 | 2.58E-01 |
| 210 | 646 | 170.076 | .10514E-01 | .86973E-02 | .18184E+10 | 1.21E+00 |
| 210 | 645 | 161.676 | .15840E+00 | .13477E+00 | .30315E+11 | 1.18E+00 |
| 210 | 644 | 146.778 | .12443E-01 | .91050E-02 | .28894E+10 | 1.37E+00 |
| 210 | 643 | 139.240 | .15912E-01 | .12733E-01 | .41058E+10 | 1.25E+00 |
| 210 | 642 | 132.813 | .23369E-02 | .20158E-02 | .66274E+09 | 1.16E+00 |
| 210 | 641 | 56.547 | .84818E-04 | .12281E-03 | .13270E+09 | 6.91E-01 |
| 654 | 209 | 173.497 | .29885E-03 | .15825E-03 | .88296E+08 | 1.89E+00 |
| 653 | 209 | 200.053 | .42288E-01 | .36309E-01 | .93973E+10 | 1.16E+00 |
| 652 | 209 | 237.531 | .37360E-03 | .61770E-03 | .58890E+08 | 6.05E-01 |
| 651 | 209 | 286.479 | .42412E-03 | .22764E-02 | .45960E+08 | 1.86E-01 |
| 209 | 650 | 224.474 | .26900E-03 | .29573E-03 | .26706E+08 | 9.10E-01 |
| 209 | 649 | 196.829 | .17618E-01 | .17581E-01 | .22749E+10 | 1.00E+00 |
| 209 | 648 | 188.279 | .17437E-01 | .16416E-01 | .24607E+10 | 1.06E+00 |
| 209 | 647 | 182.759 | .63317E-02 | .54297E-02 | .94833E+09 | 1.17E+00 |
| 209 | 646 | 180.066 | .31018E-01 | .29801E-01 | .47857E+10 | 1.04E+00 |
| 209 | 645 | 170.677 | .13661E+00 | .12399E+00 | .23460E+11 | 1.10E+00 |
| 209 | 644 | 154.159 | .40098E-01 | .31166E-01 | .84407E+10 | 1.29E+00 |
| 209 | 643 | 145.865 | .18582E-01 | .18597E-01 | .43689E+10 | 9.99E-01 |
| 209 | 642 | 138.828 | .27552E-02 | .28856E-02 | .71514E+09 | 9.55E-01 |
| 209 | 641 | 57.610 | .72181E-04 | .23593E-03 | .10880E+09 | 3.06E-01 |
| 654 | 208 | 156.348 | .37865E-03 | .27118E-03 | .13776E+09 | 1.40E+00 |
| 653 | 208 | 177.592 | .15459E-03 | .13288E-03 | .43591E+08 | 1.16E+00 |
| 652 | 208 | 206.518 | .19590E-03 | .86831E-03 | .40849E+08 | 2.26E-01 |
| 651 | 208 | 242.550 | .20728E-01 | .22032E-01 | .31336E+10 | 9.41E-01 |
| 208 | 650 | 261.599 | .77014E-03 | .76967E-03 | .56298E+08 | 1.00E+00 |
| 208 | 649 | 224.803 | .67803E-02 | .61082E-02 | .67118E+09 | 1.11E+00 |
| 208 | 648 | 213.719 | .34303E-01 | .31384E-01 | .37570E+10 | 1.09E+00 |
| 208 | 647 | 206.634 | .94068E-01 | .90471E-01 | .11021E+11 | 1.04E+00 |
| 208 | 646 | 203.198 | .13709E-02 | .11645E-02 | .16609E+09 | 1.18E+00 |
| 208 | 645 | 191.322 | .56768E-04 | .12003E-03 | .77583E+07 | 4.73E-01 |
| 208 | 644 | 170.806 | .10191E-01 | .92827E-02 | .17474E+10 | 1.10E+00 |
| 208 | 643 | 160.683 | .94474E-01 | .70493E-01 | .18305E+11 | 1.34E+00 |
| 208 | 642 | 152.185 | .77154E-01 | .55922E-01 | .16665E+11 | 1.38E+00 |
| 208 | 641 | 59.787 | .60354E-04 | .32167E-04 | .84465E+08 | 1.88E+00 |
| 654 | 207 | 151.470 | .68053E-03 | .29460E-03 | .26380E+09 | 2.31E+00 |
| 653 | 207 | 171.324 | .16700E-04 | .22212E-04 | .50601E+07 | 7.52E-01 |
| 652 | 207 | 198.091 | .11191E+00 | .80221E-01 | .25364E+11 | 1.40E+00 |
| 651 | 207 | 231.007 | .12671E-02 | .99222E-03 | .21116E+09 | 1.28E+00 |
| 207 | 650 | 276.499 | .54353E-03 | .12301E-02 | .35566E+08 | 4.42E-01 |
| 207 | 649 | 235.720 | .44954E-02 | .23176E-02 | .40473E+09 | 1.94E+00 |
| 207 | 648 | 223.561 | .85685E-02 | .81861E-02 | .85764E+09 | 1.05E+00 |
| 207 | 647 | 215.820 | .23046E-03 | .71980E-03 | .24751E+08 | 3.20E-01 |
| 207 | 646 | 212.075 | .62078E-03 | .39274E-03 | .69049E+08 | 1.58E+00 |
| 207 | 645 | 199.172 | .24520E-05 | .72921E-05 | .30921E+06 | 3.36E-01 |
| 207 | 644 | 177.036 | .24011E-02 | .10188E-02 | .38324E+09 | 2.36E+00 |
| 207 | 643 | 166.183 | .85795E-02 | .53405E-02 | .15541E+10 | 1.61E+00 |
| 207 | 642 | 157.110 | .59430E-03 | .62701E-03 | .12045E+09 | 9.48E-01 |
| 207 | 641 | 60.533 | .15819E-04 | .31896E-03 | .21597E+08 | 4.96E-02 |
| 654 | 206 | 147.954 | .15758E-01 | .12787E-01 | .64018E+10 | 1.23E+00 |
| 653 | 206 | 166.841 | .13653E-03 | .16136E-03 | .43622E+08 | 8.46E-01 |
| 652 | 206 | 192.122 | .10015E-02 | .87400E-03 | .24130E+09 | 1.15E+00 |
| 651 | 206 | 222.930 | .13700E-03 | .18725E-03 | .24516E+08 | 7.32E-01 |
| 206 | 650 | 289.035 | .86717E-01 | .10620E+00 | .51928E+10 | 8.17E-01 |
| 206 | 649 | 244.770 | .13006E-04 | .11001E-03 | .10860E+07 | 1.18E-01 |
| 206 | 648 | 231.686 | .26892E-01 | .23080E-01 | .25062E+10 | 1.17E+00 |
| 206 | 647 | 223.382 | .29506E-02 | .24829E-02 | .29581E+09 | 1.19E+00 |
| 206 | 646 | 219.372 | .11873E-01 | .83005E-02 | .12342E+10 | 1.43E+00 |
| 206 | 645 | 205.595 | .44300E-02 | .29144E-02 | .52429E+09 | 1.52E+00 |
| 206 | 644 | 182.092 | .11805E-03 | .79318E-04 | .17810E+08 | 1.49E+00 |
| 206 | 643 | 170.631 | .28721E-02 | .25659E-02 | .49349E+09 | 1.12E+00 |
| 206 | 642 | 161.080 | .21159E-03 | .16659E-03 | .40795E+08 | 1.27E+00 |
| 206 | 641 | 61.113 | .81598E-06 | .86442E-07 | .10930E+07 | 9.44E+00 |
| 654 | 205 | 137.252 | .20474E-02 | .84755E-03 | .96657E+09 | 2.42E+00 |
| 653 | 205 | 153.356 | .49497E-01 | .27581E-01 | .18718E+11 | 1.79E+00 |
| 652 | 205 | 174.456 | .77744E-04 | .11230E-03 | .22718E+08 | 6.92E-01 |
| 651 | 205 | 199.490 | .63484E-02 | .41216E-02 | .14187E+10 | 1.54E+00 |
| 205 | 650 | 340.978 | .34249E-04 | .58649E-04 | .14736E+07 | 5.84E-01 |
| 205 | 649 | 281.023 | .41020E-02 | .37933E-02 | .25984E+09 | 1.08E+00 |
| 205 | 648 | 263.912 | .16893E-02 | .12223E-02 | .12134E+09 | 1.38E+00 |
| 205 | 647 | 253.192 | .31907E-02 | .26298E-02 | .24899E+09 | 1.21E+00 |
| 205 | 646 | 248.052 | .65257E-01 | .56894E-01 | .53056E+10 | 1.15E+00 |
| 205 | 645 | 230.580 | .61795E-01 | .55711E-01 | .58144E+10 | 1.11E+00 |
| 205 | 644 | 201.423 | .46832E-02 | .46117E-02 | .57745E+09 | 1.02E+00 |
| 205 | 643 | 187.493 | .18129E-01 | .16753E-01 | .25798E+10 | 1.08E+00 |
| 205 | 642 | 176.024 | .14405E-01 | .12859E-01 | .23258E+10 | 1.12E+00 |
| 205 | 641 | 63.147 | .70322E-04 | .16305E-05 | .88223E+08 | 4.31E+01 |
| 654 | 204 | 137.003 | .26186E-03 | .18009E-03 | .12407E+09 | 1.45E+00 |
| 653 | 204 | 153.045 | .95777E-02 | .53022E-02 | .36366E+10 | 1.81E+00 |
| 652 | 204 | 174.055 | .16815E-02 | .20731E-02 | .49364E+09 | 8.11E-01 |
| 651 | 204 | 198.965 | .61566E-01 | .47276E-01 | .13831E+11 | 1.30E+00 |
| 204 | 650 | 342.523 | .14668E-03 | .48156E-03 | .62545E+07 | 3.05E-01 |
| 204 | 649 | 282.072 | .31955E-02 | .17925E-02 | .20092E+09 | 1.78E+00 |
| 204 | 648 | 264.837 | .74362E-03 | .12168E-02 | .53038E+08 | 6.11E-01 |
| 204 | 647 | 254.043 | .20258E-02 | .15474E-02 | .15703E+09 | 1.31E+00 |
| 204 | 646 | 248.869 | .63723E-02 | .60313E-02 | .51470E+09 | 1.06E+00 |
| 204 | 645 | 231.286 | .84063E-02 | .68672E-02 | .78615E+09 | 1.22E+00 |
| 204 | 644 | 201.961 | .23920E-01 | .22401E-01 | .29338E+10 | 1.07E+00 |
| 204 | 643 | 187.959 | .18384E+00 | .17468E+00 | .26032E+11 | 1.05E+00 |
| 204 | 642 | 176.435 | .13127E+00 | .11884E+00 | .21096E+11 | 1.10E+00 |
| 204 | 641 | 63.200 | .19034E-03 | .36541E-04 | .23839E+09 | 5.21E+00 |
| 654 | 203 | 126.567 | .10738E-03 | .14875E-04 | .59615E+08 | 7.22E+00 |
| 653 | 203 | 140.138 | .34282E-04 | .27034E-06 | .15525E+08 | 1.27E+02 |
| 652 | 203 | 157.551 | .70533E-03 | .13094E-02 | .25271E+09 | 5.39E-01 |
| 651 | 203 | 177.689 | .41133E-01 | .13547E-01 | .11586E+11 | 3.04E+00 |
| 203 | 650 | 431.463 | .52931E-04 | .38128E-04 | .14224E+07 | 1.39E+00 |
| 203 | 649 | 339.746 | .16959E-01 | .11468E-01 | .73500E+09 | 1.48E+00 |
| 203 | 648 | 315.051 | .32521E-02 | .26571E-02 | .16391E+09 | 1.22E+00 |
| 203 | 647 | 299.892 | .10858E-01 | .10788E-01 | .60398E+09 | 1.01E+00 |
| 203 | 646 | 292.708 | .26343E-02 | .20938E-02 | .15381E+09 | 1.26E+00 |
| 203 | 645 | 268.684 | .88990E-03 | .99286E-03 | .61667E+08 | 8.96E-01 |
| 203 | 644 | 229.904 | .12087E+00 | .12618E+00 | .11440E+11 | 9.58E-01 |
| 203 | 643 | 211.932 | .66129E-01 | .76427E-01 | .73654E+10 | 8.65E-01 |
| 203 | 642 | 197.394 | .50482E-02 | .58571E-02 | .64813E+09 | 8.62E-01 |
| 203 | 641 | 65.699 | .17126E-02 | .17229E-02 | .19849E+10 | 9.94E-01 |
| 654 | 202 | 117.682 | .50681E-04 | .21694E-04 | .32546E+08 | 2.34E+00 |
| 653 | 202 | 129.327 | .52121E-04 | .21139E-04 | .27715E+08 | 2.47E+00 |
| 652 | 202 | 144.016 | .49978E-02 | .39467E-02 | .21430E+10 | 1.27E+00 |
| 651 | 202 | 160.660 | .86716E-03 | .67281E-03 | .29878E+09 | 1.29E+00 |
| 202 | 650 | 580.997 | .11616E-02 | .32609E-02 | .17215E+08 | 3.56E-01 |
| 202 | 649 | 426.101 | .15062E-02 | .74455E-03 | .41500E+08 | 2.02E+00 |
| 202 | 648 | 387.961 | .43811E-02 | .57067E-02 | .14561E+09 | 7.68E-01 |
| 202 | 647 | 365.228 | .57048E-02 | .92118E-02 | .21395E+09 | 6.19E-01 |
| 202 | 646 | 354.629 | .65772E-04 | .12404E-04 | .26163E+07 | 5.30E+00 |
| 202 | 645 | 319.967 | .32223E-04 | .56998E-04 | .15745E+07 | 5.65E-01 |
| 202 | 644 | 266.445 | .41053E-02 | .55054E-02 | .28928E+09 | 7.46E-01 |
| 202 | 643 | 242.602 | .94895E-01 | .11501E+00 | .80658E+10 | 8.25E-01 |
| 202 | 642 | 223.740 | .26186E+00 | .33193E+00 | .26169E+11 | 7.89E-01 |
| 202 | 641 | 68.379 | .21204E-04 | .53996E-05 | .22687E+08 | 3.93E+00 |
| 654 | 201 | 58.145 | .86614E-06 | .38361E-06 | .22784E+07 | 2.26E+00 |
| 653 | 201 | 60.852 | .43748E-07 | .15317E-07 | .10507E+06 | 2.86E+00 |
| 652 | 201 | 63.920 | .51603E-07 | .15149E-07 | .11233E+06 | 3.41E+00 |
| 651 | 201 | 67.000 | .61795E-07 | .13444E-08 | .12243E+06 | 4.60E+01 |
| 650 | 201 | 143.271 | .36430E-01 | .31687E-01 | .15784E+11 | 1.15E+00 |
| 649 | 201 | 157.378 | .40863E-02 | .40966E-02 | .14673E+10 | 9.97E-01 |
| 648 | 201 | 163.308 | .10777E-01 | .11407E-01 | .35939E+10 | 9.45E-01 |
| 647 | 201 | 167.702 | .38441E-02 | .42570E-02 | .12156E+10 | 9.03E-01 |
| 646 | 201 | 170.035 | .64037E-02 | .71956E-02 | .19698E+10 | 8.90E-01 |
| 645 | 201 | 179.351 | .27057E-06 | .37793E-06 | .74807E+05 | 7.16E-01 |
| 644 | 201 | 202.107 | .83965E-04 | .10617E-03 | .18281E+08 | 7.91E-01 |
| 643 | 201 | 218.388 | .39275E-04 | .68539E-04 | .73237E+07 | 5.73E-01 |
| 642 | 201 | 236.323 | .18170E-03 | .29246E-03 | .28935E+08 | 6.21E-01 |
| 201 | 641 | 168.820 | .76849E-08 | .27800E-07 | .13489E+04 | 2.76E-01 |
| 654 | 200 | 56.395 | .14057E-05 | .33846E-06 | .39309E+07 | 4.15E+00 |
| 653 | 200 | 58.938 | .25905E-04 | .62714E-05 | .66322E+08 | 4.13E+00 |
| 652 | 200 | 61.811 | .20883E-06 | .17532E-06 | .48611E+06 | 1.19E+00 |
| 651 | 200 | 64.688 | .63147E-06 | .17759E-06 | .13421E+07 | 3.56E+00 |
| 650 | 200 | 133.096 | .30681E-04 | .24428E-04 | .15403E+08 | 1.26E+00 |
| 649 | 200 | 145.186 | .30474E-01 | .27278E-01 | .12857E+11 | 1.12E+00 |
| 648 | 200 | 150.218 | .16168E-01 | .15275E-01 | .63721E+10 | 1.06E+00 |
| 647 | 200 | 153.928 | .68183E-02 | .66173E-02 | .25593E+10 | 1.03E+00 |
| 646 | 200 | 155.891 | .55391E-01 | .54680E-01 | .20271E+11 | 1.01E+00 |
| 645 | 200 | 163.686 | .46749E-01 | .51314E-01 | .15517E+11 | 9.11E-01 |
| 644 | 200 | 182.433 | .16362E-03 | .22916E-03 | .43721E+08 | 7.14E-01 |
| 643 | 200 | 195.595 | .87047E-03 | .12165E-02 | .20235E+09 | 7.16E-01 |
| 642 | 200 | 209.860 | .13855E-03 | .21046E-03 | .27978E+08 | 6.58E-01 |
| 200 | 641 | 185.533 | .32127E-07 | .44606E-07 | .46690E+04 | 7.20E-01 |
| 654 | 199 | 55.473 | .37925E-06 | .30065E-07 | .10961E+07 | 1.26E+01 |
| 653 | 199 | 57.932 | .23987E-05 | .27280E-06 | .63565E+07 | 8.79E+00 |
| 652 | 199 | 60.706 | .68331E-05 | .32859E-07 | .16490E+08 | 2.08E+02 |
| 651 | 199 | 63.477 | .65430E-05 | .65545E-10 | .14441E+08 | 9.98E+04 |
| 650 | 199 | 128.071 | .14237E-02 | .10707E-02 | .77194E+09 | 1.33E+00 |
| 649 | 199 | 139.228 | .45977E-01 | .39166E-01 | .21094E+11 | 1.17E+00 |
| 648 | 199 | 143.849 | .87895E-02 | .77959E-02 | .37777E+10 | 1.13E+00 |
| 647 | 199 | 147.247 | .40852E-02 | .39799E-02 | .16757E+10 | 1.03E+00 |
| 646 | 199 | 149.043 | .53976E-02 | .50828E-02 | .21610E+10 | 1.06E+00 |
| 645 | 199 | 156.153 | .67912E-03 | .75015E-03 | .24770E+09 | 9.05E-01 |
| 644 | 199 | 173.124 | .13723E-01 | .14770E-01 | .40719E+10 | 9.29E-01 |
| 643 | 199 | 184.934 | .14572E-02 | .13871E-02 | .37893E+09 | 1.05E+00 |
| 642 | 199 | 197.635 | .60026E-06 | .37092E-05 | .13667E+06 | 1.62E-01 |
| 199 | 641 | 196.265 | .74156E-06 | .41580E-05 | .96306E+05 | 1.78E-01 |
| 654 | 198 | 54.486 | .50280E-07 | .83986E-08 | .15062E+06 | 5.99E+00 |
| 653 | 198 | 56.857 | .86380E-07 | .80599E-08 | .23764E+06 | 1.07E+01 |
| 652 | 198 | 59.526 | .92399E-06 | .35473E-07 | .23191E+07 | 2.60E+01 |
| 651 | 198 | 62.189 | .15657E-05 | .27540E-06 | .36005E+07 | 5.69E+00 |
| 650 | 198 | 122.932 | .91520E-04 | .78228E-04 | .53859E+08 | 1.17E+00 |
| 649 | 198 | 133.176 | .20354E-02 | .16747E-02 | .10206E+10 | 1.22E+00 |
| 648 | 198 | 137.397 | .10040E-01 | .88101E-02 | .47299E+10 | 1.14E+00 |
| 647 | 198 | 140.494 | .25858E-01 | .23173E-01 | .11651E+11 | 1.12E+00 |
| 646 | 198 | 142.128 | .77875E-03 | .69795E-03 | .34285E+09 | 1.12E+00 |
| 645 | 198 | 148.579 | .27944E-04 | .24598E-04 | .11258E+08 | 1.14E+00 |
| 644 | 198 | 163.864 | .66912E-02 | .68037E-02 | .22162E+10 | 9.83E-01 |
| 643 | 198 | 174.406 | .13387E-01 | .14029E-01 | .39142E+10 | 9.54E-01 |
| 642 | 198 | 185.658 | .22189E-01 | .25550E-01 | .57252E+10 | 8.68E-01 |
| 198 | 641 | 209.700 | .76106E-11 | .53094E-08 | .86580E+00 | 1.43E-03 |
| 654 | 197 | 52.522 | .86129E-06 | .67703E-07 | .27767E+07 | 1.27E+01 |
| 653 | 197 | 54.721 | .84591E-07 | .45643E-08 | .25123E+06 | 1.85E+01 |
| 652 | 197 | 57.190 | .18821E-04 | .50029E-06 | .51177E+08 | 3.76E+01 |
| 651 | 197 | 59.643 | .13318E-03 | .12155E-04 | .33297E+09 | 1.10E+01 |
| 650 | 197 | 113.368 | .42839E-04 | .27469E-04 | .29644E+08 | 1.56E+00 |
| 649 | 197 | 122.023 | .16191E-03 | .95560E-04 | .96709E+08 | 1.69E+00 |
| 648 | 197 | 125.558 | .68862E-03 | .54643E-03 | .38847E+09 | 1.26E+00 |
| 647 | 197 | 128.140 | .18756E-02 | .13511E-02 | .10159E+10 | 1.39E+00 |
| 646 | 197 | 129.498 | .26620E-04 | .21025E-04 | .14117E+08 | 1.27E+00 |
| 645 | 197 | 134.831 | .15193E-03 | .14020E-03 | .74327E+08 | 1.08E+00 |
| 644 | 197 | 147.299 | .45355E-01 | .40733E-01 | .18591E+11 | 1.11E+00 |
| 643 | 197 | 155.763 | .56844E-01 | .57022E-01 | .20837E+11 | 9.97E-01 |
| 642 | 197 | 164.676 | .36907E-01 | .38632E-01 | .12104E+11 | 9.55E-01 |
| 197 | 641 | 244.951 | .85670E-06 | .50841E-05 | .71427E+05 | 1.69E-01 |
| 654 | 196 | 45.955 | .21105E-05 | .19854E-05 | .88876E+07 | 1.06E+00 |
| 653 | 196 | 47.630 | .26183E-05 | .19782E-05 | .10264E+08 | 1.32E+00 |
| 652 | 196 | 49.489 | .73874E-05 | .22216E-04 | .26825E+08 | 3.33E-01 |
| 651 | 196 | 51.316 | .33653E-03 | .25135E-03 | .11365E+10 | 1.34E+00 |
| 650 | 196 | 86.643 | .13231E-08 | .86221E-08 | .15675E+04 | 1.53E-01 |
| 649 | 196 | 91.609 | .86926E-07 | .14021E-06 | .92117E+05 | 6.20E-01 |
| 648 | 196 | 93.587 | .51594E-07 | .63242E-08 | .52389E+05 | 8.16E+00 |
| 647 | 196 | 95.014 | .22981E-08 | .34426E-07 | .22639E+04 | 6.68E-02 |
| 646 | 196 | 95.759 | .25325E-07 | .69775E-09 | .24562E+05 | 3.63E+01 |
| 645 | 196 | 98.644 | .30727E-07 | .95900E-11 | .28083E+05 | 3.20E+03 |
| 644 | 196 | 105.156 | .28475E-07 | .41110E-07 | .22901E+05 | 6.93E-01 |
| 643 | 196 | 109.400 | .49699E-06 | .33159E-07 | .36931E+06 | 1.50E+01 |
| 642 | 196 | 113.723 | .15503E-06 | .48153E-08 | .10661E+06 | 3.22E+01 |
| 196 | 641 | 734.390 | .15007E+00 | .14244E+00 | .13920E+10 | 1.05E+00 |
| 569 | 269 | 179.960 | .38374E-01 | .38233E-01 | .59276E+10 | 1.00E+00 |
| 568 | 269 | 210.286 | .10173E-01 | .70924E-02 | .11509E+10 | 1.43E+00 |
| 567 | 269 | 238.462 | .14690E-01 | .11981E-01 | .12923E+10 | 1.23E+00 |
| 566 | 269 | 262.066 | .24295E-01 | .26243E-01 | .17697E+10 | 9.26E-01 |
| 565 | 269 | 273.803 | .20218E-01 | .19435E-01 | .13491E+10 | 1.04E+00 |
| 564 | 269 | 292.751 | .50930E-04 | .39602E-03 | .29729E+07 | 1.29E-01 |
| 563 | 269 | 443.432 | .33601E-02 | .41680E-02 | .85484E+08 | 8.06E-01 |
| 562 | 269 | 518.813 | .68421E-02 | .76358E-02 | .12716E+09 | 8.96E-01 |
| 269 | 561 | 182.593 | .26364E-03 | .14517E-03 | .70325E+08 | 1.82E+00 |
| 269 | 560 | 167.857 | .21558E-01 | .13574E-01 | .68045E+10 | 1.59E+00 |
| 269 | 559 | 158.724 | .20622E-01 | .15821E-01 | .72799E+10 | 1.30E+00 |
| 269 | 558 | 154.852 | .15162E-01 | .10344E-01 | .56234E+10 | 1.47E+00 |
| 269 | 557 | 150.837 | .79553E-02 | .57458E-02 | .31097E+10 | 1.38E+00 |
| 269 | 556 | 144.975 | .42784E-05 | .39618E-04 | .18103E+07 | 1.08E-01 |
| 269 | 555 | 142.182 | .61613E-02 | .52303E-02 | .27105E+10 | 1.18E+00 |
| 269 | 554 | 132.244 | .23143E-02 | .24812E-02 | .11769E+10 | 9.33E-01 |
| 269 | 553 | 126.142 | .31292E-06 | .17864E-05 | .17490E+06 | 1.75E-01 |
| 269 | 552 | 121.674 | .35456E-04 | .15219E-04 | .21299E+08 | 2.33E+00 |
| 269 | 551 | 113.815 | .55170E-05 | .39818E-03 | .37877E+07 | 1.39E-02 |
| 269 | 550 | 53.992 | .21413E-04 | .16548E-06 | .65326E+08 | 1.29E+02 |
| 569 | 268 | 175.773 | .68001E-03 | .64702E-03 | .11010E+09 | 1.05E+00 |
| 568 | 268 | 204.591 | .32385E-01 | .26260E-01 | .38705E+10 | 1.23E+00 |
| 567 | 268 | 231.165 | .71429E-01 | .65864E-01 | .66869E+10 | 1.08E+00 |
| 566 | 268 | 253.281 | .19564E-01 | .20970E-01 | .15256E+10 | 9.33E-01 |
| 565 | 268 | 264.227 | .81354E-03 | .71326E-03 | .58293E+08 | 1.14E+00 |
| 564 | 268 | 281.831 | .13212E-02 | .19581E-02 | .83210E+08 | 6.75E-01 |
| 563 | 268 | 418.849 | .15113E-03 | .29937E-03 | .43096E+07 | 5.05E-01 |
| 562 | 268 | 485.476 | .42763E-04 | .29715E-03 | .90766E+06 | 1.44E-01 |
| 268 | 561 | 187.116 | .56356E-04 | .52482E-04 | .14315E+08 | 1.07E+00 |
| 268 | 560 | 171.671 | .24476E-04 | .10287E-04 | .73861E+07 | 2.38E+00 |
| 268 | 559 | 162.130 | .16441E-03 | .25309E-03 | .55625E+08 | 6.50E-01 |
| 268 | 558 | 158.093 | .17859E-01 | .16096E-01 | .63549E+10 | 1.11E+00 |
| 268 | 557 | 153.910 | .49911E-01 | .30849E-01 | .18738E+11 | 1.62E+00 |
| 268 | 556 | 147.812 | .31447E-01 | .21446E-01 | .12801E+11 | 1.47E+00 |
| 268 | 555 | 144.909 | .43958E-02 | .15416E-02 | .18617E+10 | 2.85E+00 |
| 268 | 554 | 134.600 | .21300E-04 | .18650E-04 | .10456E+08 | 1.14E+00 |
| 268 | 553 | 128.284 | .16159E-03 | .41897E-04 | .87327E+08 | 3.86E+00 |
| 268 | 552 | 123.666 | .21917E-04 | .26308E-04 | .12745E+08 | 8.33E-01 |
| 268 | 551 | 115.556 | .70931E-05 | .42507E-05 | .47242E+07 | 1.67E+00 |
| 268 | 550 | 54.381 | .50985E-05 | .83768E-07 | .15333E+08 | 6.09E+01 |
| 569 | 267 | 160.673 | .29291E-02 | .26780E-02 | .56760E+09 | 1.09E+00 |
| 568 | 267 | 184.418 | .49256E-02 | .37878E-02 | .72451E+09 | 1.30E+00 |
| 567 | 267 | 205.737 | .48622E-01 | .40260E-01 | .57465E+10 | 1.21E+00 |
| 566 | 267 | 223.072 | .12655E-01 | .11769E-01 | .12723E+10 | 1.08E+00 |
| 565 | 267 | 231.519 | .59531E-01 | .54431E-01 | .55560E+10 | 1.09E+00 |
| 564 | 267 | 244.924 | .61106E-03 | .44672E-03 | .50959E+08 | 1.37E+00 |
| 563 | 267 | 342.211 | .14154E-02 | .93944E-03 | .60462E+08 | 1.51E+00 |
| 562 | 267 | 385.430 | .81159E-03 | .78715E-04 | .27330E+08 | 1.03E+01 |
| 267 | 561 | 207.917 | .62615E-04 | .30114E-04 | .12882E+08 | 2.08E+00 |
| 267 | 560 | 189.021 | .77535E-03 | .33175E-03 | .19300E+09 | 2.34E+00 |
| 267 | 559 | 177.518 | .11812E-04 | .16924E-05 | .33336E+07 | 6.98E+00 |
| 267 | 558 | 172.690 | .29525E-03 | .21669E-03 | .88049E+08 | 1.36E+00 |
| 267 | 557 | 167.711 | .52299E-02 | .35423E-02 | .16537E+10 | 1.48E+00 |
| 267 | 556 | 160.496 | .36670E-01 | .23520E-01 | .12661E+11 | 1.56E+00 |
| 267 | 555 | 157.079 | .66279E-01 | .54602E-01 | .23890E+11 | 1.21E+00 |
| 267 | 554 | 145.039 | .17488E-02 | .18479E-02 | .73932E+09 | 9.46E-01 |
| 267 | 553 | 137.731 | .16543E-05 | .10986E-05 | .77557E+06 | 1.51E+00 |
| 267 | 552 | 132.422 | .24632E-04 | .28728E-04 | .12493E+08 | 8.57E-01 |
| 267 | 551 | 123.165 | .58066E-04 | .12806E-03 | .34042E+08 | 4.53E-01 |
| 267 | 550 | 56.009 | .60361E-05 | .49206E-07 | .17112E+08 | 1.23E+02 |
| 569 | 266 | 158.316 | .52604E-02 | .57507E-02 | .10499E+10 | 9.15E-01 |
| 568 | 266 | 181.320 | .16246E-03 | .83448E-04 | .24720E+08 | 1.95E+00 |
| 567 | 266 | 201.888 | .42467E-02 | .41772E-02 | .52122E+09 | 1.02E+00 |
| 566 | 266 | 218.554 | .20986E-03 | .30897E-03 | .21979E+08 | 6.79E-01 |
| 565 | 266 | 226.657 | .18965E-01 | .24090E-01 | .18468E+10 | 7.87E-01 |
| 564 | 266 | 239.489 | .11075E+00 | .12642E+00 | .96599E+10 | 8.76E-01 |
| 563 | 266 | 331.694 | .34898E-01 | .29262E-01 | .15868E+10 | 1.19E+00 |
| 562 | 266 | 372.140 | .54990E-02 | .52264E-02 | .19864E+09 | 1.05E+00 |
| 266 | 561 | 212.001 | .23207E-02 | .11809E-02 | .45921E+09 | 1.97E+00 |
| 266 | 560 | 192.391 | .77680E-02 | .24991E-02 | .18664E+10 | 3.11E+00 |
| 266 | 559 | 180.487 | .19662E-01 | .96396E-02 | .53681E+10 | 2.04E+00 |
| 266 | 558 | 175.498 | .22030E-02 | .19503E-02 | .63614E+09 | 1.13E+00 |
| 266 | 557 | 170.358 | .74255E-03 | .39649E-03 | .22755E+09 | 1.87E+00 |
| 266 | 556 | 162.918 | .28828E-02 | .18971E-02 | .96591E+09 | 1.52E+00 |
| 266 | 555 | 159.399 | .73496E-03 | .63138E-03 | .25726E+09 | 1.16E+00 |
| 266 | 554 | 147.014 | .59850E-03 | .17635E-02 | .24627E+09 | 3.39E-01 |
| 266 | 553 | 139.511 | .18005E-02 | .14147E-02 | .82273E+09 | 1.27E+00 |
| 266 | 552 | 134.067 | .94007E-03 | .63749E-03 | .46515E+09 | 1.47E+00 |
| 266 | 551 | 124.587 | .34605E-03 | .44049E-04 | .19828E+09 | 7.86E+00 |
| 266 | 550 | 56.302 | .28027E-03 | .39057E-05 | .78634E+09 | 7.18E+01 |
| 569 | 265 | 131.973 | .16563E+00 | .13260E+00 | .47574E+11 | 1.25E+00 |
| 568 | 265 | 147.581 | .54457E-02 | .42108E-02 | .12508E+10 | 1.29E+00 |
| 567 | 265 | 160.926 | .17144E-01 | .15611E-01 | .33118E+10 | 1.10E+00 |
| 566 | 265 | 171.341 | .13712E-01 | .11630E-01 | .23366E+10 | 1.18E+00 |
| 565 | 265 | 176.281 | .10870E+00 | .10320E+00 | .17499E+11 | 1.05E+00 |
| 564 | 265 | 183.946 | .70138E-02 | .74768E-02 | .10370E+10 | 9.38E-01 |
| 563 | 265 | 233.883 | .18126E-01 | .13342E-01 | .16577E+10 | 1.36E+00 |
| 562 | 265 | 253.295 | .58814E-01 | .76945E-01 | .45859E+10 | 7.64E-01 |
| 265 | 561 | 289.339 | .52713E-04 | .38427E-04 | .55998E+07 | 1.37E+00 |
| 265 | 560 | 254.004 | .77545E-06 | .83967E-04 | .10689E+06 | 9.24E-03 |
| 265 | 559 | 233.658 | .63138E-03 | .61359E-03 | .10285E+09 | 1.03E+00 |
| 265 | 558 | 225.364 | .59504E-03 | .21634E-03 | .10420E+09 | 2.75E+00 |
| 265 | 557 | 216.959 | .10404E-04 | .58594E-04 | .19656E+07 | 1.78E-01 |
| 265 | 556 | 205.034 | .11289E-04 | .62854E-04 | .23881E+07 | 1.80E-01 |
| 265 | 555 | 199.491 | .10708E-03 | .83788E-06 | .23930E+08 | 1.28E+02 |
| 265 | 554 | 180.464 | .85351E-03 | .54150E-05 | .23307E+09 | 1.58E+02 |
| 265 | 553 | 169.288 | .58519E-01 | .26416E-01 | .18160E+11 | 2.22E+00 |
| 265 | 552 | 161.338 | .22676E-02 | .12084E-02 | .77476E+09 | 1.88E+00 |
| 265 | 551 | 147.804 | .20293E-01 | .11734E-01 | .82611E+10 | 1.73E+00 |
| 265 | 550 | 60.604 | .11993E-04 | .49978E-07 | .29040E+08 | 2.40E+02 |
| 569 | 264 | 128.604 | .74757E-03 | .53847E-03 | .22612E+09 | 1.39E+00 |
| 568 | 264 | 143.381 | .36159E-02 | .28154E-02 | .87989E+09 | 1.28E+00 |
| 567 | 264 | 155.944 | .54405E-01 | .49508E-01 | .11192E+11 | 1.10E+00 |
| 566 | 264 | 165.705 | .17799E+00 | .18300E+00 | .32428E+11 | 9.73E-01 |
| 565 | 264 | 170.321 | .11421E-02 | .13593E-02 | .19695E+09 | 8.40E-01 |
| 564 | 264 | 177.466 | .55849E-05 | .96926E-05 | .88711E+06 | 5.76E-01 |
| 563 | 264 | 223.506 | .63959E-04 | .66337E-04 | .64049E+07 | 9.64E-01 |
| 562 | 264 | 241.168 | .12281E-03 | .11248E-03 | .10563E+08 | 1.09E+00 |
| 264 | 561 | 306.971 | .52734E-03 | .65773E-03 | .49770E+08 | 8.02E-01 |
| 264 | 560 | 267.491 | .66887E-02 | .39662E-02 | .83137E+09 | 1.69E+00 |
| 264 | 559 | 245.023 | .10392E-01 | .55029E-02 | .15394E+10 | 1.89E+00 |
| 264 | 558 | 235.918 | .25232E-01 | .17429E-01 | .40318E+10 | 1.45E+00 |
| 264 | 557 | 226.723 | .69430E-02 | .30432E-02 | .12012E+10 | 2.28E+00 |
| 264 | 556 | 213.734 | .13543E-01 | .65347E-02 | .26365E+10 | 2.07E+00 |
| 264 | 555 | 207.717 | .34617E-01 | .17932E-01 | .71352E+10 | 1.93E+00 |
| 264 | 554 | 187.170 | .25914E-02 | .12410E-02 | .65787E+09 | 2.09E+00 |
| 264 | 553 | 175.175 | .15262E-02 | .80958E-03 | .44232E+09 | 1.89E+00 |
| 264 | 552 | 166.676 | .74194E-04 | .38404E-04 | .23752E+08 | 1.93E+00 |
| 264 | 551 | 152.272 | .16218E-03 | .88724E-04 | .62205E+08 | 1.83E+00 |
| 264 | 550 | 61.342 | .90224E-07 | .20698E-07 | .21325E+06 | 4.36E+00 |
| 569 | 263 | 112.082 | .25103E-03 | .16334E-03 | .99964E+08 | 1.54E+00 |
| 568 | 263 | 123.142 | .16275E-02 | .79747E-03 | .53690E+09 | 2.04E+00 |
| 567 | 263 | 132.296 | .73195E-04 | .55768E-04 | .20921E+08 | 1.31E+00 |
| 566 | 263 | 139.255 | .63179E-05 | .26038E-05 | .16299E+07 | 2.43E+00 |
| 565 | 263 | 142.500 | .26751E-02 | .25431E-02 | .65902E+09 | 1.05E+00 |
| 564 | 263 | 147.468 | .11830E-01 | .12826E-01 | .27213E+10 | 9.22E-01 |
| 563 | 263 | 177.923 | .41028E+00 | .44775E+00 | .64835E+11 | 9.16E-01 |
| 562 | 263 | 188.938 | .14689E-01 | .19718E-01 | .20585E+10 | 7.45E-01 |
| 263 | 561 | 473.624 | .12841E-06 | .26238E-05 | .50910E+04 | 4.89E-02 |
| 263 | 560 | 385.775 | .62614E-04 | .12889E-03 | .37418E+07 | 4.86E-01 |
| 263 | 559 | 340.717 | .26135E-04 | .81310E-04 | .20022E+07 | 3.21E-01 |
| 263 | 558 | 323.363 | .30644E-06 | .26926E-07 | .26064E+05 | 1.14E+01 |
| 263 | 557 | 306.335 | .53580E-05 | .14782E-04 | .50778E+06 | 3.62E-01 |
| 263 | 556 | 283.089 | .16002E-04 | .70269E-05 | .17758E+07 | 2.28E+00 |
| 263 | 555 | 272.630 | .65803E-04 | .24333E-04 | .78735E+07 | 2.70E+00 |
| 263 | 554 | 238.295 | .22619E-01 | .68373E-02 | .35426E+10 | 3.31E+00 |
| 263 | 553 | 219.187 | .66467E-02 | .22961E-02 | .12304E+10 | 2.89E+00 |
| 263 | 552 | 206.041 | .91138E-01 | .36974E-01 | .19093E+11 | 2.46E+00 |
| 263 | 551 | 184.470 | .25940E-03 | .19605E-03 | .67794E+08 | 1.32E+00 |

| 263 | 550 | 65.981  | .93072E-03 | .12010E-03 | .19013E+10 | 7.75E+00 |
| 569 | 262 | 105.051 | .33798E-02 | .17612E-02 | .15321E+10 | 1.92E+00 |
| 568 | 262 | 114.708 | .19639E-03 | .80266E-04 | .74668E+08 | 2.45E+00 |
| 567 | 262 | 122.611 | .98325E-03 | .45451E-03 | .32719E+09 | 2.16E+00 |
| 566 | 262 | 128.565 | .23560E-06 | .12159E-05 | .71307E+05 | 1.94E-01 |
| 565 | 262 | 131.326 | .83049E-04 | .13847E-04 | .24089E+08 | 6.00E+00 |
| 564 | 262 | 135.534 | .30513E-03 | .70043E-04 | .83097E+08 | 4.36E+00 |
| 563 | 262 | 160.836 | .41870E-01 | .35294E-01 | .80971E+10 | 1.19E+00 |
| 562 | 262 | 169.784 | .46444E+00 | .45730E+00 | .80598E+11 | 1.02E+00 |
| 262 | 561 | 660.375 | .84647E-05 | .21511E-04 | .17262E+06 | 3.94E-01 |
| 262 | 560 | 501.230 | .11436E-02 | .55895E-04 | .40482E+08 | 2.05E+01 |
| 262 | 559 | 427.734 | .33137E-02 | .34285E-04 | .16108E+09 | 9.67E+01 |
| 262 | 558 | 400.736 | .13251E-03 | .41511E-05 | .73386E+07 | 3.19E+01 |
| 262 | 557 | 374.909 | .41391E-02 | .23207E-03 | .26190E+09 | 1.78E+01 |
| 262 | 556 | 340.672 | .66730E-03 | .75369E-04 | .51135E+08 | 8.85E+00 |
| 262 | 555 | 325.639 | .43127E-04 | .12862E-03 | .36170E+07 | 3.35E-01 |
| 262 | 554 | 277.824 | .14562E-03 | .17464E-03 | .16778E+08 | 8.34E-01 |
| 262 | 553 | 252.193 | .18006E-01 | .42793E-02 | .25178E+10 | 4.21E+00 |
| 262 | 552 | 234.945 | .10699E-02 | .76285E-03 | .17238E+09 | 1.40E+00 |
| 262 | 551 | 207.304 | .78697E-01 | .34485E-01 | .16286E+11 | 2.28E+00 |
| 262 | 550 | 68.687  | .38064E-05 | .28061E-06 | .71753E+07 | 1.36E+01 |
| 569 | 261 | 54.899  | .10536E-07 | .10975E-07 | .17488E+05 | 9.60E-01 |
| 568 | 261 | 57.425  | .85258E-05 | .17453E-05 | .12934E+08 | 4.89E+00 |
| 567 | 261 | 59.340  | .29190E-05 | .52177E-06 | .41469E+07 | 5.59E+00 |
| 566 | 261 | 60.701  | .11788E-04 | .22049E-05 | .16005E+08 | 5.35E+00 |
| 565 | 261 | 61.309  | .33118E-06 | .54574E-07 | .44075E+06 | 6.07E+00 |
| 564 | 261 | 62.211  | .86749E-10 | .41528E-08 | .11213E+03 | 2.09E-02 |
| 563 | 261 | 67.053  | .26879E-06 | .11700E-06 | .29906E+06 | 2.30E+00 |
| 562 | 261 | 68.559  | .60602E-08 | .44648E-07 | .64498E+04 | 1.36E-01 |
| 561 | 261 | 139.241 | .61873E-04 | .51711E-04 | .15965E+08 | 1.20E+00 |
| 560 | 261 | 149.232 | .22444E-02 | .20022E-02 | .50416E+09 | 1.12E+00 |
| 559 | 261 | 157.278 | .99342E-02 | .97547E-02 | .20091E+10 | 1.02E+00 |
| 558 | 261 | 161.273 | .22947E-02 | .25509E-02 | .44137E+09 | 9.00E-01 |
| 557 | 261 | 165.871 | .13498E-01 | .14397E-01 | .24543E+10 | 9.38E-01 |
| 556 | 261 | 173.590 | .44260E-01 | .52658E-01 | .73478E+10 | 8.41E-01 |
| 555 | 261 | 177.772 | .17809E+00 | .21727E+00 | .28191E+11 | 8.20E-01 |
| 554 | 261 | 196.206 | .52438E-03 | .72500E-03 | .68142E+08 | 7.23E-01 |
| 553 | 261 | 211.378 | .13038E-03 | .19583E-03 | .14598E+08 | 6.66E-01 |
| 552 | 261 | 225.237 | .16839E-03 | .27942E-03 | .16605E+08 | 6.03E-01 |
| 551 | 261 | 258.248 | .40710E-03 | .93252E-03 | .30536E+08 | 4.37E-01 |
| 261 | 550 | 170.569 | .35077E-09 | .11100E-09 | .10723E+03 | 3.16E+00 |
| 569 | 260 | 53.580  | .34533E-06 | .22644E-06 | .60176E+06 | 1.53E+00 |
| 568 | 260 | 55.984  | .31524E-05 | .38968E-06 | .50316E+07 | 8.09E+00 |
| 567 | 260 | 57.802  | .23144E-06 | .44369E-07 | .34653E+06 | 5.22E+00 |
| 566 | 260 | 59.092 | .53890E-06 | .10290E-06 | .77204E+06 | 5.24E+00 |
| 565 | 260 | 59.669 | .85665E-05 | .20461E-05 | .12036E+08 | 4.19E+00 |
| 564 | 260 | 60.523 | .19493E-04 | .57129E-05 | .26621E+08 | 3.41E+00 |
| 563 | 260 | 65.096 | .21685E-03 | .43503E-04 | .25601E+09 | 4.98E+00 |
| 562 | 260 | 66.514 | .10346E-04 | .16315E-05 | .11699E+08 | 6.34E+00 |
| 561 | 260 | 131.058 | .20377E-03 | .15621E-03 | .59348E+08 | 1.30E+00 |
| 560 | 260 | 139.871 | .28083E-02 | .22123E-02 | .71810E+09 | 1.27E+00 |
| 559 | 260 | 146.916 | .30035E-02 | .25324E-02 | .69611E+09 | 1.19E+00 |
| 558 | 260 | 150.396 | .78297E-03 | .71304E-03 | .17317E+09 | 1.10E+00 |
| 557 | 260 | 154.387 | .19479E-03 | .20648E-03 | .40883E+08 | 9.43E-01 |
| 556 | 260 | 161.053 | .21589E-03 | .23124E-03 | .41638E+08 | 9.34E-01 |
| 555 | 260 | 164.646 | .69595E-03 | .78621E-03 | .12843E+09 | 8.85E-01 |
| 554 | 260 | 180.339 | .12309E+00 | .15828E+00 | .18933E+11 | 7.78E-01 |
| 553 | 260 | 193.076 | .14164E-01 | .20576E-01 | .19007E+10 | 6.88E-01 |
| 552 | 260 | 204.574 | .17410E+00 | .26118E+00 | .20811E+11 | 6.67E-01 |
| 551 | 260 | 231.445 | .14667E-03 | .27052E-03 | .13698E+08 | 5.42E-01 |
| 260 | 550 | 184.696 | .30405E-06 | .23477E-06 | .79268E+05 | 1.30E+00 |
| 569 | 259 | 51.805 | .32735E-06 | .11091E-04 | .61019E+06 | 2.95E-02 |
| 568 | 259 | 54.049 | .57418E-06 | .38787E-06 | .98327E+06 | 1.48E+00 |
| 567 | 259 | 55.741 | .11843E-05 | .96579E-06 | .19067E+07 | 1.23E+00 |
| 566 | 259 | 56.940 | .10211E-06 | .33022E-06 | .15755E+06 | 3.09E-01 |
| 565 | 259 | 57.476 | .79778E-07 | .28824E-05 | .12081E+06 | 2.77E-02 |
| 564 | 259 | 58.267 | .26793E-05 | .11841E-05 | .39478E+07 | 2.26E+00 |
| 563 | 259 | 62.494 | .65993E-04 | .75424E-05 | .84531E+08 | 8.75E+00 |
| 562 | 259 | 63.800 | .19754E-03 | .17943E-04 | .24278E+09 | 1.10E+01 |
| 561 | 259 | 120.922 | .33019E-04 | .24386E-04 | .11296E+08 | 1.35E+00 |
| 560 | 259 | 128.387 | .34712E-03 | .25410E-03 | .10535E+09 | 1.37E+00 |
| 559 | 259 | 134.298 | .55608E-02 | .44780E-02 | .15424E+10 | 1.24E+00 |
| 558 | 259 | 137.200 | .17919E-02 | .14490E-02 | .47622E+09 | 1.24E+00 |
| 557 | 259 | 140.514 | .17399E-03 | .17061E-03 | .44085E+08 | 1.02E+00 |
| 556 | 259 | 146.014 | .26633E-03 | .22834E-03 | .62493E+08 | 1.17E+00 |
| 555 | 259 | 148.961 | .21479E-03 | .18775E-03 | .48424E+08 | 1.14E+00 |
| 554 | 259 | 161.691 | .75683E-04 | .78107E-04 | .14482E+08 | 9.69E-01 |
| 553 | 259 | 171.856 | .46274E-01 | .56407E-01 | .78379E+10 | 8.20E-01 |
| 552 | 259 | 180.906 | .19279E-02 | .22891E-02 | .29470E+09 | 8.42E-01 |
| 551 | 259 | 201.604 | .24533E+00 | .36272E+00 | .30195E+11 | 6.76E-01 |
| 259 | 550 | 209.435 | .33471E-07 | .21296E-07 | .67865E+04 | 1.57E+00 |
| 654 | 269 | 218.018 | .16997E-01 | .15559E-01 | .23852E+10 | 1.09E+00 |
| 653 | 269 | 261.666 | .71328E-02 | .92953E-02 | .69486E+09 | 7.67E-01 |
| 652 | 269 | 329.710 | .18291E-01 | .18589E-01 | .11223E+10 | 9.84E-01 |
| 651 | 269 | 432.217 | .23860E-04 | .62707E-05 | .85193E+06 | 3.80E+00 |
| 269 | 650 | 177.561 | .10590E-02 | .74673E-03 | .22405E+09 | 1.42E+00 |
| 269 | 649 | 159.807 | .27491E-01 | .21939E-01 | .71800E+10 | 1.25E+00 |
| 269 | 648 | 154.125 | .23303E-02 | .18969E-02 | .65433E+09 | 1.23E+00 |
| 269 | 647 | 150.405 | .44917E-02 | .39709E-02 | .13244E+10 | 1.13E+00 |
| 269 | 646 | 148.577 | .47986E-03 | .86968E-03 | .14499E+09 | 5.52E-01 |
| 269 | 645 | 142.126 | .31970E-01 | .24282E-01 | .10557E+11 | 1.32E+00 |
| 269 | 644 | 130.484 | .27689E-01 | .23699E-01 | .10847E+11 | 1.17E+00 |
| 269 | 643 | 124.492 | .16922E-01 | .13793E-01 | .72827E+10 | 1.23E+00 |
| 269 | 642 | 119.329 | .24446E-02 | .15619E-02 | .11451E+10 | 1.57E+00 |
| 269 | 641 | 53.952 | .12861E-04 | .48466E-07 | .29472E+08 | 2.65E+02 |
| 654 | 268 | 211.903 | .63222E-01 | .56973E-01 | .93913E+10 | 1.11E+00 |
| 653 | 268 | 252.907 | .73392E-02 | .86299E-02 | .76535E+09 | 8.50E-01 |
| 652 | 268 | 315.923 | .36352E-02 | .50881E-02 | .24294E+09 | 7.14E-01 |
| 651 | 268 | 408.829 | .61253E-04 | .11122E-03 | .24444E+07 | 5.51E-01 |
| 268 | 650 | 181.835 | .21612E-03 | .39372E-04 | .43599E+08 | 5.49E+00 |
| 268 | 649 | 163.260 | .12509E-01 | .12460E-01 | .31304E+10 | 1.00E+00 |
| 268 | 648 | 157.334 | .13235E+00 | .11851E+00 | .35662E+11 | 1.12E+00 |
| 268 | 647 | 153.460 | .18124E-01 | .16634E-01 | .51333E+10 | 1.09E+00 |
| 268 | 646 | 151.557 | .29705E-01 | .25034E-01 | .86260E+10 | 1.19E+00 |
| 268 | 645 | 144.851 | .22480E-02 | .26579E-02 | .71462E+09 | 8.46E-01 |
| 268 | 644 | 132.777 | .33590E-02 | .27899E-02 | .12709E+10 | 1.20E+00 |
| 268 | 643 | 126.577 | .90222E-05 | .21938E-04 | .37561E+07 | 4.11E-01 |
| 268 | 642 | 121.244 | .52237E-03 | .47891E-03 | .23702E+09 | 1.09E+00 |
| 268 | 641 | 54.340 | .98773E-06 | .59490E-09 | .22312E+07 | 1.66E+03 |
| 654 | 267 | 190.338 | .32259E-01 | .24741E-01 | .59393E+10 | 1.30E+00 |
| 653 | 267 | 222.782 | .33696E-01 | .30574E-01 | .45284E+10 | 1.10E+00 |
| 652 | 267 | 270.270 | .37085E-02 | .23393E-02 | .33864E+09 | 1.59E+00 |
| 651 | 267 | 335.493 | .89572E-04 | .30401E-03 | .53081E+07 | 2.95E-01 |
| 267 | 650 | 201.417 | .24931E-03 | .31514E-03 | .40990E+08 | 7.91E-01 |
| 267 | 649 | 178.875 | .23957E-02 | .30482E-02 | .49942E+09 | 7.86E-01 |
| 267 | 648 | 171.785 | .83898E-03 | .48285E-03 | .18963E+09 | 1.74E+00 |
| 267 | 647 | 167.178 | .55241E-02 | .44800E-02 | .13184E+10 | 1.23E+00 |
| 267 | 646 | 164.921 | .14428E-01 | .89826E-02 | .35383E+10 | 1.61E+00 |
| 267 | 645 | 157.011 | .10535E+00 | .91301E-01 | .28505E+11 | 1.15E+00 |
| 267 | 644 | 142.923 | .77715E-02 | .72641E-02 | .25376E+10 | 1.07E+00 |
| 267 | 643 | 135.766 | .85119E-02 | .78689E-02 | .30802E+10 | 1.08E+00 |
| 267 | 642 | 129.649 | .30738E-03 | .26717E-03 | .12197E+09 | 1.15E+00 |
| 267 | 641 | 55.966 | .28372E-06 | .51126E-07 | .60421E+06 | 5.55E+00 |
| 654 | 266 | 187.040 | .18198E-02 | .10373E-02 | .34697E+09 | 1.75E+00 |
| 653 | 266 | 218.276 | .79365E-05 | .38354E-06 | .11111E+07 | 2.07E+01 |
| 652 | 266 | 263.667 | .38064E-02 | .29271E-02 | .36520E+09 | 1.30E+00 |
| 651 | 266 | 325.379 | .39565E-02 | .45623E-02 | .24927E+09 | 8.67E-01 |
| 266 | 650 | 205.247 | .22156E-02 | .30957E-02 | .35080E+09 | 7.16E-01 |
| 266 | 649 | 181.889 | .34816E-02 | .40572E-02 | .70194E+09 | 8.58E-01 |
| 266 | 648 | 174.564 | .14023E-01 | .12996E-01 | .30695E+10 | 1.08E+00 |
| 266 | 647 | 169.808 | .83482E-01 | .76849E-01 | .19311E+11 | 1.09E+00 |
| 266 | 646 | 167.480 | .32104E-02 | .29922E-02 | .76342E+09 | 1.07E+00 |
| 266 | 645 | 159.329 | .41846E-03 | .38848E-03 | .10995E+09 | 1.08E+00 |
| 266 | 644 | 144.841 | .81084E-02 | .87266E-02 | .25780E+10 | 9.29E-01 |
| 266 | 643 | 137.495 | .17849E-01 | .15175E-01 | .62975E+10 | 1.18E+00 |
| 266 | 642 | 131.225 | .32733E-01 | .29379E-01 | .12679E+11 | 1.11E+00 |
| 266 | 641 | 56.257 | .40294E-04 | .32166E-06 | .84921E+08 | 1.25E+02 |
| 654 | 265 | 151.348 | .61897E-03 | .54900E-03 | .18024E+09 | 1.13E+00 |
| 653 | 265 | 171.170 | .76582E-03 | .30172E-03 | .17434E+09 | 2.54E+00 |
| 652 | 265 | 197.884 | .86809E-01 | .85992E-01 | .14787E+11 | 1.01E+00 |
| 651 | 265 | 230.726 | .70582E-01 | .67344E-01 | .88437E+10 | 1.05E+00 |
| 265 | 650 | 276.904 | .11089E-02 | .10105E-02 | .96467E+08 | 1.10E+00 |
| 265 | 649 | 236.013 | .16582E-01 | .98770E-02 | .19856E+10 | 1.68E+00 |
| 265 | 648 | 223.826 | .11812E-02 | .58798E-03 | .15727E+09 | 2.01E+00 |
| 265 | 647 | 216.067 | .15571E-01 | .12819E-01 | .22247E+10 | 1.21E+00 |
| 265 | 646 | 212.313 | .33591E-06 | .61267E-04 | .49705E+05 | 5.48E-03 |
| 265 | 645 | 199.381 | .12220E-02 | .17645E-02 | .20504E+09 | 6.93E-01 |
| 265 | 644 | 177.201 | .76106E-01 | .77048E-01 | .16167E+11 | 9.88E-01 |
| 265 | 643 | 166.329 | .30934E-01 | .30907E-01 | .74582E+10 | 1.00E+00 |
| 265 | 642 | 157.241 | .38143E-03 | .13073E-02 | .10290E+09 | 2.92E-01 |
| 265 | 641 | 60.552 | .17532E-03 | .35367E-05 | .31893E+09 | 4.96E+01 |
| 654 | 264 | 146.934 | .96744E-01 | .78022E-01 | .29889E+11 | 1.24E+00 |
| 653 | 264 | 165.544 | .12499E-01 | .12630E-01 | .30421E+10 | 9.90E-01 |
| 652 | 264 | 190.404 | .64228E-03 | .73567E-03 | .11817E+09 | 8.73E-01 |
| 651 | 264 | 220.621 | .12864E-03 | .49674E-03 | .17628E+08 | 2.59E-01 |
| 264 | 650 | 293.010 | .21669E-01 | .24026E-01 | .16835E+10 | 9.02E-01 |
| 264 | 649 | 247.615 | .43810E-02 | .43676E-02 | .47660E+09 | 1.00E+00 |
| 264 | 648 | 234.233 | .22018E-01 | .16331E-01 | .26768E+10 | 1.35E+00 |
| 264 | 647 | 225.750 | .51017E-02 | .41142E-02 | .66771E+09 | 1.24E+00 |
| 264 | 646 | 221.655 | .44019E-01 | .27795E-01 | .59762E+10 | 1.58E+00 |
| 264 | 645 | 207.598 | .21394E-01 | .12823E-01 | .33111E+10 | 1.67E+00 |
| 264 | 644 | 183.662 | .27384E-03 | .24127E-03 | .54150E+08 | 1.13E+00 |
| 264 | 643 | 172.009 | .19178E-02 | .12944E-02 | .43235E+09 | 1.48E+00 |
| 264 | 642 | 162.307 | .25048E-03 | .19061E-03 | .63422E+08 | 1.31E+00 |
| 264 | 641 | 61.289 | .66093E-06 | .14294E-07 | .11736E+07 | 4.62E+01 |
| 654 | 263 | 125.754 | .88969E-04 | .31266E-04 | .37526E+08 | 2.85E+00 |
| 653 | 263 | 139.141 | .15943E-04 | .17821E-04 | .54926E+07 | 8.95E-01 |
| 652 | 263 | 156.293 | .87916E-07 | .50403E-04 | .24006E+05 | 1.74E-03 |
| 651 | 263 | 176.090 | .74425E-02 | .89306E-02 | .16010E+10 | 8.33E-01 |
| 263 | 650 | 441.190 | .17741E-04 | .41031E-04 | .60792E+06 | 4.32E-01 |
| 263 | 649 | 345.748 | .66684E-04 | .61919E-04 | .37208E+07 | 1.08E+00 |
| 263 | 648 | 320.206 | .63552E-02 | .34553E-02 | .41343E+09 | 1.84E+00 |
| 263 | 647 | 304.559 | .24413E-01 | .12256E-01 | .17555E+10 | 1.99E+00 |
| 263 | 646 | 297.153 | .30398E-03 | .19299E-03 | .22963E+08 | 1.58E+00 |
| 263 | 645 | 272.424 | .51041E-04 | .24930E-04 | .45873E+07 | 2.05E+00 |
| 263 | 644 | 232.638 | .37655E-02 | .53165E-03 | .46409E+09 | 7.08E+00 |
| 263 | 643 | 214.252 | .66888E-02 | .91149E-02 | .97192E+09 | 7.34E-01 |
| 263 | 642 | 199.406 | .57015E-01 | .52192E-01 | .95642E+10 | 1.09E+00 |
| 263 | 641 | 65.920 | .21900E-04 | .31063E-05 | .33615E+08 | 7.05E+00 |
| 654 | 262 | 116.971 | .62189E-03 | .31827E-03 | .30317E+09 | 1.95E+00 |
| 653 | 262 | 128.468 | .11680E-02 | .51788E-03 | .47205E+09 | 2.26E+00 |
| 652 | 262 | 142.953 | .96042E-01 | .74949E-01 | .31348E+11 | 1.28E+00 |
| 651 | 262 | 159.337 | .51815E-01 | .40339E-01 | .13613E+11 | 1.28E+00 |
| 262 | 650 | 598.980 | .35029E-03 | .37560E-04 | .65124E+07 | 9.33E+00 |
| 262 | 649 | 435.694 | .12278E-01 | .49765E-02 | .43143E+09 | 2.47E+00 |
| 262 | 648 | 395.898 | .39901E-03 | .40670E-05 | .16981E+08 | 9.81E+01 |
| 262 | 647 | 372.254 | .18483E-02 | .71695E-03 | .88969E+08 | 2.58E+00 |
| 262 | 646 | 361.248 | .29003E-03 | .66526E-04 | .14824E+08 | 4.36E+00 |
| 262 | 645 | 325.346 | .13558E-03 | .74296E-04 | .85438E+07 | 1.82E+00 |
| 262 | 644 | 270.165 | .72231E-02 | .90358E-02 | .66008E+09 | 7.99E-01 |
| 262 | 643 | 245.682 | .30075E-01 | .18014E-01 | .33235E+10 | 1.67E+00 |
| 262 | 642 | 226.357 | .52898E-01 | .45874E-01 | .68863E+10 | 1.15E+00 |
| 262 | 641 | 68.621 | .22459E-03 | .22899E-04 | .31813E+09 | 9.81E+00 |
| 654 | 261 | 57.987 | .10321E-04 | .27766E-05 | .20473E+08 | 3.72E+00 |
| 653 | 261 | 60.679 | .13824E-05 | .32055E-06 | .25044E+07 | 4.31E+00 |
| 652 | 261 | 63.729 | .64434E-07 | .11236E-07 | .10582E+06 | 5.73E+00 |
| 651 | 261 | 66.791 | .18490E-07 | .38280E-08 | .27646E+05 | 4.83E+00 |
| 650 | 261 | 142.317 | .63362E-04 | .52352E-04 | .20867E+08 | 1.21E+00 |
| 649 | 261 | 156.228 | .20220E-01 | .20166E-01 | .55259E+10 | 1.00E+00 |
| 648 | 261 | 162.070 | .12564E+00 | .13288E+00 | .31906E+11 | 9.46E-01 |
| 647 | 261 | 166.397 | .18471E-01 | .20337E-01 | .44496E+10 | 9.08E-01 |
| 646 | 261 | 168.694 | .47454E-01 | .53040E-01 | .11123E+11 | 8.95E-01 |
| 645 | 261 | 177.859 | .39711E-02 | .47321E-02 | .83732E+09 | 8.39E-01 |
| 644 | 261 | 200.215 | .12549E-03 | .18580E-03 | .20881E+08 | 6.75E-01 |
| 643 | 261 | 216.180 | .10623E-03 | .18733E-03 | .15162E+08 | 5.67E-01 |
| 642 | 261 | 233.739 | .34344E-04 | .81125E-04 | .41929E+07 | 4.23E-01 |
| 261 | 641 | 170.163 | .92764E-11 | .29177E-09 | .21369E+01 | 3.18E-02 |
| 654 | 260 | 56.517 | .11988E-05 | .15130E-06 | .25033E+07 | 7.92E+00 |
| 653 | 260 | 59.072 | .14725E-09 | .15900E-12 | .28146E+03 | 9.26E+02 |
| 652 | 260 | 61.958 | .26717E-05 | .37019E-06 | .46422E+07 | 7.22E+00 |
| 651 | 260 | 64.849 | .26110E-05 | .52508E-06 | .41414E+07 | 4.97E+00 |
| 650 | 260 | 133.779 | .21640E-02 | .16813E-02 | .80652E+09 | 1.29E+00 |
| 649 | 260 | 146.000 | .15562E-02 | .13383E-02 | .48696E+09 | 1.16E+00 |
| 648 | 260 | 151.089 | .29083E-01 | .27483E-01 | .84979E+10 | 1.06E+00 |
| 647 | 260 | 154.842 | .11451E+00 | .11372E+00 | .31856E+11 | 1.01E+00 |
| 646 | 260 | 156.830 | .25668E-02 | .25876E-02 | .69610E+09 | 9.92E-01 |
| 645 | 260 | 164.721 | .14416E-03 | .15466E-03 | .35439E+08 | 9.32E-01 |
| 644 | 260 | 183.720 | .69165E-02 | .89657E-02 | .13668E+10 | 7.71E-01 |
| 643 | 260 | 197.075 | .60849E-04 | .84111E-04 | .10450E+08 | 7.23E-01 |
| 642 | 260 | 211.563 | .16793E-01 | .25733E-01 | .25025E+10 | 6.53E-01 |
| 260 | 641 | 184.221 | .47378E-08 | .10401E-09 | .93117E+03 | 4.56E+01 |
| 654 | 259 | 54.546 | .17175E-05 | .25526E-06 | .38503E+07 | 6.73E+00 |
| 653 | 259 | 56.921 | .39314E-05 | .45819E-06 | .80934E+07 | 8.58E+00 |
| 652 | 259 | 59.597 | .73176E-05 | .36711E-06 | .13742E+08 | 1.99E+01 |
| 651 | 259 | 62.266 | .91784E-04 | .19215E-04 | .15790E+09 | 4.78E+00 |
| 650 | 259 | 123.235 | .53127E-03 | .38686E-03 | .23333E+09 | 1.37E+00 |
| 649 | 259 | 133.532 | .10072E-01 | .84555E-02 | .37676E+10 | 1.19E+00 |
| 648 | 259 | 137.776 | .12733E-02 | .10540E-02 | .44743E+09 | 1.21E+00 |
| 647 | 259 | 140.891 | .18647E-03 | .20970E-03 | .62658E+08 | 8.89E-01 |
| 646 | 259 | 142.534 | .94236E-03 | .90029E-03 | .30939E+09 | 1.05E+00 |
| 645 | 259 | 149.022 | .43669E-03 | .44623E-03 | .13116E+09 | 9.79E-01 |
| 644 | 259 | 164.403 | .14056E+00 | .15490E+00 | .34686E+11 | 9.07E-01 |
| 643 | 259 | 175.017 | .48079E-03 | .95618E-03 | .10470E+09 | 5.03E-01 |
| 642 | 259 | 186.350 | .33463E-02 | .36213E-02 | .64275E+09 | 9.24E-01 |
| 259 | 641 | 208.823 | .23130E-06 | .46327E-07 | .35380E+05 | 4.99E+00 |
| 704 | 269 | 230.997 | .18659E-02 | .21572E-02 | .29156E+09 | 8.65E-01 |
| 269 | 703 | 171.712 | .21196E-01 | .20263E-01 | .38361E+10 | 1.05E+00 |
| 269 | 702 | 153.942 | .35384E-01 | .28781E-01 | .79675E+10 | 1.23E+00 |
| 269 | 701 | 149.451 | .21305E-02 | .23928E-02 | .50898E+09 | 8.90E-01 |
| 269 | 700 | 142.694 | .20501E-02 | .17174E-02 | .53726E+09 | 1.19E+00 |
| 269 | 699 | 120.883 | .24973E-01 | .14786E-01 | .91193E+10 | 1.69E+00 |
| 704 | 268 | 224.144 | .43840E-01 | .45334E-01 | .72754E+10 | 9.67E-01 |
| 268 | 703 | 175.705 | .54786E-01 | .47829E-01 | .94694E+10 | 1.15E+00 |
| 268 | 702 | 157.143 | .37414E-02 | .25194E-02 | .80847E+09 | 1.49E+00 |
| 268 | 701 | 152.467 | .25330E-03 | .68111E-04 | .58144E+08 | 3.72E+00 |
| 268 | 700 | 145.441 | .45291E-02 | .28342E-02 | .11425E+10 | 1.60E+00 |
| 268 | 699 | 122.848 | .66200E-03 | .47656E-03 | .23407E+09 | 1.39E+00 |
| 704 | 267 | 200.156 | .52187E-01 | .43975E-01 | .10861E+11 | 1.19E+00 |
| 267 | 703 | 193.923 | .13022E-01 | .14731E-01 | .18477E+10 | 8.84E-01 |
| 267 | 702 | 171.558 | .15475E-02 | .18661E-02 | .28056E+09 | 8.29E-01 |
| 267 | 701 | 166.000 | .67725E-01 | .64457E-01 | .13115E+11 | 1.05E+00 |
| 267 | 700 | 157.705 | .99934E-02 | .91277E-02 | .21441E+10 | 1.09E+00 |
| 267 | 699 | 131.485 | .44602E-02 | .35358E-02 | .13767E+10 | 1.26E+00 |
| 704 | 266 | 196.512 | .21987E-02 | .17375E-02 | .47472E+09 | 1.27E+00 |
| 266 | 703 | 197.471 | .20312E-03 | .25240E-03 | .27795E+08 | 8.05E-01 |
| 266 | 702 | 174.329 | .33414E-01 | .27808E-01 | .58670E+10 | 1.20E+00 |
| 266 | 701 | 168.593 | .14555E-01 | .11541E-01 | .27325E+10 | 1.26E+00 |
| 266 | 700 | 160.043 | .17071E-01 | .12627E-01 | .35564E+10 | 1.35E+00 |
| 266 | 699 | 133.106 | .29792E-02 | .13287E-02 | .89727E+09 | 2.24E+00 |
| 704 | 265 | 157.491 | .26416E-03 | .16002E-03 | .88796E+08 | 1.65E+00 |
| 265 | 703 | 262.935 | .40687E-03 | .41642E-03 | .31404E+08 | 9.77E-01 |
| 265 | 702 | 223.440 | .60963E-01 | .62735E-01 | .65157E+10 | 9.72E-01 |
| 265 | 701 | 214.103 | .94866E-02 | .10410E-01 | .11043E+10 | 9.11E-01 |
| 265 | 700 | 200.501 | .45650E-01 | .40236E-01 | .60594E+10 | 1.13E+00 |
| 265 | 699 | 159.949 | .18565E+00 | .14253E+00 | .38721E+11 | 1.30E+00 |
| 704 | 264 | 152.717 | .85741E-02 | .74710E-02 | .30652E+10 | 1.15E+00 |
| 264 | 703 | 277.415 | .92035E-01 | .10929E+00 | .63814E+10 | 8.42E-01 |
| 264 | 702 | 233.811 | .13516E-01 | .11768E-01 | .13193E+10 | 1.15E+00 |
| 264 | 701 | 223.607 | .22356E-01 | .18088E-01 | .23859E+10 | 1.24E+00 |
| 264 | 700 | 208.812 | .61107E-03 | .37058E-03 | .74783E+08 | 1.65E+00 |
| 264 | 699 | 165.194 | .20428E-02 | .17559E-02 | .39945E+09 | 1.16E+00 |
| 704 | 263 | 129.966 | .42789E-07 | .63687E-07 | .21121E+05 | 6.72E-01 |
| 263 | 703 | 406.760 | .70383E-04 | .23485E-03 | .22700E+07 | 3.00E-01 |
| 263 | 702 | 319.417 | .27305E-01 | .49541E-01 | .14280E+10 | 5.51E-01 |
| 263 | 701 | 300.673 | .31211E-01 | .49094E-01 | .18422E+10 | 6.36E-01 |
| 263 | 700 | 274.519 | .11250E+00 | .14624E+00 | .79656E+10 | 7.69E-01 |
| 263 | 699 | 203.782 | .14309E-01 | .14990E-01 | .18387E+10 | 9.55E-01 |
| 704 | 262 | 120.607 | .19598E-04 | .48505E-05 | .11234E+08 | 4.04E+00 |
| 262 | 703 | 537.240 | .15783E-02 | .52311E-02 | .29180E+08 | 3.02E-01 |
| 262 | 702 | 394.693 | .70255E-02 | .11381E-01 | .24065E+09 | 6.17E-01 |
| 262 | 701 | 366.464 | .12097E-03 | .60115E-03 | .48066E+07 | 2.01E-01 |
| 262 | 700 | 328.337 | .14449E-01 | .22422E-01 | .71520E+09 | 6.44E-01 |
| 262 | 699 | 232.012 | .31820E+00 | .40446E+00 | .31543E+11 | 7.87E-01 |
| 704 | 261 | 58.867 | .48707E-06 | .24083E-06 | .11719E+07 | 2.02E+00 |
| 703 | 261 | 146.312 | .37206E-01 | .33299E-01 | .14491E+11 | 1.12E+00 |
| 702 | 261 | 162.273 | .13406E-02 | .13862E-02 | .42449E+09 | 9.67E-01 |
| 701 | 261 | 167.580 | .85783E-02 | .94589E-02 | .25468E+10 | 9.07E-01 |
| 700 | 261 | 176.978 | .14725E-02 | .17147E-02 | .39197E+09 | 8.59E-01 |
| 699 | 261 | 228.000 | .42219E-03 | .67705E-03 | .67714E+08 | 6.24E-01 |
| 704 | 260 | 57.353 | .33442E-07 | .52189E-08 | .84766E+05 | 6.41E+00 |
| 703 | 260 | 137.303 | .53620E-05 | .30925E-05 | .23714E+07 | 1.73E+00 |
| 702 | 260 | 151.265 | .12669E-01 | .11158E-01 | .46163E+10 | 1.14E+00 |
| 701 | 260 | 155.867 | .70389E-02 | .62840E-02 | .24157E+10 | 1.12E+00 |
| 700 | 260 | 163.965 | .21473E-02 | .16423E-02 | .66593E+09 | 1.31E+00 |
| 699 | 260 | 206.851 | .17603E-03 | .26679E-03 | .34301E+08 | 6.60E-01 |
| 704 | 259 | 55.324 | .20845E-08 | .41273E-10 | .56784E+04 | 5.05E+01 |
| 703 | 259 | 126.220 | .37722E-04 | .40239E-04 | .19741E+08 | 9.37E-01 |
| 702 | 259 | 137.923 | .13835E-01 | .12261E-01 | .60637E+10 | 1.13E+00 |
| 701 | 259 | 141.738 | .30199E-02 | .26103E-02 | .12533E+10 | 1.16E+00 |
| 700 | 259 | 148.403 | .10659E-01 | .10489E-01 | .40353E+10 | 1.02E+00 |
| 699 | 259 | 182.684 | .42176E-01 | .47566E-01 | .10537E+11 | 8.87E-01 |
| 654 | 302 | 204.842 | .36174E-01 | .27901E-01 | .46003E+10 | 1.30E+00 |
| 653 | 302 | 242.912 | .64885E-01 | .63648E-01 | .58677E+10 | 1.02E+00 |
| 652 | 302 | 300.479 | .16018E-01 | .27980E-01 | .94670E+09 | 5.72E-01 |
| 651 | 302 | 383.333 | .22247E-02 | .13611E-02 | .80786E+08 | 1.63E+00 |
| 302 | 650 | 187.378 | .54419E-03 | .47693E-03 | .12923E+09 | 1.14E+00 |
| 302 | 649 | 167.715 | .10506E-01 | .50933E-02 | .31140E+10 | 2.06E+00 |
| 302 | 648 | 161.467 | .38638E-02 | .32946E-02 | .12356E+10 | 1.17E+00 |
| 302 | 647 | 157.390 | .97091E-08 | .18794E-04 | .32679E+04 | 5.17E-04 |
| 302 | 646 | 155.388 | .54886E-01 | .39899E-01 | .18953E+11 | 1.38E+00 |
| 302 | 645 | 148.347 | .66696E-03 | .24686E-04 | .25269E+09 | 2.70E+01 |
| 302 | 644 | 135.708 | .19353E-05 | .16069E-03 | .87613E+06 | 1.20E-02 |
| 302 | 643 | 129.239 | .24342E-04 | .16599E-05 | .12151E+08 | 1.47E+01 |
| 302 | 642 | 123.684 | .20401E-03 | .14880E-03 | .11119E+09 | 1.37E+00 |
| 302 | 641 | 54.824 | .63281E-04 | .91299E-06 | .17554E+09 | 6.93E+01 |
| 654 | 301 | 181.298 | .16308E-02 | .14822E-02 | .26475E+09 | 1.10E+00 |
| 653 | 301 | 210.496 | .95089E-02 | .86932E-02 | .11452E+10 | 1.09E+00 |
| 652 | 301 | 252.399 | .12037E+00 | .14915E+00 | .10082E+11 | 8.07E-01 |
| 651 | 301 | 308.388 | .30468E-01 | .16907E-01 | .17095E+10 | 1.80E+00 |
| 301 | 650 | 212.637 | .91112E-03 | .55497E-03 | .16801E+09 | 1.64E+00 |
| 301 | 649 | 187.669 | .15165E-01 | .65738E-02 | .35902E+10 | 2.31E+00 |
| 301 | 648 | 179.881 | .49257E-02 | .23620E-02 | .12692E+10 | 2.09E+00 |
| 301 | 647 | 174.835 | .52554E-04 | .20031E-03 | .14335E+08 | 2.62E-01 |
| 301 | 646 | 172.369 | .19762E-01 | .14597E-01 | .55458E+10 | 1.35E+00 |
| 301 | 645 | 163.747 | .60260E-04 | .10724E-03 | .18738E+08 | 5.62E-01 |
| 301 | 644 | 148.483 | .12202E-03 | .11299E-02 | .46145E+08 | 1.08E-01 |
| 301 | 643 | 140.773 | .36318E-04 | .45149E-03 | .15280E+08 | 8.04E-02 |
| 301 | 642 | 134.207 | .13849E-02 | .83945E-03 | .64107E+09 | 1.65E+00 |
| 301 | 641 | 56.798 | .26120E-03 | .33159E-05 | .67506E+09 | 7.88E+01 |
| 654 | 300 | 145.677 | .40130E-04 | .36896E-04 | .10090E+08 | 1.09E+00 |
| 653 | 300 | 163.951 | .24726E+00 | .24972E+00 | .49084E+11 | 9.90E-01 |
| 652 | 300 | 188.300 | .56935E-05 | .20160E-04 | .85684E+06 | 2.82E-01 |
| 651 | 300 | 217.800 | .28951E-03 | .64541E-03 | .32566E+08 | 4.49E-01 |
| 300 | 650 | 298.138 | .64851E-03 | .69760E-03 | .60831E+08 | 9.30E-01 |
| 300 | 649 | 251.267 | .18920E-01 | .10916E-01 | .24986E+10 | 1.73E+00 |
| 300 | 648 | 237.499 | .26999E-01 | .16055E-01 | .39909E+10 | 1.68E+00 |
| 300 | 647 | 228.781 | .98103E-02 | .57050E-02 | .15627E+10 | 1.72E+00 |
| 300 | 646 | 224.576 | .13627E-01 | .53346E-02 | .22527E+10 | 2.55E+00 |
| 300 | 645 | 210.159 | .35584E-01 | .19151E-01 | .67174E+10 | 1.86E+00 |
| 300 | 644 | 185.663 | .62611E-03 | .29111E-03 | .15144E+09 | 2.15E+00 |
| 300 | 643 | 173.763 | .11312E-02 | .51489E-03 | .31237E+09 | 2.20E+00 |
| 300 | 642 | 163.868 | .42752E-03 | .22982E-03 | .13274E+09 | 1.86E+00 |
| 300 | 641 | 61.510 | .10567E-05 | .11175E-06 | .23287E+07 | 9.46E+00 |
| 654 | 299 | 125.629 | .33684E-02 | .17819E-02 | .11389E+10 | 1.89E+00 |
| 653 | 299 | 138.989 | .15555E-02 | .11088E-02 | .42967E+09 | 1.40E+00 |
| 652 | 299 | 156.101 | .14222E-02 | .32718E-02 | .31144E+09 | 4.35E-01 |
| 651 | 299 | 175.846 | .42072E+00 | .45305E+00 | .72603E+11 | 9.29E-01 |
| 299 | 650 | 442.731 | .27323E-04 | .22874E-05 | .11622E+07 | 1.19E+01 |
| 299 | 649 | 346.694 | .14277E-02 | .84703E-03 | .99037E+08 | 1.69E+00 |
| 299 | 648 | 321.016 | .28677E-02 | .12481E-02 | .23202E+09 | 2.30E+00 |
| 299 | 647 | 305.293 | .37362E-02 | .13603E-02 | .33423E+09 | 2.75E+00 |
| 299 | 646 | 297.851 | .25678E-03 | .61534E-04 | .24133E+08 | 4.17E+00 |
| 299 | 645 | 273.011 | .22842E-04 | .25872E-04 | .25551E+07 | 8.83E-01 |
| 299 | 644 | 233.065 | .67164E-01 | .22628E-01 | .10309E+11 | 2.97E+00 |
| 299 | 643 | 214.615 | .30800E-01 | .16214E-01 | .55753E+10 | 1.90E+00 |
| 299 | 642 | 199.720 | .97822E-02 | .29696E-02 | .20447E+10 | 3.29E+00 |
| 299 | 641 | 65.954 | .92939E-03 | .12196E-03 | .17814E+10 | 7.62E+00 |
| 654 | 298 | 57.814 | .78767E-05 | .14850E-05 | .12575E+08 | 5.30E+00 |
| 653 | 298 | 60.490 | .13777E-04 | .22910E-05 | .20091E+08 | 6.01E+00 |
| 652 | 298 | 63.520 | .71938E-07 | .44040E-08 | .95139E+05 | 1.63E+01 |
| 651 | 298 | 66.561 | .11120E-05 | .25804E-06 | .13393E+07 | 4.31E+00 |
| 650 | 298 | 141.278 | .41119E-03 | .34854E-03 | .10993E+09 | 1.18E+00 |
| 649 | 298 | 154.978 | .90640E-03 | .90903E-03 | .20138E+09 | 9.97E-01 |
| 648 | 298 | 160.724 | .16510E-02 | .17285E-02 | .34104E+09 | 9.55E-01 |
| 647 | 298 | 164.979 | .56239E-04 | .52000E-04 | .11026E+08 | 1.08E+00 |
| 646 | 298 | 167.237 | .17810E-02 | .21996E-02 | .33980E+09 | 8.10E-01 |
| 645 | 298 | 176.240 | .24967E+00 | .30073E+00 | .42893E+11 | 8.30E-01 |
| 644 | 298 | 198.165 | .13338E-03 | .17506E-03 | .18124E+08 | 7.62E-01 |
| 643 | 298 | 213.793 | .22036E-03 | .32853E-03 | .25726E+08 | 6.71E-01 |
| 642 | 298 | 230.951 | .66147E-04 | .11165E-03 | .66175E+07 | 5.92E-01 |
| 298 | 641 | 171.672 | .38237E-09 | .45966E-10 | .10818E+03 | 8.32E+00 |
| 654 | 297 | 56.472 | .18388E-05 | .20833E-06 | .30767E+07 | 8.83E+00 |
| 653 | 297 | 59.022 | .98642E-07 | .11467E-07 | .15110E+06 | 8.60E+00 |
| 652 | 297 | 61.903 | .19409E-04 | .52993E-05 | .27027E+08 | 3.66E+00 |
| 651 | 297 | 64.788 | .11437E-03 | .19790E-04 | .14539E+09 | 5.78E+00 |
| 650 | 297 | 133.523 | .12486E-02 | .96707E-03 | .37372E+09 | 1.29E+00 |
| 649 | 297 | 145.695 | .25091E-04 | .12186E-04 | .63075E+07 | 2.06E+00 |
| 648 | 297 | 150.762 | .82222E-03 | .63687E-03 | .19303E+09 | 1.29E+00 |
| 647 | 297 | 154.500 | .85234E-05 | .24956E-04 | .19054E+07 | 3.42E-01 |
| 646 | 297 | 156.478 | .74060E-03 | .73277E-03 | .16140E+09 | 1.01E+00 |
| 645 | 297 | 164.333 | .11009E-03 | .12579E-03 | .21753E+08 | 8.75E-01 |
| 644 | 297 | 183.237 | .95139E-02 | .13213E-01 | .15120E+10 | 7.20E-01 |
| 643 | 297 | 196.520 | .31865E+00 | .45799E+00 | .44027E+11 | 6.96E-01 |
| 642 | 297 | 210.924 | .92697E-02 | .15622E-01 | .11118E+10 | 5.93E-01 |
| 297 | 641 | 184.709 | .93852E-07 | .41088E-07 | .22936E+05 | 2.28E+00 |
| 654 | 296 | 56.033 | .18921E-05 | .26182E-06 | .32157E+07 | 7.23E+00 |
| 653 | 296 | 58.542 | .18658E-06 | .28448E-07 | .29050E+06 | 6.56E+00 |
| 652 | 296 | 61.376 | .15455E-04 | .41572E-05 | .21892E+08 | 3.72E+00 |
| 651 | 296 | 64.211 | .93881E-04 | .17734E-04 | .12150E+09 | 5.29E+00 |
| 650 | 296 | 131.095 | .53108E-04 | .38520E-04 | .16490E+08 | 1.38E+00 |
| 649 | 296 | 142.808 | .30166E-03 | .28272E-03 | .78928E+08 | 1.07E+00 |
| 648 | 296 | 147.674 | .64096E-03 | .58004E-03 | .15684E+09 | 1.11E+00 |
| 647 | 296 | 151.257 | .11401E-01 | .10516E-01 | .26592E+10 | 1.08E+00 |
| 646 | 296 | 153.153 | .61727E-04 | .59258E-04 | .14042E+08 | 1.04E+00 |
| 645 | 296 | 160.670 | .99760E-06 | .83803E-06 | .20621E+06 | 1.19E+00 |
| 644 | 296 | 178.694 | .14412E-01 | .18971E-01 | .24084E+10 | 7.60E-01 |
| 643 | 296 | 191.304 | .34468E-03 | .33809E-03 | .50257E+08 | 1.02E+00 |
| 642 | 296 | 204.927 | .31233E+00 | .48415E+00 | .39686E+11 | 6.45E-01 |
| 296 | 641 | 189.566 | .11744E-06 | .10132E-06 | .27248E+05 | 1.16E+00 |
| 704 | 302 | 216.258 | .12081E+00 | .11313E+00 | .17230E+11 | 1.07E+00 |
| 302 | 703 | 180.875 | .11707E-02 | .20607E-02 | .23868E+09 | 5.68E-01 |
| 302 | 702 | 161.266 | .82279E-01 | .75261E-01 | .21102E+11 | 1.09E+00 |
| 302 | 701 | 156.346 | .11610E+00 | .10478E+00 | .31682E+11 | 1.11E+00 |
| 302 | 700 | 148.966 | .78300E-05 | .40586E-05 | .23535E+07 | 1.93E+00 |
| 302 | 699 | 125.354 | .32619E-02 | .30493E-02 | .13846E+10 | 1.07E+00 |
| 704 | 301 | 190.184 | .25664E-01 | .18186E-01 | .47327E+10 | 1.41E+00 |
| 301 | 703 | 204.302 | .25997E-02 | .32684E-02 | .41544E+09 | 7.95E-01 |
| 301 | 702 | 179.631 | .19359E-01 | .17938E-01 | .40017E+10 | 1.08E+00 |
| 301 | 701 | 173.547 | .34997E-01 | .36111E-01 | .77505E+10 | 9.69E-01 |
| 301 | 700 | 164.501 | .80820E-01 | .72865E-01 | .19921E+11 | 1.11E+00 |
| 301 | 699 | 136.175 | .35384E-01 | .35428E-01 | .12727E+11 | 9.99E-01 |
| 704 | 300 | 151.360 | .95745E-01 | .82036E-01 | .27876E+11 | 1.17E+00 |
| 300 | 703 | 282.007 | .20770E-01 | .21200E-01 | .17420E+10 | 9.80E-01 |
| 300 | 702 | 237.064 | .16032E-01 | .10924E-01 | .19027E+10 | 1.47E+00 |
| 300 | 701 | 226.581 | .73764E-01 | .52518E-01 | .95836E+10 | 1.40E+00 |
| 300 | 700 | 211.403 | .80840E-02 | .59071E-02 | .12065E+10 | 1.37E+00 |
| 300 | 699 | 166.812 | .25486E-02 | .17928E-02 | .61092E+09 | 1.42E+00 |
| 704 | 299 | 129.833 | .19366E-05 | .34212E-05 | .76629E+06 | 5.66E-01 |
| 299 | 703 | 408.069 | .10414E-03 | .18739E-04 | .41712E+07 | 5.56E+00 |
| 299 | 702 | 320.224 | .24216E-01 | .15489E-01 | .15752E+10 | 1.56E+00 |
| 299 | 701 | 301.388 | .73563E-02 | .59958E-02 | .54018E+09 | 1.23E+00 |
| 299 | 700 | 275.114 | .55439E-02 | .63247E-03 | .48856E+09 | 8.77E+00 |
| 299 | 699 | 204.110 | .63523E-01 | .64172E-01 | .10170E+11 | 9.90E-01 |
| 704 | 298 | 58.688 | .13849E-04 | .37528E-05 | .26820E+08 | 3.69E+00 |
| 703 | 298 | 145.214 | .30427E-02 | .26914E-02 | .96243E+09 | 1.13E+00 |
| 702 | 298 | 160.924 | .65945E-01 | .68904E-01 | .16985E+11 | 9.57E-01 |
| 701 | 298 | 166.142 | .14665E+00 | .16053E+00 | .35437E+11 | 9.14E-01 |
| 700 | 298 | 175.374 | .63974E-02 | .75877E-02 | .13874E+10 | 8.43E-01 |
| 699 | 298 | 225.347 | .30565E-03 | .57279E-03 | .40147E+08 | 5.34E-01 |
| 704 | 297 | 57.306 | .13997E-05 | .18543E-06 | .28430E+07 | 7.55E+00 |
| 703 | 297 | 137.033 | .31346E-02 | .25698E-02 | .11134E+10 | 1.22E+00 |
| 702 | 297 | 150.938 | .82495E-01 | .77859E-01 | .24153E+11 | 1.06E+00 |
| 701 | 297 | 155.519 | .39116E-01 | .38872E-01 | .10787E+11 | 1.01E+00 |
| 700 | 297 | 163.580 | .97921E-02 | .10581E-01 | .24409E+10 | 9.25E-01 |
| 699 | 297 | 206.240 | .23246E-02 | .31909E-02 | .36453E+09 | 7.29E-01 |
| 704 | 296 | 56.854 | .68565E-06 | .86571E-07 | .14149E+07 | 7.92E+00 |
| 703 | 296 | 134.477 | .81989E-03 | .63201E-03 | .30241E+09 | 1.30E+00 |
| 702 | 296 | 147.842 | .38144E-03 | .44853E-03 | .11640E+09 | 8.50E-01 |
| 701 | 296 | 152.235 | .34784E-02 | .31467E-02 | .10011E+10 | 1.11E+00 |
| 700 | 296 | 159.950 | .11676E+00 | .12057E+00 | .30440E+11 | 9.68E-01 |
| 699 | 296 | 200.503 | .12742E-01 | .18501E-01 | .21141E+10 | 6.89E-01 |
| 302 | 726 | 172.393 | .68554E-01 | .57828E-01 | .12822E+11 | 1.19E+00 |
| 302 | 725 | 144.515 | .11463E-01 | .59953E-02 | .30510E+10 | 1.91E+00 |
| 301 | 726 | 193.546 | .10777E-01 | .11386E-01 | .15992E+10 | 9.47E-01 |
| 301 | 725 | 159.091 | .49156E-01 | .33882E-01 | .10795E+11 | 1.45E+00 |
| 300 | 726 | 261.915 | .12373E+00 | .13663E+00 | .10026E+11 | 9.06E-01 |
| 300 | 725 | 202.551 | .24566E-04 | .32665E-04 | .33283E+07 | 7.52E-01 |
| 299 | 726 | 367.299 | .46762E-05 | .17976E-04 | .19267E+06 | 2.60E-01 |
| 299 | 725 | 260.309 | .19168E+00 | .24531E+00 | .15723E+11 | 7.81E-01 |
| 726 | 298 | 151.186 | .29897E-01 | .28007E-01 | .10469E+11 | 1.07E+00 |
| 725 | 298 | 181.972 | .15911E-03 | .17103E-03 | .38459E+08 | 9.30E-01 |
| 726 | 297 | 142.339 | .34555E-03 | .30423E-03 | .13651E+09 | 1.14E+00 |
| 725 | 297 | 169.306 | .15318E-01 | .14982E-01 | .42772E+10 | 1.02E+00 |
| 726 | 296 | 139.583 | .12724E-03 | .11240E-03 | .52274E+08 | 1.13E+00 |
| 725 | 296 | 165.420 | .20972E-02 | .18522E-02 | .61345E+09 | 1.13E+00 |
| 704 | 311 | 56.875 | .83286E-07 | .18679E-07 | .14311E+06 | 4.46E+00 |
| 703 | 311 | 134.596 | .13386E-02 | .10404E-02 | .41072E+09 | 1.29E+00 |
| 702 | 311 | 147.987 | .93548E-03 | .84579E-03 | .23743E+09 | 1.11E+00 |
| 701 | 311 | 152.388 | .32389E-02 | .32459E-02 | .77526E+09 | 9.98E-01 |
| 700 | 311 | 160.120 | .24320E-02 | .22728E-02 | .52726E+09 | 1.07E+00 |
| 699 | 311 | 200.769 | .36072E+00 | .54358E+00 | .49742E+11 | 6.64E-01 |
| 726 | 311 | 139.712 | .15992E-03 | .16113E-03 | .54649E+08 | 9.92E-01 |
| 725 | 311 | 165.601 | .89808E-01 | .96797E-01 | .21843E+11 | 9.28E-01 |